\documentclass[aps,rmp,twocolumn,groupedaddress,amsfonts,amsmath,amssymb,showpacs]{revtex4}
\usepackage{graphicx}

\allowdisplaybreaks[1]
\begin{document}

\title{Ultrashort filaments of light in weakly-ionized, 
optically-transparent media}

\author{L. Berg\'e}
\email{luc.berge@cea.fr}
\affiliation{D\'epartement de Physique Th\'eorique et Appliqu\'ee, 
CEA-DAM/Ile
de France, B.P. 12, 91680 Bruy\`eres-le-Ch\^atel, France}
\author{S. Skupin}
\affiliation{D\'epartement de Physique Th\'eorique et Appliqu\'ee, 
CEA-DAM/Ile
de France, B.P. 12, 91680 Bruy\`eres-le-Ch\^atel, France}
\author{R. Nuter}
\affiliation{D\'epartement de Physique Th\'eorique et Appliqu\'ee, 
CEA-DAM/Ile
de France, B.P. 12, 91680 Bruy\`eres-le-Ch\^atel, France}
\author{J. Kasparian}
\affiliation{Laboratoire de Spectroscopie Ionique et Mol\'eculaire,
Universit\'e Claude Bernard Lyon 1,
43 bd du 11 Novembre, 69622 Villeurbanne Cedex, France}
\author{J.-P. Wolf}
\affiliation{GAP-Biophotonics, Universit\'e de Gen\`eve, 20  Rue de l'Ecole de 
M\'edecine,
1211 Gen\`eve 4, Switzerland}

\date{\today}

\begin{abstract}
Modern laser sources nowadays deliver ultrashort light pulses reaching
few cycles in duration, high energies beyond the Joule level and 
peak powers exceeding several terawatt (TW). When such 
pulses propagate through optically-transparent 
media, they first self-focus in space and grow in intensity, until 
they generate a tenuous plasma by photo-ionization. For free electron 
densities and beam intensities below their breakdown limits, these pulses 
evolve as self-guided objects, resulting from successive equilibria between the Kerr 
focusing process, the chromatic dispersion of the medium, and the defocusing action of 
the electron plasma. Discovered one decade ago, this self-channeling mechanism reveals 
a new physics, widely extending the frontiers of nonlinear optics. Implications include 
long-distance propagation of TW beams in the atmosphere, supercontinuum emission, pulse 
shortening as well as high-order harmonic generation. This review presents the landmarks 
of the 10-odd-year progress in this field. Particular emphasis is laid to the theoretical 
modeling of the propagation equations, whose physical ingredients are discussed from 
numerical simulations. The dynamics of single 
filaments created over laboratory scales in various materials such 
as noble gases, liquids and dielectrics reveal new perspectives 
in pulse shortening techniques. Differences 
between femtosecond pulses propagating in gaseous or 
condensed materials are underlined. 
Attention is also paid to the multifilamentation instability of
broad, powerful beams, breaking up the energy distribution into 
small-scale cells along the optical path. 
The robustness of the resulting filaments in adverse weathers, their 
large conical emission exploited for multipollutant remote sensing, 
nonlinear spectroscopy, and the possibility to guide electric 
discharges in air are finally
addressed on the basis of experimental results.
\end{abstract}


\maketitle

\tableofcontents

\section{Introduction}

Over the past two decades, ultrafast laser sources producing ultrashort pulses have come of age. Technological advances in this field have permitted the generation of light wave packets comprising only a few oscillation cycles of the electric field. In space, the extent of the pulses becomes focusable to a spot size comparable to the laser wavelength. In time, mode-locking and chirped-pulse amplification (CPA) technologies allow to access smaller and smaller durations and optical intensities locally exceeding hundreds of terawatt (TW) per cm$^2$ at moderate (sub-mJ) energies. Power levels nowadays approach the petawatt range with femtosecond pulse durations (1 fs = 10$^{-15}$ sec.). Ultrashort light pulses enable researchers to probe ultrafast relaxation processes on never-before-accessed time scales and study light-matter interactions at unprecedented intensity levels. Due to extreme temporal and spatial confinements, the pulse strength exceeds that of the Coulomb field which binds electrons at their nucleus. It becomes strong enough to overcome the Coulomb barrier and triggers optical-field ionization. Pushed to ultrahigh intensities $> 10^{18}$ cm$^2$, the availability of CPA lasers has extended the horizon of laser physics from atomic and optics studies to relativistic plasmas, nuclear and high-energy physics \cite{Mourou:rmp:78:309}.

\begin{figure}
\includegraphics[width=\columnwidth]{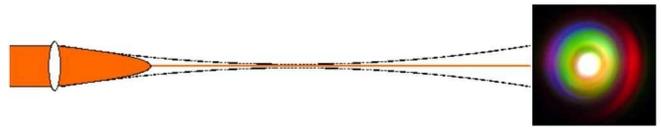}
\caption{Principle of producing femtosecond filaments in air at the laser wavelength of 800 nm. Photograph at the right-hand side shows a transverse cut of the filament profile.}
\label{fig0}
\end{figure}

Before reaching such extreme intensities, progress in ultrashort 
laser systems has led to the ability to observe tunnel or multiphoton ionization in several spectral ranges, before the ionization process reaches saturation. In this regime, an intriguing phenomenon was discovered by Braun {\it et al.}  \cite{Braun:ol:20:73} in the middle of the nineties. By launching infrared pulses with femtosecond durations and gigawatt (GW) powers in the atmosphere, the beam became confined to a long-living, self-confined tube of light capable of covering several tens of meters, i.e., many linear diffraction lengths, along the propagation axis. The mechanism supporting this ''light bullet'' results from the balance between Kerr focusing, which increases the local optical index with the wave intensity, and self-induced ionization. When an ultrashort pulse self-focuses and couples with a self-induced plasma channel, its spatial profile exhibits a narrow extent along the optical path. Spectra broaden due to self-phase modulation (SPM), which is sustained by the mechanism of high-intensity clamping. This picture classically refers to what is commonly called a ''femtosecond filament''. In the diffraction plane, this filament is characterized by a white-light spot, surrounded by concentric ''rainbows'' with colors ranging from red to green. The high nonlinearities competing through the filamentation process produce an impressive supercontinuum leading to white-light emission. They also affect the beam divergence through an apparent conical emission, as illustrated in Fig.\ \ref{fig0}

Femtosecond filaments have, for the last decade, opened the route to a 
fascinating physics. Their highly nonlinear dynamics 
sparked broad interest, first because of the capability of 
femtosecond pulses to convey high intensities over spectacular 
distances, second because of the white light emitted by the filaments 
transforming infrared lasers into ''white-light lasers'' 
\cite{Chin:jjap:38:L126,Kasparian:sc:301:61}. For appropriate beam configurations, long fs filaments can be created not only in the atmosphere, but also in noble gases, liquids and dielectric solids, as long as the pulse intensity does not reach the limit of optical breakdown, which implies the electron plasma density to remain at subdense levels. These optical structures are subject to strong modifications of their temporal profile, triggered by SPM together with the ionization front and the chromatic dispersion of the medium. The dynamical balance between nonlinear focusing and ionization can result in a drastic shortening of the pulse duration, down to the optical cycle limit. This property opens quite promising ways to generically deliver light with durations of a few fs only, which should further impact the fields of high-order harmonic generation and sub-fs pulse production. Besides, femtosecond pulses with broad spatial extents create several filaments, whose mutual interactions support the self-guiding of the beam envelope and can preserve a confined state upon several kilometers. This recently led to develop ultrashort LIght Detection And Ranging (Lidar) facilities, that exploit the white light emitted by these filaments, in order to detect and identify several pollutants within a single laser shot. 

Many applications have been inspired by this ultrafast ''light bullet'', which justifies the present review. Before commenting on these, an accurate understanding of the filamentation phenomenon requires a rigorous derivation of the propagation equations together with basic tools for capturing the underlying physics. For this purpose, Section \ref{sec2} addresses the model describing the long range propagation of ultrashort laser pulses in any optically-transparent medium. Section \ref{sec3} reviews the basic phenomenon of wave self-focusing and its limitation by potential players such as plasma generation, chromatic dispersion and optical nonlinearities. Emphasis is given to semi-analytic (e.g., variational) methods providing qualitative information about these effects. Section \ref{sec4} lists the major phenomena driving femtosecond filaments, whatever the propagation medium may be. Section \ref{sec5} is devoted to pulse shortening that can be achieved by letting femtosecond pulses ionize atom gases at appropriate pressures and to high-order harmonic generation. Section \ref{sec6} addresses different propagation regimes in condensed materials (water, silica glasses), from the laser-induced breakdown limit to $X$-shaped nonlinear waves self-guided with no plasma generation. Section \ref{sec7} concentrates on the atmospheric applications of the white-light supercontinuum emitted by ultrashort filaments. Current techniques for changing their onset distance and self-channeling length are discussed, together with the diagnostics used for plasma and optical-field measurements. Attention is paid to novel ultrashort LIDAR-based setups and their use in remotely analyzing aerosols, biological agents, dense targets through remote filament-induced breakdown spectroscopy. Their ability to trigger and guide electric discharges over several meters is also discussed. Section \ref{sec8} finally summarizes the principal aspects of this review and presents future prospects.

\section{Propagation equations\label{sec2}}

To start with, we derive the model describing the 
propagation of ultrashort optical pulses in 
transparent media. Using conventional description of nonlinear 
optics, straightforward combination of the Maxwell's equations yields 
\cite{Agrawal:NFO:01,He:PNO:99,Shen:PNO:84}
\begin{subequations}
\label{wave_equation}
\begin{align}
\label{wave_equation1}
\nabla^2 \vec{E} - \vec{\nabla} (\vec{\nabla} \cdot \vec{E}) - 
c^{-2} \partial_t^2 \vec{E} & = \mu_0 (\partial_t^2 \vec{P} + 
\partial_t \vec{J}),\\
\label{wave_equation2}
{\vec \nabla} \cdot \vec{E} & = (\rho - {\vec \nabla} \cdot \vec{P})/\epsilon_0,
\end{align}
\end{subequations}
where $\epsilon_0$, $\mu_0$ and $c$ denote the electric permittivity, magnetic permeability and the speed of light in vacuum, respectively. The optical electric field $\vec{E}$, the 
polarization vector $\vec{P}$, the 
carrier density $\rho$ and the current density $\vec{J}$ are real 
valued. For further convenience, we introduce standard 
Fourier transforms applied to the fields $(\vec{E}, \vec{P}, 
\vec{J})^{\dagger}$ as 
\begin{equation}
\label{TF}
(\widehat{\vec{E}}, \widehat{\vec{P}}, 
\widehat{\vec{J}})^{\dagger}(\vec{r},\omega) \equiv \frac{1}{2\pi} 
\int (\vec{E}, \vec{P}, \vec{J})^{\dagger}(\vec{r},t) \mbox{e}^{i 
\omega t} dt.
\end{equation}
The current density $\vec{J}$ describes the motions of the free 
electrons created by ionization of the ambient atoms, for which ion dynamics is discarded. The 
polarization vector $\vec{P}$ describes the bounded electron response 
driven by the laser 
radiation. It is usually 
decomposed into a linear part $\vec{P}_L \equiv \vec{P}^{(1)}$ 
related 
to the first-order susceptibility tensor 
$\overset{\leftrightarrow}{\chi}
\raisebox{1ex}{\scriptsize(1)}$ and a nonlinear one $\vec{P}_{\rm 
NL}$ 
satisfying $|\vec{P}^{(1)}| \gg |\vec{P}_{\rm NL}|$. For isotropic, homogeneous, non magnetizable media and spectral ranges far 
from any material resonance, $\vec{P}$ can be expressed as a power 
series in 
$\vec{E}$:
\begin{equation}
\label{polarization}
\widehat{\vec{P}} = 
\widehat{\vec{P}}\raisebox{1ex}{\scriptsize(1)}(\vec{r},\omega) + 
\widehat{\vec{P}}
\raisebox{1ex}{\scriptsize(3)}(\vec{r},\omega) + 
\widehat{\vec{P}}\raisebox{1ex}{\scriptsize(5)}(\vec{r},\omega) + 
\ldots 
\end{equation}
with scalar components
\begin{equation}
\label{polarization_2}
\begin{split}
\widehat{P}_{\mu}^{(j)} & = \epsilon_0 
\sum_{\alpha_1 \ldots \alpha_j} \int ... \int \chi_{\mu \alpha_1 
\ldots 
\alpha_j}^{(j)}(-\omega_{\sigma};\omega_1,\ldots,\omega_j)\\
& \quad \times \widehat{E}_{\alpha_1}(\vec{r},\omega_1) \ldots 
\widehat{E}_{\alpha_j}(\vec{r},\omega_j) 
\delta(\omega-\omega_{\sigma}) 
d\omega_1 \ldots d\omega_j,
\end{split}
\end{equation}
where $\omega_{\sigma} = \omega_1 + \ldots + \omega_j$. All 
susceptibility tensors 
$\overset{\leftrightarrow}{\chi}\raisebox{1ex}{(j)}$ with 
even index $j$ vanish due to inversion symmetry. The subscript $\mu$ indicates the field vector 
component in Cartesian coordinates and the indices $\alpha_j$ have to be summed up over $x$, $y$, and $z$. The tensor 
$\overset{\leftrightarrow}{\chi}\raisebox{1ex}{\scriptsize(1)}$ is 
diagonal with $\displaystyle{\chi_{\mu 
\alpha}^{(1)} 
= \chi^{(1)} \delta_{\mu \alpha}}$, so that
\begin{equation}
\widehat{\vec{P}}\raisebox{1ex}{\scriptsize(1)}(\vec{r},\omega) = 
\epsilon_0 \chi^{(1)}(\omega) \widehat{\vec{E}}(\vec{r},\omega), 
\label{linpol}
\end{equation}
and the scalar dielectric function, defined by
\begin{equation}
\epsilon(\omega) = 1 + \chi^{(1)}(\omega),
\label{epsilon}
\end{equation}
enters the wave number of the electromagnetic field $k(\omega) = 
\sqrt{\epsilon(\omega)} \omega/c$. Since $\chi^{(1)}$ is 
complex-valued, the dielectric function 
$\epsilon(\omega)$ contains every information not only
about the material dispersion, but also about the linear 
losses given by the imaginary part of $\chi^{(1)}(\omega)$. 
When losses are negligible, $\epsilon(\omega)$ is real and reduces to $\epsilon(\omega) = n^2(\omega)$, where $n(\omega)$ here denotes the 
linear refractive index of the medium, which can be described in certain frequency ranges (far from resonances) by, e.g., a Sellmeier formula. By convention, $n_0 \equiv n(\omega_0)$ for the central frequency 
$\omega_0 = 2 \pi 
c/\lambda_0$ of a laser operating at the wavelength $\lambda_0$ and 
$k_0 \equiv k(\omega_0)$. Without any specification, $\omega = 2 \pi c/\lambda$, $w_{\perp} = 2 \pi/k_{\perp}$ is the waist of the 
optical wave packet in the plane $(x,y)$ and $k_{\perp}$ is the 
corresponding extent in the transverse Fourier space.

\subsection{Helmoltz equation}

By taking its Fourier transform, Eqs.\ (\ref{wave_equation1}) expresses as the Helmoltz equation
\begin{subequations}
\label{k_equations}
\begin{align}
\label{k_equation}
\left[\partial_z^2 + k^2(\omega) + \nabla_{\perp}^2\right] 
\widehat{\vec{E}} & = - \mu_0 \omega^2 \widehat{\vec{\cal{F}}}_{\rm NL} 
+ \vec{\nabla} (\vec{\nabla} \cdot \widehat{\vec{E}}),\\
\label{gauge}
\vec{\nabla} \cdot \widehat{\vec{E}} & = (\epsilon_0 \epsilon)^{-1} 
(\widehat{\rho} - 
\vec{\nabla} \cdot \widehat{\vec{P}}_{\rm NL}),
\end{align}
\end{subequations}
where $\nabla_{\perp}^2 \equiv \partial_x^2 + \partial_y^2$ stands 
for transverse diffraction and
\begin{equation}
\label{NL}
\widehat{\vec{\cal{F}}}_{\rm NL} \equiv \widehat{\vec{P}}_{\rm NL} 
+ i\widehat{\vec{J}}/\omega \equiv \widehat{F} \otimes \widehat{\vec{E}},
\end{equation}
gathers all nonlinear contributions through the function $F$ that depends on $\vec{E}$ and can be viewed as the
effective nonlinear refractive index change of the medium ($\otimes$ is the convolution operator). 
Because Eqs.\ (\ref{k_equations}) are usually difficult to integrate in the full space-time 
domain, assumptions 
are requested to simplify them into more tractable form. 
The most fundamental of those consists in supposing that the 
wavefield keeps a transverse extension always fulfilling
\begin{equation}
\label{basic}
k_{\perp}^2/k^2(\omega) \ll 1,
\end{equation}
i.e., for $k(\omega)$ located around $\omega_0$, the transverse waist of the beam has dimensions larger than the central wavelength. The second one assumes small nonlinearities, i.e., 
\begin{equation}
\label{weaknl}
\frac{\widehat{F}}{\epsilon_0 \epsilon(\omega)} \ll 1.
\end{equation}

\subsubsection{From vectorial to scalar description}

The previous conditions make vectorial effects negligible for, e.g, a transversally-polarized light field $E_{\perp} = (E_x,E_y)$. Indeed, Eqs.\ (\ref{k_equations}) can be combined, with the help of the continuity equation $\partial_t \rho + \vec{\nabla} \cdot \vec{J} = 0$ expressed in Fourier variable, into the form  
\begin{equation}
\label{k_equation_perp}
\left[\partial_z^2 + k^2(\omega) + \nabla_{\perp}^2\right] 
\widehat{\vec{E}} = - \mu_0 \omega^2 \left[ \widehat{\vec{\cal{F}}}_{\rm 
NL} + \frac{\vec{\nabla} (\vec{\nabla} 
\cdot \widehat{\vec{\cal{F}}}_{\rm NL})}{k^2(\omega)}\right],
\end{equation}
whose last term scrambles nonlinear vectorial components. When we 
project the vectors $\vec{E} = (\vec{E}_{\perp}, E_z)^{\dagger}$, 
$\vec{P}_{\rm NL} = (\vec{P}_{\rm NL}^{\perp},P_{\rm 
NL}^{z})^{\dagger}$, $\vec{J} = (\vec{J}_{\perp},J_{z})^{\dagger}$, $\vec{\nabla} = (\vec{\nabla}_{\perp}, 
\partial_z)^{\dagger}$ onto the transverse and longitudinal axis with 
unit vectors $\vec{e}_{\perp}$ and $\vec{e}_z$, respectively, 
$\widehat{E}_z$ is found to scale as $O(k_{\perp}/k$). This follows from a direct Fourier transform of Eq.\ (\ref{gauge}) for weak nonlinearities [Eq.\ (\ref{weaknl})]. Expressed in Fourier space, the 
nonlinear coupling of transverse/longitudinal components described by the scrambling term behave as $O(k_{\perp}^2/k^2$) \cite{Milsted:pra:53:3536,Fibich:ol:26:840,Fibich:pd:157:112}. So, these effects become important in the limit 
$k_{\perp} \rightarrow k(\omega)$ only. Reversely, if we postpone to further demonstrations that nonlinear compression processes are stopped {\it before} $k_{\perp}$ becomes 
comparable with $k$ (by, e.g., chromatic dispersion or plasma 
generation), the last term in the right-hand side (RHS) of 
(\ref{k_equation_perp}) is 
close to zero, implying thereby $E_z \simeq 0$. The field remains 
transversally polarized along the propagation axis, making the 
influence of $\vec{\nabla} (\vec{\nabla} \cdot \vec{\cal{F}}_{\rm NL})/k^2$ 
negligible. Hence, as long as the nonlinear polarization and current density preserve the conditions (\ref{basic}) and (\ref{weaknl}), vectorial effects 
can be ignored for purely optical or weakly-ionized materials as well. This property justifies the use of a scalar description for linearly-polarized beams having, e.g., $E_y = 0$.

\subsubsection{Weak backscattering}

The question of evaluating backscattering waves may be crucial in 
several areas, such as remote sensing experiments 
\cite{Yu:ol:26:533,Kasparian:sc:301:61}, which spectrally analyze the photons returning 
towards the laser source. The amount of backscattered photons, 
however, constitutes a weak percentage of those traveling in the 
forward 
direction through a transparent medium. The reason can be seen from 
the scalar version of Eq.\ (\ref{k_equation_perp}) expressed as
\begin{equation}
\label{D_equation}
D_+(\omega) D_-(\omega) \widehat{E} = - \nabla_{\perp}^2 
\widehat{E} - \mu_0 \omega^2 \widehat{F} \otimes \widehat{E},
\end{equation}
where $D_{\pm}(\omega) \equiv \partial_z \mp ik(\omega)$.
By substituting the solution $\widehat{E} = \widehat{U}_+ \mbox{e}^{i 
k(\omega) z} + \widehat{U}_- \mbox{e}^{-i 
k(\omega) z}$, Eq.\ (\ref{D_equation}) expands as
\begin{equation}
\label{back_eq1}
\begin{split}
\mbox{e}^{2 i k(\omega) z} [ \partial_z^2 + 2 i k(\omega) 
\partial_z + \nabla_{\perp}^2 + \mu_0 \omega^2 \tilde{F} ] \widehat{U}_+ & \\
+ [\partial_z^2 - 2 i k(\omega) \partial_z + \nabla_{\perp}^2 + \mu_0 
\omega^2 \tilde{F} ] \widehat{U}_- & = 0.
\end{split}
\end{equation}
Here, $\widehat{U}_+$ and $\widehat{U}_-$ represent the Fourier components of the forward and 
backward running fields, for which we a priori assume $|\partial_z U_{\pm}| \ll |k(\omega)U_{\pm}|$ and $U_+ \gg U_-$.
For technical convenience, we assume $\widehat{F}=\tilde{F}\delta(\omega)$.
Following Fibich {\it et al.} \cite{Fibich:jsc:17:351}, we can 
integrate Eq.\ (\ref{back_eq1}) over the interval $z - \pi/2 
k \leq z \leq z + \pi/2 k$ (one fast oscillation) and Taylorize $\widehat{U}_{\pm}(z)$ to evaluate 
\begin{equation}
\label{back_eq3}
2ik(\omega) \partial_z \widehat{U}_- \sim - \frac{\mbox{e}^{2 i k(\omega) z}}{2ik(\omega)} \partial_z [\nabla_{\perp}^2 + \mu_0 \omega^2 \tilde{F} ] \widehat{U}_+.
\end{equation}
Since $\nabla_{\perp}^2 \sim - k_{\perp}^2$ in Fourier space, the backscattered component has a weak influence on the beam dynamics if $k_{\perp}^2 \ll k^2(\omega)$ and as long as the longitudinal variations of the nonlinearities remain small.

\subsubsection{Unidirectional pulse propagation}

The limit (\ref{basic}) moreover implies that the wave components forming the 
angle $\theta = \arcsin{(k_{\perp}/k)}$ between the transverse and 
longitudinal directions mostly propagate forwards since $\theta \ll 
\pi/2$ \cite{Feit:josab:5:633}. Because the 
propagation physics is mainly brought by the forward component, one has $\widehat{U}_- \rightarrow 0$ and $\widehat{E} \simeq \widehat{U}_+ \mbox{e}^{i k(\omega) z}$. With the above approximations, the operator $D_-(\omega)$ for backscattering mainly applies to the most rapid variations of the field, expected to be driven by the complex exponential, i.e., $D_-(\omega) \widehat{E} \approx 2 i k(\omega) 
\widehat{E}$. The so-called Unidirectional Pulse Propagation Equation (UPPE)
\begin{equation}
\label{UPPE_1}
\partial_z \widehat{E} = \frac{i}{2 k(\omega)} \nabla_{\perp}^2 
\widehat{E} + i k(\omega) \widehat{E} + \frac{i \mu_0 \omega^2}{2 
k(\omega)} \widehat{\cal F}_{\rm NL}
\end{equation}
then naturally emerges from Eq.\ (\ref{D_equation}). Validity of this model explicitly requires that the second-order derivative in $z$ of the envelope function $\widehat{U}_+$ must be small compared with $|k(\omega) \partial_z \widehat{U}_+|$, since 
$D_+(\omega) D_-(\omega) \widehat{\mathcal{E}} = \mbox{e}^{i k(\omega) z} 
[\partial_z^2 + 2 i k(\omega) \partial_z] \widehat{U}_+$ 
\cite{Geissler:prl:83:2930,Husakou:prl:87:203901}. This approximation, 
usually expressed as $|\partial_z \widehat{U}_+| \ll |k(\omega) \widehat{U}_+|$, refers to the ''paraxiality'' assumption. It holds if the field envelope $U_+$ does not significantly change over propagation distances of the order of $\lambda$, for all wavelengths under consideration. Paraxiality is again linked to the weakness of both the ratio 
$k_{\perp}/k$ and the nonlinearities. Let us indeed assume the nonlinear function $F$ clamped at a maximal constant level, $F_{\rm max}$. The forward component of Eq.\ (\ref{back_eq1}) then goes like
\begin{equation}
\label{forward}
\widehat{U}_+ \sim \mbox{e}^{i k(\omega) z [\sqrt{1 - \frac{k_{\perp}^2}{k^2(\omega)} + \frac{\mu_0 \omega^2}{k^2(\omega)} F_{\rm max}} - 1]}. 
\end{equation}
It is seen right away that this solution fits that of the paraxial model discarding second derivatives in $z$, i.e., $\widehat{U}_+ \sim \mbox{e}^{-i (k_{\perp}^2 - \mu_0 \omega^2 F_{\rm max}) z/2k(\omega)}$, as long as the two constraints (\ref{basic}) and (\ref{weaknl}) apply.

Recently, Kolesik {\it et al.} 
\cite{Kolesik:prl:89:283902,Kolesik:pre:70:036604} proposed an UPPE 
model based on the exact linear modes of the Helmoltz equation. The 
key idea is to preserve the linear solutions, 
$\widehat{E}_{\rm lin} = \mbox{e}^{\pm i \sqrt{k^2(\omega) + \nabla_{\perp}^2} z}$, as long as possible along the manipulation of Eq.\ (\ref{k_equation}). 
Rewriting Eq.\ (\ref{D_equation}) as
\begin{equation}
\label{D_perp_equation}
D_+^{\perp}(\omega) D_-^{\perp} (\omega) \widehat{E} = - \mu_0 
\omega^2 \widehat{\cal F}_{\rm NL},
\end{equation}
with $D_{\pm}^{\perp}(\omega) \equiv (\partial_z \mp i \sqrt{k^2(\omega) + \nabla_{\perp}^2})$, we may retain the forward running component $\widehat{E} = \widehat{U}_+ 
\mbox{e}^{i \sqrt{k^2(\omega) + \nabla_{\perp}^2} z}$ 
constrained to the ''paraxial'' limit $D_-^{\perp} (\omega) 
\widehat{E} \approx 2 i \sqrt{k^2(\omega) + \nabla_{\perp}^2} 
\widehat{E}$. 
Repeating the previous procedure immediately yields
\begin{equation}
\label{UPPE_2}
\partial_z \widehat{E} = i \sqrt{k^2(\omega) + 
\nabla_{\perp}^2} \widehat{E} + \frac{i \mu_0 \omega^2 
\widehat{\cal F}_{\rm NL}}{2 \sqrt{k^2(\omega) + \nabla_{\perp}^2}}.
\end{equation}
Eq.\ (\ref{UPPE_2}) allows to formally describe DC-field components ($\omega = 
0$), whereas Eq.\ (\ref{UPPE_1}) is limited to non-zero frequencies 
strictly. Despite these minor differences, UPPE models (\ref{UPPE_1}) and (\ref{UPPE_2}) become quite analogous when the condition 
$k_{\perp}^2/k^2 \ll 1$ applies. The major advantage of the UPPE models is to elude the formal use of a central optical frequency and 
correctly describe the complete spectrum of pulses in nonlinear 
regimes, even when they develop very large bandwidths.

For practical use, it is convenient to introduce the complex version of the electric field
\begin{equation}
\label{complex}
E = \sqrt{c_1}(\mathcal{E} + \mathcal{E}^*),\,\,\,\mathcal{E} = \frac{1}{\sqrt{c_1}} \int \Theta(\omega) \widehat{E} \mbox{e}^{- i \omega t} d\omega,
\end{equation}
where $c_1 \equiv \omega_0 \mu_0/2 k_0$ and $\Theta(x)$ denotes the Heaviside function. Because $\mathcal{E}$ satisfies $\widehat{\mathcal{E}^*}(\omega) = \widehat{\mathcal{E}}(-\omega)^*$ ($^*$ means complex conjugate), it is then sufficient to treat the UPPE model (\ref{UPPE_1}) in the frequency domain $\omega > 0$ only. The field intensity can be defined by $E^2$ averaged over an optical period at least, for a given central frequency $\omega_0$. This quantity usually follows from the modulus of the time averaged Poynting vector. It is expressed in W/cm$^2$ and with the above normalization factor $c_1$ it is simply given by the classical relation $I = |\mathcal{E}|^2$.

\subsubsection{Envelope description}

When a central frequency $\omega_0$ is imposed, 
Eq.\ (\ref{UPPE_1}) restitutes the Nonlinear Envelope Equation (NEE), 
earlier derived by Brabec and Krausz \cite{Brabec:prl:78:3282}.
We can make use of the Taylor expansion
\begin{equation}
\label{taylor}
k(\omega) = k_0 + k' \bar{\omega} + \widehat{\cal 
D},\,\,\, \widehat{\cal D} \equiv \sum_{n \geq 2}^{+\infty} 
\frac{k^{(n)}}{n!} \bar{\omega}^n,
\end{equation}
where $\bar{\omega} = \omega - \omega_0$, $k' = \partial k/\partial 
\omega|_{\omega = \omega_0}$ and $k^{(n)} = \partial^n k/\partial 
\omega^n|_{\omega = \omega_0}$ and develop Eq.\ (\ref{UPPE_1}) as
\begin{equation}
\label{NEE_1}
\begin{split}
\partial_z E & = \int \left[ \frac{i \nabla_{\perp}^2}{2 k(\omega)} + i(k_0 + k' 
\bar{\omega} + \widehat{\cal D}) \right] \\
 & \quad \times \widehat{E}(\omega) \mbox{e}^{- i 
\omega t} d\omega + i \frac{\mu_0}{2} \int 
\frac{\omega^2}{k(\omega)} \widehat{\cal F}_{\rm NL}(\omega) 
\mbox{e}^{- i \omega t} d\omega.
\end{split}
\end{equation}
We apply the property following which $\widehat{E}(\omega)$ is the Fourier transform
of $E(t)\mbox{e}^{i \omega_0 t}$ in $\bar{\omega}$, so that  
$\bar{\omega}$ corresponds to $i \partial_t$ by inverse Fourier transform.
Terms with $k(\omega)$ in their denominator are expanded up to first order in $\bar{\omega}$ only.
Furthermore, we introduce the complex-field representation
\begin{equation}
\label{envelope}
\mathcal{E} = U \mbox{e}^{i k_0 z - i \omega_0 t},
\end{equation}
involving the novel envelope function $U$. Next, the new time variable $t \rightarrow t - 
z/v_g$ can be utilized to replace the pulse into the frame moving with the group velocity 
$v_g= k'^{-1}$. Eq.\ (\ref{NEE_1}) then restores the NEE model 
\begin{equation}
\label{NEE_3}
(i \partial_{z} + {\cal D}) U \simeq - \frac{T^{-1}}{2 
k_0} (\nabla_{\perp}^2 U) - \frac{\mu_0 \omega_0^2}{2 k_0 \sqrt{c_1}} T {\cal 
F}_{\rm NL}^{\rm env}(U),
\end{equation}
where
\begin{equation} 
\label{approx_1}
{\cal D} \equiv \sum_{n \geq 2}^{+\infty} 
(k^{(n)}/n!) (i \partial_{t})^n,\,\,\, T = (1 + \frac{i}{\omega_0} \partial_{t}),
\end{equation}
whenever $|k_0 - \omega_0 k'|/k_0 \ll 1$. This condition is met 
if the difference between group and phase velocity relative to the 
latter is small, which is fulfilled in a wide range of propagation 
phenomena. The operator $T^{-1}$ introduces {\it space-time focusing} 
in front of the diffraction term $[T^{-1}(\nabla_{\perp}^2 U)]$. On the other hand, nonlinearities with the envelope function ${\cal F}_{\rm NL}^{\rm 
env}(U)$ are also affected by the operator $T$, which refers to {\it 
self-steepening}. For dispersion relations truncated at some finite orders $n < +\infty$, the NEE model applies to optical fields with sufficiently narrow spectral bandwidths. With full chromatic dispersion, it holds for describing light pulses down to the single cycle.

\subsection{Nonlinear optical responses}

We henceforth assume a linearly polarized field (along, e.g., $\vec{e}_x$) and treat nonlinear effects within a scalar description. For centro-symmetric 
materials, only one relevant component of the tensor remains in the cubic contribution $P^{(3)}$, e.g., $\chi^{(3)}=\chi_{xxxx}^{(3)}$ \cite{Agrawal:NFO:01}. For 
simplicity, we may first consider $\chi^{(3)}$ as 
keeping a constant value for a spectral domain centered around 
$\omega_0$. Eq.\ (\ref{polarization_2}) then simplifies with a single 
component, noted $\chi_{\omega_0}^{(3)}$, and in time 
domain one finds $P^{(3)}(\vec{r},t) = \epsilon_0 
\chi_{\omega_0}^{(3)} E^3$. This expression holds whenever we suppose an 
instantaneous response of 
the medium, which ignores the contribution of molecular 
vibrations and rotations to 
$\chi^{(3)}$. Strictly speaking, however, the phenomenon of 
Raman scattering comes into play when the laser field interacts with 
anisotropic molecules. This interaction can be schematized by a three-level system built from the rotational states of a 
molecule. The molecular scatterer has two rotational eigenstates, the ground state (level 1) with energy $\hbar \Omega_1$ and an excited one (level 2) with energy $\hbar \Omega_2$ where $\hbar = 1.06 \times 10^{-34}$ Js and $\Omega_m$ denotes the frequency of the state $m=1,2,3$. Far above lies an electronic (or translational) state with energy $\hbar \Omega_3 \gg \hbar \Omega_2 - \hbar \Omega_1$. This molecule interacts with the laser field whose photon frequency $\omega_0$ fulfills $\Omega_{13},\Omega_{23} \gg \omega_0 \gg \Omega_{21}$, [$\Omega_{nm} \equiv \Omega_n - \Omega_m$], such that state $|3\rangle$ cannot be populated. Because of the definite parity of these molecular states, the dipole matrix element $\mu_{12}$ associated with the transition $|1\rangle \rightarrow |2\rangle$ via a single photon is null, so that the rotational state $|2\rangle$ can only be excited via transition through a virtual state $|3\rangle$ [$\mu_{13} \simeq \mu_{23} \equiv \mu \neq 0$]. Following this path, a Stokes photon with energy $\hbar \omega_s = \hbar \omega_0 - \hbar \Omega_{21}$ is emitted and the corresponding polarization vector involves the density matrix element associated with the states $|1\rangle$ and $|2\rangle$ as $P_{\rm Raman} = \chi^{(1)} [\rho_{12} \mbox{e}^{i \omega_R t} + c.c.] E$. Here, $\omega_R = \Omega_{21}$ is the fundamental rotational frequency and $\rho_{12}$ is found to satisfy \cite{Penano:pre:68:056502}
\begin{equation}
\label{Raman}
\partial_t \rho_{12} \simeq - \frac{\rho_{12}}{\tau_2} - i \frac{\mu^2 E^2}{\hbar^2 \Omega_{31}} \mbox{e}^{- i t/\tau_1},
\end{equation}
where $\tau_1 = 1/\omega_R$ and $\tau_2$ is the dipole dephasing time. Eq.\ (\ref{Raman}) provides the Raman response
\begin{equation}
\label{PRaman}
P_{\rm Raman} 
= \frac{2 \chi^{(1)} \mu^2}{\Omega_{31} \hbar^2} E \int_{-\infty}^t 
\mbox{e}^{-\frac{t-t'}{\tau_2}} \sin(\frac{t-t'}{\tau_1}) E^2(t') dt',
\end{equation}
which originates from nonresonant, nonlinear couplings. 

Expressed in terms of the rescaled complex field $\mathcal{E}$ [Eq.\ (\ref{complex})] and with appropriate normalizations \cite{Sprangle:pre:66:046418}, it completes the cubic polarization as
\begin{subequations}
\label{Raman_123}
\begin{align}
\label{Raman_1}
\begin{split}
P^{(3)} & = 2 n_0 n_2 \epsilon_0 \sqrt{c_1} \int_{-\infty}^{+\infty} {\bar R}(t-t') |\mathcal{E}(t')|^2 dt' \mathcal{E} \\
& \quad  + 2 n_0 n_2 \epsilon_0 \sqrt{c_1} (1 - x_K) \mathcal{E}^3/3+c.c.,
\end{split} \\
\label{Raman_3}
{\bar R}(t) & = (1 - x_K) \delta(t) + x_K \Theta (t) h(t), \\
\label{Raman_2}
h(t) & = \frac{2}{3} \frac{\tau_1^2 + \tau_2^2}{\tau_1 \tau_2^2} 
\mbox{e}^{-t/\tau_2} \sin(t/\tau_1),
\end{align}
\end{subequations}
with the definition of the nonlinear refractive index $n_2= 3 
\chi_{\omega_0}^{(3)}/(4 n_0^2 c \epsilon_0)$. Here, contributions in O($\mathcal{E}^3$) 
are retained to further describe third-harmonic generation. Expression (\ref{Raman_1}) possesses both retarded and instantaneous components in the ratio $x_K$. 
The instantaneous part $\sim \delta(t)$ describes the response from 
the bound electrons upon a few 
femtoseconds or less. The retarded part 
$\sim h(t)$ accounts for nuclear responses, namely, the 
Raman contribution, in which fast oscillations in $E^2$ give negligible contributions, as $\tau_1$ and $\tau_2$ currently far exceed the optical period $\sim \omega_0^{-1}$.

The fraction of delayed Kerr depends on the molecular species under consideration. For air at 800 nm, Sprangle \cite{Sprangle:pre:66:046418} suggests $\tau_1 \simeq 62$ fs, $\tau_2 \simeq 77$ fs and 
$x_K=1/2$. This choice is 
consistent with that proposed in experimental papers 
\cite{Nibbering:josab:14:650,Ripoche:oc:135:310}. When 
$\tau_1 \sim \tau_2$, the function $h(t) \simeq (1/\tau_1) 
\mbox{e}^{-t/\tau_1}$ can also be used in the ratio $x_K=1/2$ 
\cite{Chiron:epjd:6:383}. For condensed materials, the parameter ranges $\tau_2/\tau_1 = 2-4,\,\tau_2 = 30-50$ fs with $x_{K} = 0.15-0.18$ have been suggested \cite{Zozulya:prl:82:1430,Agrawal:NFO:01}. Values of the 
nonlinear Kerr index $n_2$ can be found in the literature 
\cite{Hellwarth:pra:41:2766,Nibbering:josab:14:650,Lehmeier:oc:56:67,Gong:cpl:15:30,Luo:jms:336:61}. Comprised between $10^{-19}$ cm$^2$/W for gases and $10^{-16}$ cm$^2$/W in dense media, they may, however, vary by a factor of the order of unity, depending on the procedure used for their evaluation (polarization spectroscopy, self- or cross-phase modulated spectra, 
time-resolved interferometry), together with the laser 
wavelength and pulse durations at which measurements are performed.

Besides, the susceptibility tensor has nonlinear 
components $\chi^{(j>3)}$ that satisfy the ordering
\cite{Boyd:NO:92,Shen:PNO:84}
\begin{equation}
\label{chi_law}
\frac{P^{(k+2)}}{P^{(k)}} = \frac{\chi^{(k+2)}}{\chi^{(k)}} \cdot \frac{E^{k+2}}{E^k} \approx \frac{|E|^2}{|E_{\rm at}|^2},
\end{equation}
where $E_{\rm at} \simeq 3 \times 10^{10}$ V/m is the characteristic atomic electric field strength with intensity $I_{\rm at} > 10^{14}$ W/cm$^2$. Typically, the evaluation $\chi^{(5)}/\chi^{(3)} \sim 
10^{-12}$ holds for nonresonant interactions in, e.g., gases. Despite 
the lack of knowledge 
on the sign of $\chi^{(5)}$ \cite{Pan:josab:7:509}, the quintic 
susceptibility is often expected to saturate Kerr focusing and has, 
therefore, a negative sign \cite{Nurhuda:rr:48:40}. Since the  
ordering (\ref{chi_law}) suggests 
that $\chi^{(j)}$ is rapidly decreasing with the order $j$, the 
Taylor series with respect to the electric field is truncated at the 
5th order.
Quintic polarization can be derived following the same procedure as 
above, with $\chi^{(5)}$ assumed constant in the frequency domain. By developing $E^5$ in terms of 
$(\mathcal{E},\mathcal{E}^*)$ [Eq.\ (\ref{complex})], the quintic 
contribution of the polarization vector then expands as 
\begin{equation}
\label{quintic}
P^{(5)} = - 2 n_0 n_4 \epsilon_0 
\sqrt{c_1}  
\left(\left|\mathcal{E}\right|^4 + \frac{1}{2} \left|\mathcal{E}\right|^2 
\mathcal{E}^2 + \frac{1}{10} \mathcal{E}^4\right) \mathcal{E} + c.c. 
\end{equation}
where $n_4 = 5 |\chi_{\omega_0}^{(5)}|/(4 n_0^3 c^2 \epsilon_0^2)$.
On the whole, the total nonlinear polarization vector reads as
\begin{equation}
\label{nlpoltotal}
P_{\rm NL} = P^{(3)} + P^{(5)}.
\end{equation}

\subsection{Plasma generation for singly-charged ionization}

When free electrons are created, they induce a current density $\vec{J} = q_e 
\rho \vec{v}_e$. This quantity depends on the electron charge $q_e=-1.6\times10^{-19}$ C,
the electron density $\rho$ and the electron velocity $\vec{v}_e$. $\vec{J}$ is 
computed from the fluid equations \cite{Esarey:ieeejqe:33:1879,Sprangle:pre:54:4211}
\begin{subequations}
\begin{align}
\label{plasma_1}
\partial_t \rho + \vec{\nabla} \cdot (\rho \vec{v}_e) & = {\cal S}, \\
\label{plasma_2}
\partial_t \vec{v}_e + (\vec{v}_e \cdot \vec{\nabla}) \vec{v}_e & = 
\frac{q_e}{m_e} \left(\vec{E} + \frac{\vec{v}_e \times \vec{B}}{c}\right) - 
\nu_e \vec{v}_e - {\cal S} \vec{v}_e/\rho.
\end{align}
\end{subequations}
Here, ${\cal S}$ represents external plasma sources and $\nu_e$ is 
the effective electron collision frequency. These equations can be 
combined to yield
\begin{equation}
\label{plasma_3}
\partial_t \vec{J} + \nu_e \vec{J} = \frac{q_e^2 \rho}{m_e} \vec{E} + 
\vec{\Pi},
\end{equation}
where 
\begin{equation}
\label{plasma_4}
\vec{\Pi} = \frac{q_e}{m_e c} \vec{J} \times \vec{B} - \frac{\vec{J}}{\rho 
q_e} (\vec{\nabla} \cdot \vec{J}) - (\vec{J} \cdot \vec{\nabla}) \vec{v}_e 
\end{equation}
represents ponderomotive forces acting on slowly-varying 
time scales. For linearly-polarized electromagnetic fields 
$(\vec{E},\vec{B})$ oscillating at the high frequency $\omega_0$, 
the driving term $\vec{\Pi}$ admits envelope components containing gradients of the field 
intensity, radiation 
pressure due to electron collisions and changes in the electron 
density. Ponderomotive forces induce low plasma currents, which can in 
turn generate electromagnetic pulses (EMP) 
\cite{Cheng:prl:87:213001,Cheng:prl:89:139302} and provide sources of coherent sub-THz 
radiation \cite{Tzortzakis:ol:27:1944}. Numerical simulations \cite{Sprangle:pre:69:066415,Penano:pop:11:2865} have 
shown, however, that for 100-fs pulses reaching intensities of 
$10^{14}$ W/cm$^2$ and free electron densities $\rho \sim 10^{16}$ 
cm$^{-3}$ in the atmosphere, the efficiency conversion to EMP is of 
the order of $10^{-9}$ with local intensities 
attaining only 10 kW/cm$ ^2$. In dielectrics, the plasma generates 
EMP intensities remaining about $\sim$ MW/cm$^2$ for 
peak laser intensities of $\sim 10^{13}$ W/cm$^2$. These ponderomotive terms can thus be ignored, as long as peak intensities are below $10^{15}$ W/cm$^2$. In this range, plasma density perturbations due to Langmuir wave oscillations and relativistic increase of the electron mass have also a negligible 
influence. Therefore, the equation for the current density reduces to 
Eq.\ (\ref{plasma_3}) in which $\Pi = 0$. At the lowest order in $v_e$, 
the growth of the electron density is only governed by the source 
term ${\cal S}$, i.e., 
\begin{equation}
\label{rho1}
\partial_t \rho = {\cal S} = W(I) (\rho_{\rm nt} - \rho) + \frac{\sigma}{U_i} \rho I - 
f(\rho),
\end{equation}
that involves photo-ionization processes with rate $W(I)$, 
collisional ionization with cross-section $\sigma$, and 
a function describing electron recombination or attachment with 
neighboring ions denoted by $f(\rho)$. Here, $\rho_{\rm nt}$ and $U_i$ are the density of neutral species and the ionization potential, respectively, while $\rho \ll \rho_{\rm nt}$. Typically, the recombination function in gases has a quadratic 
dependency on $\rho$, so that $f(\rho) = \beta_{\rm recomb} \rho^2$ 
with $\beta_{\rm recomb} 
[\mbox{cm}^3/\mbox{s}] \sim 2 \times 10^{-8}$ at electron temperatures 
$T_e = 1$ eV 
\cite{Tzortzakis:oc:181:123,Sprangle:pre:66:046418,Mlejnek:ol:23:382}. Recombination times belong to the nanosecond scale. In dielectrics, much 
shorter recombination times are involved ($\tau_{\rm 
recomb} = 50-150$ fs) and the density linearly decreases like $f(\rho) = \rho/\tau_{\rm recomb}$ 
\cite{Audebert:prl:73:1990,Tzortzakis:prl:87:213902,Penano:pre:72:036412}.

Besides, the electron collisional rate depends on
the electron energy distribution function and temperature versus the 
ionization potential $U_i$. Assuming a Maxwellian distribution function for the electron 
velocity, this rate linearly varies like $\sigma 
\left|\mathcal{E}\right|^2/U_i$, as long as the electron thermal 
energy is small compared with $U_i$. Here, $\sigma$ is 
the inverse bremsstrahlung cross-section \cite{Lotz:zp:206:205,Lotz:josab:57:873}. If we omit 
Ohmic heating, solving for the current density (\ref{plasma_3}) 
leads by Fourier transformation to
\begin{equation}
\label{current1}
\widehat{\vec{J}} = \frac{q_e^2}{m_e (\nu_e^2 + \omega^2)} (\nu_e + i 
\omega) \widehat{(\rho \vec{E})}.
\end{equation}
The current 
density term in Eq.\ (\ref{wave_equation1}) transforms as
\begin{equation}
\label{current2}
\mu_0 \partial_t \vec{J} \rightarrow [- i \frac{\omega n_0 
\sigma(\omega)}{c} + \frac{\omega_0^2}{c^2 \rho_c (1 + 
\nu_e^2/\omega^2)}] \widehat{(\rho \vec{E})},
\end{equation}
after introducing the critical plasma density
\begin{equation}
\label{rhocrit}
\rho_c \equiv \frac{\omega_0^2 m_e \epsilon_0}{q_e^2} \simeq 
\frac{1.11 \times 10^{21}}{\lambda_0^2[\mu\mbox{m}]} \mbox{cm}^{-3},
\end{equation}
at which the laser wave number vanishes. The cross-section
\begin{equation}
\label{sigma}
\sigma(\omega) = \frac{q_e^2}{m_e \epsilon_0 n_0 c \nu_e (1 + 
\omega^2/\nu_e^2)}
\end{equation}
then provides the frequency-dependent collisional rate. This expression 
was earlier 
derived in the limit of zero elastic collisions with ions by 
Kennedy \cite{Kennedy:ieeejqe:31:2241}, Yablonovitch and Bloembergen 
\cite{Yablonovitch:prl:29:907} and by Feit and Fleck 
\cite{Feit:apl:24:169}. Often linked to what is 
called the ''Drude'' model, it determines energy losses 
through plasma (cascade) ionization.

In Eq.\ (\ref{rho1}), $W(I)$ denotes the rate for photo-ionization. 
It is evaluated from perturbative theories valid as long as the electric field $E$ is weaker than the atom field strength $E_{\rm at}$. This rate has been rederived in Appendix \ref{appA}, following Keldysh's and Perelomov, Popov and Terent'ev (PPT)'s theories applying to atoms or dielectrics (crystals) \cite{Keldysh:spjetp:20:1307,Perelomov:spjetp:25:336,Perelomov:spjetp:23:924,
Perelomov:spjetp:24:207}. To describe the ionization of complex atoms, 
PPT formula usually 
includes the so-called ADK coefficients [for Ammosov, Delone and Krainov 
\cite{Ammosov:spjetp:64:1191}], originally established in 
the limit of high intensities. Optical field ionization theories stress two major limits bounded by the ''adiabaticity'' Keldysh parameter,
\begin{equation}
\label{keldyshparam}
\gamma = \omega_0 \frac{\sqrt{2 m_e U_i}}{|q_e| E_p},
\end{equation}
namely, the limit for multiphoton ionization (MPI, $\gamma \gg 1$) 
concerned with rather low intensities and the tunnel limit ($\gamma 
\ll 1$) concerned with high intensities, from which the Coulomb barrier becomes low enough to let the electron tunnel out. Here, $E_p$ denotes the peak optical amplitude ($E_p = \sqrt{2 c_1 I}$). For simplicity, we ignore the phenomenon of above-threshold ionization 
\cite{Agostini:prl:42:1127,Corkum:prl:62:1259}, through which the electron embarks more kinetic energy than $K \hbar \omega_0 - U_i$. For laser 
intensities $I = |\mathcal{E}|^2 < 10^{13}$ W/cm$^2$, MPI characterized by the limit
\begin{equation}
\gamma \gg 1 \Longrightarrow W(I) \rightarrow W_{\rm 
MPI} = \sigma_K I^K
\label{mpi}
\end{equation}
dominates, where $K=\textrm{mod}(U_i/\hbar\omega_0)+1$ is the number 
of photons necessary to liberate one electron.  For higher intensities, tunnel ionization starts to contribute, i.e., electrons tunnel out within one optical cycle. Despite the complexity of ionization formulas, all 
of them exhibit common dependencies on the laser field strength \cite{Perry:prl:60:1270,Reiss:pra:22:1786}. By 
producing charged ions of noble gases as a function of the laser 
intensity, quantitative measurements confirmed the validity of 
Keldysh theory for 
singly-charged ions only \cite{Perry:pra:37:747}. This theory basically applies to hydrogen atoms and discards the Coulomb field of the residual ion on the outgoing electron. The PPT/ADK or 
Krainov's models \cite{Krainov:josab:14:425}, instead, better 
reproduce experimental ionization rates of atoms in, e.g., the tunnel regime and may even suit for higher-charged states \cite{Augst:josab:8:858,Cornaggia:pra:62:023403}. On this basis, ionization of diatomic molecules can be 
predicted through semi-empirical theories using the PPT rate in which 
the electron tunnels through a barrier with effective potential 
$Z_{\rm eff}/r$. Here, $Z_{\rm eff}$ is determined by fitting the theoretical slope of the PPT formula onto measured ion signals. For ion signals collected 
from the interaction of O$_2$ and N$_2$ molecules with a 800 nm laser 
pulse, $Z_{\rm eff}$ is inferred from PPT applied to a corresponding 
atom (Xe atom) with ionization potential close to that of dioxygen 
molecules ($U_i^{O_2} \simeq 12.1$ eV, $Z_{\rm eff} = 0.53$) and to 
Ar atoms with $U_i$ close to the energy gap of $N_2$ molecules 
($U_i^{N_2} \simeq 15.6$ eV, $Z_{\rm eff} = 0.9$) 
\cite{Talebpour:oc:163:29}. The PPT/ADK formula 
can alternatively be extended to O$_2$ molecules by selecting the 
molecular coefficients of Tong {\it 
et al.} \cite{Tong:pra:66:033402} within the so-called ''ADK 
Molecular'' ionization model \cite{Nuter:josab:23:874}.

\begin{figure}
\includegraphics[width=0.99\columnwidth]{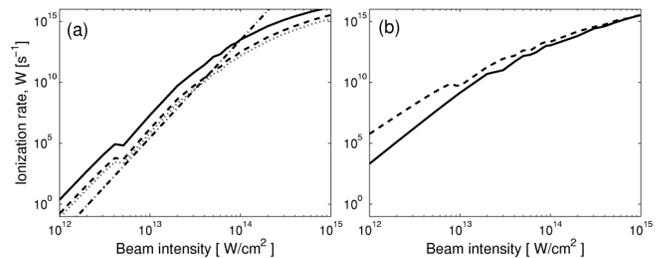}
\caption{(a) Ionization rate of O$_2$ molecules vs. laser intensity 
obtained from the PPT theory (solid curve), ADK molecular model 
(dashed curve), the fitting curve from PPT with $Z_{\rm eff} = 0.53$ 
(dotted curve) \cite{Talebpour:oc:163:29} and the MPI-like 
formulation (dash-dotted curve) used in \cite{Couairon:josab:19:1117} 
at 800 nm. (b) Keldysh ionization rates for crystals applied to fused silica 
(solid curve, $U_i = 7.8$ eV) and water (dashed curve, $U_i = 7$ eV) 
at  the same laser wavelength.}
\label{fig1}
\end{figure}

Figure \ref{fig1} illustrates ionization rates for some of the previous 
theories applied to O$_2$ molecules [Fig.\ \ref{fig1}(a)] or to fused 
silica [Fig.\ \ref{fig1}(b)]. PPT and ADK molecular ionization 
rates plotted in solid and dashed curves differ 
by only one decade. The dotted curve indicates the 
experimental fit from O$_2^+$ signals with the PPT rate using $Z_{\rm 
eff} = 0.53$. A good agreement is achieved between this fit and the 
ADK molecular ionization curve. Note that as long as 
the beam saturates below $10^{13}$ W/cm$^2$, the MPI limit alone can be retained. Beyond this intensity, the photo-ionization rate departs from the law $\sigma_K I^K$ and must include tunnel contributions. To get qualitative behaviors only, the MPI rate may, however, be used at higher intensities, while keeping the interaction physics valid. 

Because free carriers are first generated by photo-ionization, we must take the 
corresponding losses into account. Via energy conservation law, 
self-consistent expressions for 
these losses can be established. The temporal evolution of 
the energy density $w$ is determined by a local version of the 
Poynting theorem, i.e.,
\begin{equation}
\frac{d}{dt}w(\vec{r},t) = \vec{J}(\vec{r},t) \cdot 
\vec{E}(\vec{r},t),
\label{poynting}
\end{equation}
from which we can compute the energy lost by the 
pulse when it extracts electrons through single ionization process \cite{Rae:pra:46:1084,Kandidov:apb:77:149}. The amount of energy per time and volume units is then given by 
$\vec{J} \cdot \vec{E} = U_i \partial_t \rho_{\rm PI}$ where $\partial_t \rho_{\rm PI} \equiv W(I)(\rho_{\rm nt} - \rho)$. Using complex-valued 
fields, the current associated with 
photo-ionization losses is merely found to be
\begin{equation}
\label{losses}
\vec{J}_{\rm loss} = \sqrt{\frac{k_0}{2\omega_0 \mu_0}} U_i 
\frac{W(I)}{I} (\rho_{\rm nt} - \rho)
(\mathcal{\vec{E}}+\mathcal{\vec{E}}^*),
\end{equation}
where fast oscillations are canceled in the denominator. Geissler 
{\it et al.} \cite{Geissler:prl:83:2930} 
derived similar losses by introducing the overall polarization vector for free electrons $\vec{\cal P} = q_e \rho_{\rm PI} 
\vec{x}$, so that $\partial_t \vec{\cal P} = q_e \dot{\rho}_{\rm PI} \vec{x} + 
\vec{J}$. Free electrons arise with zero velocity at the position $\vec{x}(t) \simeq U_i \vec{E}/(2 |q_e| I)$, yielding the same loss current.

As a final result, the propagation equation within the UPPE 
description (\ref{UPPE_1}) reads in Fourier space as
\begin{equation}
\label{finalUPPE1}
\begin{split}
\frac{\partial}{\partial z} \widehat{\cal E} & = \left[ \frac{i}{2 
k(\omega)} 
\nabla_{\perp}^2 + i k(\omega) \right] \widehat{\cal E} + \frac{i \mu_0 
\omega^2}{2 k(\omega) \sqrt{c_1}} \Theta(\omega) \widehat{P}_{\rm NL} \\
 & - \frac{i k_0^2 \Theta(\omega)}{2 
\epsilon(\omega_0) k(\omega)(1+\frac{\nu_e^2}{\omega^2})} \left(\widehat{\frac{\rho 
\mathcal{E}}{\rho_c}}\right) - \frac{\Theta(\omega)}{2} \sqrt{\frac{\epsilon(\omega_0)}{\epsilon(\omega)}} {\cal L}(\omega),
\end{split}
\end{equation}
where 
\begin{equation}
\label{TFlosses}
{\cal L}(\omega) = \frac{U_i}{2\pi} \int \mathcal{E} \left[ \frac{W(I)}{I}(\rho_{\rm nt} - \rho) + \frac{\sigma(\omega)}{U_i} \rho \right] \mbox{e}^{i \omega t} dt.
\end{equation}
$P_{\rm NL}(\vec{r},t)$ [Eq.\ (\ref{nlpoltotal})] and the expression containing the
electron density $\rho(\vec{r},t)$ [Eq.\ (\ref{rho1})] must be transformed to Fourier space. Treating the complex field $\mathcal{E}$ only for positive frequencies is sufficient because of the symmetry $\widehat{\mathcal{E}^*}(\omega) = \widehat{\mathcal{E}}(-\omega)^*$, which imposes to select the parts of the nonlinear terms belonging to the frequency range $\omega > 0$.

\begin{figure}
\includegraphics[width=\columnwidth]{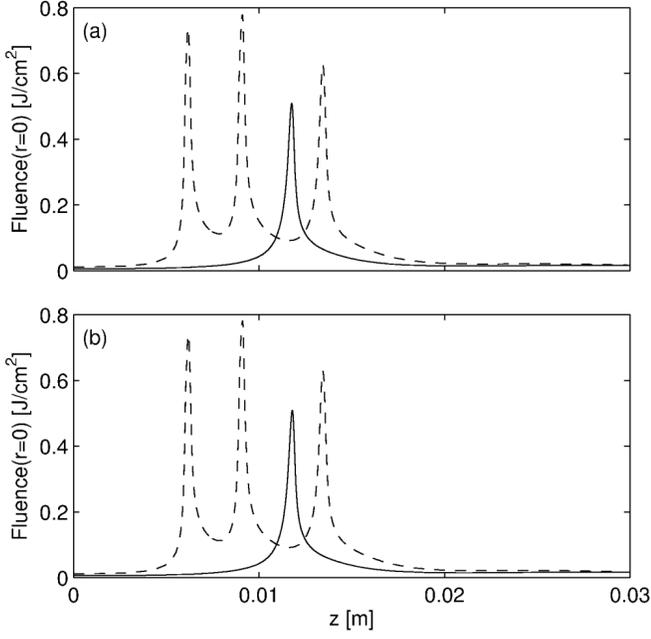}
\caption{Maximum fluence for pulses propagating in silica glass with 0.45 
$\mu$J (solid curve) and 0.9 $\mu$J (dashed line) energies simulated 
from (a) the UPPE model, (b) the NEE model with dispersion limited to fifth order.}
\label{fig2}
\end{figure}

For practical use, the 
collision cross-section $\sigma$ is 
stated for a central frequency $\omega_0$. We furthermore assume $\nu_e^2/\omega_0^2 \ll 1$ and $\sqrt{\epsilon(\omega_0)/\epsilon(\omega)} \approx 1$. The link to the NEE model is then straightforward, whenever 
the dispersion relation supports a Taylor expansion around $\omega_0$. By retaining only waveforms beating at 
$\omega_0$, the nonlinear envelope 
equation for the forward component $U$ is directly inferred from 
(\ref{NEE_3}) as
\begin{widetext}
\begin{subequations}
\label{modeleq}
\begin{align}
\label{1}
\begin{split}
\frac{\partial}{\partial z} U & =  \frac{i}{2k_0} {T}^{-1}\nabla_{\perp}^2 U + i \mathcal{D} U
+ i \frac{\omega_0}{c} n_2 T \left[ \left(1-x_K \right)|U|^2 
+ x_K \int_{-\infty}^t h(t-t') \left|U(t^{\prime})\right|^2 
dt^{\prime} \right] U \\
 & \quad - i\frac{\omega_0}{c} n_4 T |U|^4 U - i \frac{k_0}{2n_0^2 \rho_c} {T}^{-1} \rho U - \frac{\sigma}{2} \rho U - \frac{\beta_{\rm MPA}(|U|)}{2} U,
\end{split} \\
\label{2} 
\frac{\partial}{\partial t} \rho & = W(I) (\rho_{\rm nt} - \rho)
+ \frac{\sigma(\omega_0)}{U_i} \rho |U|^2 - 
f(\rho)
\end{align}
\end{subequations}
\end{widetext}
where $t$ stands for the retarded time variable $t - 
z/v_g$. The function $\beta_{\rm MPA}(|U|) = (\rho_{\rm nt} - \rho) U_i W(I)/|U|^2$ accounts for losses caused by photo-ionization. In the MPI limit (\ref{mpi}), this dissipative function takes the form $\beta_{\rm MPA}(|U|) \rightarrow \beta^{(K)} |U|^{2K-2}$ where $\beta^{(K)} \equiv K \hbar \omega_0 \sigma_K \rho_{\rm nt}$ is the coefficient for multiphoton 
absorption (MPA). The first term of the operator $\mathcal{D}$ corresponds 
to group-velocity dispersion (GVD) with coefficient $k''= \partial^2 
k/\partial \omega^2|_{\omega = \omega_0}$. Equations (\ref{modeleq}) 
describe wave diffraction, Kerr focusing response, 
plasma generation, chromatic dispersion with a self-consistent action 
of deviations from the classical slowly-varying envelope 
approximation 
through space-time focusing and self-steepening operators [$({T}^{-1} 
\nabla_{\perp}^2 \mathcal{E})$ and $(T |\mathcal{E}|^2 
\mathcal{E})$, respectively]. They are usually integrated numerically by using, e.g.,
initially singly-humped pulses taken with a super-Gaussian beam shape
\begin{equation}
\label{incond}
U(x,y,z=0,t) = U_0 \mbox{e}^{-\frac{r^{2N}}{w_0^{2N}} - i k_0 
\frac{r^2}{2f} - \frac{t^2}{t_p^2} - i C \frac{t^2}{t_p^2}},
\end{equation}
which may be focused through a lens of focal length $f$ and be 
temporally chirped if $C \neq 0$. Here, $r=\sqrt{x^2+y^2}$. For Gaussian beams ($N=1$), $U_0 = 
\sqrt{2P_{\rm in}/\pi w_0^2}$ involves the input power $P_{\rm in}$, $w_0$ is the beam waist and 
$t_p$ the $1/e^2$ pulse half-width, such that its 
full-width-at-half-maximum (FWHM) is $\Delta t = \sqrt{2 \ln{2}} 
t_p$. The initial level ($t = -\infty$) of plasma density is zero. 
In linear propagation, such Gaussian pulses diffract over the distance 
\begin{equation}
\label{zf}
z_f = (f^2/z_0)/(1 + f^2/z_0^2),
\end{equation}
where $z_0 = \pi n_0 w_0^2/\lambda_0$ is the diffraction 
range of the collimated beam ($f = +\infty$). In 
nonlinear propagation, the interplay between all competitors in Eqs.\ 
(\ref{modeleq}) severely alters beam diffraction and can maintain the 
pulse in self-guided state over longer distances. As an example, Fig.\ 
\ref{fig2} shows the dynamics of the maximum fluence (${\bar F} = 
\int_{-\infty}^{+\infty} |\mathcal{E}|^2 dt$) for an unchirped, collimated Gaussian pulse 
with 0.9 and 0.45 $\mu$J energies, input waist $w_0 = 71$ $\mu$m, 
half-width duration $t_p = 42.5$ fs propagating in a 3-cm long silica sample. This beam is numerically simulated from 
the UPPE model [Fig.\ \ref{fig2}(a)] and the NEE including higher-order dispersion up to $n = 5$ [Fig.\ \ref{fig2}(b)]. The ionization rate is 
delivered by the Keldysh's formula for 
crystals [Fig.\ \ref{fig1}(b)]. Silica promotes strong chromatic 
dispersion, which is approximated by the NEE model, but not by UPPE 
that describes the full linear dispersion relation. Despite this difference, 
we can observe that the beam fluence remains identical according to both 
descriptions (\ref{finalUPPE1}) and (\ref{modeleq}). This property is true in several media (dielectrics, liquids and air as well), 
whenever Eq.\ (\ref{1}) includes 
dispersive contributions $k^{(n)}$ with $n > 3$ \cite{Kolesik:prl:89:283902,Kolesik:apb:77:185}.

Earlier studies on femtosecond pulse propagation in the atmosphere 
started with simpler models than Eqs.\ (\ref{finalUPPE1}) or 
(\ref{modeleq}) by, e.g., ignoring time dispersion 
\cite{Kandidov:qe:24:905}. Their derivation was 
improved later
\cite{Feng:ol:20:1958,Sprangle:pre:54:4211,Mlejnek:ol:23:382,Mlejnek:pre:58:4903}, 
before being finalized into
NEE and UPPE formulations. 

\section{Optical Ultrashort filaments: A few tools for their 
analytical description\label{sec3}} 

\subsection{Kerr focusing and wave collapse}

\begin{figure}
\includegraphics[width=0.95\columnwidth]{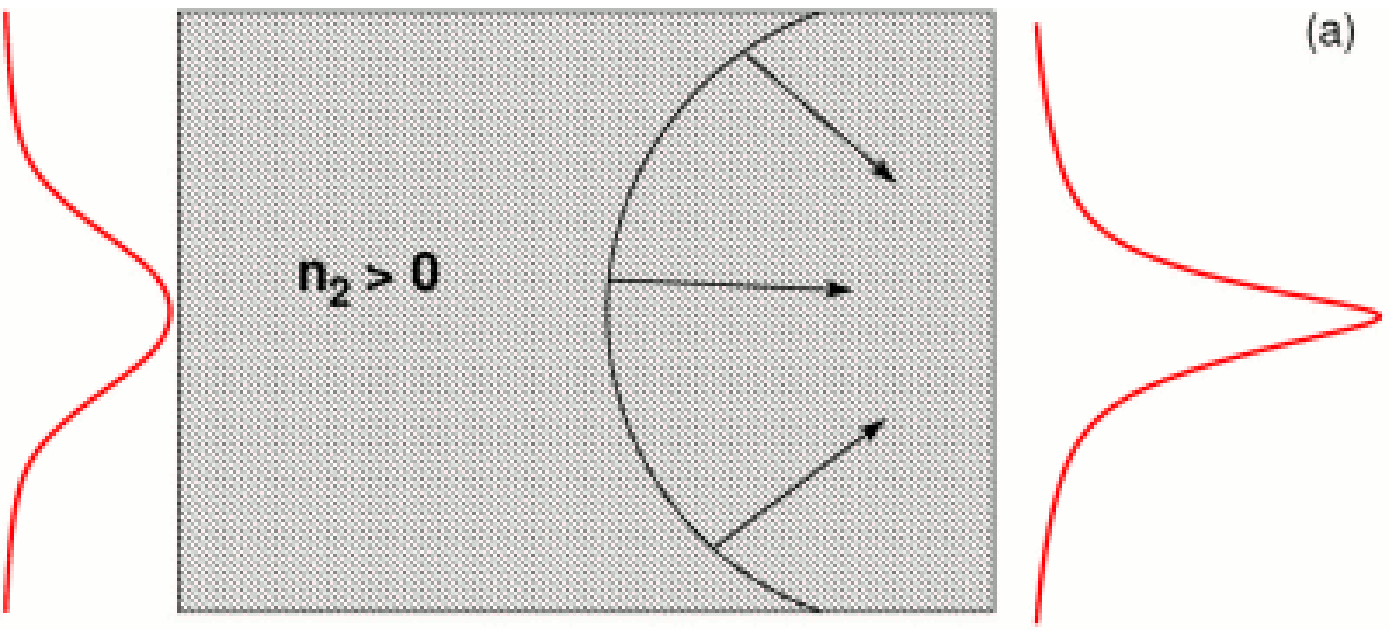}
\includegraphics[width=0.95\columnwidth]{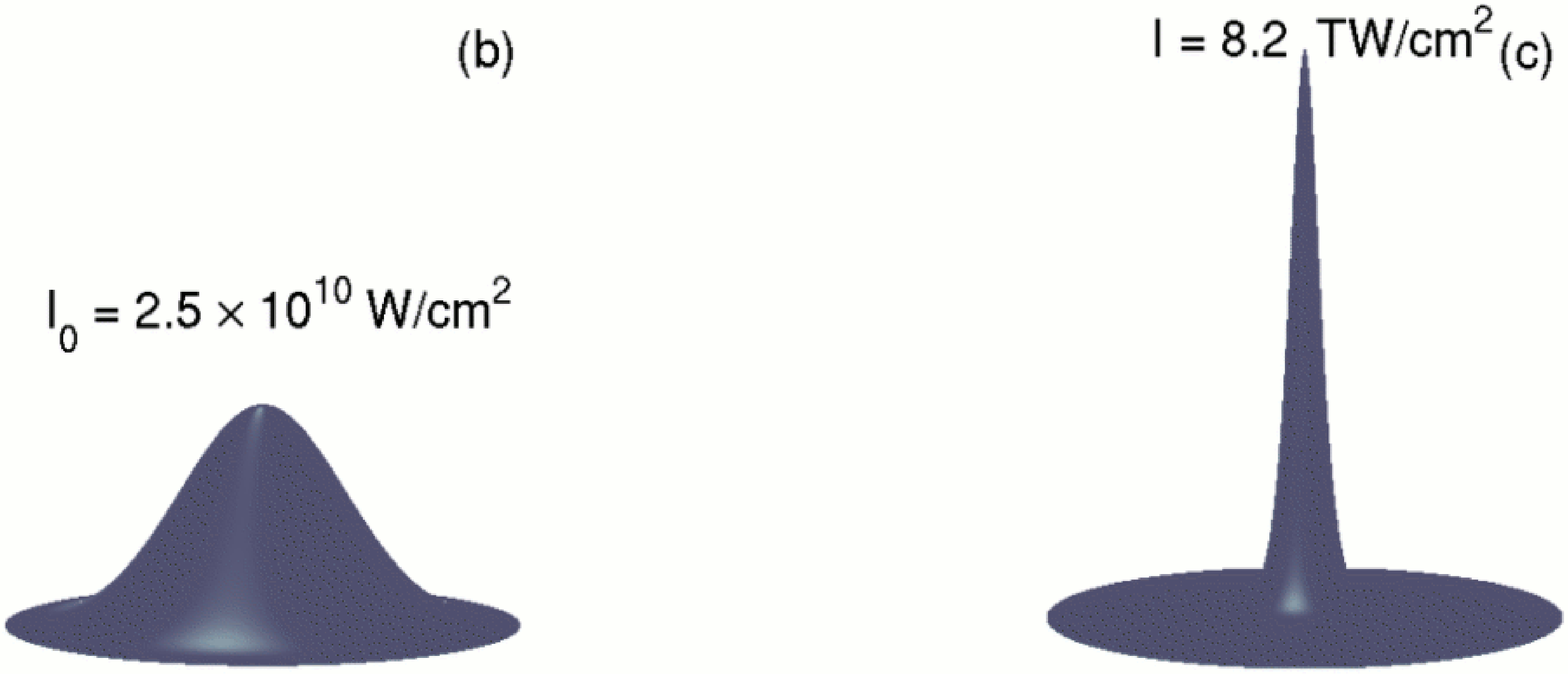}
\caption{(a) Principle of wave self-focusing. Insets (b) and (c) detail 
some intensity profiles of the solution to Eq.\ (\ref{cubicNLS}), applied to silica ($n_2 = 3.2 \times 10^{-16}$ cm$^2$/W). The initial condition is 
the Gaussian pulse (\ref{incond}) where $f=+\infty$, $w_0 = 130$ 
$\mu$m, $P_{\rm in}/P_{\rm cr} = 3$, $t_p = 85$ fs and $C = 0$.}
\label{fig3}
\end{figure}

For beams with no temporal dispersion, Eq.\ (\ref{1}) with no other 
nonlinear response than the instantaneous cubic Kerr term $(x_K = 
0)$, i.e., 
\begin{equation}
\label{cubicNLS}
\frac{\partial}{\partial z} U = \frac{i}{2k_0} \nabla_{\perp}^2 U + i 
\frac{\omega_0}{c} n_2 |U|^2 U = 0,
\end{equation}
describes the self-focusing of optical wave-packets with $n_2 > 0$
\cite{Askaryan:spjetp:15:1088,Marburger:pqe:4:35}. Self-focusing is a 
nonlinear phenomenon common to several branches of physics. In 
optics, it intervenes through the refractive optical index $N_{\rm 
opt} = n_0 + n_2 I$ that increases with the field intensity and 
forces light rays to refract onto the axis. This causes a compression 
of the beam in the diffraction plane, which leads to ''wave 
collapse'' when the Kerr nonlinearity is not saturated. A necessary condition for the collapse is that the input power $P_{\rm in} = \int |\mathcal{E}|^2
d{\vec r}$ exceeds some critical value,
\begin{equation}
\label{Pcrit}
P_{\rm cr} \simeq \frac{3.72 \lambda_0^2}{8\pi 
n_0 n_2},
\end{equation}
computed on the Townes mode (see below). The beam waist decreases more and more, as the field amplitude $|U|$ diverges.
Figure \ref{fig3} illustrates the self-focusing principle taking 
place in purely Kerr media. The two insets detail the collapsing 
evolution of a Gaussian wave-packet with initial condition 
(\ref{incond}) at $z = 0$ and near the collapse distance $z_c$. This distance, also termed as ''nonlinear focus'', locates the point at which the beam 
amplitude diverges along the optical path. It is given by the well-known semi-empirical Marburger formula 
\cite{Marburger:pqe:4:35}
\begin{equation}
\label{marburger}
z_c = \frac{0.367 z_0}{\sqrt{\left(\sqrt{\frac{P_{\rm in}}{P_{\rm cr}}} - 
0.852\right)^2 - 0.0219} + \frac{z_0}{f}}.
\end{equation}

Below we briefly review properties of wave collapse and how 
self-focusing is affected by temporal dispersion and plasma 
generation. In order to argue on the relevant mechanisms 
participating in this process, we discard
non-resonant terms causing harmonic generation and consider an 
instantaneous cubic response ($x_K = 0$). We apply the approximation $T^{-1} \simeq 1 - (i/\omega_0) \partial_t$ supposing $\omega_0 t_p \gg 1$, neglect recombination and treat the plasma coupling term driven by MPI only and subject to the limit $T^{-1} \rightarrow 1$. Equations (\ref{modeleq}) are then rescaled in dimensionless form by using the substitutions $r \rightarrow w_0 r$, $t \rightarrow t_p t$, $z \rightarrow 4 z_0 z$, $U \rightarrow \sqrt{c_2} \psi$ 
and $\rho \rightarrow (n_0^2 \rho_c/2 z_0 k_0) \rho$, where $c_2 \equiv \lambda_0^2/8\pi^2 n_0 n_2 w_0^2$. They result into the extended nonlinear Schr{\"o}dinger (NLS) equation
\begin{subequations}
\label{adim}
\begin{gather}
   \label{adim_1}
   i \partial_z \psi + \nabla_{\perp}^2 \psi +
   |\psi|^2 \psi + {\cal F}(\psi) = 0, \\
\begin{split}
{\cal F} & = - \delta \partial_t^2 \psi - \rho \psi - \epsilon |\psi|^4 \psi + i \nu |\psi|^{2K-2} \psi \\
 &  \quad + \frac{i(|\psi|^2
   \psi - \nabla_{\perp}^2 \psi)_t}{t_p \omega_0},\quad \partial_t \rho = \Gamma |\psi|^{2K},
\end{split} 
\end{gather}
\end{subequations}
where ${\cal F}(\psi)$ includes ''perturbations'' of a critical collapse, such as GVD, MPI, 
nonlinear saturation, MPA and pulse steepening. GVD and MPA have 
the normalized coefficients $\delta \equiv 2 z_0
k''/t_p^2$ and $\nu = 2 z_0 \beta^{(K)}c_2^{K-1}$, respectively. Quintic saturation is 
taken into account through $\epsilon = n_4 c_2/n_2$. The rescaled MPI 
coefficient
reads as $\Gamma = (2 z_0 k_0/n_0^2 \rho_c) \sigma_K \rho_{\rm nt} t_p 
c_2^K$. For Gaussian beams, the incident 
amplitude (\ref{incond}) reduces to
\begin{equation}
\label{adim_0}
|\psi(z=0)| = \sqrt{\frac{16 \pi n_0 n_2 P_{\rm in}}{\lambda_0^2}} 
\mbox{e}^{-r^2 - t^2}.
\end{equation}

\subsubsection{Principles of wave self-focusing}

With a purely cubic nonlinearity, solutions to the Cauchy problem 
(\ref{adim_1}) can blow-up (or collapse) at finite distance
\cite{Kelley:prl:15:1005,Sulem:NLS:99}. To describe this singular 
process, let us consider 
the NLS equation (\ref{adim_1}) with ${\cal F} = \partial_t^2 \psi$ ($\delta = -1$), so that the 
Laplacian ${\vec \nabla}_{\perp}^2 \rightarrow {\vec \nabla}^2 = 
\partial_x^2 + \partial_y^2 + \partial_t^2 +...$ formally accounts 
for the dispersion of a wave-packet along $D$ orthogonal spatial axes 
[${\vec r} = 
(x,y,t,...)$]. The wavefunction $\psi$ evolves from the 
spatially-localized initial datum
$\psi({\vec r},0) \equiv \psi_0({\vec r})$, assumed to belong to the 
Hilbert space $H^1$ with finite norm $\|\psi\|_{H^1} = (\|\psi\|_2^2 
+ 
\|{\vec \nabla} \psi\|_2^2)^{1/2}$, where $\|f\|_p \equiv (\int |f|^p 
d{\vec 
r})^{1/p}$. Two invariants are associated with $\psi$, namely, the 
$L^2$ norm (power) $P$ and Hamiltonian $H$:
\begin{equation}
\label{NLS_2}
\displaystyle{P = \|\psi\|_2^2\,,\,\,\,H = \|{\vec \nabla} \psi 
\|_2^2 -
\frac{1}{2} \|\psi\|_4^4.}
\end{equation}
The following ``virial'' equality can be 
established \cite{Vlasov:rqe:14:1062,Glassey:jmp:18:1794}
\begin{equation}
\label{NLS_3}
\displaystyle{P d_z^2 \langle r^2 \rangle = 4 \{2H + 
(1-D/2)\|\psi\|_4^4\},}
\end{equation}
where $\langle r^2 \rangle = \int r^2 |\psi|^2 d{\vec r}/P$ denotes 
the mean-squared radius of the solution $\psi$. By a double 
integration 
in $z$, Eq.\ (\ref{NLS_3})
shows that, whenever $D \geq 2$, there exist initial conditions for 
which $\langle r^2 \rangle$ vanishes at
finite distance, which is the sufficient signature of a wave collapse. For 
finite norms $P$,
the inequality $P \leq (4/D)^2 \langle r^2 \rangle
\times \|{\vec \nabla} \psi\|_2^2$ thus implies that the gradient
norm diverges in collapse regimes. As $H$ is finite, the collapse 
dynamics makes the $L^4$ norm $\|\psi\|_4^4$ blow up in turn and 
max$_r|\psi|$ diverges accordingly, by virtue of the mean-value theorem $\int 
|\psi|^4d{\vec r} \leq \mbox{max}_r |\psi|^2 \times P$ 
\cite{Kuznetsov:chaos:6:381}. This leads to a 
finite-distance blow-up, at which the solution $\psi$ stops to exist 
in $H^1$ \cite{Rasmussen:ps:33:481}. This mathematical singularity 
reflects the ultimate issue of the nonlinear self-focusing in the 
absence of saturation of the Kerr response and under the paraxiality assumption.

While $H < 0$ arises from Eq.\ (\ref{NLS_3}) as a sufficient condition 
for collapse, sharper requirements can be derived by means of the Sobolev 
inequality
\begin{equation}
\label{NLS_3bis}
\displaystyle{\|\psi\|_4^4 \leq C \|{\vec \nabla} \psi\|_2^D
\times \|\psi\|_2^{4-D}.}
\end{equation}
In the critical case $D=2$,
this inequality can be used to bound $H$ from below, so that the 
gradient norm blows up only if $P$ fulfills the constraint $P > P_c$. 
The best constant in Eq.\ (\ref{NLS_3bis}) is exactly $C_{\rm best} = 2/P_c$ 
and it involves the quantity $P_c = \int R^2 d{\vec r} = 11.68$, where $R$ 
is the radially-symmetric soliton solution, called the ''Townes mode'', 
of $- R + 
r^{-1} \partial_r r \partial_r R + R^3 = 0$ 
\cite{Chiao:prl:13:479,Weinstein:cmp:87:567}. $P_c$ justifies the 
existence of a critical 
power for the 2D self-focusing of optical beams in nonlinear Kerr 
media. In the supercritical case $D = 3$, a criterion for collapse, 
sharper than $H < 0$, can be established from a combination of Eqs.\ 
(\ref{NLS_3}) and (\ref{NLS_3bis}) as $H < P_c^2/P$ for gradient 
norms initially above $3P_c^2/P$ \cite{Kuznetsov:pd:87:273}. Here, 
$P_c$ again corresponds to the mass of the 3D soliton 
satisfying $- R + r^{-2} \partial_r r^2 \partial_r R + R^3 = 0$.

Once collapse is triggered, the solution focuses self-similarly 
near the singularity point $z_c$ as \cite{Rypdal:pd:16:339}
\begin{equation}
\label{NLS_4}
\displaystyle{\psi({\vec r},z) = L^{-1}(z) \phi({\vec \xi},\zeta)
\mbox{e}^{i\lambda \zeta + i L L_z \xi^2/4},}
\end{equation}
where ${\vec \xi} = {\vec r}/L(z)$, $\zeta(z) \equiv \int_0^z
du/L^2(u)$ and the parameter $\lambda$
is positive for making the new wavefunction $\phi$
localized. The function $L(z)$ represents the scale length
that vanishes as collapse develops, and $\phi$ converges to an exactly
self-similar form $\phi({\vec \xi})$ fulfilling $\partial_{\zeta} 
\phi \rightarrow 0$. For radial 
solutions Eq.\ 
(\ref{adim_1}) transforms into
\begin{equation}
\label{NLS_5}
\displaystyle{i\partial_{\zeta} \phi + \xi^{1-D} \partial_{\xi} 
\xi^{D-1} \partial_{\xi} \phi + |\phi|^2 \phi +
\beta [\xi^2 - \xi_{T}^2]\phi = 0,}
\end{equation}
where $\xi_{T}^2 \equiv
\beta^{-1}[\lambda - i L L_z (\frac{D}{2}-1)]$ is viewed as a complex turning
point, with $\beta \equiv -\frac{1}{4}L^3 L_{zz}$. As $L(z) \rightarrow 0$, 
$\phi$ can be treated by means of quasi-self-similar
techniques \cite{Berge:pr:303:259}. The solution $\phi$ is split into
a nonlinear core, $\phi_c$, extending in the range $\xi \ll \xi_T$, 
and a linear tail, $\phi_{T} \sim \mbox{e}^{- \lambda 
\pi/\beta}/\xi^{1+i \lambda/\beta}$, defined in
the complementary spatial domain $\xi \gg \xi_{T}$ where
the nonlinearity vanishes. The length $L(z)$ is then identified
from the continuity equation describing the mass exchanges between 
the core and tail parts of $\psi$. The dynamics of self-similar 
collapses 
vary with the space dimension number as follows.

\begin{itemize}

\item For $D = 2$, $L(z)$ has
a twice-logarithmic correction: $L(z) \simeq L_0
\sqrt{z_c-z}/\sqrt{\ln{\ln{[1/(z_c-z)]}}}$  
\cite{Fraiman:spjetp:61:228,Landman:pra:38:3837,Malkin:pla:151:285}. 
As $z \rightarrow z_c$, the exponential contribution of the tail 
decreases to zero, while the core
converges to the Townes 
mode $R$. The power $P$ relaxes to the critical
value $P_c$ and stays mostly located around the
center. Self-similar relaxation of self-focusing beams to the Townes 
mode has been reported experimentally \cite{Moll:prl:90:203902}.

\item For $D = 3$, $\beta$ attains a fixed point
$\beta_0 \neq 0$, leading to the scaling law $L(z) \simeq L_0 
\sqrt{z_c - z}$. The power is no longer preserved self-similarly in space,
since $P = L(z) \int |\phi|^2 d{\vec \xi}$. This integral behaves as 
$P \simeq P_{\rm core}(z) + P_{\rm tail}(z)$, where $P_{\rm core}(z) 
\sim L(z)$ vanishes, while $P_{\rm tail}(z)$ contains almost all 
the initial mass as $L(z) \rightarrow 0$. A 3D collapse is thus 
accompanied by an expulsion of mass towards the large distances 
where it keeps a stationary density $r^2 |\psi|^2 \rightarrow {\rm const}$ 
\cite{Zakharov:spjetp:64:773,Vlasov:spjetp:68:1125,Kosmatov:pd:52:16,Lemesurier:pd:138:334}.

\end{itemize}

\subsubsection{Variational approaches}

Because the cubic NLS equation is not integrable at high dimension 
numbers $D \geq 2$, approximation methods describing the fate of 
singular solutions may be employed. Among those, ''variational 
approaches'' consist in building a set of dynamical equations 
governing the size, amplitude and phase of the beam. These beam 
parameters are evaluated from functional integrals (Action or 
Lagrangian integrals) computed from a given trial function, which the 
pulse is supposed to keep along the optical path. Such methods 
give global, qualitative behaviors within a pedestrian way. Their 
principal drawback is that, as they need to fulfill the main 
conservation laws (e.g., preservation of the $L^2$ norm), they 
capture the entire initial mass of the solution and do not let the 
latter evacuate radiation to the boundaries or relax to an exact 
ground state. This discrepancy may partly be cured by accounting for 
corrective damping of the nonlinear core 
\cite{Kath:pre:51:1484,Arevalo:pre:72:026605}, but improvements 
become rapidly tricky and limited. Variational principles applied to 
NLS solutions were extensively studied 
\cite{Anderson:pf:22:1838,Anderson:pra:27:3135,Bondeson:ps:20:479, 
Anderson:pf:22:105,Desaix:josab:8:2082} and further exploited in 
different contexts 
\cite{Vidal:pre:55:3571,Manassah:ol:13:589,Karlsson:ol:17:22,
Manassah:ol:17:1259,Pietsch:epl:15:173,Silberberg:ol:15:1282,Cerullo:ol:21:65}. 
They were also widely worked out by several authors in the present scope \cite{Sprangle:pre:54:4211,Esarey:ieeejqe:33:1879,Lehner:pre:61:1996,Couairon:epjd:27:159}.

\begin{figure}
\includegraphics[width=\columnwidth]{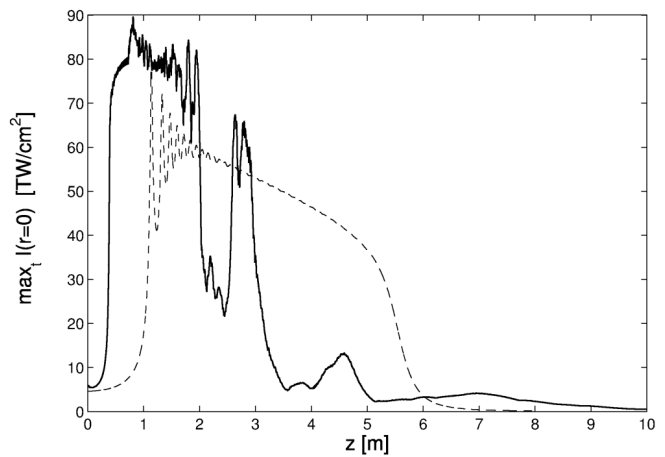}
\caption{Peak intensities for a 1 mm-waisted, 100-fs collimated Gaussian pulse with 22 critical powers. Dashed line: Results from Eqs.\ (\ref{VA}) in the limits $T, T^{-1} \rightarrow 1$; Solid line: $(3+1)$-dimensional numerical simulations (initial conditions are perturbed by a $10 \%$ amplitude random noise).}
\label{fig4}
\end{figure}

For example, considering the shape (\ref{incond}) with $N=1$, a Gaussian trial function in the form
\begin{equation}
\label{adim_7a}
\psi = \frac{\sqrt{{\cal J}(z)}}{L(z) \sqrt{T(z)}} \phi(\xi_{\perp},\eta) 
\mbox{e}^{i\frac{L_z L}{4} \xi_{\perp}^2 - i 
\frac{T_z T}{4 \delta} \eta^2}
\end{equation}
with $\phi = \mbox{e}^{-\xi_{\perp}^2/2 - \eta^2/2 + i \theta(z)}$, 
$\xi_{\perp} = r/L(z)$, $\eta = t/T(z)$ can be employed in the virial 
identities governing the transverse and temporal mean squared extents 
\cite{Berge:pop:7:210,Sprangle:pre:66:046418}. Separate dynamical 
systems for the transverse width [$L(z)$], the temporal duration 
[$T(z)$] and the intensity factor [${\cal J}(z)$] of the pulse are then analytically established as
\begin{subequations}
\label{VA}
\begin{align}
\label{VA1}
\frac{L_{zz}}{4} & = \frac{1}{L^3} - \frac{{\cal J}T(0)}{4 T 
L^3} + \frac{\Gamma \sqrt{\pi K} {\cal J}^K}{2 (K+1)^2 L^{2K+1} T^{K-1}} + \dotsb \\
\label{VA2}
\frac{T_{zz}}{4} & = \frac{\delta^2}{T^3} + \frac{\delta {\cal J} T(0)}{4 
L^2 T^2} + \dotsb, \\
{\cal J}_z & = - \frac{2 \nu}{K^{3/2}} \frac{{\cal J}^K}{(L^2T)^{K-1}} + \dotsb
\end{align}
\end{subequations}
where final dots formally omit contributions from MPA, Raman-delayed, quintic and steepening effects.
With ${\cal J}(0) = 2 \sqrt{2} P_{\rm in}/P_{\rm cr}$ imposed by the initial power, Eqs.\ (\ref{VA}) yield qualitative behaviors 
for the propagation of a single femtosecond pulse preserving its 
cylindrical symmetry. It, of course, cannot depict any temporal or 
spatial splitting phenomenon, but offers a global evolution of the 
beam intensity. This evolution fairly supports the 
comparison with $(3+1)$-dimensional numerical simulations 
\cite{Champeaux:pre:71:046604}, apart from an overestimated nonlinear focus. Figure 
\ref{fig4} shows an example of peak intensity reached by Gaussian pulses with parameters $P_{\rm 
in}/P_{\rm cr} = 22$, $w_0 = 1$ mm, $t_p = 100$ fs at the wavelength 
$\lambda_0 = 800$ nm in the limit $T \rightarrow 1$ and dispersion 
limited to GVD. Collapse is arrested by plasma generation. The dashed curve corresponds to the solution 
reconstructed from the two-scale variational method. The solid one shows the result obtained from numerical 
integration of the propagation equations.

Besides, alternative procedures, such as perturbative methods, may provide 
qualitative information about the changes in the scale $L(z,t)$ induced by 
contributions that can stop the collapse. Here, 
the radial size $L(z,t)$ tends to zero in the diffraction plane as $z 
\rightarrow z_c$, but it now depends on time. The principle is to 
perturb a collapsing state that naturally tends to the Townes mode $R$, by means of $L^2$ orthogonal perturbative modes. These will provide an integral relation for the $z$-dependent function 
$\beta$ [Eq.\ (\ref{NLS_5})], directly depending on ${\cal F}$ \cite{Fibich:siamjam:60:183}. At the critical dimension, the zeroth-order collapsing solution, 
whose core converges as 
\begin{equation}
\psi \rightarrow \psi_s = L^{-1} R({\vec r}/L) \exp{(i\zeta + 
iL_z r^2/4L),}
\end{equation}
has its mean radius $L(z,t)$ modified by the perturbation $|{\cal F}| 
\ll 1$. Appropriate functions ${\cal F}$ are then capable of turning the sign 
of $\beta = - \frac{1}{4} L^3 L_{zz}$ from positive to negative, 
which predicts the arrest of collapse. This method was applied to 
many ''perturbations'' of NLS, such as normal GVD, steepening and 
non-paraxial deviations \cite{Fibich:prl:76:4356,Fibich:ol:22:1379} 
as well as for NLS with attractive potentials 
\cite{Schjoedteriksen:pre:64:066614,Lemesurier:pre:70:046614}. 
It was later extended to a variational system  
formally avoiding the constraints $\epsilon, \beta, {\cal F} \ll 1$ by introducing an
amplitude factor accounting for dissipative losses, $\psi \rightarrow \sqrt{{\cal J}(z,t)} \psi_s$ \cite{Berge:pd:152:752}. Applied to Eqs.\ (\ref{adim}) for pulses having a Gaussian distribution in time, this procedure involves the substitution
\begin{equation}
\label{subs}
\psi \rightarrow \sqrt{{\cal J}(z,t)} \psi_s;\,\,\,{\cal J}(z=0,t) = P_{\rm in} 
\mbox{e}^{-2t^2}/P_{\rm cr},
\end{equation}
and leads to the dynamical equations
\begin{subequations}
\label{adim_56}
\begin{align}
\label{adim_5}
\frac{M}{4P_c} L^3 L_{zz} & = 1 - {\cal J} + 4 \epsilon \frac{{\cal J}^2}{L^2} 
- \frac{L^2}{2 P_c} \int R^2 \xi \partial_{\xi} \rho d{\vec \xi} \\
   \label{adim_6}
   \frac{{\cal J}_z}{{\cal J}} & \simeq 2
   \delta \zeta_{tt} - 2 \nu A \frac{{\cal J}^{K-1}}{L^{2K - 2}} - 
\frac{B (L^{-2})_t}{t_p \omega_0},
\end{align}
\end{subequations}
where $M = \int \xi^2 R^2(\xi) d{\vec \xi}$, $A = \int R^{2K} d{\vec 
\xi}/P_c$, $B = 3 {\cal J} + 1$, and $C = \int R^{2K+2} d{\vec \xi}/P_c$.
Equation (\ref{adim_6}) originates from the power variations, assuming ${\cal J}_z/{\cal J} \ll 1$. 
The first term refers to GVD, the second one to MPA and the last term 
corresponds to pulse steepening.

\begin{figure}
\includegraphics[width=\columnwidth]{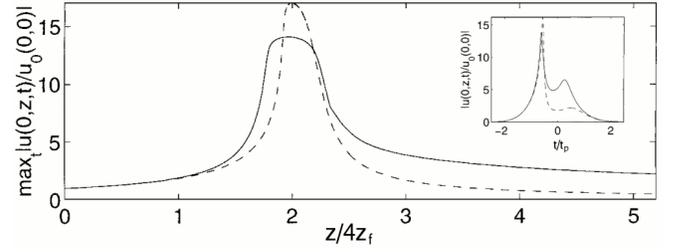}
\caption{Normalized maximum intensity of Gaussian pulses in argon with power of 1.85 $P_{\rm cr}$ integrated from Eqs.\ 
(\ref{adim_56}) (dashed curves) and from a numerical integration of Eqs.\ 
(\ref{modeleq}) (solid curves). Inset shows temporal profiles at $z = 2.4$ cm.}
\label{fig5}
\end{figure}

Figure \ref{fig5} depicts the normalized maximum intensity and 
temporal pulse distortions driven by MPI alone at $z = 2.4$ cm for the solution to Eqs.\ (\ref{adim_56}) in the absence of 
dissipation (${\cal J}_z = 0$). The Townes mode is here approached 
by a Gaussian profile for a beam containing 1.85 critical powers. The 
temporal dependencies in $L(z,t)$ alter the pulse profile. Note that 
discrepancies occur in the maximal amplitude, in the location of 
the first focus and in the temporal shapes. Because they rely on 
trial functions that capture the same power and shape as initial, 
variational procedures cannot accurately describe fluctuations in the wave 
envelope. They should therefore be employed for what they are, i.e., 
as qualitative procedures yielding first indications on the 
beam dynamics.

\subsection{Saturation by temporal dispersion}

\begin{figure}
\includegraphics[width=0.85\columnwidth]{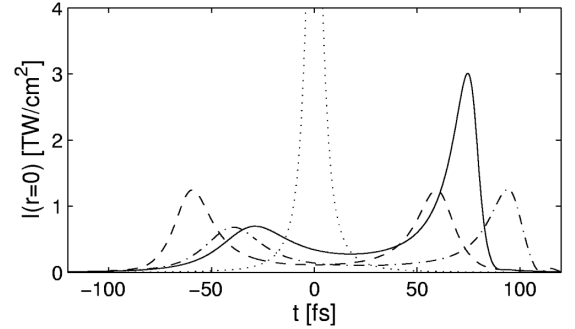}
\caption{Temporal profiles of Gaussian pulses ($w_0 = 130\,\mu$m, $t_p = 85$ fs) near the nonlinear focus with power $P_{\rm in} = 3 P_{\rm cr}$ for the Kerr response alone (dotted curve), for Kerr + normal GVD (dashed curve), for Kerr + GVD + $T, T^{-1}$ (solid curve), for Kerr + GVD + $T,T^{-1}$ + higher-order dispersion (dash-dotted curve). In all cases, plasma response is omitted.}
\label{fig6}
\end{figure}

By mixing GVD and spatial diffraction when $T \simeq 1$, the 
Kerr nonlinearity causes defocusing in time for $k'' > 0$ (normal 
dispersion) and temporal compression for $k'' < 0$ (anomalous 
dispersion), besides wave focusing in the transverse
direction. The interplay of these processes results
in the symmetric splitting of the pulse along the time axis with 
normal GVD \cite{Chernev:ol:17:172,Chernev:oc:87:28} and
to a 3D spatiotemporal collapse with anomalous GVD 
\cite{Berge:pr:303:259}. In addition, when ultrashort pulses develop 
sharp temporal gradients, the operator $T$ in front of the Kerr 
term (self-steepening) induces a shock 
dynamics: The field develops a singular 
profile with $|\psi_{t}| \rightarrow +\infty$ in the trail ($t > 0$) 
of the pulse \cite{Anderson:pra:27:1393}. This dynamics is reinforced 
by space-time focusing \cite{Rothenberg:ol:17:1340}. Figure \ref{fig6} 
depicts temporal profiles of pulses undergoing steepening effects in 
normally dispersive regime (silica glass at 790 nm), together with a 
transverse collapse. 

To understand these dynamics, we can consider ultrashort 
pulses as being stacked along the temporal direction into different 
time slices having each their own power, e.g., $P(t) = P_{\rm in} 
\mbox{e}^{-2 t^2}$ for Gaussian profiles. Slices located at times $t 
< 0$ correspond to the front (or leading) pulse; those at $t > 0$ 
constitute the back (trailing) pulse. Each time slice self-focuses at 
its respective singularity point, $z_c(t)$, according to Eq.\ 
(\ref{marburger}) in which the ratio $P_{\rm in}/P_{\rm cr}$ must be 
replaced by $P_{\rm in}\mbox{e}^{- 2 t^2}/P_{\rm cr}$. This scenario 
is known as the ''moving-focus'' model 
\cite{Lugovoi:spu:16:658,Shen:rmp:1:48} and yields simple 
comprehension elements to figure out the pulse distortions. Here, the central time slice focuses 
at the shortest distance $z_c(t=0)$. Furthermore, $\dot{z}_c(t)$ is 
positive for
$t > 0$ and negative for $t < 0$ [$\dot{z}_c(0) = 0$], whereas
$\ddot{z}_c(t)$ always remains positive. Near the collapse 
distance, we can replace all time derivatives by $\partial_t = - 
\dot{z}_c \partial_z$, $\partial_t^2 = -
\ddot{z}_c \partial_z + \dot{z}_c^2 \partial_z^2$ 
\cite{Luther:pd:74:59, Fibich:ol:22:1379}. Power fluctuations 
(\ref{adim_6}) are then given by
\begin{equation}
    \label{13}
    \displaystyle{\frac{{\cal J}_z^{\rm disp}}{{\cal J}} = 2 \delta [- 
\ddot{z}_c/L^2 +
    \dot{z}_c^2 (1/L^2)_z] + \frac{2 B}{\omega_0 t_p} \dot{z_c} 
(L^{-2})_z,}
\end{equation}
where $L_z < 0$ in compression regime. With $\delta > 0$, Eq.\ 
(\ref{13}) describes defocusing around the central slice: Normal 
GVD transfers power towards non-zero instants, symmetrically located 
with respect to $t = 0$. Self-steepening and space-time focusing 
moreover produce a transfer of power from the leading ($\dot{z}_c < 0$) to 
the trailing portion of the pulse ($\dot{z}_c > 0$). This asymmetrizes the temporal profile, which was retrieved by direct experiments \cite{Ranka:ol:23:534}. Normal GVD alone ''splits'' a focusing pulse into two regular, symmetric spikes at powers $< 2 P_c$. For higher powers, the peak edges develop shock profiles and 
disintegrate into ripplelike cells 
\cite{Germaschewski:pd:151:175,Fibich:pre:67:056603}. Because of the 
hyperbolicity of the
operator $\nabla_{\perp}^2 - \delta 
\partial_t^2$, one splitting event transforms the optical field
in the $(r,t)$ plane into an $X$-shaped waveform 
\cite{Litvak:pre:61:891,Litvak:jetp:91:1268,Zharova:jetp:96:643,Berge:prl:89:153902,
Ditrapani:prl:91:093904,Conti:prl:90:170406,Christodoulides:ol:29:1446}.

Normal GVD $(k''>0)$ and plasma formation 
compete at powers moderately above critical to halt the wave 
collapse. The stronger the GVD coefficient, the larger the power 
interval in which the collapse is arrested by pulse splitting. By 
solving the cubic NLS equation with normal GVD, a boundary 
$\delta_{\rm crit}({\bar p})$, function of the ratio of input power over 
critical, ${\bar p} = P_{\rm in}/P_{\rm cr}$, can be calculated in such a 
way that initial conditions fulfilling $\delta > \delta_{\rm 
crit}({\bar p})$ will limit the Kerr self-focusing through GVD splitting 
instead of letting the solution diverge into a singular state 
\cite{Luther:ol:19:862}. Higher-order dispersion together with 
steepening effects $(T, T^{-1})$ modify this curve in the sense of 
''delaying'' the self-focusing threshold to higher powers 
\cite{Skupin:pd:220:14,Skupin:pra:74:043813}. In particular, third-order dispersion tends to delocalize the pulse by pushing the temporal centroid
to the back
\cite{Fibich:ol:29:887}.

In contrast, for anomalous GVD $(k'' < 0)$ power is transferred to 
center, as seen from Eq.\ (\ref{13}) with $\delta < 0$. Ultrashort 
pulses thus collapse both in space and time. A mapping $|\delta| > 
\delta_{\rm crit}({\bar p})$ can again be constructed on the basis of virial-type arguments 
\cite{Kuznetsov:pd:87:273,Berge:pre:71:065601}. It predicts pulse 
spreading when the dispersion length $\sim t_p^2/|k''|$ is short 
enough to prevail over diffraction and Kerr nonlinearity. 
As for normal GVD, the theoretical boundaries are modified by higher-order dispersion and steepening terms to some extent.

\begin{figure}
\includegraphics[width=0.85\columnwidth]{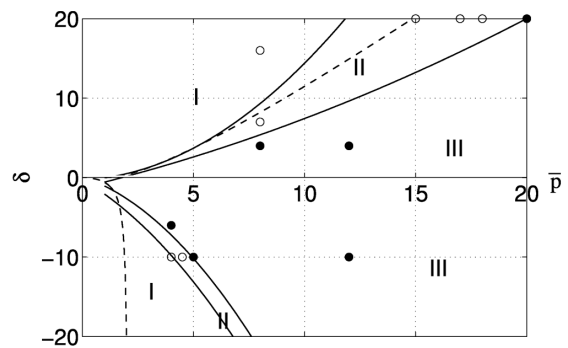}
\caption{Collapse regions in the plane $(\delta,{\bar p})$. The solid curves 
extrapolate results from the propagation equations. The dashed curves 
represent theoretical limits estimated from Eqs.\ (\ref{adim}), above 
which dispersion inhibits the self-focusing. Regions I, II and III are explained in the text.}
\label{fig7}
\end{figure}

Figure \ref{fig7} summarizes the zone of collapse and no-collapse in 
the plane $(\delta,{\bar p})$ for both normally and anomalously dispersive 
media computed from Eqs.\ (\ref{modeleq}). Open circles represent initial conditions that do not 
collapse. Close ones result in a strong divergence of the beam 
intensity. Three zones clearly occur: (I) A dispersion-dominated 
domain leading to pulse spreading; (II) A transition zone in which GVD 
and steepening operators inhibit the self-focusing; (III) A 
Kerr-dominated region, in which chromatic dispersion is unable to 
stop the wave blow-up. Here, the field intensity can increase by several decades before 
reaching the ionization threshold. Note that for anomalous GVD, the collapse 
process, which encompasses spatial and temporal compressions, makes 
the intermediate zone II almost meaningless.

\subsection{Saturation by plasma defocusing}

The action of MPI is to deplete the pulse temporal profile
through the sudden emergence of an ionization front near the focus 
point $z_c$. In the absence of time dispersion and nonlinear losses, 
the plasma equation $\partial_t \rho = \Gamma |\psi|^{2 K}$ can be integrated 
with the {\it ansatz} (\ref{subs}) as
\begin{equation}
\label{MPI_1}
\rho(r,t,z) \simeq \sqrt{\frac{\pi}{8K}} \Gamma 
\left(\frac{P_{\rm in}}{P_{\rm cr}} R^2\right)^K \frac{\mbox{Erf}(\sqrt{2K}
   t) + 1}{L^{2K}(z,t)},
\end{equation}
where Erf$(x)=(2/\sqrt{\pi}) \int_0^x \mbox{e}^{-u^2} du$ denotes 
the error function. The last 
term of Eq.\ (\ref{adim_5}) then scales as $L^{2(1-K)} \partial_{\xi} 
\rho$ where
$\partial_{\xi} \rho \sim \partial_{\xi} R^{2K}$ is negative. As $L
\rightarrow 0$, this term efficiently competes with the Kerr 
contribution ${\cal J} \sim P_{\rm in} \mbox{e}^{-2t^2}/P_{\rm cr}$. With 
the step function Erf$(\sqrt{2K} t) +1$, the plasma response arises like a defocusing plateau, as the field intensity reaches its maximum near the focus point $z_c$. All time slices belonging to the interval 
$t > 0$ are defocused. At negative times, the pulse
continues to self-focus and feeds plasma defocusing until
forming a single time slice located near $t = t^* \simeq - 
[\ln{\sqrt{P_{\rm in}/P_{\rm cr}}}]^{1/2}$ at powers close to 
critical \cite{Berge:prl:86:1003,Henz:pra:59:2528}.
Thus, plasma generation is
accompanied by a sharp duration shortening of the pulse (see, e.g., inset of Fig.\ \ref{fig5}).

\begin{figure}
\includegraphics[width=0.49\columnwidth]{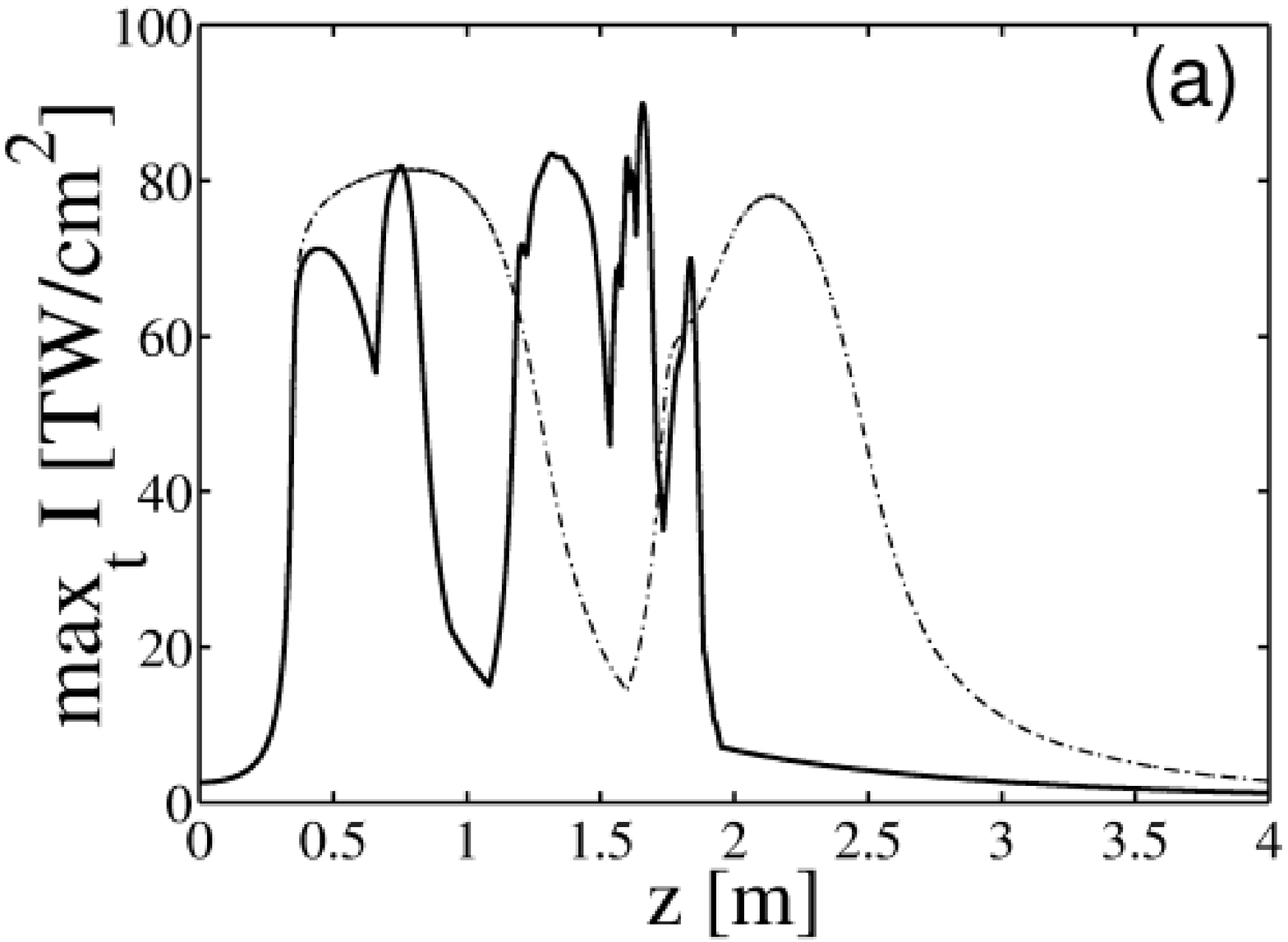}
\includegraphics[width=0.49\columnwidth]{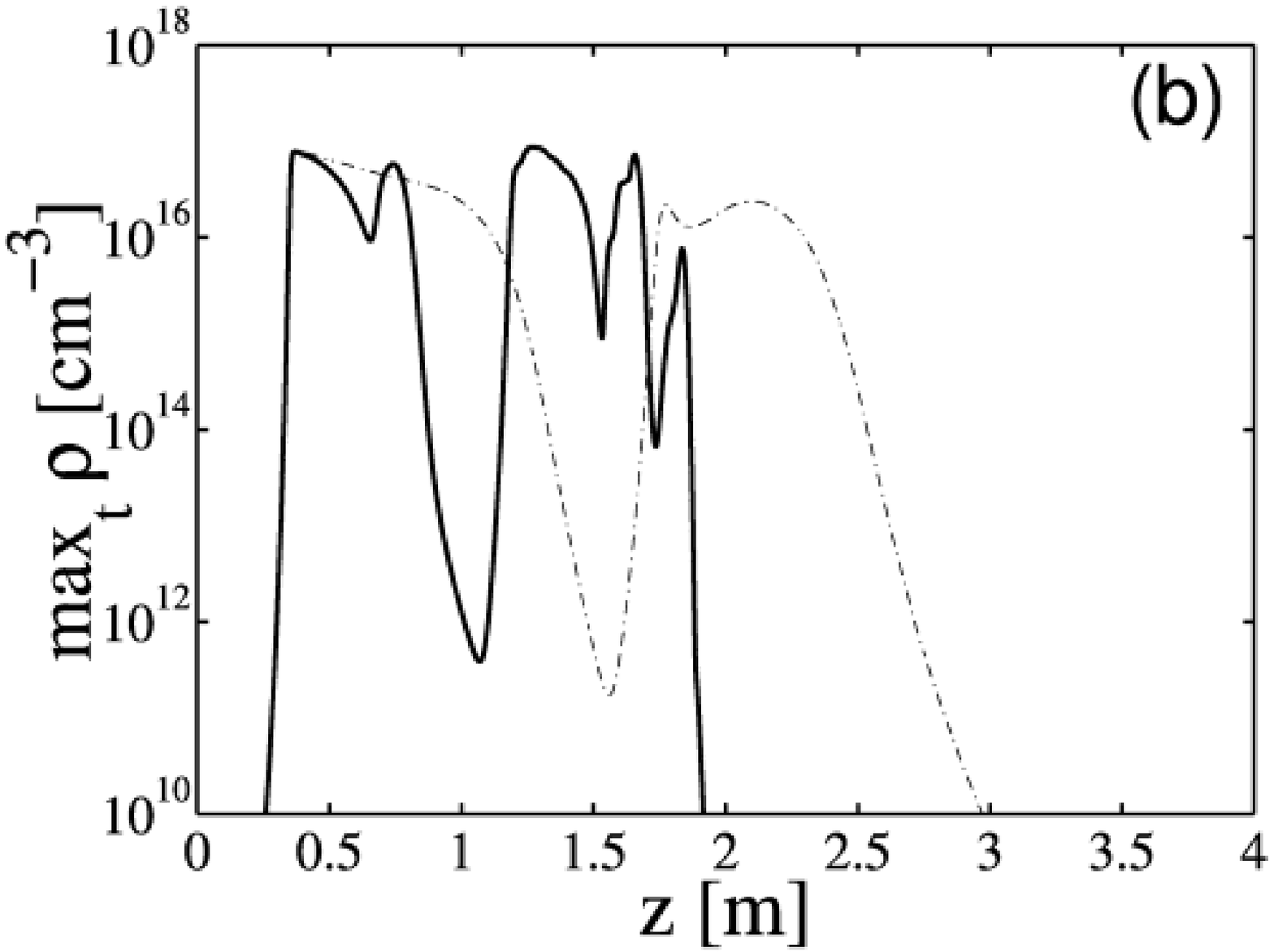}
\includegraphics[width=0.49\columnwidth]{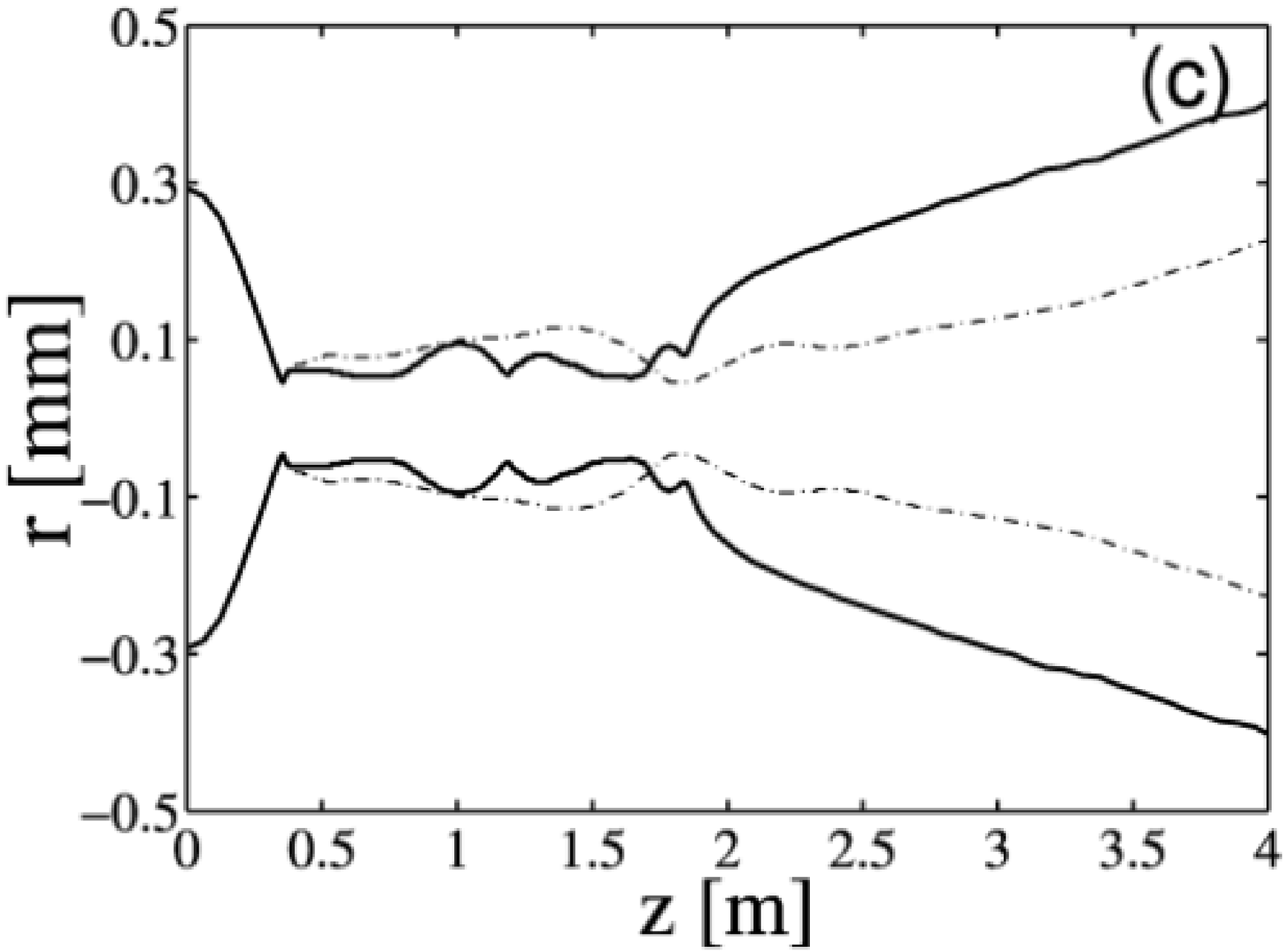}
\includegraphics[width=0.49\columnwidth]{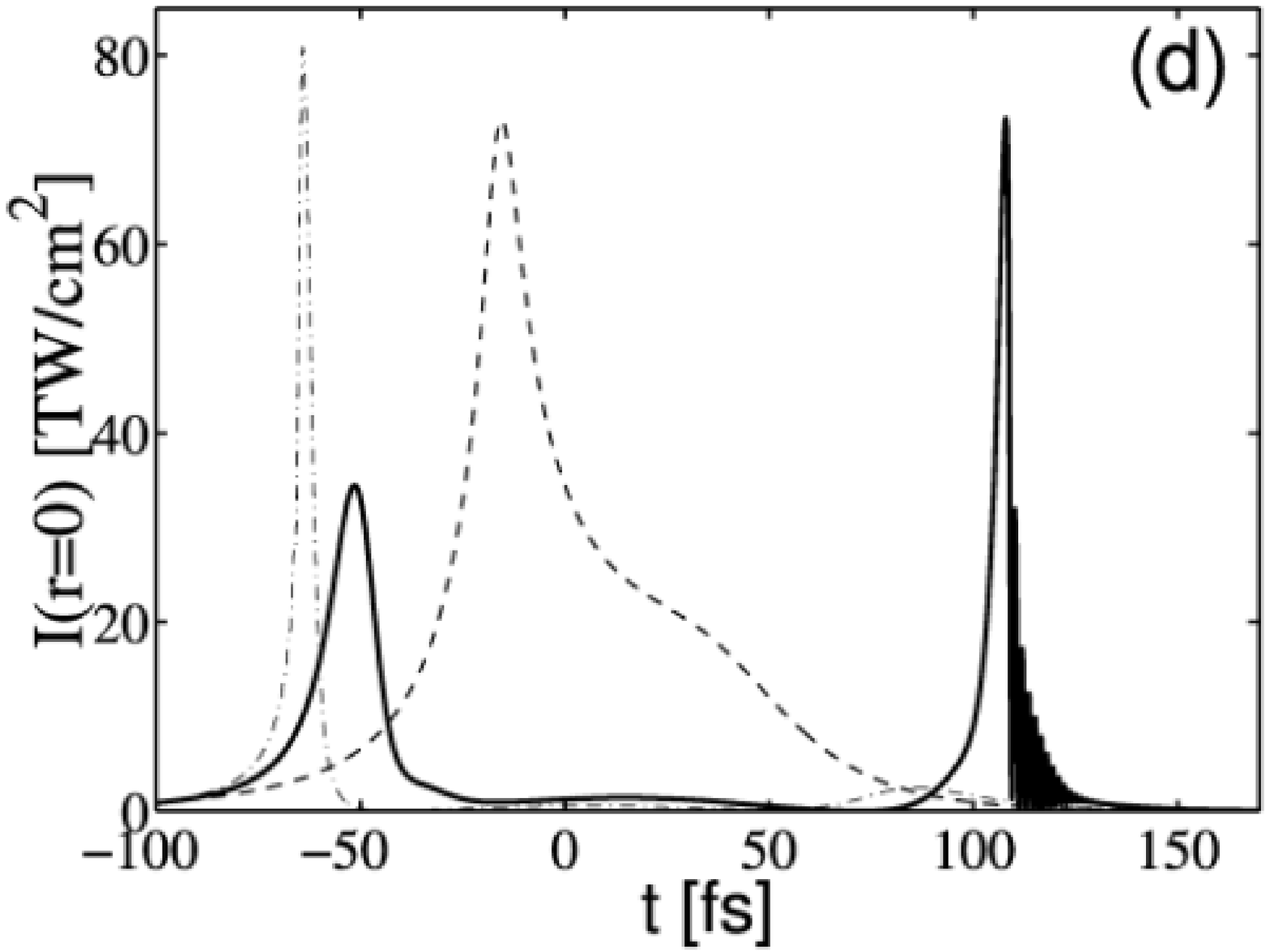}
\includegraphics[width=0.49\columnwidth]{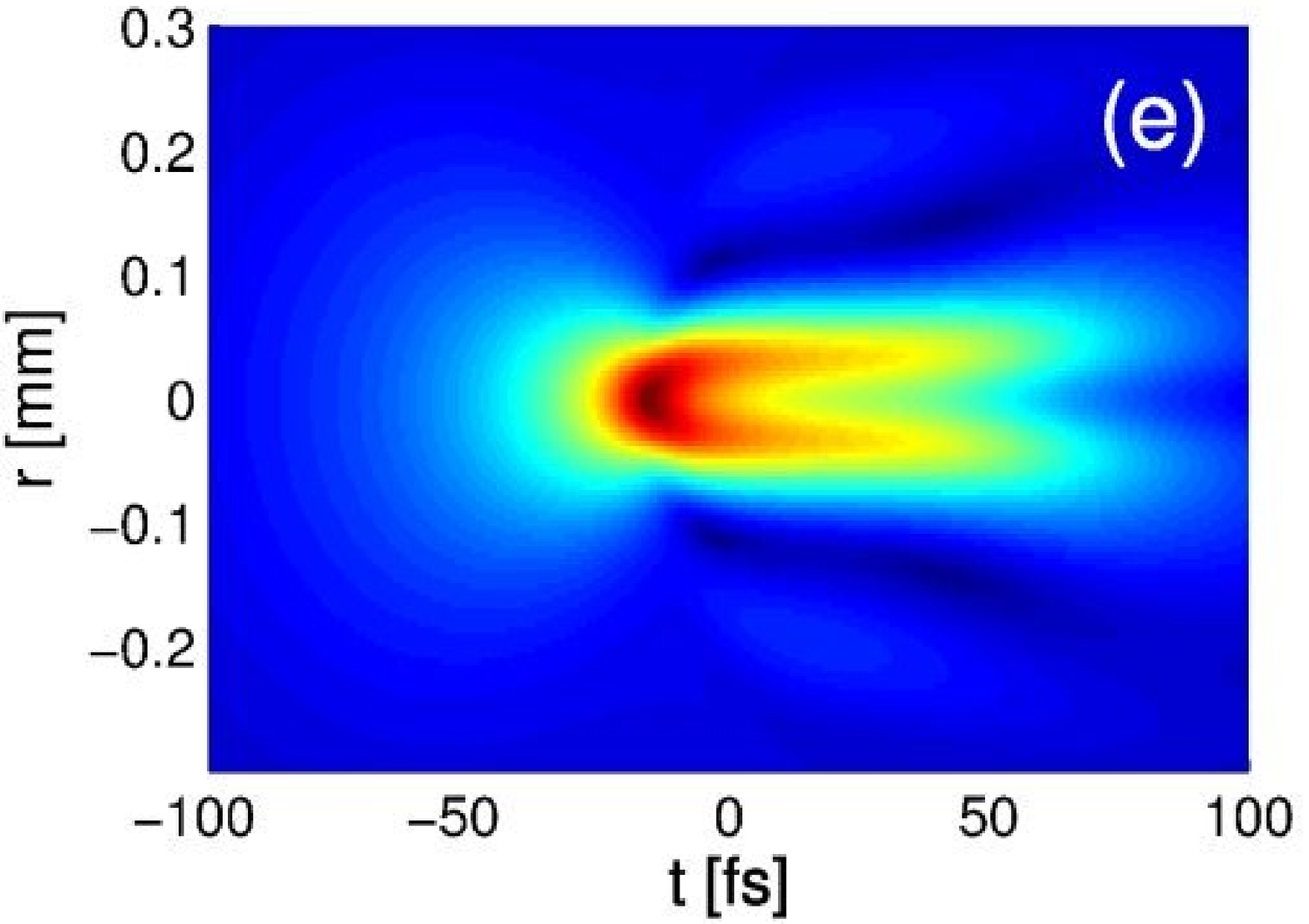}
\includegraphics[width=0.49\columnwidth]{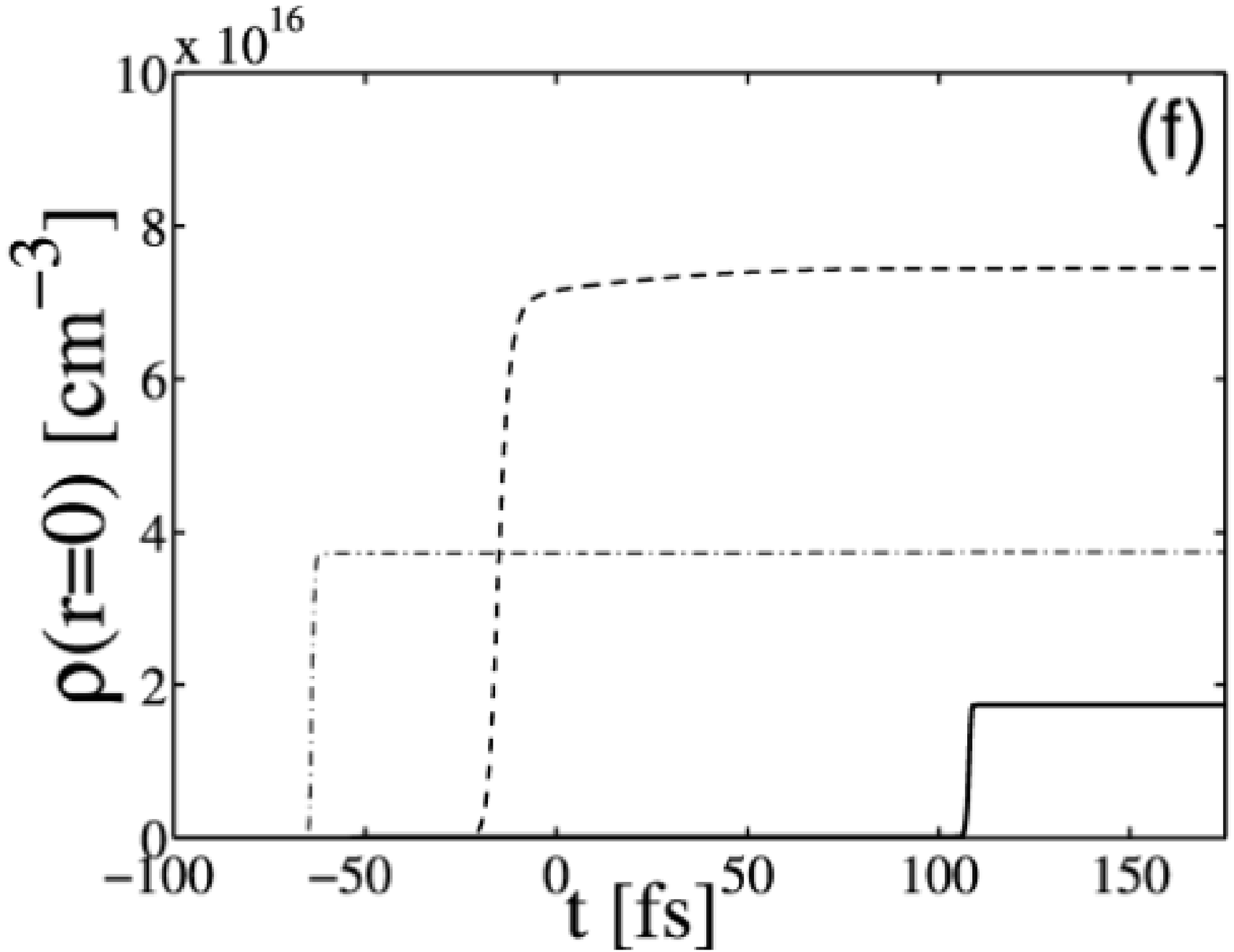}
\caption{(a) Maximum intensity, (b) peak electron density and (c) mean radius (FWHM of the fluence)
of a 150-fs, 0.5 mm-waisted Gaussian pulse with $4P_{\rm cr}$ in air at $\lambda_0 = 800$ nm. Dash-dotted curves refer to the case when $T, T^{-1} = 1$ and chromatic dispersion is limited to GVD ($k'' = 0.2$ fs$^2$/cm). Solid curves refer to the full model (\ref{modeleq}). (d) Temporal profiles at $z = 0.8$ m. The dashed curve refers to the distribution at $z = 0.4$ m (identical for both models). (e) Radial profile vs time near $z_c$. (f) Plasma responses associated with (d).}
\label{fig9}
\end{figure}

When GVD and MPA come into play, the leading peak becomes unstable 
and refocusing of the trail occurs at high enough input powers. With 
MPA, the leading component is partly damped in the
intensity ratio ${\cal J}_z/{\cal J} \simeq - 2 \nu A ({\cal J}/L^2)^{K-1}$, as the peak
electron plasma forms a density plateau [see Eq.\ (\ref{adim_6})]. 
Consequently, the electron density
attains much lower levels and permits the emergence of a trail. 
This complex dynamics leads to two following signatures for plasma 
defocusing: (i) MPI shortens the pulse duration near the nonlinear 
focus; (ii) Because $\rho$ scales as $R^{2K}$, the spatial zone of 
plasma defocusing takes place inside a narrow region ($\sim 1/\sqrt{K}$ times the beam width) of the intensity distribution, which creates spatial rings. This ''self-guiding'' mechanism is therefore not static. Certain pulse components are defocused to the 
benefit of the others. To clear it up, Figure \ref{fig9} depicts 
focusing/defocusing cycles affecting the peak intensity, electron density, beam radius and the distribution in space 
and time of a 150-fs pulse for atmospheric propagation. Note 
the rings formed just at the stage of plasma defocusing [Fig.\ \ref{fig9}(e)] 
and the occurrence of a sharp trail when steepening effects $(T, T^{-1})$ are included [Fig.\ \ref{fig9}(d)]. 

\subsection{Saturation by optical nonlinearities}

Additional defocusing quintic nonlinearities produce stable 
solitons in continuous-wave (cw) media \cite{Trillo:SS:01}. This property follows from 
a balance between diffraction, Kerr focusing and nonlinear 
saturation. When a quintic contribution is added (${\cal F} = 
-\epsilon |\psi|^4 \psi$), the beam starts to self-focus and evacuates part of power into diffracting components to reach a stable ground state of Eqs.\ (\ref{adim}) 
\cite{Gatz:josab:14:1795,Vakhitov:rqe:16:783,Kolokolov:rqe:17:1016,Malkin:pd:64:251,Vidal:prl:77:1282}. This property is partly reflected by Eqs.\ (\ref{adim_56}) emphasizing an 
arrest of collapse like $\beta = - \frac{1}{4} L^3 L_{zz} \sim 1/L^2$, which induces 
regular oscillations of the size $L(z)$ beyond the focus point.

\begin{figure}
\includegraphics[width=\columnwidth]{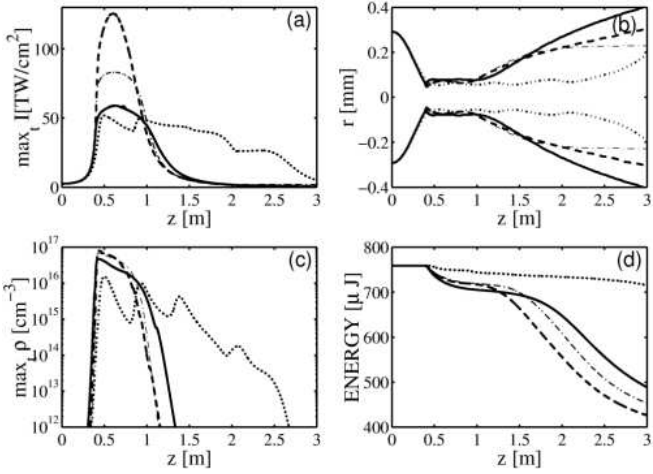}
\caption{(a) Peak intensities, (b) pulse radii, (c) electron density 
and (d) energy losses for Gaussian pulses with $w_0 = 0.5$ mm, 
$t_p = 70$ fs, $T = T^{-1} = 1$ and different ionization models: PPT 
(solid curves), ADK molecular (dashed curves), ADK molecular + 
$\chi^{(5)}$ susceptibility (dotted curves) and the MPI-like 
approximation of Fig.\ \ref{fig1} (dash-dotted curves).}
\label{fig8}
\end{figure}

Reframed in the present context, several papers addressed the 
question of limiting self-focusing by nonlinear optical saturations, 
instead of relying on plasma generation alone 
\cite{Koprinkov:prl:84:3847,Koprinkov:prl:87:229401,Gaeta:prl:87:229401}. 
In fact, it appears that the weight of $\chi^{(5)}$ susceptibility depends 
on the actual pulse length, on the ratio of instantaneous over delayed Kerr 
responses and on the ionization sources, that influence the maximum 
intensity attained by the beam 
\cite{Akozbek:oc:191:353,Couairon:pra:68:015801,Vincotte:pra:70:061802}. 
This property can be seen from 
the zeroes of the nonlinear refraction index
\begin{equation}
\label{sat}
\Delta n = [(1-x_K) + x_K G(t)]|\psi|^2 - \epsilon |\psi|^4 - \rho,
\end{equation}
in which the 
Raman-delayed response has been introduced by 
changing the Kerr term $|\psi|^2$ of Eqs.\ (\ref{adim}) into 
$[(1-x_K) + x_K G(t)]|\psi|^2$. Here, the function $G(t)$ follows 
from computing the integral $\int h(t-t') |\psi(t')|^2 dt' \sim G(t) 
|\psi|^2$ over the initial pulse profile (\ref{adim_0}). High-order saturation 
prevails over ionization and induces a solitonlike dynamics at 
high enough values of $n_4$, before the occurrence of 
an electron plasma. Relevance of higher-order optical 
nonlinearities above all depends on the ionization rate: For example in air, if the laser intensity saturates above $10^{14}$ W/cm$^2$ by plasma 
generation alone, then even weak values of $n_4$ can soften this peak 
intensity. At lower intensities, quintic saturation has a 
more limited role. For instance, Fig.\ \ref{fig8} displays the peak 
intensities, beam radius (FWHM of the fluence distribution), maximum 
electron density and energy losses of a 70-fs, 0.5 mm-waisted 
unchirped Gaussian pulse propagating in parallel geometry in air. 
Limitation of the pulse growth depends on the ionization model employed.
One of them involves the ADK molecular rate without (dashed 
curve) and with (dotted curve) a weak quintic saturation ($n_4 = 2.5 
\times 10^{-33}$ cm$^4$/W$^2$). The measured intensity peak of $5 
\times 10^{13}$ W/cm$^2$ is refound in the latter configuration. 
$\chi^{(5)}$ susceptibility lowers the peak intensity and density, 
which stays close to $10^{16}$ cm$^{-3}$. Energy losses are 
weaker, which increases the self-guiding range.

Other high-order effects may soften the Kerr focusing of nonlinear waves. 
For instance, the Raman 
response function (\ref{Raman_123}) becomes all the weaker as the pulse 
is short, which affects the critical power for self-focusing like 
$P_{\rm cr} \rightarrow P_{\rm cr}/[1 - x_K + x_K G(t)]$ 
\cite{Ting:pop:12:056705}. Unlike long pulses, for which $G(t) 
\rightarrow 1$, the non-instantaneous nonlinearity reduces the 
effective value of the Kerr index and results in delaying the 
nonlinear focus from which a filament emerges 
\cite{Akozbek:oc:191:353,Chiron:epjd:6:383,Liu:oe:13:5750}. In connection, 
modulational instabilities have their growth rate decreasing as the 
ratio $x_K$ is augmented 
\cite{Wyller:pd:157:90}.

Besides, the cubic polarization $\sim E^3$ generates third-order 
harmonics as
\begin{equation}
E \rightarrow U_{\omega} \mbox{e}^{i k(\omega) z - i 
\omega t} + {\tilde U}_{3\omega} \mbox{e}^{i k(3 \omega) z - 3 i \omega 
t} + c.c.
\end{equation}
at $\omega = \omega_0$ \cite{Agrawal:NFO:01}. Two coupled 
equations can then be derived under the basic assumptions of 
slowly-varying envelopes and narrow-spectral bandwidths $\Delta 
\omega_j/\omega_j \ll 1$ ($j=\omega,3\omega$). The coupling is 
insured by cross-phase modulation (XPM) and four-wave mixing (FWM) 
induced by the cubic nonlinearity. By substituting ${\tilde 
U}_{3\omega} \rightarrow U_{3\omega} \mbox{e}^{i \Delta k z}$, where 
$\Delta k = 3 k(\omega) - k(3\omega)$ is the linear wavevector 
mismatch parameter, the propagation equations read like Eq.\ (\ref{adim}) 
where the function ${\cal F}$ for the two 
components $U_{\omega}$ and $U_{3 \omega}$ must be adapted as
\begin{subequations}
\label{TH}
\begin{align}
\mathcal{F}_{\omega} & \rightarrow \mathcal{F}_{\omega} + 
2 |\psi_{3 \omega}|^2 \psi_{\omega} + (\psi_{\omega}^{*})^2 \psi_{3 
\omega}, \\
\mathcal{F}_{3 \omega} & \rightarrow \mathcal{F}_{3 \omega} + 2 
|\psi_{\omega}|^2 \psi_{3 \omega} + \frac{\psi_{\omega}^3}{3} + 
(\frac{i}{\Delta v} \partial_t - \frac{\Delta k}{(4 z_0)^{-1}}) 
\psi_{3 \omega},
\end{align}
\end{subequations}
that includes the group-velocity mismatch responsible for temporal 
walk-off, $\Delta v = [v_g(3\omega)^{-1} - v_g(\omega)^{-1}]^{-1}$. 
In self-focusing regimes, the third-harmonic intensity usually 
contributes by a little percentage to the overall beam fluence 
\cite{Akozbek:prl:89:143901}. Despite the smallness of the TH field, this 
component may act as a saturable nonlinearity for the pump component 
when the term containing $|\Delta k|$ has an order of magnitude comparable with the FWM and 
XPM terms, $\psi_{3\omega} \sim \psi_{\omega}^3/3(4z_0 
|\Delta k| + 2 |\psi_{\omega}|^2)$, and all other contributions driving the TH component are neglected. Once inserted into the pump wave 
equation, this so-called ''cascading'' limit \cite{Buryak:ol:24:1859} 
introduces a saturable quintic nonlinearity that lowers the peak 
intensity and enhances the propagation over longer distances \cite{Berge:pre:71:016602}.

So far, we have been dealing with linearly-polarized waves. However, by writing the electric field as $\mathcal{E} = 
(\mathcal{E}_x,\mathcal{E}_y)$, each polarization component undergoes optical nonlinearities as
\begin{equation}
\label{TH3}
P_{\rm NL} \sim |\psi_j|^2 \psi_j + \frac{2}{3} |\psi_k|^2 \psi_j + 
\frac{1}{3} \psi_j^2 \psi_k^*,
\end{equation}
with $j,k=x,y$ ($j \neq k$). A 
linearly-polarized state is described by the single-component 
NLS equation with, e.g., $\psi_y = 0$. A circularly-polarized state corresponds 
to the configuration $\psi_y=\pm i\psi_x$. In between, 
elliptically-polarized states facilitate energy transfers through the 
phase-dependent terms $\psi_j^2 \psi_k^*/3$ from one orthogonal 
component to the other \cite{Schjoedteriksen:ol:26:78}. These energy 
transfers increase the power threshold for self-focusing 
\cite{Berge:pd:176:181}. For circular polarization, the collapse 
threshold for the total power $P = \int(|\psi_x|^2 + 
|\psi_y|^2)d{\vec r}$ is increased by a factor $3/2$ when passing 
from linearly- to circularly-polarized geometries. In addition, circular 
polarizations on femtosecond filaments self-channeling in the 
atmosphere produce stabler patterns than those issued from linear 
polarization \cite{Kolesik:pre:64:046607}. They 
moreover weaken the MPI efficiency and decrease ion signals by a couple
 of decades in the fluorescence of $N_2$ molecules collected from 
infrared pulses \cite{Petit:oc:175:323}.

Compared with plasma formation and chromatic dispersion, quintic saturation and polarization 
effects have a limited impact on the nonlinear 
dynamics. Therefore, we shall 
henceforth focus on the first two players mainly.

\subsection{Self-phase modulation and supercontinuum generation}

The Kerr effect creates spectral broadening through self-phase 
modulation (SPM). The solving for $i \partial_{z} \psi = - |\psi|^2 
\psi$ indeed yields the exact solution $\psi = \psi_0 \mbox{e}^{i  
|\psi_0|^2 z}$ [$\psi_0 \equiv \psi(z=0)$], which describes 
a self-induced phase modulation 
experienced by the optical field during its
propagation. This intensity-dependent phase shift is responsible for spectral broadening by 
virtue of the relation $\Delta \omega = - \partial_{t}
\mbox{arg}(\psi)$ \cite{Agrawal:NFO:01,Shen:PNO:84}. Because the frequency spectrum is expanded by the nonlinearity, SPM leads to supercontinuum generation and white-light emission, as 
the wave intensity strongly increases through the self-focusing 
process. With plasma generation, this phenomenon can be 
described by employing the following approximations. First, we assume 
that the 
MPI response behaves like a static density plateau [see, e.g., Fig.\ 
\ref{fig9}(f)], so that the ratio 
$\rho/|\psi|^2$ is either zero or close to unity with $\partial_z 
(\rho/|\psi|^2) \simeq 
0$. Second, omitting GVD and MPA for simplicity we retain 
self-steepening
to the detriment of space-time focusing, i.e., we neglect diffraction
in Kerr-dominated regimes. Phase and amplitude of solutions $\psi = 
A \mbox{e}^{i\varphi}$ 
to Eq.\ (\ref{adim_1}) satisfy $\varphi_z + \frac{A^2}{t_p \omega_0} 
\partial_t
    \varphi = A^2 - \rho$ and $A_z + \frac{3}{t_p \omega_0} A^2 
\partial_t A = 0$,
respectively. If we model the pulse initially 
located at $t = t_0$ by $A^2 = 
A_0^2/\mbox{cosh}(\tau)$, $\tau =
(t-t_0)/\tau_0$, and suppose that the amplitude does not change too 
much, then
\begin{subequations}
\begin{align}
    A^2 & \simeq A_0^2/\mbox{cosh}(\tau - 3 Q A^2/A_0^2), \\
\begin{split}
    \varphi & \simeq t_p \omega_0 \tau_0 
\int_{-\infty}^{\tau} \left(1 -
    \frac{\rho}{A^2}\right) d\tau' \\ 
 & \quad - t_p \omega_0\mbox{sinh}^{-1}[\mbox{sinh}(\tau) - Q],
\end{split}
\end{align}
\end{subequations}
where $Q \equiv z A_0^2/t_p \omega_0 \tau_0$ is linked to self-steepening. These expressions yield the spectral broadening 
\begin{equation}
    \label{22}
    \frac{\Delta \lambda}{\lambda_0} \simeq 1 -
    \frac{\rho}{A^2} - \left[1 + \frac{Q^2 - 2Q
    \mbox{sinh}(\tau)}{\mbox{cosh}^2(\tau)}\right]^{-1/2},
\end{equation}
valid under the basic assumption $\Delta \omega/\omega_0 = - \Delta \lambda/\lambda_0 \ll 1$. As long as $Q \ll 1$ and in the absence of MPI, 
variations in wavelength $\Delta \lambda/\lambda_0 \approx - Q
\mbox{sinh}(\tau)/\mbox{cosh}^2(\tau)$ starting with $t_0 = 0$ represent the
early symmetric broadening through SPM \cite{Yang:ol:9:510}. When MPI forms a 
defocusing plateau,
$\Delta \lambda/\lambda_0$ then becomes larger in the region where
$\rho = 0$ (non-defocused leading pulse), than when $\rho
\rightarrow A^2$ (defocused trail).
Consequently, as the beam reaches the first focus point $z_c$, the 
dominant 
part of the pulse is the front edge ($\rho = 0$) and MPI creates a 
primary redshift \cite{Lehner:pre:61:1996}. In the trail, full 
plasma coupling ($\rho/A^2 \rightarrow 1$) 
limits the spectral enlargement to the opposite side. However, 
self-steepening 
induced at increasing $Q$ inhibits the MPI redshifting 
and instead displaces more the spectrum to the blue
side. As a result, this creates an asymmetric spectral broadening
with a prominent blueshift $\Delta \lambda < 0$ for $\tau > 0$ 
\cite{Gaeta:prl:84:3582,Yang:ol:9:510,Ward:prl:90:053901,Berkovsky:pra:72:043821,Champeaux:pre:68:066603}.

\begin{figure}
\includegraphics[width=0.49\columnwidth]{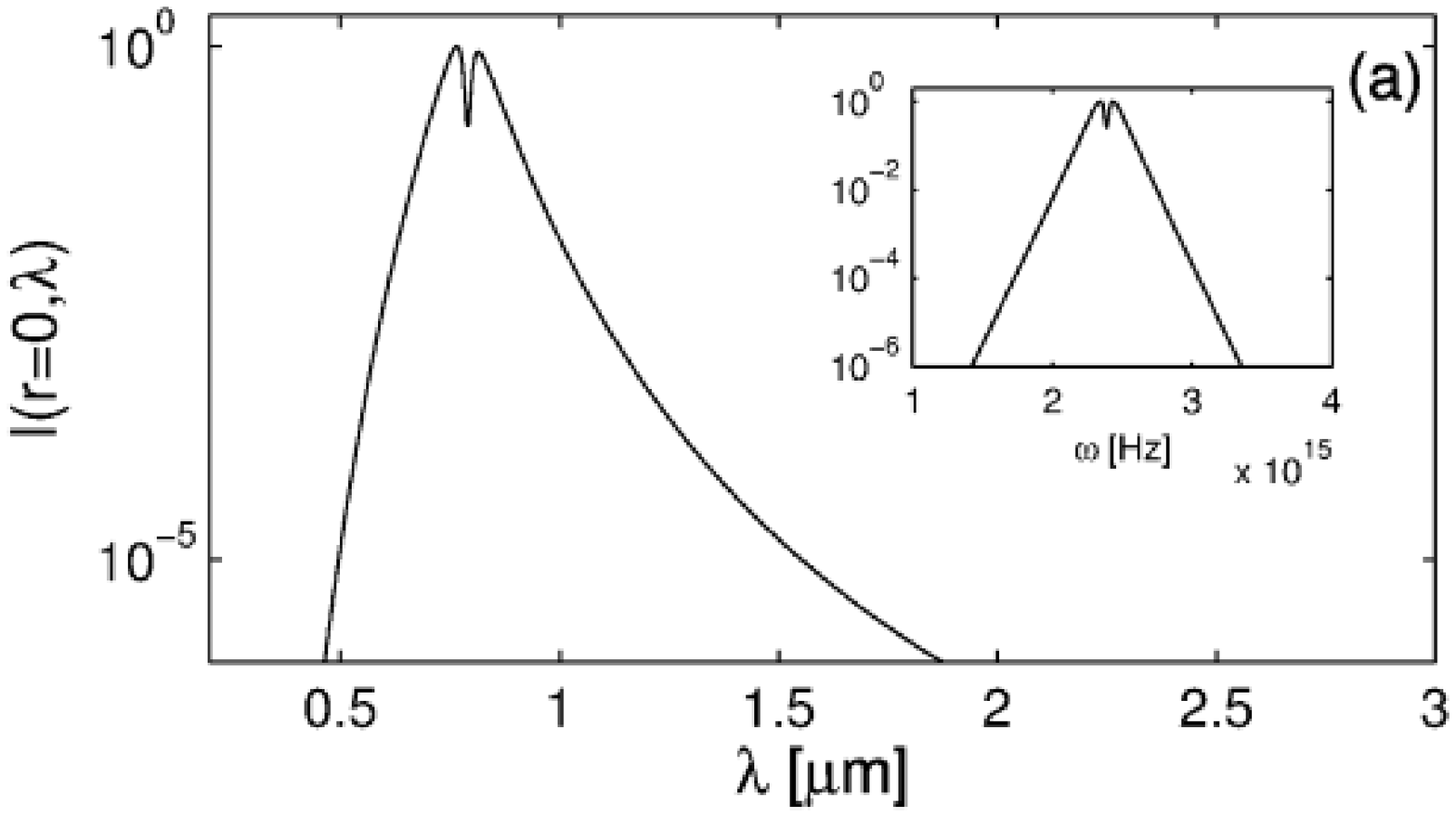}
\includegraphics[width=0.49\columnwidth]{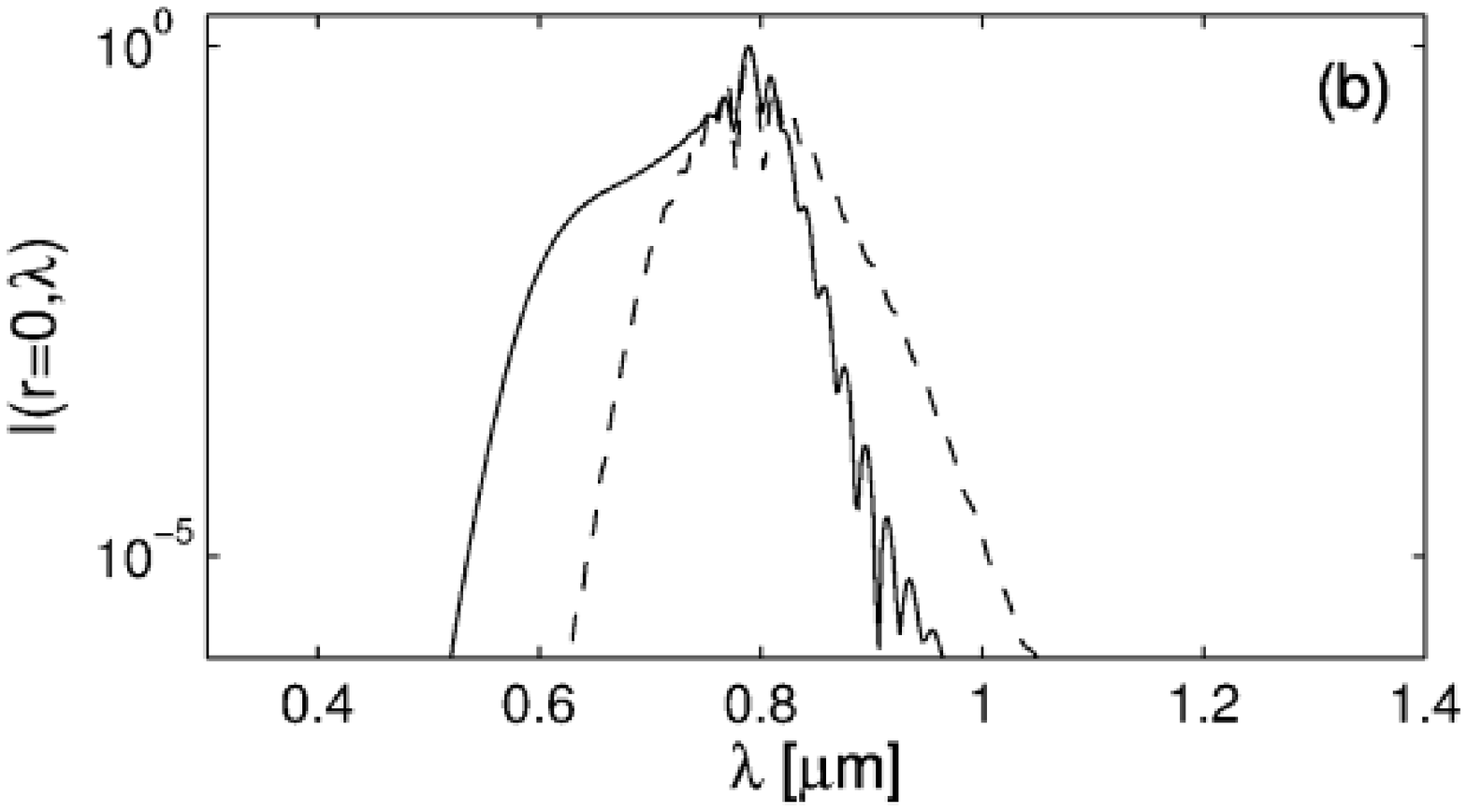}
\includegraphics[width=0.49\columnwidth]{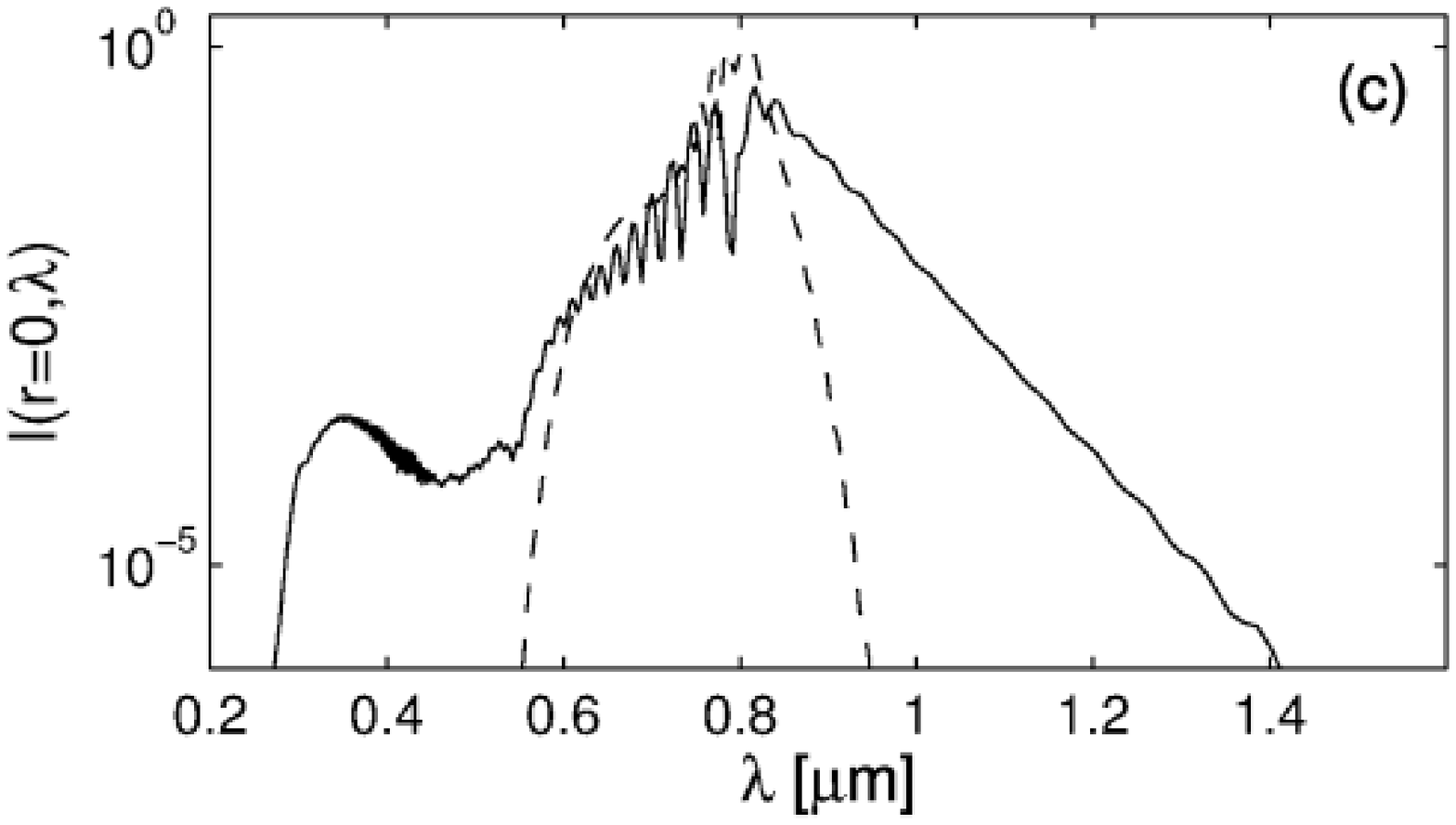}
\includegraphics[width=0.49\columnwidth]{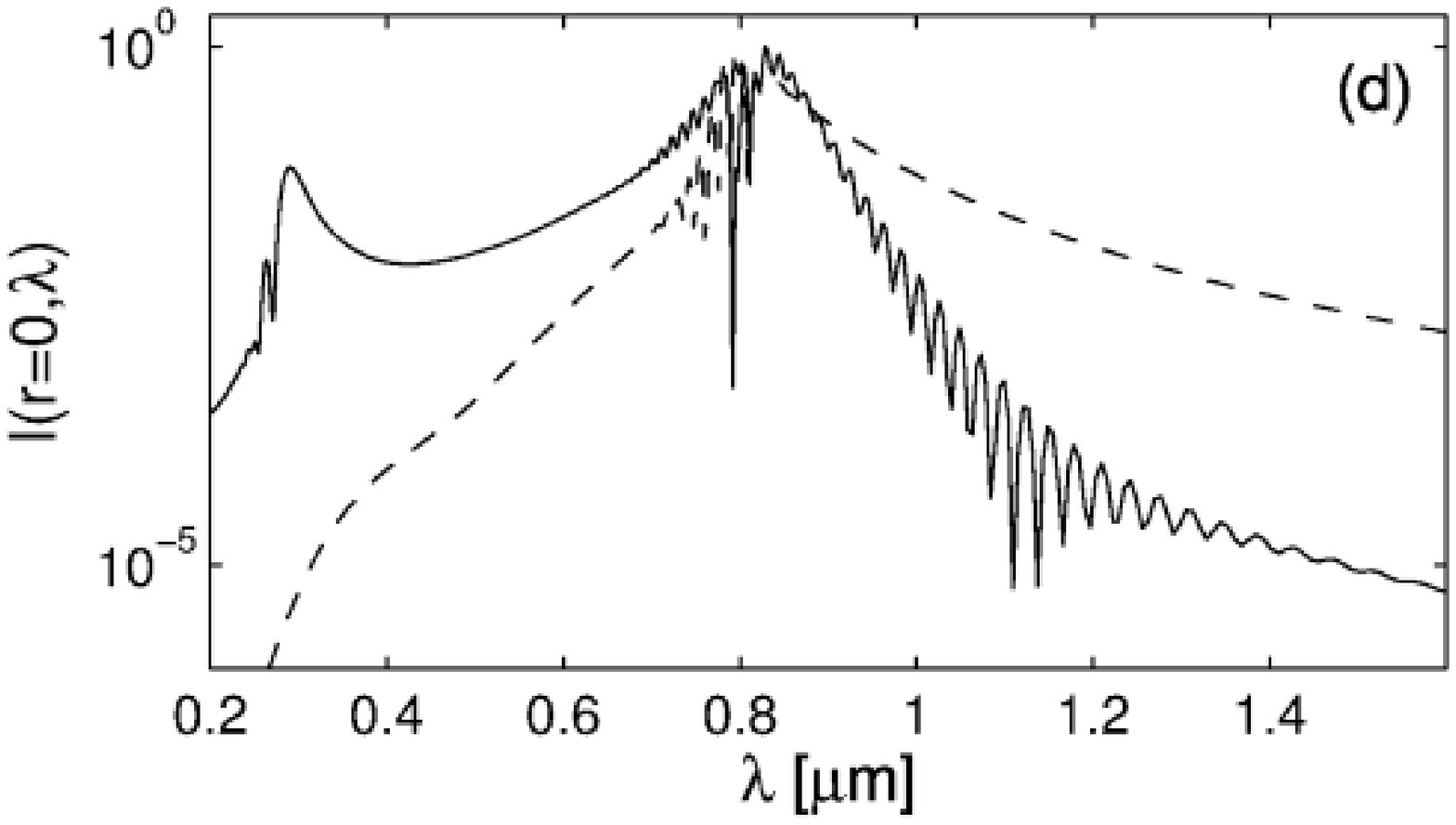}
\caption{Spectral broadening induced by SPM and steepening effects 
in fused silica: (a) with purely cubic nonlinearity (inset shows the spectrum in frequency for comparison), (b) with Kerr $+$ GVD only (dashed curve) and when adding space-time focusing and self-steepening (solid curve) for $P_{\rm in} = 3 P_{\rm cr}$, (c) same with plasma generation that remains inefficient at this power level (dashed curve, see Fig.\ \ref{fig7}), but strongly affects the spectrum when $P_{\rm in} = 6 P_{\rm cr}$ (solid curve). For comparison, 
\ref{fig10b}(d) shows supercontinuum with (solid) and without 
(dash-dotted) $T, T^{-1}$ in air for the initial data of Fig.\ \ref{fig9}.}
\label{fig10b}
\end{figure}

Figure \ref{fig10b} shows spectral modifications around 790 nm induced by SPM 
affected by space-time focusing and self-steepening in the presence 
(or not) of ionization in fused silica. The properties stressed by Eq.\ 
(\ref{22}) can be refound. Note the asymmetry in the SPM profile when the Fourier transform of the on-axis field intensity is expressed in wavelengths [Fig.\ \ref{fig10b}(a)], since $\Delta \omega = - 2 \pi c \Delta \lambda/\lambda^2$. For comparison, Fig.\ \ref{fig10b}(d) illustrates supercontinuum reached in air with or without pulse steepening at 800 nm. In all cases, we can observe a clear amplification of UV/blue wavelengths due to steepening terms. Because the spectral component rapidly decreases outside the central wavelength, the principal source of ionization (scaling as $I^K$ in MPI regime), however, remains given by that 
operating at $\lambda_0$.  

\subsection{Modulational instabilities: The route to multiple 
filamentation}

In the context of nonlinear optics, an intense wave propagating in a 
focusing Kerr medium can break up into several spots
from the inhomogeneities affecting its initial distribution. 
This phenomenon was first evidenced by self-focusing experiments in 
liquids \cite{Bespalov:jetp:3:307,Campillo:apl:23:628} and the 
resulting small-scale structures are 
usually named ''filaments''. Because it mostly concerns spatial distortions, 
this process of ''multifilamentation''
can be understood from Eqs.\ (\ref{adim}), where temporal 
variations are discarded. $(2+1)$-dimensional models freezing 
temporal dependencies were used to simulate the 
early filamentation stage of high-power beams in various transparent 
media, yielding rather good qualitative agreements with experimental 
measurements 
\cite{Kandidov:qe:29:911,Fibich:ol:29:1772,Dubietis:ol:29:1126,Dubietis:prl:92:253903,
Porras:prl:93:153902,Kandidov:apb:80:267,Fibich:oe:13:5897}. 
In plasma regimes, 2D models can also simulate complex multifilamentation patterns 
created from terawatt laser pulses
\cite{Berge:prl:92:225002,Skupin:pre:70:046602}. For instance, by 
assuming that MPI mainly counterbalances Kerr focusing at a dominant 
time slice $t = t_c(z)$, the field envelope can be decomposed as 
$\psi = {\tilde \psi}(x,y,z) \times \chi[t,t_c(z)]$ 
where the temporal distribution for the highest-intensity peak is 
modeled by the Gaussian $\chi[t,t_c(z)] = \mbox{e}^{- 
[t-t_c(z)]^2/\tau^2}$ with constant extent $\tau$. Plugging the 
above expression into Eqs.\ (\ref{adim}), computing the expression 
of $\rho$ and averaging the result over the entire 
time domain supplies the extended NLS equation for the spatial profile ${\tilde 
\psi}$:
\begin{subequations}
\label{2D_1}
\begin{align}
\label{2D_1a}
i \partial_z {\tilde \psi} & = - \nabla_{\perp}^2 {\tilde \psi} -
f(|{\tilde \psi}|^2) {\tilde \psi}\\
f(s) & = \alpha s - \frac{\epsilon s^2}{\sqrt{3}} - \sqrt{\frac{\pi}{8K}} \tau \Gamma s^K + i \frac{\nu s^{K-1}}{\sqrt{K}},
\end{align}
\end{subequations}
where the coefficient $\alpha$ averages the Raman delayed response.
This model does not formally depend on the longitudinal 
location of the time slice $t = t_c(z)$. The only arbitrariness is 
the choice of the peak duration $\tau$. Because MPI shortens pulses to mean duration reaching $1/10$ of their 
initial values, the value $\tau = 0.1$ was found to 
provide the best approximations of fluence 
patterns developed by $(3+1)$-dimensional fs pulses.

Modeling filament formation then requires a perturbation 
theory involving a steady-state solution, expressed as 
${\tilde \psi}_s({\vec r},z) = \phi({\vec r}) \mbox{e}^{i \lambda z}$. 
Defined in the limit of no dissipation ($\nu \rightarrow 0$), $\phi$ satisfies the differential equation
\begin{equation}
\label{MF_2}
- \lambda \phi + {\vec \nabla}_{\perp}^2 \phi + 
f(\phi^2) \phi = 0
\end{equation}
and $\lambda ={\rm const}$. Stability of $\phi$ is investigated 
from perturbations $v+iw$ with small real-valued components $(v,w)$ 
acting against this stationary mode. Linearizing Eqs.\ (\ref{2D_1}) 
yields the
eigenvalue problem \cite{Kuznetsov:pr:142:103}
\begin{subequations}
\label{MF_4}
\begin{align}
\partial_z v &= L_0 w,&
L_0 & = \lambda - {\vec \nabla}_{\perp}^2 - f(\phi^2), \\
- \partial_z w &= L_1 v,&
L_1 &= \lambda - {\vec \nabla}_{\perp}^2 - [f(\phi^2) + 2 f'(\phi^2) \phi^2],
\end{align}
\end{subequations}
where $L_0$ and $L_1$ are self-adjoint operators
with $f'(\phi^2) = \partial f(u)/\partial u|_{u = \phi^2}$. 
Combining Eqs.\ (\ref{MF_4}), we then obtain 
$\partial_z^2 v = - L_0 L_1 v$, from which different kinds of 
instabilities may be investigated.

{\it Modulational instabilities}: Originally proposed by Bespalov and 
Talanov
\cite{Bespalov:jetp:3:307}, the modulational instability (MI) 
theory involves oscillatory perturbations with an
exponential growth rate, that split the beam envelope
approximated by a background uniform solution. Perturbative modes are 
chosen 
as $v,w \sim \cos{(k_x x)} \cos{(k_y y)} \mbox{e}^{\gamma z}$ and 
they apply 
to a plane wave $\phi$ which satisfies ${\vec \nabla}_{\perp}^2 
\phi = 0$ and $\lambda = f(\phi^2)$. The growth rate $\gamma$ is 
then given by
\begin{equation}
\label{MF_6}
\gamma^2 = k_{\perp}^2(2 A - k_{\perp}^2),\quad A 
\equiv u f'(u)|_{u = \phi^2}.
\end{equation}
Plane waves are unstable with $\gamma^2 > 0$ in the range $0 < 
k_{\perp} < 
\sqrt{2A}$ and the maximum growth rate $\gamma_{\rm max} = A$ is 
attained 
for $k_{\perp} = k_{\rm max} = \sqrt{A}$.
This instability promotes the beam breakup into arrays of
small-scale filaments periodically distributed in the diffraction
plane with the transversal spacing $\lambda_{\rm mod} \simeq 2 
\pi/k_{\rm max}$ and longitudinal length $\sim \gamma_{\rm 
max}^{-1}$. The number 
of filaments is close to the
ratio $P_{\rm in}/P_{\rm fil}$, where $P_{\rm fil}$ is the power enclosed in
one filament. Considering each filament with
radial symmetry, the evaluation $P_{\rm fil} \simeq 2 \pi
\int_0^{\lambda_{\rm mod}/2} r |\phi|^2 dr \simeq 2.65 P_c$ holds for 
unsaturated Kerr media [$f(s) = s$].

Because they constitute rough representations 
of physical beams, plane waves can be replaced by the soliton modes of the NLS
equation (\ref{MF_2}). The resulting instability appears when a 
soliton $\phi$ is perturbed by oscillatory modulations
developing along one transverse axis. 
Perturbations are, e.g., local in $x$ and they
promote the formation of bunches periodically distributed over the 
$y$ axis.
The operators $L_0$ and $L_1$ in Eqs.\ (\ref{MF_4}) are
transformed as $L_0 = \lambda - \partial_x^2 - f(\phi^2) + k_y^2$ and 
$L_1 =
L_0 - 2 f'(\phi^2)\phi^2$. Numerical computations 
are then often required for solving the spectral problem (\ref{MF_4}) 
\cite{Rypdal:ps:40:192,Zakharov:spjetp:38:494,Akhmediev:ol:17:393}.

{\it Azimuthal instabilities}: For broad beams, 
self-focusing takes place as a regular distribution of dots located upon 
ring diffraction patterns \cite{Feit:josab:5:633}. To model this
instability, the Laplacian in Eq.\ (\ref{2D_1a}) must be rewritten as 
${\vec \nabla}_{\perp}^2 = r^{-1}
\partial_r r \partial_r + r^{-2} \partial_{\theta}^2$, where 
${\theta}$ denotes the azimuthal angle. Within a first approximation, 
unstable modes $v,w \sim 
\mbox{cos}(M\theta)
\mbox{e}^{\gamma_M z}$ with azimuthal index
number $M$ break up a spatial ring, which is modeled by a uniform 
background solution
$\phi$ lying on a circular path with length $s = {\bar r} \theta$, 
and mean radius ${\bar r}$. Eqs.\ (\ref{MF_4}) 
then yield
\cite{Atai:pra:49:3170,Sotocrespo:pra:45:3168}
\begin{equation}
\label{MF_7}
\gamma_M^2 = \left(\frac{M}{{\bar r}}\right)^2 \left(2 A - 
\frac{M^2}{{\bar r}^2}\right),
\end{equation}
and the maximum number of modulations on the ring is provided by the 
integer part of $M_{\rm max} = {\bar r} \sqrt{A}$.

In this regards, solitary waveforms having a ring-shaped distribution 
may also exhibit a phase 
singularity with an integer number of windings, $m$ (topological 
charge). Such structures are termed ''optical vortices''. They convey 
a constant orbital angular momentum 
\cite{Kruglov:pl:111A:401,Kruglov:jmo:39:2277} and are  
experimentally designed by means of phase masks and holographic 
techniques 
\cite{Desyatnikov:po:47:291,Tikhonenko:prl:76:2698,Petrov:ol:23:1444}. 
For cubic media, they undergo azimuthal MI, which make vortex solitons decay into $M_{\rm 
max} \sim 2|m|+1$ filaments 
\cite{Vincotte:prl:95:193901,Vuong:prl:96:133901}. For cubic-quintic 
nonlinearities, optical vortices may be linearly stable for $|m| = 1$
\cite{Quigora-Teixeiro:josab:14:2004,Berezhiani:pre:64:057601,Towers:pla:288:292,Skarka:pla:319:317,Desyatnikov:pre:61:3107}, but they decay into $M_{\rm max} \sim 2|m|$ filaments otherwise
\cite{Skryabin:pre:58:1252,Firth:prl:79:2450}. The orbital motion 
confers robustness to vortex solitons, which can propagate powerful beams beyond several tens of Rayleigh distances 
\cite{Michinel:jobqso:3:314}, before they break up into filaments.

\begin{figure}
\includegraphics[width=\columnwidth]{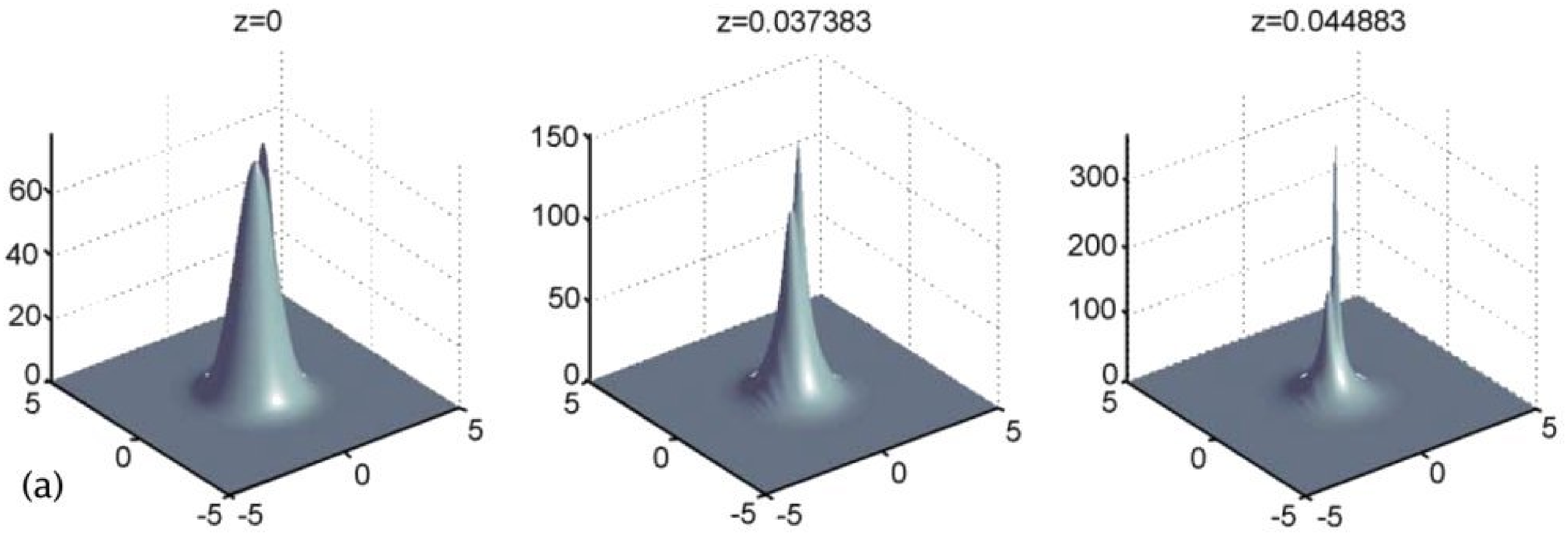}
\includegraphics[width=\columnwidth]{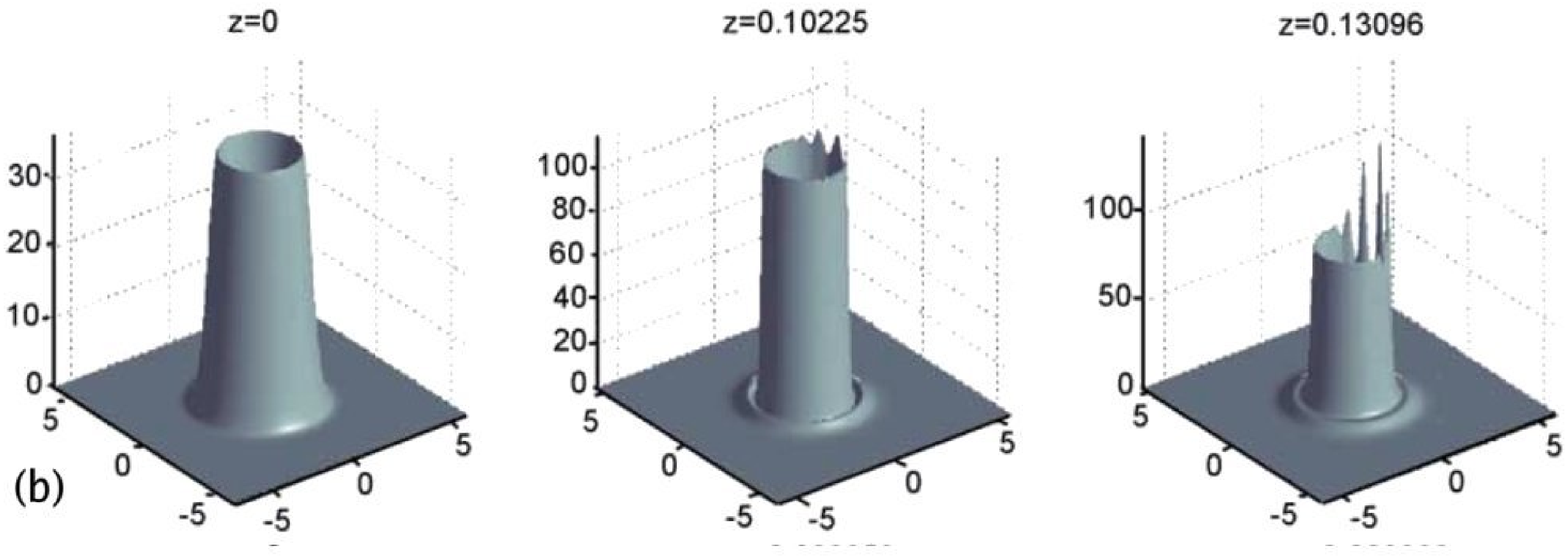}
\includegraphics[width=\columnwidth]{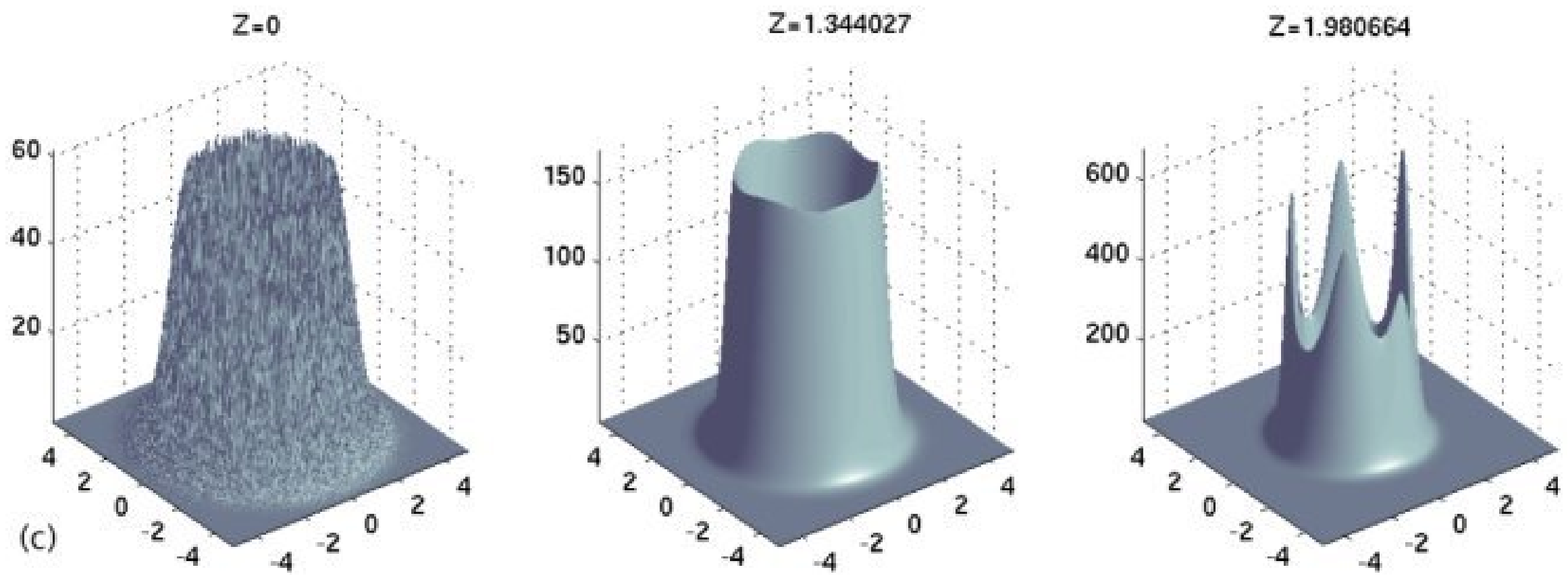}
\caption{Multifilamentation patterns in Kerr focusing regime 
at powers above critical ($P_{\rm in} \le 30 P_{\rm cr}$) for (a) a Gaussian beam ($x_0 = 0.35,y_0 = 
0,\epsilon = 0.1$) and (b) an annular ring with flat phase ($x_0 = 1.7,y_0 = 0,\epsilon = 
0.01$). These two beams undergo MI from a local defect with relative amplitude $\epsilon$ 
located at the coordinates $(x_0,y_0)$. (c) Vortex-shaped beams 
with charge $|m| = 2$ initially perturbed by a $10 \%$ random noise.}
\label{fig11}
\end{figure}

Figure \ref{fig11} shows the multifilamentation of (a) a Gaussian beam, (b) a flat-phase, ring-shaped waveform and (c) an $m = 2$ vortex breaking up into $2|m|+1$ filaments in a cubic medium. MI is triggered either from an initial defect or from random noise \cite{Berge:pd:176:181}. Note the robustness of the 
Gaussian profile, whose shape definitively differs from that of a 
plane wave.

Modulational instabilities rather affect the early Kerr stage in 
the beam propagation. At later stages, the resulting filaments become fully 
nonlinear and they may interact mutually. From the interplay between 
diffraction and nonlinearity, two filaments can fuse, whenever their separation distance is below a critical 
value, function of their individual powers 
\cite{Berge:josab:14:2550,Mckinstrie:prl:61:2929}. With no saturation, 
each filament whose 
power is above critical creates its own attractor, at which it freely 
collapses. By including 
saturation, filaments 
with powers above critical are able to coalesce into an intense 
central 
lobe and develop various patterns 
\cite{Konno:ps:20:382}. Without any control, 
multifilamentation happens as an harmful
instability destroying the homogeneity of the energy distribution. 
It has detrimental consequences on the quality of the beam shape.

\section{Universal features of femtosecond filamentation\label{sec4}}

For an efficient propagation, the beam intensity and electron density must both remain below their so-called 
''laser-induced breakdown'' (LIB) limits. LIB means the 
breakdown of the beam that results in damage or tremendous energy 
absorption in dielectrics as well as in liquids 
\cite{Kennedy:ieeejqe:31:2241}. It involves three major 
processes, namely, (i) ionization, (ii) heating of the ''free'' electrons by the radiation, 
and (iii) transfer of plasma energy to the medium. Once generated, 
the electron plasma absorbs and scatters the remaining pulse energy. 
Assuming free 
electrons seeded by avalanche ionization, $\partial_t \rho 
\simeq (\sigma/U_i) \rho I$, an incident pulse launched in tightly focused geometry will be significantly attenuated by energy absorption at density levels above a critical value, $\rho_{\rm cr}$. An estimate of the LIB intensity then follows as $I_{\rm LIB} \sim (U_i/\sigma t_p) 
\mbox{ln}(\rho_{\rm cr})$ \cite{Feng:ol:20:1958}. For 
subpicosecond pulses, multiphoton processes mostly provide the 
dominant breakdown mechanism, such that the estimate $I_{\rm LIB} 
\sim (\rho_{\rm cr}/\rho_{\rm nt}\sigma_K t_p)^{1/K}$ applies 
\cite{Feng:ieeejqe:33:127}. Ranges of values for $\rho_{\rm cr}$ are often close to $10 \%$ of $\rho_{\rm nt}$ (e.g., $\rho_{\rm cr} = 10^{21}$ cm$^{-3}$ in silica or water) when using femtosecond pulses \cite{Lenzner:prl:80:4076,Noack:ieeejqe:35:1156}.

In what follows, characteristics of ultrashort pulse propagation are mostly
depicted for peak electron densities less than the 
LIB threshold. Density levels maintain the pulse in self-channeled state without significant conversion of beam energy into the plasma.

\subsection{Radial self-focusing and temporal splittings\label{sec4a}}

The self-channeling mechanism relies on the balance between Kerr 
focusing and plasma generation. The most illuminating example is given by 
''femtosecond filaments'' in air, as presented in Fig.\ \ref{fig0}. Atmospheric filaments form microstructures exhibiting $\sim 100-150\,\mu$m in FWHM diameter and $\sim 1$ mJ in energy at infrared wavelengths (800 nm). They are accompanied by a strong SPM-induced spectral 
broadening and cover several tens of meters, whenever the input pulse power exceeds $\sim 10$ GW \cite{Braun:ol:20:73,Nibbering:ol:21:62,Brodeur:ol:22:304}. Propagation distances 
in excess of 200 m were even reported for 60 
fs, 795 nm pulses with 4 mm diameter, traveling thus upon a few Rayleigh distances $z_0$. 
These experiments revealed, after a possible short 
stage of multiple filamentation, the emergence of one robust filament 
coupled to electron densities of a few 
times $10^{16}$ cm$^{-3}$ \cite{LaFontaine:pop:6:1615}. Further 
measurements of infrared filaments specified a peak intensity of 
about $5 \times 10^{13}$ W/cm$^2$ \cite{Kasparian:apb:71:877}. Similar behaviors involving slightly higher peak intensities were also reported in ${\rm N}_2$ molecular gases at different pressures \cite{Becker:apb:73:287}.

The ''moving-focus'' model was first
revisited to yield qualitative explanations of this phenomenon 
\cite{Brodeur:ol:22:304,Chin:jnop:8:121}. Though pleasant and rather 
intuitive, this model, however, failed at describing the 
self-guiding of light that persists beyond the linear focus of 
convergent beams \cite{Lange:ol:23:120}. The appropriate scenario for 
long distance propagation, called ''dynamical spatial replenishment'', 
was specified by numerical simulations. During the focusing 
stage, the beam generates a narrow plasma that strongly defocuses the 
trailing part of the pulse and creates one leading peak. Once the 
intensity decreases enough (via, e.g., MPA), plasma generation turns 
off. The back pulse can then refocus, which produces a two-spiked 
temporal profile \cite{Mlejnek:ol:23:382,Mlejnek:pre:58:4903}. In space, plasma defocuses only the 
inner part of the beam intensity, so that the trailing edge decays 
into spatial rings \cite{Kandidov:qe:24:905}. With increasing 
propagation distances and as the front pulse intensity decreases, the 
spatial rings coalesce under Kerr compression and allow refocusing of 
the trail. Although the beam radius may look ''quasi-static'', the 
temporal and spatial distributions of the pulse strongly fluctuate (see, e.g., Figs.\ \ref{fig9} and \ref{fig8}). At high enough powers, self-channeling is supported by several focusing/defocusing cycles 
triggered from a dynamical interplay between multiple, stringent peaks alternating in the temporal pulse profile. ''Pushed'' by steepening 
effects, these peaks are responsible for the formation of optical 
shocks in the medium. This complex dynamics was reported by several authors, dealing with pulse propagation either in air 
\cite{Berge:pop:7:210,Chiron:epjd:6:383,Couairon:josab:19:1117,Akozbek:oc:191:353}, 
in argon cells \cite{Nurhuda:pra:66:023811,Champeaux:pre:68:066603}, 
or even in fused silica \cite{Tzortzakis:prl:87:213902}.

Because the fundamental scenario for self-channeling is that aforementioned, 
estimates for peak intensities ($I_{\rm max}$), electron densities ($\rho_{\rm max}$) and filament 
radius ($L_{\rm min}$) can be deduced from equating diffraction, Kerr and ionization 
responses. This yields the simple relations
\begin{subequations}
\label{estimate13}
\begin{gather}
\label{estimate1}
I_{\rm max} \simeq \frac{\rho_{\rm max}}{2 \rho_c n_0 {\bar 
n}_2}, \quad
\rho_{\rm max} \simeq t_p \rho_{\rm nt} W(I_{\rm max}), \\
\label{estimate3}
L_{\rm min} \sim \pi (2 k_0^2 {\bar n}_2 I_{\rm max}/n_0)^{-1/2},
\end{gather}
\end{subequations}
where
\begin{equation}
\label{n2bar}
{\bar n}_2 = n_2 (1 - x_K) + n_2 x_K \mbox{max}_t G(t),
\end{equation}
represents the effective Kerr index when the Raman-delayed 
response is maximum over the initial pulse profile and $W(I_{\rm 
max})$ reduces to $W(I_{\rm max}) = \sigma_K I_{\rm max}^K$ in MPI 
regime. Some parameter values for different material used throughout this review have been indicated in Table I. From these values the above relations supply filament diameters of the order of $150 \mu$m in air at 800 nm and $10-20$ $\mu$m in dense transparent media (dielectrics, water), which agrees with current numerical data. Note that for laser wavelengths about $1 \mu$m, the requirement (\ref{basic}) is always fulfilled. As inferred from Table I, the critical power for self-focusing exceeds the GW level in gases, but remains limited to a few MW in dense materials.

\begin{table}
\caption{Parameter values for dioxygen molecules (air) at 800 nm ($U_i = 12.1$ eV, $\rho_{\rm nt} = 5.4 \times 10^{18}$ cm$^{-3}$), fused silica at  $\lambda_0 = 790$ nm and 1550 nm ($U_i = 7.8$~eV, $\rho_{\rm nt} = 2.1 \times 10^{22}$~cm$^{-3}$), and water at $527$ nm ($U_i = 7$~eV, $\rho_{\rm nt} = 3.32 \times 10^{22}$~cm$^{-3}$).}
\vspace{1mm}

\begin{tabular}{|r|r|r|r|r|}
\hline $\lambda_0$ [nm] & $n_2$ [cm$^2$/W] & 
$k''$ [fs/cm$^{2}$] & $\sigma_K$ 
[s$^{-1}$cm$^{2K}$/W$^K$] & K \tabularnewline
\hline 800/O$_2$ & 4 $\times 10^{-19}$ & 0.2 & 2.9 $\times 10^{-99}$ & 8 \tabularnewline
\hline 248/O$_2$ & 8 $\times 10^{-19}$ & 1.21 & 2 $\times 10^{-28}$ & 3 \tabularnewline
\hline 790/SiO$_2$ & 3.2 $\times 10^{-16}$ & 370 & $6.8 \times 10^{-56}$ & 5 \tabularnewline
\hline 1550/SiO$_2$ & 2.2 $\times 10^{-16}$ & -280 & $1.5 \times 10^{-119}$ & 10 \tabularnewline
\hline 527/H$_2$O & 2.7 $\times 10^{-16}$ & 500 & 1.07 $\times 10^{-28}$ & 3 \tabularnewline
\hline
\end{tabular}
\end{table}

Plasma generation is mostly expected to stop the divergence of the beam caused by self-focusing, except for specific values of the parameters $P_{\rm in}/P_{\rm cr}$ and $\delta \sim k''$ that privilege collapse arrest by dispersion (see Fig.\ \ref{fig7}). For $\lambda_0 = 800$ nm, nonlinear fluorescence techniques \cite{Liu:pra:72:053817} recently revealed that small-scale filaments created in water at powers lower than $8 P_{\rm cr}$ did not need plasma saturation. With 38-fs, 110-$\mu$m pulses and $k'' \simeq 560$ 
fs$^2$/cm, the normalized dispersion length is $\delta \sim 4.9$, for which the plasma response may indeed be inhibited, following Fig.\ \ref{fig7}. Pulses are then subject to space-frequency coupling from the interplay among SPM, 
dispersion and phase mismatching, that produces conical emission (see 
below). The energy loss caused by the conical wave depends on the 
energy contained in the colored part of the spectrum. Liu {\it et al.} \cite{Liu:pra:72:053817} suggests that if the phase matching width of light waves whose frequencies experience self-focusing is small enough compared with the spectral width of the on-axis laser pulse, the spectral energy going into the self-focusing components is small and the energy loss caused by conical emission becomes large. This phase matching width is all the smaller as GVD is large and the spread of energy can stop the Kerr focusing without the help of plasma defocusing at powers moderately above critical.

\begin{figure}
\includegraphics[width=\columnwidth]{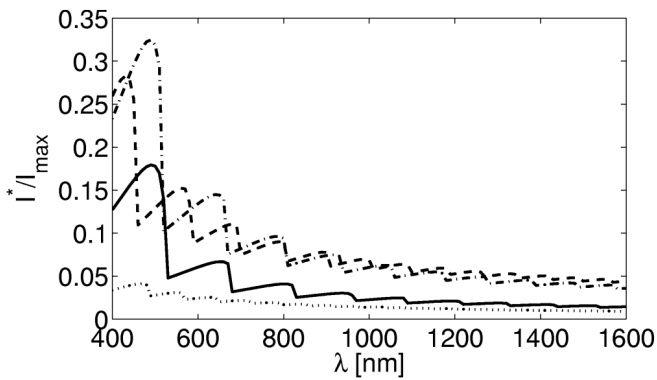}
\caption{$I^*/I_{\rm max}$ versus $\lambda_0$ for different pulse durations 
$t_p$ in air (70 fs, dotted curve), in fused silica (20 fs, dashed 
curve), and in water (170 fs, solid curve; 50 fs, dash-dotted curve).}
\label{fig10}
\end{figure}

Nonetheless, there exist 
different scenarios having attracted attention in the past few years. 
For instance, research groups led by P. Di Trapani and A. Dubietis 
\cite{Porras:prl:93:153902,Dubietis:prl:92:253903,Dubietis:ol:28:1269} 
emphasized the apparent absence of free electron emission in the 
self-guiding of 527 nm, 200 fs pulses with $\sim 100$ $\mu$m waist 
and powers up to 10 $P_{\rm cr}$, propagating in water in loosely 
focused geometry. Numerical simulations outlined the filament 
reshaping into a nonlinear $X$ (biconical) wave, driven by the 
combination between linear diffraction, normal GVD, self-focusing and 
nonlinear losses \cite{Kolesik:prl:89:283902}. Whereas GVD cannot 
halt the beam collapse even in these configurations ($\delta$ is too small), normal 
dispersion plays an important role by dispersing the most intense time slices of the pulse. Instead of an extensive 
production of free electrons, the nonlinear losses (MPA) inherent to 
this process moreover dominate at laser wavelengths as low as 527 nm. From Eqs.\ (\ref{estimate1}), it is always possible to compute numerically 
the ratio $I^* / I_{\rm max}$ 
that dictates the MPA efficiency [$\sim U_i \rho_{\rm nt} W(I)/I$] compared 
with plasma defocusing [$\sim i k_0 \rho/\rho_c$] in Eq.\ (\ref{finalUPPE1}). This ratio involves the quantity
\begin{equation}
\label{Istar}
I^* = n_0^2 \rho_c U_i/(k_0 t_p),
\end{equation}
that scales like $n_0/(\lambda_0 t_p)$. The resulting 
curve is displayed
in Fig.\ \ref{fig10} for fused silica, 
water and air at different pulse durations. Keldysh 
ionization rate for crystals is used for dense media, while the PPT rate is employed for air (O$_2$ molecules). This figure shows that the percentage of 
MPA over the emission of free electrons augments all the 
more as the laser wavelength is low and the pulse duration is short. 
In air, MPA becomes negligible compared with MPI. For dense materials, Fig.\ \ref{fig10} 
points out to a sudden change in the 
influence of MPA whenever $\lambda_0 < 600$ nm. This mainly follows 
from the dependences over $\lambda_0$ of both the critical plasma 
density (\ref{rhocrit}) and the photo-ionization rate $W(I)$. As a result, self-focusing is stopped at lower intensities and the 
free electron density is maintained at lower levels 
\cite{Skupin:pra:74:043813}.

\subsection{Robustness and multifilamentation}

Femtosecond filaments are remarkably robust, in the sense that they 
can propagate over several meters and even reform after hitting an 
obstacle. In the atmosphere, aerosol particles, like water 
droplets or dust, may have dimensions comparable with the filament diameter, which could seriously harm the 
dynamical balance required to propagate filaments. However, even opaque droplets as large as 100 $\mu$m in diameter do not 
block filamentation. More intriguing, energy losses are limited to $15 \%$ of the filament energy \cite{Courvoisier:apl:83:213}. The filament rapidly self-heals and 
seems unaffected by the droplet, so that an energy balance between 
the filament core and the unfocused part of the beam was conjectured 
to explain the rebuilding of the pulse. Unfocused parts of the beam 
refer to the ''photon bath'', i.e., the 
low-amplitude background surrounding the filament core.

\begin{figure}
\includegraphics[width=\columnwidth]{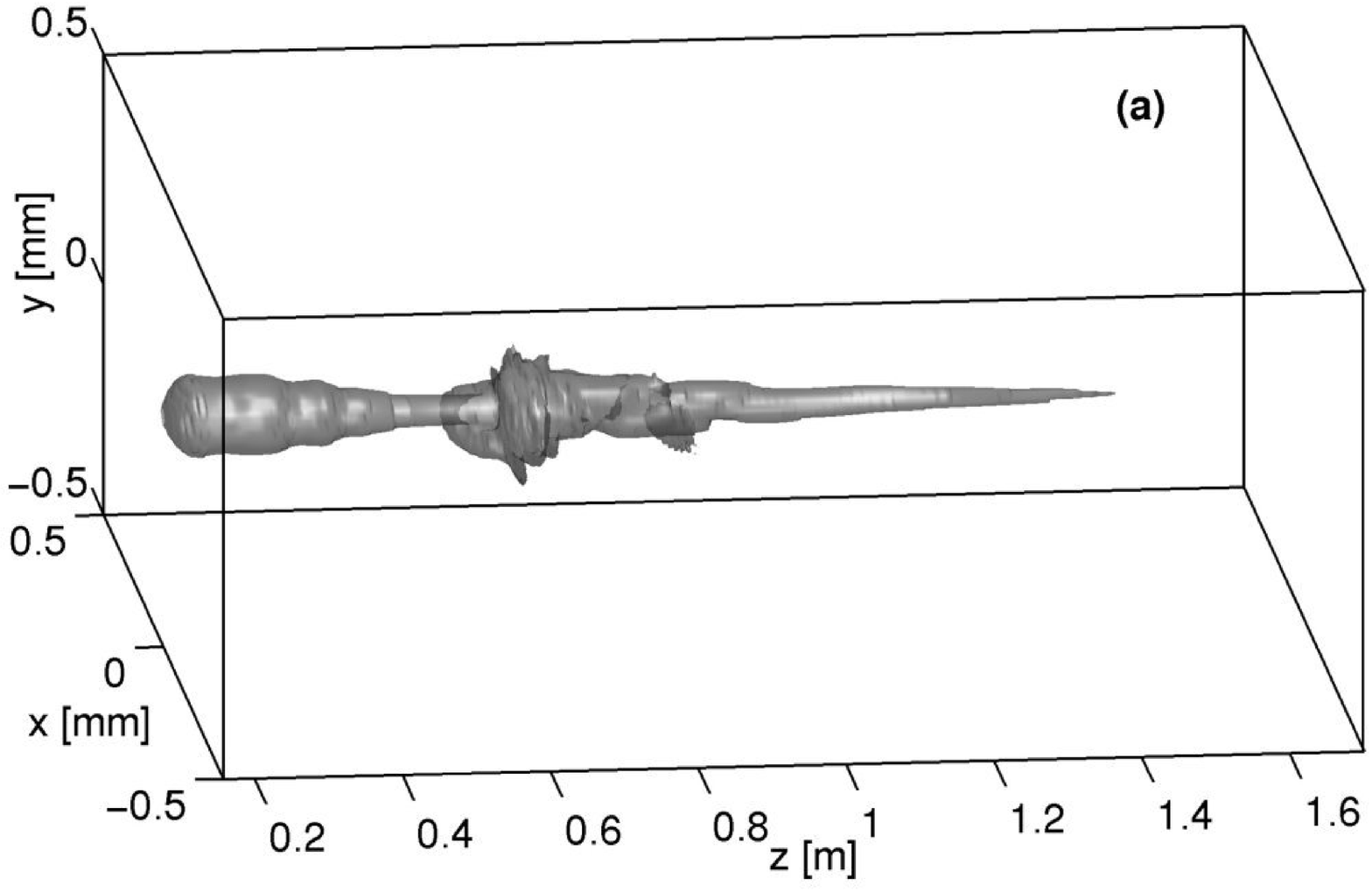}
\includegraphics[width=\columnwidth]{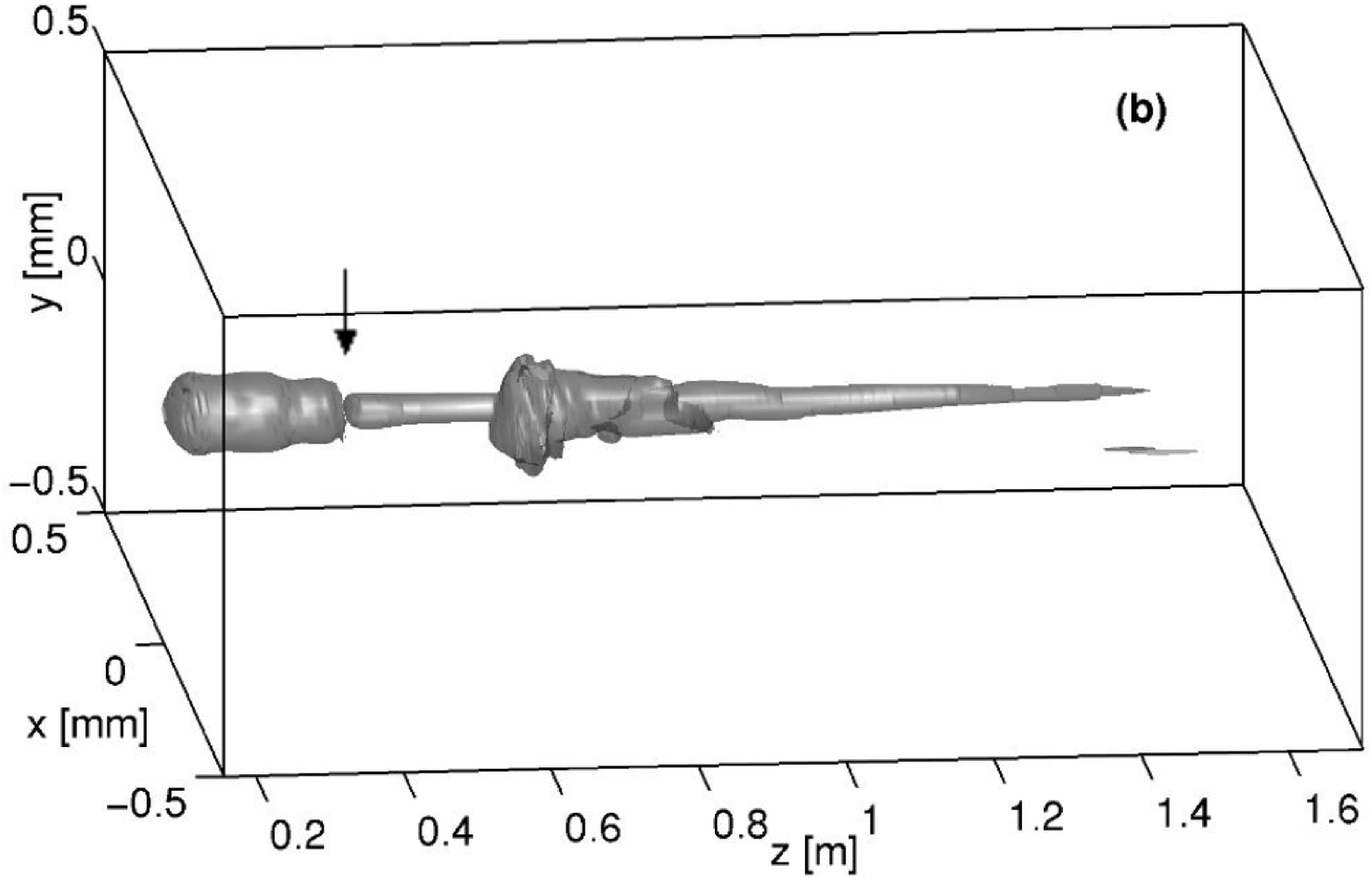}
\includegraphics[width=\columnwidth]{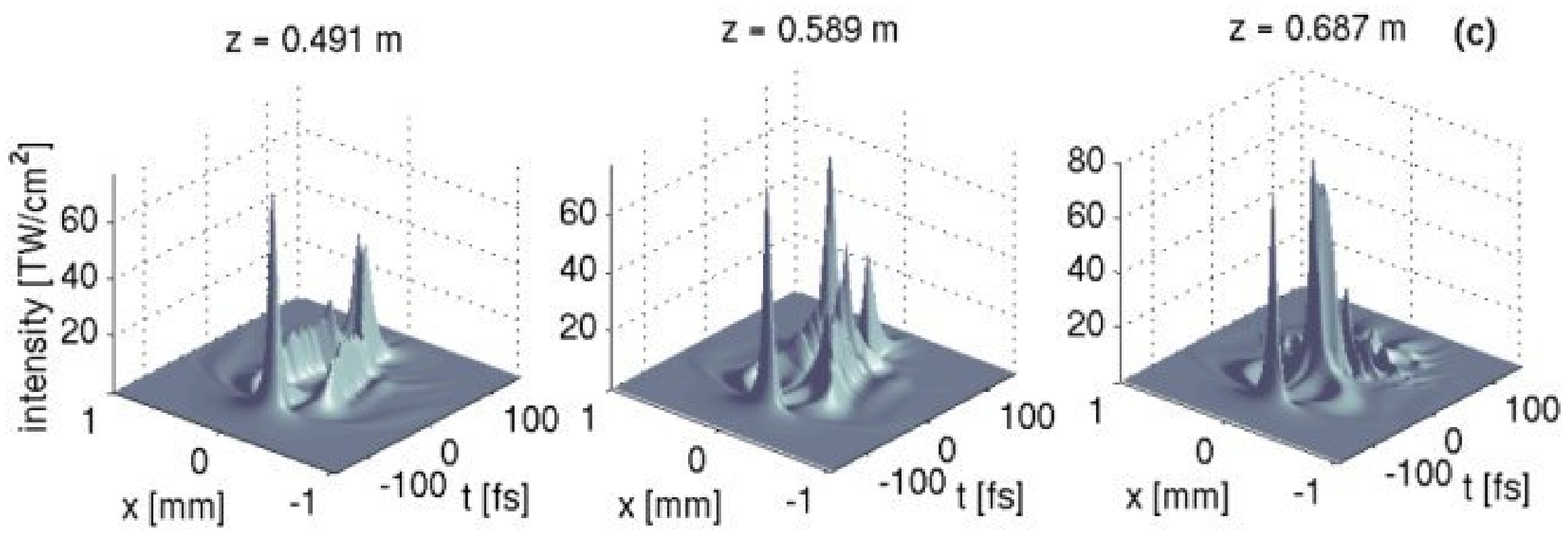}
\caption{(a,b) 3D surface plots 
of plasma strings in cases of (a) free propagation or (b) an impact with a 95 $\mu$m large 
droplet located at $z = 0.4$ m (see arrow), computed from $(3+1)$-dimensional numerical simulations. Note the convected components at $z \sim 0.65$ m, 
corresponding to the second focusing sequence when the trail peak 
takes over the front pulse and develops azimuthal instabilities in 
space. (c) Temporal profiles occupying an active zone of $\sim 300$ $\mu$m in diameter.}
\label{fig13}
\end{figure}

Numerical simulations constrained to the radial symmetry reproduced 
the smallness of the immediate energy losses in air and the interaction 
dynamics was similar whatever the pulse self-channels in the first 
(leading edge) or in the second (trailing edge) focusing stage 
\cite{Kolesik:ol:29:590}. From water experiments 
\cite{Dubietis:ol:29:2893}, analogous conclusions were drawn, which 
insisted more on the linear contribution from the outer conical waves 
to the filament reconstruction. The ''active zone'' of the photon 
bath must, however, keep time slices with power above critical to 
re-localize the beam at center. The filament must 
also be somehow delocalized in space, in order to justify the 
smallness of energy losses during the collision. Answering 
this point, simulations revealed that 
the breakup of rotational symmetry of pulsed beams can push
the focusing components, still untouched by the plasma, out of axis. 
The defocused time slices diffract to the low-intensity background, 
but the focusing ones exit out of center 
\cite{Skupin:ol:27:1812,Skupin:oqe:35:573}. A femtosecond filament 
thus consists of a ''spatially-extended'' structure, whose most 
intense components can move within a zone of several hundreds of 
$\mu$m around the origin. Through a collision, just a few 
components are involved, which explains the smallness 
of the losses. Three-dimensional simulations enclosing the filament inside a 
circular disk of 300 $\mu$m in diameter clearly showed that over the 
2 cm of self-healing the beam components outside this zone play no 
significant role in the filament reconstruction 
\cite{Skupin:prl:93:023901}. Some of these results are summarized in 
Fig.\ \ref{fig13}. Recent experiments using pinholes to block the 
photon bath confirmed these theoretical predictions. The energy 
reservoir actively feeding the filament core was measured within a 
diameter of $220-440\,\mu$m and was found to contain up to $50 \%$ of 
the pulse energy \cite{Liu:ol:30:2602}.

Whereas a single filament appears as a robust object, broad 
pulses having many critical powers become subject to modulational 
instability. The background energy reservoir favors energy 
exchanges between the different optical cells, that focus under Kerr 
compression and defocus by plasma generation. A sea of collapsing 
spots is thus nucleated, forming an 
''optically turbulent femtosecond light guide'' along which nonlinear 
dissipation consumes small energy per filament 
\cite{Mlejnek:prl:83:2938}. 
Inside the bundle, the filaments are able to merge and 
relaunch recurrent collapse events at further distances. 
They gather into long-range clusters composed of secondary cells that 
occur around the primary filaments created from the initial beam defects 
\cite{Berge:prl:92:225002,Skupin:pre:70:046602}. For broad pulses ($w_0 > 1$ cm) enclosing enough power (e.g., more than 100 $P_{\rm cr}$), the self-focusing distance is close to 
the MI filamentation distance $z_{\rm fil} \sim 1/n_2 I_0$, which varies like $1/P_{\rm in}$ whenever $I_0 \ll I_{\rm max}$ \cite{Campillo:apl:24:178,Fibich:oe:13:5897}.

\begin{figure}
\includegraphics[width=\columnwidth]{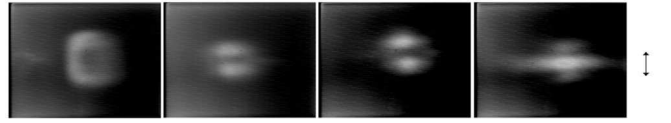}
\caption{Merging of two filaments created from mm-waisted, fs pulses 
with 5 mJ energy. Arrow indicates a scale length of 2 mm.}
\label{fig14}
\end{figure}

For narrower beams, experiments exploring mJ, mm-waisted 
pulses \cite{Hosseini:pra:70:033802,Liu:njp:6:1,Tzortzakis:prl:86:5470} 
reported the beam breakup into two filaments that continue to 
increase in 
intensity by Kerr focusing. Depending on their separation distance 
and individual powers, 
these filaments can merge. An example of this fusion 
mechanism is shown in Fig \ref{fig14} for two (plasma-free) optical spots, 
resulting from the modulational instability of a 42-fs, 3 mm-waisted 
collimated pulse in air \cite{Tzortzakis:prl:86:5470}. In the 
presence of plasma 
defocusing, ''parent'' filaments develop spatial rings whose radius 
increases with $z$. Two systems of rings can thus interfere to yield 
''child'' filaments. Each focusing spot 
triggers a plasma sequence that defocuses the trail pulse and locates 
the cells at specific negative instants. The number of ''filaments'' is therefore higher in intensity than in the beam fluence, which sums 
up all the time slices \cite{Champeaux:pre:71:046604}.

Since multifilamentation emerges from the initial 
beam defects, several techniques to control this process in 
shot-to-shot, meter-range experiments have been 
proposed. Among those, we can mention the introduction of beam 
ellipticity \cite{Dubietis:ol:29:1126,Grow:oe:13:4594} or of a tilted 
lens 
\cite{Fibich:ol:29:1772} along the optical path. Strong 
field gradients or forcing aberrations through phase  
masks can also be used to control high intensities over long 
scales \cite{Mechain:prl:93:035003}. Other techniques consist in 
monitoring the filamentation distance by changing the beam divergence 
angle with a deformable mirror \cite{Jin:oe:13:10424}. By this method, a single filament has been 
produced with powers as high as 420 GW, usually yielding multiple 
filaments in collimated geometry. Alternative methods consist in tuning the focal distance of a telescope setup composed of divergent and convergent lenses \cite{Fibich:oe:14:4946,Liu:apb:85:55}. Interesting 
features have recently been published concerning the fusion, 
repulsion and spiral co-propagation of two filaments, depending on their 
relative phase shift and crossing angle, which may further be used 
to optimize the propagation range \cite{Xi:prl:96:025003}. In the 
same 
spirit, arrays of diffractive microlenses \cite{Cook:apl:86:021105} 
or 
periodic meshes \cite{Kandidov:apb:80:267} into the propagation path 
helps in making the multifilamentation ''more deterministic''. As a 
matter of fact, these methods amount to altering the 
background surface of the beam from which small-scale filaments can 
randomly emerge from modulational instability. Alternative ways 
may be provided by femtosecond 
optical vortices \cite{Vincotte:prl:95:193901} that keep the 
filaments rotating 
along a ring and are nowadays designed experimentally on fs time 
scales \cite{Mariyenko:oe:13:7599,Vuong:prl:96:133901}. Further explorations should 
concern optical smoothing techniques breaking the laser spatial or temporal 
coherence \cite{Marklund:ol:31:1884}. These techniques are currently 
employed in laser facilities devoted to the inertial confinement fusion, in order to homogenize the beam distribution 
\cite{Donnat:ol:17:331,Labaune:pfb:4:2224}.

\subsection{White light generation and Conical emission}

Spectral broadening has been the topic of intense investigations for several 
decades \cite{Alfano:prl:24:584}. In 1995, Gil'denburg {\it et al.} 
\cite{Gildenburg:pla:203:214} performed 2D numerical simulations of 
ultrashort focused electromagnetic waves creating breakdown plasma 
due to tunnel gas ionization. High blueshifts of the pulse spectrum 
up to $\Delta \omega/\omega_0 \equiv \omega/\omega_0 - 1 > 40 \%$ were reported. At that time, 
blueshift was attributed to plasma generation, as the growth of free 
electrons increases the plasma frequency by $\omega_{\rm pe}^2 = 
q_e^2 \rho(I)/m_e \epsilon_0$ and implies a positive frequency shift 
$\Delta \omega_+ \sim \int_0^z \partial_t \rho dl$ 
\cite{Yablonovitch:pra:10:1888,Penetrante:josab:9:2032,Rae:pra:46:1084}. Later, experimental spectra revealed 
an ''ultrafast white-light continuum''. This occurs in a wide variety of 
condensed and gaseous media, whenever the input beam power exceeds 
the self-focusing threshold, which fixes a striking ''band 
gap'' for supercontinuum generation 
\cite{Brodeur:josab:16:637,Brodeur:prl:80:4406}. Around the central 
wavelength of 800 nm, blueshifts ($\Delta \omega_+$) down to $0.5$ $\mu$m in wavelength enlarged the spectra at intensities lower than 
$10^{-2}$ times the central one. Spectral enlargements appeared 
asymmetric, with a limited redshift, which could not be 
explained by SPM theory alone. Continua emitted from gases were found 
narrower $(\Delta \omega_+ \simeq 0.5 \,\omega_0$) than those for 
dense media $(\Delta \omega_+ \simeq 0.8-1.6 \,\omega_0)$. 

Because this process mixes all spectral components in the visible 
range, the core of the filament evolves like a white spot, 
which gives rise to a ''white-light laser'' \cite{Chin:jjap:38:L126}. 
Noting by $\varphi(\vec{r},t)$ the 
phase of the field envelope, frequency variations are dictated by
\begin{equation}
\label{SPMChin}
\Delta \omega = - \partial_t \varphi \sim - k_0 \Delta z \partial_t ({\bar 
n_2} I - \rho/2 n_0 \rho_c), 
\end{equation}
which varies with the superimposed actions of the Kerr and plasma 
responses. Near the focus point, only the front leading edge survives 
from this interplay and a redshift is enhanced by plasma 
generation. At later distances, second focusing/defocusing sequences 
relax the spectra to the blue side while the redshifted components 
decrease in intensity. A salient blueshift around the central 
frequency follows from self-steepening that creates a shock edge at 
the back of the pulse and amplifies a ''blue shoulder'' in the spectrum 
\cite{Akozbek:oc:191:353}. This spectral dynamics readily follows 
from Eq.\ (\ref{22}). Asymmetries caused by the focusing events 
are important: They design the pulse spectrum which will be preserved after the nonlinear stage of the beam.

\begin{figure}
\includegraphics[width=0.6\columnwidth]{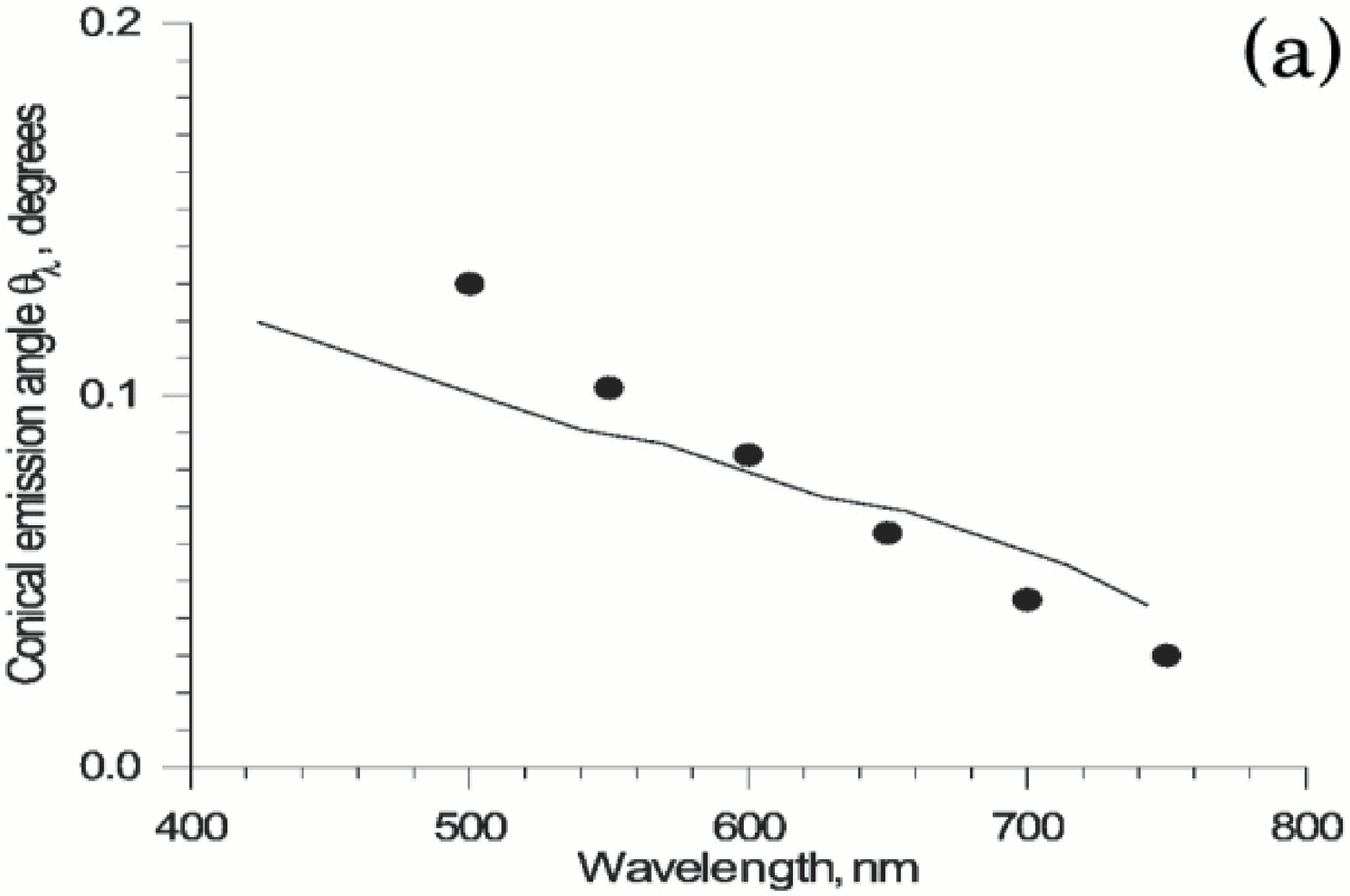}
\includegraphics[width=\columnwidth]{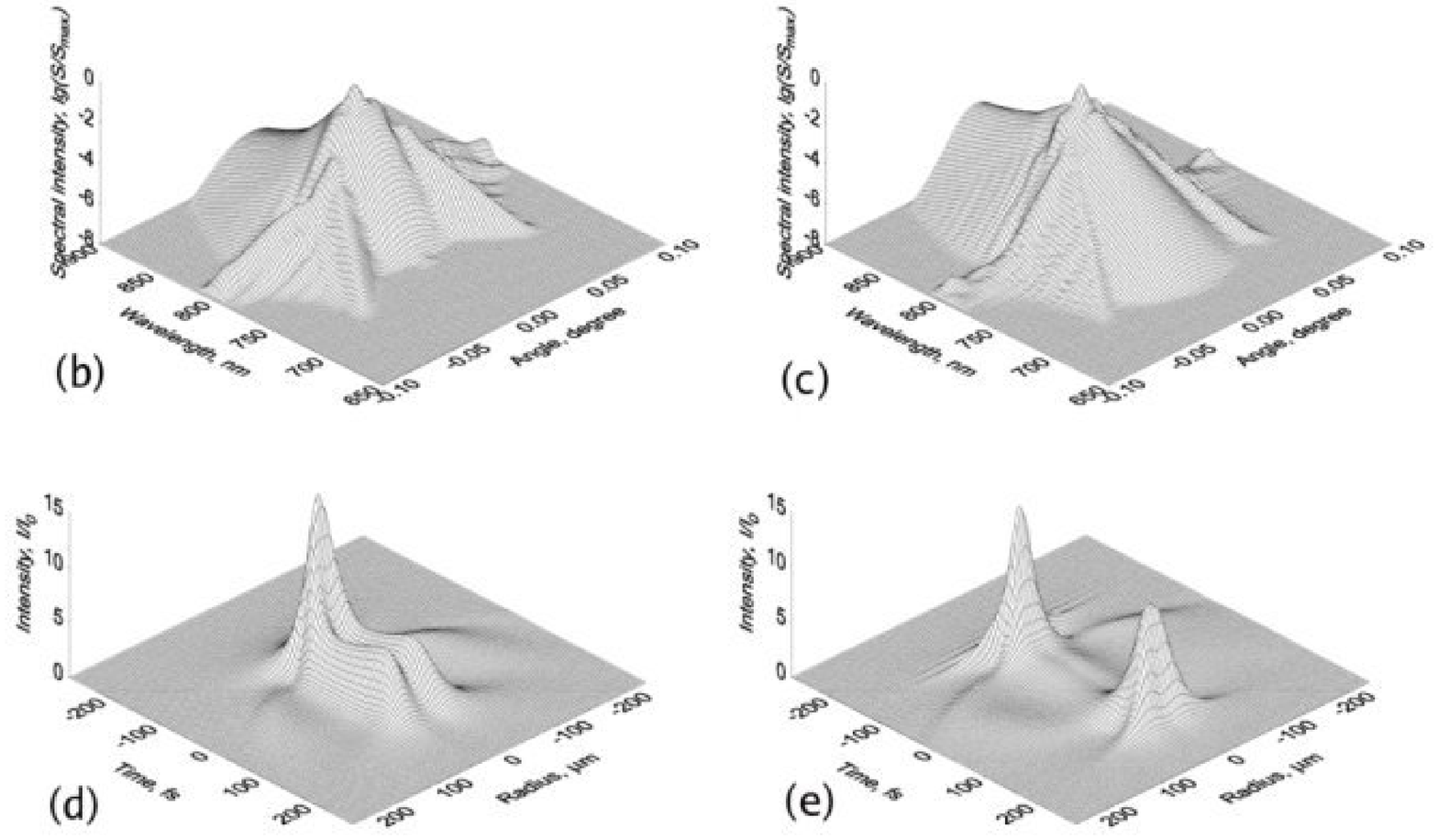}
\caption{(a) Measured (symbols) and calculated (curve) CE angles in 
air. (b,c) Frequency angular spectra and (d,e) spatio-temporal 
distributions of an atmospheric, fs filament starting from 
infrared pulses with 6.3 critical powers, 250 fs FWHM duration and 
0.18-mm radius at two different propagation distances (b,d) $z = 0.43 z_0$; (c,e) $z = 0.6 z_0$ \cite{Kandidov:apb:77:149}.
Courtesy of O.\ Kosareva.}
\label{fig15}
\end{figure}

The pulse spectrum is usually computed from the Fourier transform of 
the field intensity, either at $r = 0$, or being averaged over the 
filament radius, 
$S_{\rm fil}(\omega) = 2\pi \int_0^{r_{\rm fil}} 
|\widehat{\mathcal{E}}(r, \omega)|^2 rdr$. More insights can be 
obtained by looking at the frequency-angular spectral function $S(\theta_x,\theta_y,\omega,z) = |Q(\theta_x,\theta_y,\omega,z)|^2$,
defining
\begin{equation}
\label{spectrum2}
Q(\theta_x,\theta_y,\omega,z) = \int dx dy dt U(x,y,z,t) \mbox{e}^{i \omega t - i \theta_x kx - i \theta_y k y},
\end{equation}
where $\theta_x = k_x/k,\,\theta_y = k_y/k$ are the angles at which different frequency components $\omega$ propagate in the medium. The dependence of the angle $\theta = \theta_x = \theta_y$ on wavelength defines the supercontinuum cone emission and is obtained by means of the expression $\lambda = 2 \pi c/\omega$. Surface 
plots of $\ln{S(\lambda,\theta)}$ compared with the spatio-temporal 
intensity distributions figure out a conical emission (CE) driven by 
''wings'' directed to non-zero angles and connected with the 
plasma-induced spatial rings \cite{Kandidov:qe:34:348,Kandidov:apb:77:149}. Nibbering, Kosareva 
{\it et al.} \cite{Nibbering:ol:21:62,Kosareva:ol:22:1332} first 
revealed the existence of this conical emission accompanying 
femtosecond filamentation in air at 800 nm. Recalled in Fig.\ \ref{fig15}(a), CE 
increases in the interval $0 \leq \theta \leq 0.12^{\circ}$ with 
decreasing wavelengths $800 \geq \lambda \geq 500$ nm and is 
independent of the position along the filament. This ''conical 
emission'' corresponds to the angular divergence 
of spectral components. The (half-) angle $\theta$ at which the 
radiation propagates is determined by the spatial gradient
\begin{equation}
\label{CE}
\theta({\vec r}, t) = \mbox{arctan}[-k_0^{-1} 
\frac{\partial}{\partial r} \varphi({\vec r},t)].
\end{equation}
This angle depends on the spatial intensity distribution of the 
pulse, which strongly evolves in self-channeling regime, as 
illustrated in Figs.\ \ref{fig15}(b,c,d,e). By comparing the radial 
dependences of the divergence angle 
$\theta(r)$, low-frequency 
components are located nearest the optical axis. High-frequency 
components lie near the periphery rings and their radiation is 
directed out of the axis.

Besides Kerr and plasma effects, the material dispersion can play a 
relevant role in condensed media, for which the GVD coefficient is 
usually high. Kolesik {\it et al.} 
\cite{Kolesik:apb:77:185,Kolesik:prl:91:043905} compared the 
extension of supercontinuum wings generated in air ($k'' = 0.2$ 
fs$^2$/cm at 800 nm) and in water ($k'' = 500$ fs$^2$/cm at 400 nm). 
Numerical simulations accounting for the complete linear dispersion 
relation displayed evidence of that chromatic dispersion becomes a 
key player (for water above all) in determining the spectral extent 
of supercontinuum generation: Full chromatic dispersion included in 
$k(\omega)$ encompasses the $T, T^{-1}$ operators when passing to an 
envelope description (see Sec.\ \ref{sec2}). These operators induce 
steepening of the pulse profile and strongly ''blueshift'' the 
spectra.

On the other hand, spectral broadening becomes enhanced by harmonic 
generation. The question of coupling ultrafast IR components with 
self-induced third-harmonics was raised ten years ago 
\cite{Backus:ol:21:665}. While conversion efficiency as high as $0.1 
\%$ was first reported for third-harmonic (TH) generation in air, 
higher efficiencies up to $0.2 \%$ were next achieved by using 
focused pulses filamenting over only $\sim 10$ cm. At $800$ nm, the 
coupling of TH with the IR pump produces a ''two-colored'' 
femtosecond filament from the threshold intensity $I(\omega) \geq 2 
\times 10^{13}$ W/cm$^2$ 
\cite{Yang:pre:67:015401,Akozbek:prl:89:143901,Alexeev:ol:30:1503}. 
Along this process, the pump wave injects part of its energy into the 
third harmonics. 
The amount of pump energy depends on the linear wave vector mismatch 
parameter $\Delta k = [3k(\omega)- k(3\omega)]^{-1}$ fixing the 
coherence length $L_c = \pi/|\Delta k|$. The smaller the coherence 
length, the weaker TH fields, since most of the pump energy is 
periodically injected and depleted by the pump. In 
self-channeling regime, the balance between TH, pump wave 
nonlinearities and the linear mismatch parameter makes the phase 
difference $\Delta \varphi = 3 \varphi(\omega) - \varphi(3 \omega)$ 
be clamped at a constant value equal to $\pi$ 
\cite{Akozbek:prl:89:143901}. Along meter-range distances, the 
two-colored filament propagates over longer scales than an IR pulse 
alone, as the TH component stabilizes the pump wave and achieves $0.5 
\%$ conversion efficiency \cite{Berge:pre:71:016602}. Experimental 
identification of the TH component reveals some central spot 
surrounded by ring structures embarking most of the TH energy and 
having a half-divergence angle of 0.5 mrad 
\cite{Theberge:oc:245:399}. This contributes to enhance the total 
conical emission of the beam.

\begin{figure}
\includegraphics[width=\columnwidth]{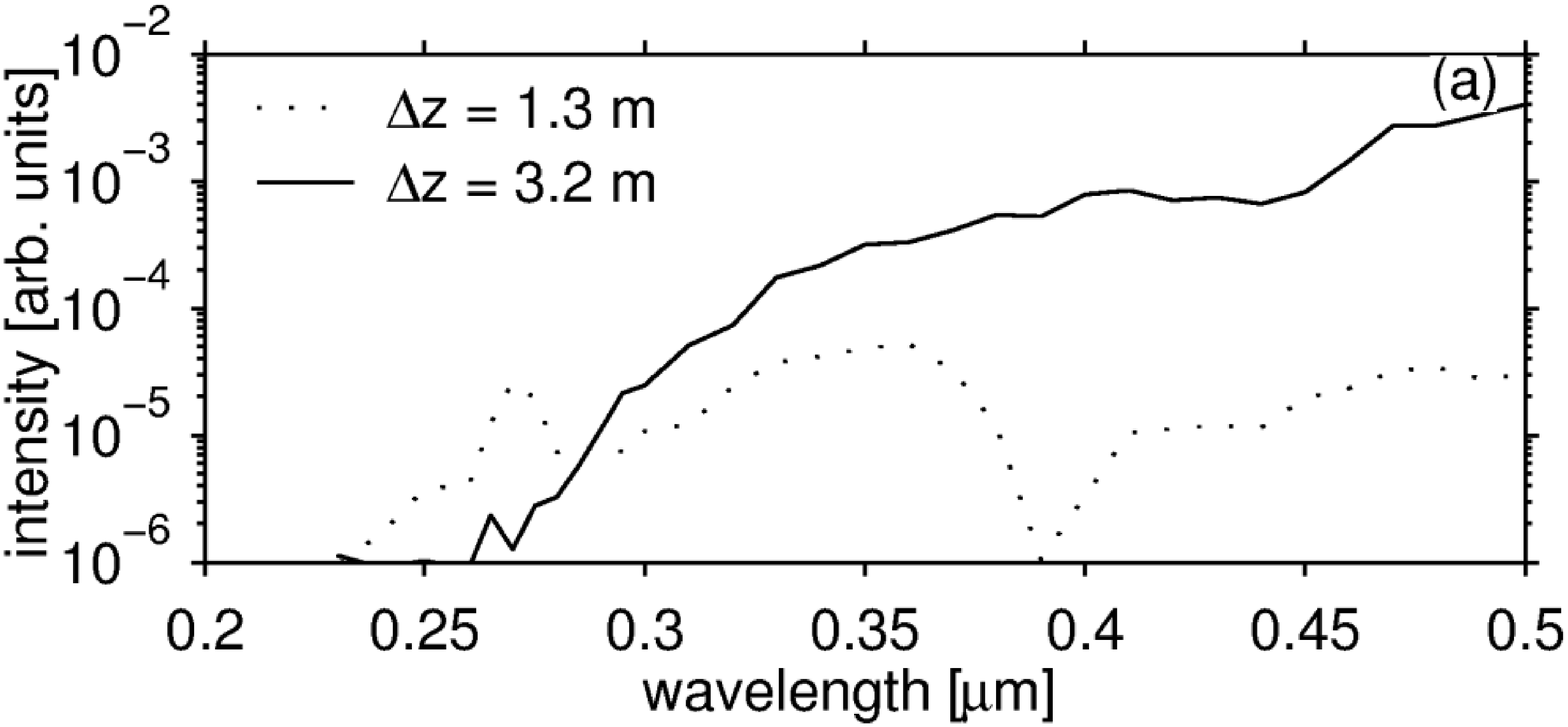}
\includegraphics[width=\columnwidth]{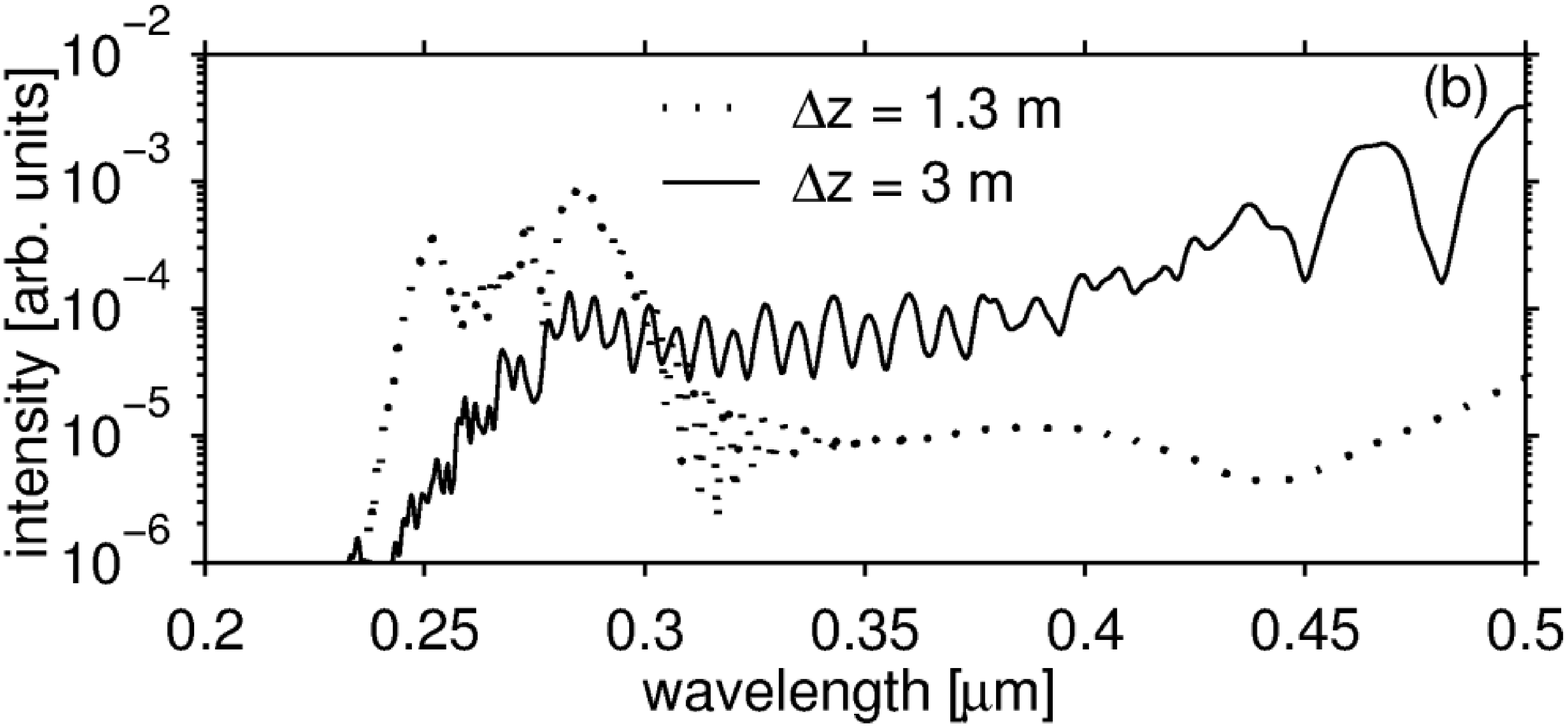}
\includegraphics[width=\columnwidth]{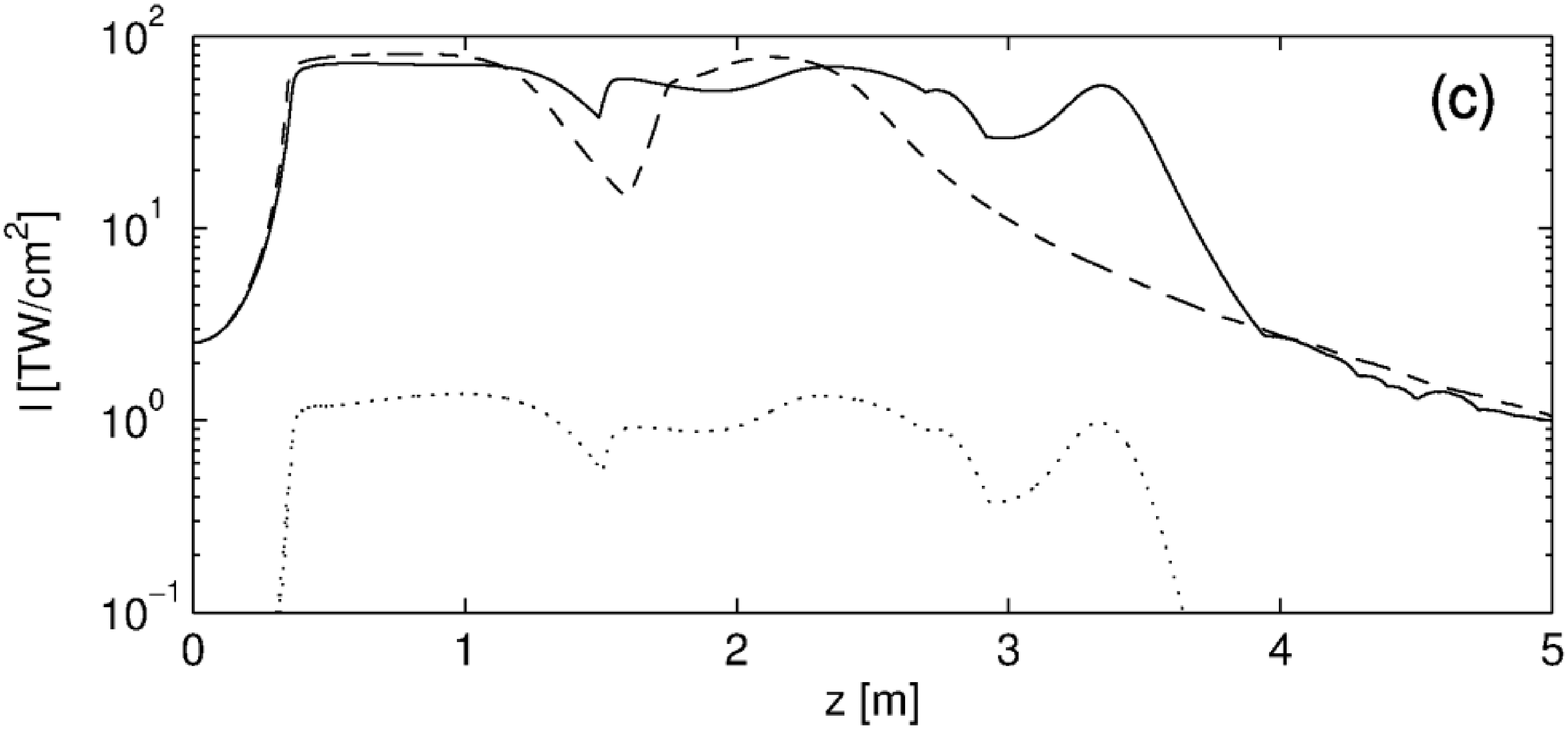}
\caption{UV-visible supercontinua (a) measured in laboratory, (b) 
computed numerically. (c) TH (dotted curve) and pump (solid curve) 
intensities integrated numerically. The dashed curve represents the same quantity computed using the pump component alone.}
\label{fig16}
\end{figure}

Harmonic generation permits the occurrence of amazing phenomena, such 
as, e.g., the emergence of new wavelengths with non-trivial relative spectral 
intensities. Manipulating intense, powerful filaments centered at 800 
nm results in supercontinuum generation extending down to 230 nm in air 
and yielding a continuous spectral band of UV-visible wavelengths 
\cite{Theberge:apb:80:221}. This phenomenon was observed over laboratory scales as well as over LIDAR 
propagation distances $> 200$ m \cite{Berge:pre:71:016602,Mejean:apb:82:341}. Numerical 
simulations evidenced that UV-visible spectral broadening is 
created by the overlap of redshift from the TH component and 
blueshift from the IR pump, $\Delta \omega_j = - \partial_t 
\varphi_j$ ($j=\omega,3\omega$), fulfilling the phase-locking 
constraint $\Delta \varphi = \pi$ mentioned above. Also, TH-induced 
saturation reinforces the defocusing action of the electron plasma 
and enlarges the conical emission to $\sim 0.25^{\circ}$ at 0.26 $\mu$m.
UV-visible supercontinuum generation in air is plotted in Figs.\ 
\ref{fig16}(a,b) for a meter-range propagation of one filament. 
Numerical results reproduce the experimental data in the 
wavelength domain $200 \leq \lambda \leq 500$ nm at the same 
distances after the nonlinear focus ($\Delta z = z - z_c$). Fig.\ 
\ref{fig16}(c) illustrates the maximum TH and pump intensities, calculated from a propagation model accounting for TH generation [see Eqs.\ (\ref{TH})] without steepening terms. Resembling spectral dynamics 
have also been reported from 45-fs, 1-mJ 
infrared pulses propagating in argon at atmospheric pressure, after 
subsequent compression by chirped mirrors \cite{Trushin:apb:80:399}.
Simulations of these experiments \cite{Akozbek:njp:8:177} were performed from the one-component NEE model (\ref{modeleq}) in the limit of weak THG. It was found that temporal gradients inherent to the $T, T^{-1}$ operators are sufficient to amplify UV shifts and cover the THG bandwidth for very short pulses ($\leq 10$ fs) over short conversion lengths that involve a limited number of focusing/defocusing cycles. Huge UV shifts have already been displayed for air propagation in Fig.\ \ref{fig10b}(d). The issue of knowing under which conditions steepening effects or THG prevail in the supercontinuum generation remains open.

\subsection{Role of the laser wavelength}

Whereas emphasis was put so far onto optical pulses delivered by 
Ti:Sa oscillators 
operating around 800 nm, one could wonder whether the 
propagation dynamics changes at different laser wavelengths. Besides 
the Rayleigh length $z_0$, several physical parameters 
become modified when varying $\lambda_0$, namely,

\begin{itemize}

\item The Kerr index $n_2$, function of the cubic susceptibility 
$\chi_{\omega_0}^{(3)}$, decreases inversely proportional to 
$\lambda_0$ \cite{Agrawal:NFO:01,Nibbering:josab:14:650}.

\item The photo-ionization rate and the number of photons consumed by 
MPI transitions $\sim U_i/\hbar \omega_0$ vary with $\lambda_0$. 
Even though MPI gain and MPA losses cannot be separated from each 
other, their 
respective influence changes with $\lambda_0$ and stresses more 
losses 
at low wavelengths (see Fig.\ \ref{fig10}).

\item The value and sign of the GVD parameter, $k'' = 
\partial^2 k/\partial \omega^2|_{\omega = \omega_0}$, are modified by 
varying $\lambda_0$. Depending on the material 
considered, $k''$ can turn from positive (normal GVD) to negative 
(anomalous GVD), which deeply modifies the spatio-temporal structure 
of the 
beam.

\end{itemize}

On this topic, propagation studies mostly concerned 
the differences between infrared pulses at 800 nm in the atmosphere and ultraviolet ones at, e.g., 248 nm, created by an hybrid, 
frequency-tripled Ti:Sa/KrF excimer chain. Long distance propagation experiments employing ps, 248-nm pulses \cite{Schwarz:apb:72:343,Schwarz:oc:180:383} 
reported that after a transient stage of multiple filamentation, the 
beam relaxed to a single filament upon propagation 
ranges of more than 10 m. The filament length, defined as 
the spatial region where the plasma couples 
with the light channel, only covered $\sim 1$ m, where the electron density, 
evaluated from the energy consumption per filament, attained $3 
\times 10^{15}$ cm$^{-3}$. Computed from the MPI rate, the peak intensity 
was deduced to be about $10^{12}$ W/cm$^2$. Also associated with this 
filament, a narrow spectral bandwidth of $\Delta \lambda \sim 10$ nm was measured around 248 nm. Later, Mysyrowicz's group 
\cite{Tzortzakis:ol:25:1270,Tzortzakis:oc:197:131} observed the 
filamentation of UV pulses with durations from 5 ps down to 450 fs. 
Long, meter-range ($\sim 2-4$ m) filaments coupled with 
$10^{15}-10^{16}$ cm$^{-3}$ peak electron densities at intensities of 
$\sim 10^{11}$ W/cm$^2$ were identified. The ''apparent'' 
differences between UV and IR peak values follow from the pulse 
duration between UV and IR experiments, the 
plasma critical density $\rho_c \sim 1/\lambda_0^2$ together with the 
number of photons involved in  multiphoton transitions. This varies 
from $K = 8$ at 800 nm to $K = 3$ at 248 nm for ${\rm O}_2$ molecules. 
Numerical simulations, however, showed that, for 50-fs FWHM input pulses, the 
filament dynamics and length ($< 5$ m), starting from the same 
nonlinear focus, were identical in both cases with peak intensities and electron densities attaining $4-8 \times 
10^{13}$ W/cm$^2$ and $10^{17}$ cm$^{-3}$, respectively. Even though 
the ionization parameters vary 
with $\lambda_0$, the peak values are fixed by the balance 
$\rho_{\rm max}/2 \rho_c \sim n_2 I_{\rm max}$, which remains similar 
at the two wavelengths. Averaging the peak quantities over a given 
filament width can justify the different measurements made 
in this field \cite{Couairon:prl:88:135003}.

Besides, changes in supercontinuum generation were numerically 
addressed by Ak{\"o}zbek {\it et al.} \cite{Akozbek:apb:77:177}, who 
accounted for THG at other laser wavelengths such as 1064 nm and 
1550 nm. The latter corresponds to eye-safety and is 
particularly important for LIDAR applications. For these 
wavelengths, the GVD parameters for both pump and TH frequencies are 
close to each other. However, the wavevector mismatch strongly 
decreases from $- 5$ cm$^{-1}$ (800 nm) to $-0.65$ cm$^{-1}$ (1550 
nm), 
whereas the temporal walk-off parameter $\Delta v = 
[v_g^{-1}(3\omega) 
- v_g^{-1}(\omega)]^{-1}$ increases from 0.4 to 2.0 cm/s, 
respectively. Consequently, 
the coherence length significantly augments at large wavelengths, 
which 
allows higher TH intensities and energy conversion efficiency.

\begin{figure}
\includegraphics[width=\columnwidth]{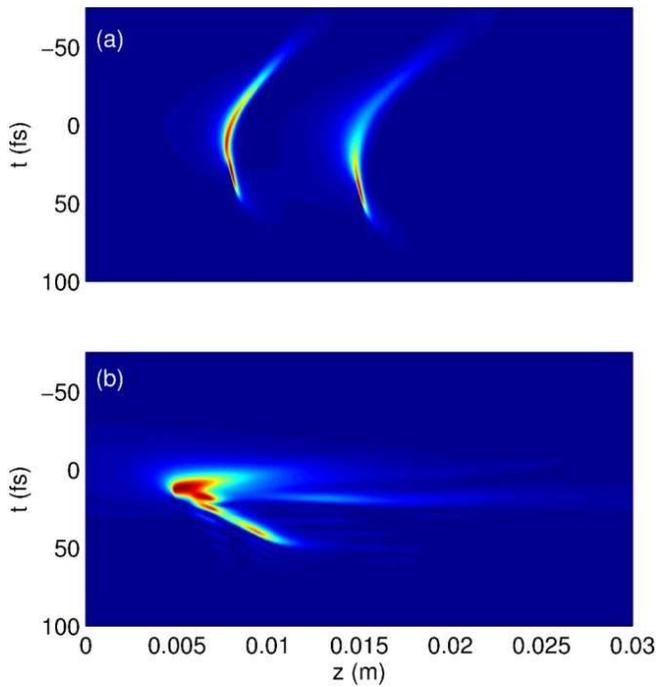}
\caption{Temporal dynamics of 42.5-fs pulses with waist $w_0 = 71$ 
$\mu$m propagating in silica at the laser wavelength (a) 
$\lambda_0 = 790$ nm with power $P_{\rm in} = 6 P_{\rm cr}$, (b) $\lambda_0 = 1550$ 
nm with $P_{\rm in} = 3 P_{\rm cr}$. Note that the physical value of $P_{\rm cr}$ changes with $\lambda_0$.}
\label{fig17}
\end{figure}

More surprising features occur when the laser wavelength is selected in 
such a way that $k''$ becomes negative and leads to anomalous dispersion. 
In that case, the pulse ceases to be dispersed in time and undergoes a 
temporal compression, in addition to the spatial Kerr focusing. This happens, 
e.g., in fused silica at $\lambda_0 =1550$ nm, for which $k'' \simeq 
-280$ fs$^2$/cm (see Table I). Recent experiments at this wavelength have shown 
that 
collapse events looked ''extended'' along the $z$ axis, unlike the 
localized events promoted by normal GVD. Plasma
 halts the collapse at powers above critical, but anomalous 
GVD continues to transfer energy into the collapse region, resulting 
in 
the formation of longer filaments before the beam eventually 
defocuses 
\cite{Moll:ol:29:995}. The pulse can thus remain confined along 
several 
diffraction lengths and develops very narrow, isolated temporal 
peaks. A strong temporal compression produces optical spots whose 
duration is shrunk to the few-cycle limit \cite{Berge:pre:71:065601}. 
Sharp shock profiles emerge from this 3D collapsing 
dynamics and pulse steepening along each focusing/defocusing event 
tremendously amplifies the blue part of the spectrum. Fig.\ \ref{fig17} displays 
the temporal distributions of a femtosecond filament created in fused 
silica at 790 nm ($k''= 370$ fs$^2$/cm $>0$) and at 1550 nm 
($k''<0$). At 790 nm, normal GVD stretches the 
pulse along the time direction. Reversely, at 1550 nm, anomalous GVD 
compresses it temporally and makes it shift to positive 
instants through third-order dispersion and self-steepening that 
push the pulse centroid towards the region $t > 0$.

\section{Ultrashort filaments in gases\label{sec5}}

This section is devoted to novel trends opened by the self-guiding of 
femtosecond pulses over small distances ($<1$ m), such as pulse 
shortening and high-harmonic generation. Most of these 
applications are conducted in pressurized gas cells, 
in which the raise of plasma excitations can be controlled by changing the medium pressure.

\subsection{Novel perspectives for pulse shortening}

In the last decade, an important challenge has consisted in testing 
powerful techniques of spectral broadening and ultrabroadband 
dispersion control, in order to compress 
pulses to durations of a few optical cycles only ($\tau_{\rm o.c.} = 
\lambda_0/c$). Nisoli and co-workers 
\cite{Nisoli:ol:22:522,Nisoli:apb:65:189,Nisoli:ieeejstqe:4:414} 
achieved the shortest pulse durations from $\sim 1$ mJ, 20-fs pulses 
by manipulating the spectral 
broadening attained along a 60-cm long fused-silica hollow fiber 
filled with atomic (Ar, Kr) or molecular (N$_2$) gases. Conditions 
for optimum pulse compression outlined an appropriate combination of 
SPM and gas dispersion managed by external chirped compensation 
systems. Pulses as short as 4.5 fs with output energies up to 70 
$\mu$J could be delivered at 800 nm and 1-kHz repetition rate after 
using a chirped-mirror delay line. The potential scalability of this 
system to higher pulse energies was claimed to hold provided that two conditions are 
fulfilled: (i) The laser peak power must 
be smaller than $P_{\rm cr}$; 
(ii) The peak intensity must not exceed the ionization 
threshold, to preserve a flat spectral phase for an 
efficient recompression. These two requirements could at that time be 
satisfied by a delicate tuning of the pressure parameter. For a 
gaseous medium with pressure $p$, the coefficients affecting Eq.\ 
(\ref{1}) indeed vary as follows
\begin{equation}
\label{pressure}
k'' \sim p,\quad\beta^{(K)} \sim p,\quad P_{\rm cr} \sim 1/n_2 \sim 
1/p,\quad\rho_{\rm nt} \sim p.
\end{equation}
An optimal compression length can be estimated by $L_{\rm opt} 
\approx (6L_{\rm NL} L_{\rm disp})^{1/2}$ where $L_{\rm disp} \sim 
t_p^2/k''$ and $L_{\rm NL} \simeq n_2 \omega_0 P_{\rm in}/c S_{\rm 
eff}$ are the dispersion and nonlinearity lengths, respectively, with 
$S_{\rm eff}$ being the effective area of the beam mode \cite{Nisoli:ol:22:522}. This distance is tuned by the cell length and the local pressure $p$, in order to reach the smallest possible pulse duration. Stability of the beam was insured by selecting the EH$_{11}$ hybrid mode with 
intensity $I(r) = I_0 J_0^2(2.405 r/a)$, where $I_0$ is the peak 
intensity, $J_0$ is the zero-order Bessel function and $a$ is the 
capillary radius \cite{Tempea:ol:23:762}. Keeping the beam stable 
inside the waveguide requires to limit the excitation of higher-order 
modes owing to the nonlinear spatio-temporal dynamics. The spectral 
broadening thus inherently depends on how close the initial peak 
power of the beam is to $P_{\rm cr}$ \cite{Homoelle:ol:25:761}. Sub-10 fs pulses are
routinely produced through these techniques \cite{Steinmeyer:sc:286:1507}. They can even be applied to bulk (BK7 glass) media, from which 
pulse shortening by a factor 3-5 and output energies of 220 $\mu$J have been achieved \cite{Mevel:josab:20:105}. 

In this scope, plasma generation may not 
constitute a drawback for pulse compression. Tempea and 
Brabec \cite{Tempea:ol:23:1286} mentioned the possibility to produce 
few-cycle optical pulses by compensating the spectral chirp generated 
by the plasma nonlinearity with a dispersive line delay for 
intense pulses propagating in filled-gas hollow 
fibers. This scenario was experimentally confirmed by a JILA team 
\cite{Wagner:prl:93:173902}, who succeeded in compressing 30 fs 
pulses to 13 fs in hollow argon capillaries, even without any need of 
external dispersion compensation. It was recently improved to deliver 
5 fs, high energy (1 mJ) pulses following the same scheme 
supplemented by amplifier systems and chirped-mirror compressors 
\cite{Verhoef:apb:82:513}. In Sec.\ \ref{sec4}, Fig.\ \ref{fig9}(d) 
recalled the basic property of pulse compression through the combined 
effects of Kerr nonlinearity and self-induced ionization. Chromatic dispersion and 
nonlinear losses, however, saturate the defocusing action of free 
electrons on the leading edge and enable the rear pulse to refocus at 
later distances. This produces a two-peaked temporal distribution 
whose time extent is often of the order of the input pulse duration. 
Achieving an efficient pulse compression, instead, requires
to isolate one (at least dominant) peak shrunk in time  
\cite{Champeaux:pre:68:066603}.

\begin{figure}
\includegraphics[width=\columnwidth]{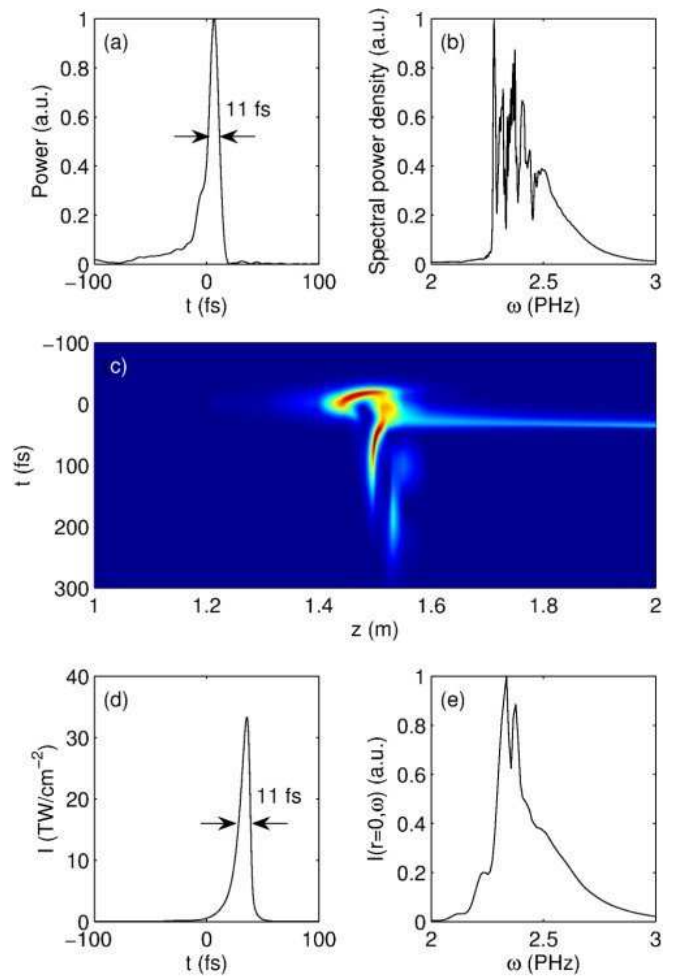}
\caption{(a) Experimental pulse shape for the self-compression mechanism in a 50 k-Pa argon cell, (b) 
associated spectrum. (c) On-axis numerically-computed temporal evolution of 45-fs pulses in the $(t,z)$ plane for a similar setup. (d) On-axis intensity and (e) spectral intensity at $z = 1.6$ m (dashed 
curve) and $z = 2$ m (solid curve, exit window of the argon cell). 
} 
\label{fig20}
\end{figure}

To realize this challenge, Hauri {\em et al.} used a configuration of 
two cascaded gas-filled
cells with intermittent dispersion compensation for 
producing
sub-mJ light pulses with durations down to 5.7 fs at $\sim 800$ nm 
\cite{Hauri:apb:79:673}. The
cells were operated at different pressures. It was later suggested 
that this scheme could be improved when a single cell with a pressure 
gradient is used instead, in order to monitor the refocusing stage of the 
trail and create waveforms with a single-cycle temporal
signature \cite{Couairon:ol:30:2657,Couairon:jmo:53:75}. Despite 
first results mainly directed to higher pulse energies
\cite{Suda:apl:86:11116}, further efforts are needed to adjust the 
pressure gradient
for optimum compression to extremely short pulse duration. 
Alternatively, Stibenz {\em et al.} recently demonstrated an efficient
pure self-compression obliterating the need for any kind of dispersion compensation, pressure
gradients, or capillaries for beam guiding \cite{Stibenz:ol:31:274}. 
The experimental setup 
involves beams in convergent geometry with energy up to 5 mJ, $w_0 = 
11$ mm, $f = 1.5$ m, FWHM durations of 45 fs and input powers equal 
to 5 times critical. 
The beam is initially focused and propagates inside an 1-m long, 50 k-Pa argon 
cell whose center is positioned at the location of the geometrical focus. Output pulses reach durations down to about 10 fs and develop a strong blue-shift, as reported in Figs.\ \ref{fig20}(a,b). Figure \ref{fig20}(c) details this self-compression mechanism, reproduced numerically from Eqs.\ (\ref{modeleq}). First, near
$z_c \simeq 1.45$ m, the front pulse focuses 
and plasma depletes its rear part. Second, at the 
linear focus ($z \simeq 1.5$ m), the plasma density decreases, and the 
back pulse rises again. Finally, from $z \simeq 1.6$ m, only the 
components close to center, keeping intensity levels above $10^{13}$ 
W/cm$^2$, are effectively trapped in the filament, whereas the 
temporal wings diffract rapidly. A robust, 
temporally-compressed
structure of 11 fs forms in the core region of the filament [Fig.\ \ref{fig20}(d)]. The pulse center relaxes to a narrow ''waveguide'',
preserving its energy at low plasma density levels ($< 10^{14}$ 
cm$^{-3}$) after the linear focus.
Fig.\ \ref{fig20}(e) shows the on-axis intensity spectra,
corresponding to this temporal
compression. Not shown here, the spectral phase at the cell exit ($z = 2$ cm) becomes nearly flat in the blue wavelengths with tiny variations comprised
in the same proportions as those measured by spectral phase 
interferometry for direct electric-field reconstruction (SPIDER). The robustness for this self-compression mechanism was numerically found limited to rather long $f/w_0$ ratios and powers around 5 
$P_{\rm cr}$, in order to 
avoid another focusing sequence \cite{Skupin:pre:74:056604}.

To end with this aspect, let us mention that Chen {\it et al.} 
\cite{Chen:oc:259:331} also succeeded in achieving pulse 
self-compression from 50 to 20 fs, by making 800 nm pulses with 
energy $<$ 1 mJ pass through a thin BK7 glass plate, with no 
subsequent dispersion compensation. SPIDER and spectral 
measurements revealed that self-compression takes place as long as the front pulse focuses 
while plasma cuts off the trail part. At this stage, the spectrum is 
marked by a strong redshift. By increasing the input peak intensity, 
the back pulse refocuses and blueshifts the spectrum. Accounting on the balance between plasma-induced 
negative chirps and SPM-induced positive chirps, the spectral output 
phase was, again, observed to be flat. All these methods represent 
new promising techniques to produce high-energy few-cycle laser 
sources in the future.

\subsection{High-order harmonic generation}

The possibility to reduce pulse durations to a single optical cycle 
provides ideal conditions for high-harmonic generation and 
laser-assisted X-ray photo-ionization \cite{Drescher:sc:291:1923}. 
Several fundamental atomic processes such as inner-shell electronic 
relaxation or ionization take place within a fraction of the 
oscillation period of visible or near-infrared radiation and they 
require very short probes for being investigated. For this purpose, bursts of 
attosecond pulses (1 as = 10$^{-18}$ sec.) need to be isolated. 
This is now feasible by combining single bursts emitted at 
extreme ultraviolet (XUV) wavelengths and ultrashort lasers, in order 
to preserve the coherence properties at so short time 
scales. In this microcosm, the ionization processes become 
crucial, and more particularly the transient stage of about half 
laser period along which an electron is liberated. Developments in 
attosecond X-ray science have been addressed in several 
excellent reviews 
\cite{Brabec:rmp:72:545,Pfeifer:rpp:69:443,Scrinzi:jpb:39:1}. 
Subfemtosecond light pulses can be experimentally obtained by 
superposing several high harmonics and making them emit 
simultaneously. By controlling their synchronization, pulses of 130 as in duration 
have been achieved \cite{Mairesse:sc:302:1540}.

High-Harmonic Generation (HHG) describes the process by which laser 
light at central frequency $\omega_0$ is converted into integer 
multiples of $\omega_0$ during the highly nonlinear interaction with 
the medium. In 1993, Corkum \cite{Corkum:prl:71:1994} proposed a 
quasiclassical theory for this process, which can be divided into 
three steps:

(i) The electron is freed by ionization and driven away 
from the parent ion.

(ii) Since the laser field changes its sign over 
times about $\tau_{o.c.}/2$, this electron slows down, stops at 
a certain position, then re-accelerates towards the ion.

(iii) When the electron recombines with the nucleus, a photon with energy equal to $U_i$ plus the electron kinetic energy is emitted. This gives rise to 
very high harmonic orders.

\begin{figure}
\includegraphics[width=\columnwidth]{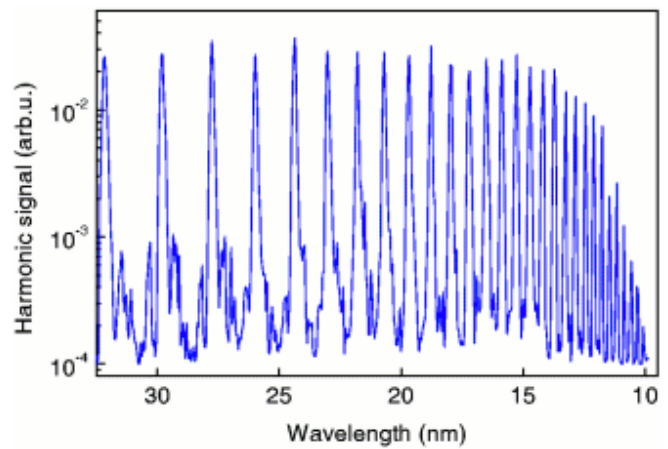}
\caption{Harmonic spectrum generated in neon by a 800 nm Ti:Sa laser. 25th to 81st harmonics are displayed
\cite{Salieres:mst:12:1818}.}
\label{fig21}
\end{figure}

Figure \ref{fig21} illustrates a typical HHG 
spectrum. The harmonic intensity 
is rapidly decreasing after an almost flat plateau ($\lambda<15$ nm). It is 
terminated by a cut-off, signaling the highest harmonics that can be 
generated. Among all possible electron trajectories, there indeed exists a maximum kinetic energy corresponding to the 
maximum photon energy embarked by this process. The numerical solving 
for the electron trajectories with various phases leads this maximum kinetic energy to 
be $\sim 3.17 \times U_p$ \cite{Corkum:prl:71:1994,Zeng:pra:67:013815}.
Here, $U_p=q_e^2 E_p^2/4m_e \omega_0^2$ denotes the ponderomotive energy of the electron in the wave field (see Appendix \ref{appA}).
Electrons thus produce 
harmonic photons up to the cut-off energy 
\begin{equation}
\label{HHGcutoff}
\hbar \omega_c = 3.17 \times U_p + U_i.
\end{equation}
The laser pulses employed in HHG experiments are generally short ($< 40 $ fs), with the 
elementary profile $E(t) = E_0 \cos{(\omega_0 t + \varphi_{\rm 
CE})}$, where the carrier-envelope phase $\varphi_{\rm CE}$ 
becomes an important parameter for the 
conversion efficiency into HHG when approaching the optical cycle \cite{Scrinzi:jpb:39:1}.

In ionized media with symmetry inversion, harmonic 
peaks only exist at odd integers multiple of $\omega_0$. The above 
three-step process repeats every half-cycle of the laser field, so 
that the Fourier sum in the spectra makes even harmonics cancel each 
other along one cycle. The analytical description of this process requires a 
quantum-mechanical approach of the dipole moment associated with the 
harmonics. To this aim, Lewenstein {\it et al.} 
\cite{Lewenstein:pra:49:2117} derived the time-dependent dipole moment $\vec{d}(t) = q_e \langle \psi({\vec 
r},t)|{\vec r}|\psi({\vec r}, t) \rangle$ that represents the 
expected position of the electron in the quantum state $|\psi({\vec 
r}, t) \rangle$. Details for computing $\vec{d}(t)$ have been given in Appendix \ref{appB}. The 
time-dependent dipole takes into account ionization at time $t'$, 
energy gain computed from a phase integrand and recombination at time 
$t$. Stationary phase conditions yield the information about the 
electron trajectories. By Fourier transforming the 
dipole moment, the harmonic spectrum can be calculated and it restores 
the cut-off law (\ref{HHGcutoff}).

\begin{figure}
\includegraphics[width=\columnwidth]{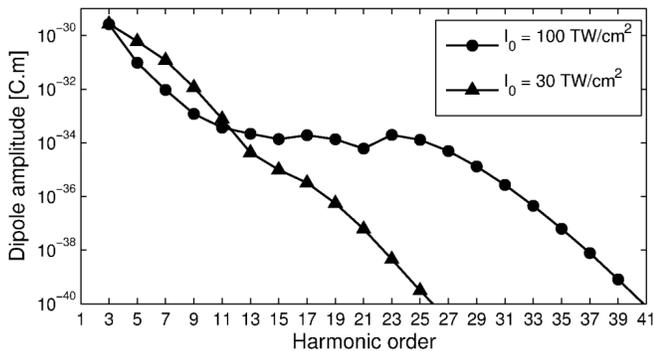}
\caption{Harmonic spectrum computed from the dipole moment ${\vec d}(t)$ in pressurized argon for two different pump intensities.} 
\label{fig22}
\end{figure}

The challenging issue is then to describe HHG by accounting for the 
propagation physics of ultrashort pulses having initially a few fs 
durations. In 1998, Lange {\it et al.} \cite{Lange:prl:81:1611} 
employed self-guided fs pulses to produce HHG up to the 15th harmonics 
in noble gases and maintain the longitudinal coherence of the generated harmonics. 
Tamaki {\it et al.} \cite{Tamaki:prl:82:1422} demonstrated 
experimentally efficient HHG through the phase-matched 
propagation of laser beams in pressurized Ne with peak 
intensities up to $10^{15}$ W/cm$^2$. By changing the 
propagation length, phase-matching magnifies the 
conversion efficiency around the 49th harmonics by 40 times near the 
cut-off region. Tosa {\it et al.} \cite{Tosa:pra:67:063817} reported 
the H13 harmonics triggered by the 
self-guided propagation of a focused pulse inside a 14-cm long cell 
filled with Xe, in which the Kerr response was negligible.
From these results, we easily guess that the next step in 
''attosecond'' investigations should concern the optimization of HHG 
spectra by means of femtosecond filaments. The promising feature of pulse shortening induced by the balance Kerr/plasma could become an excellent technique to produce an isolated XUV pulse. Moreover, the ability to keep the laser pump clamped at sufficient high intensity over several centimeters may magnify the harmonic signal.
For instance, figure \ref{fig22} illustrates the amplitude of the atomic dipole
computed numerically for a plane-wave pulse traversing an argon cell pressurized at 60 kPa
for two different pump intensities. The HHG spectrum becomes magnified at intensity levels close to 100 TW/cm$^2$.
This preliminary result keeps us confident with obtaining harmonic signals with high enough energy level.

\section{Ultrashort filaments in dense media\label{sec6}}

Below, we review major physical aspects
of ultrashort filaments in solid dielectrics and in liquids.

\subsection{Pulse propagation in dielectrics and damages}

Transparent dielectrics such as SiO$_2$ samples were routinely 
examined during the 90s, in the
framework of the optical breakdown on the one hand (see Sec.\ \ref{sec4}) and of 
the carrier trapping dynamics
of band-gap crystals on the other hand. One of the key parameters 
being the potential gap
(which is the energy difference between valence and conduction band), several investigations were led to identify 
this gap in silicon dioxide and to measure the ultrafast
excitation and relaxation of an electron gas pumped into the 
conduction 
band by intense fs laser pulses 
\cite{Tohmon:prb:39:1337,Audebert:prl:73:1990,Martin:prb:55:5799}.

Meanwhile, other researches concerned laser-induced breakdown (LIB) 
in dielectrics. The threshold 
damage fluence varying as $\sqrt{t_p}$ for
long pulses in thermal conduction regimes was seen to deviate from this scaling with fs pulses, for which electrons have 
no time to efficiently couple to the 
lattice. Du {\it et al.} \cite{Du:apl:64:3071} reported a damage 
threshold 
fluence higher than the $\sqrt{t_p}$ prediction rule for
pulse durations $< 10$ ps and they furthermore underlined that 
short-pulse damage exhibits a deterministic nature, unlike long
pulses. A theoretical model accounting for MPI, avalanche ionization
and Joule heating stressed the dominant role of photo-ionization at 
fs time scales \cite{Stuart:prb:53:1749}. However, both 
photo-ionization and avalanche come 
into play during the occurrence of damage. Although the
role of the former is initially dominant 
\cite{Rayner:oe:13:3208}, it can be masked by
avalanche from an initially-high electron density $(> \rho_c/100)$ 
\cite{Tien:prl:82:3883}. To complete these two processes, the
rapid electron decay over $\sim 100$ fs scales through 
recombination softens the maximum 
electron density for LIB \cite{Li:prl:82:2394}.

\begin{figure}
\includegraphics[width=0.45\columnwidth]{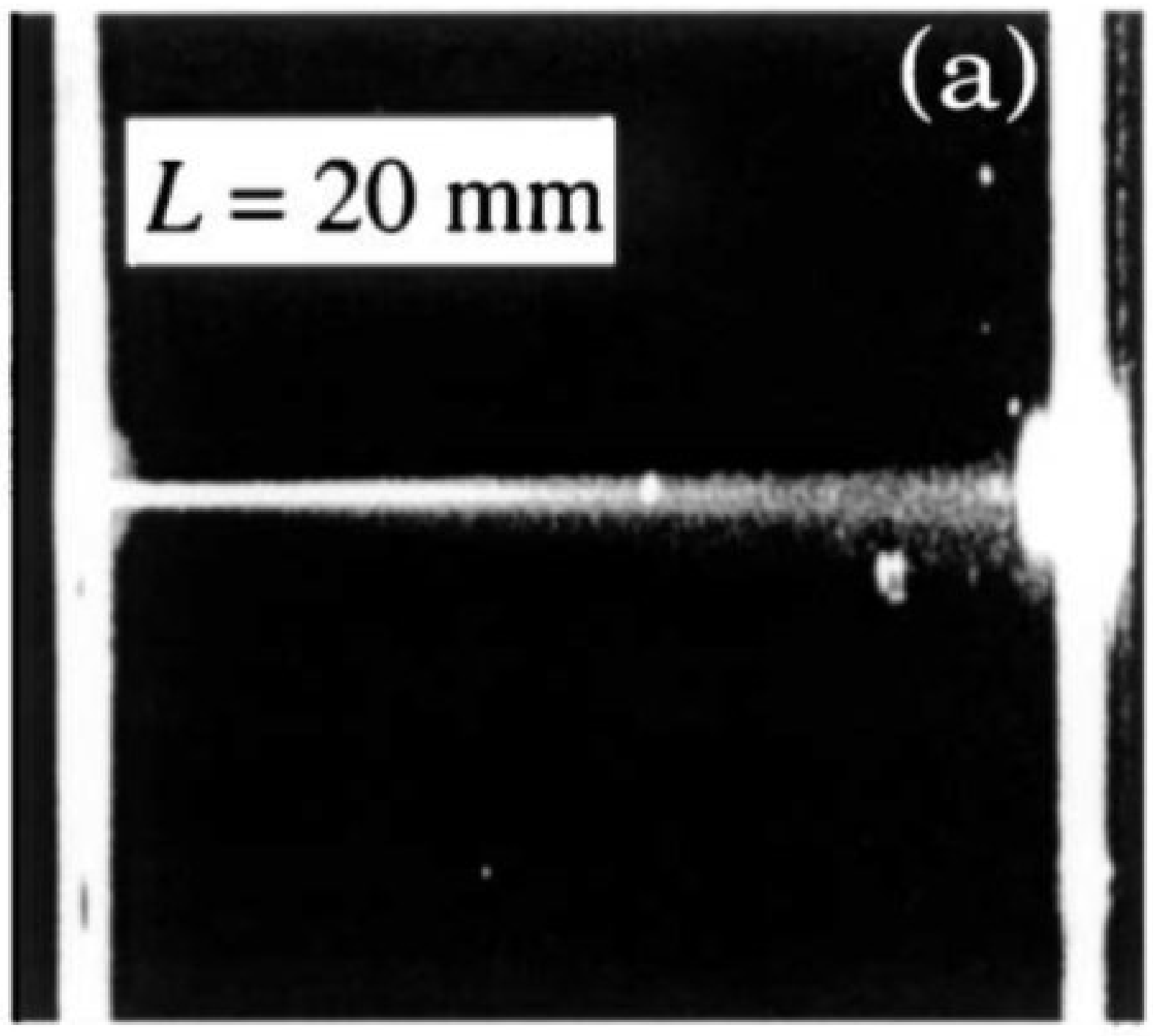}
\includegraphics[width=0.53\columnwidth]{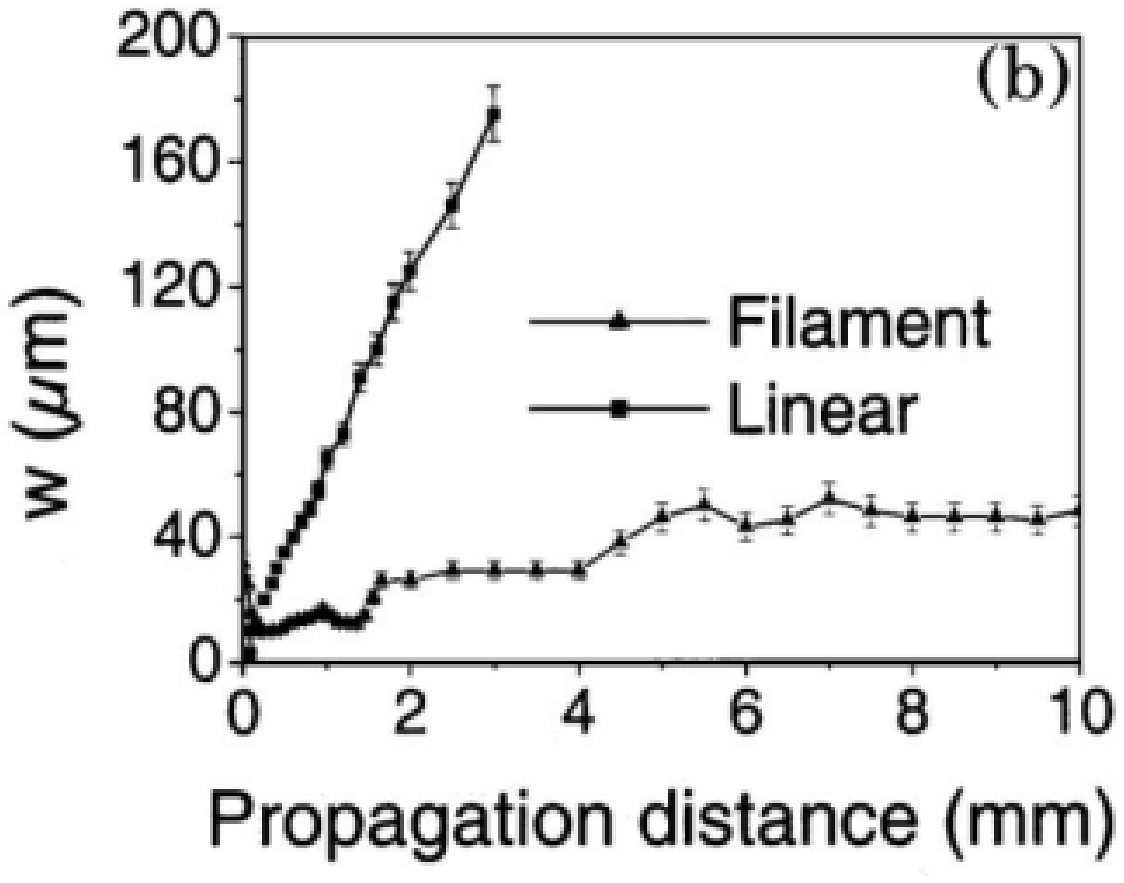}
\includegraphics[width=\columnwidth]{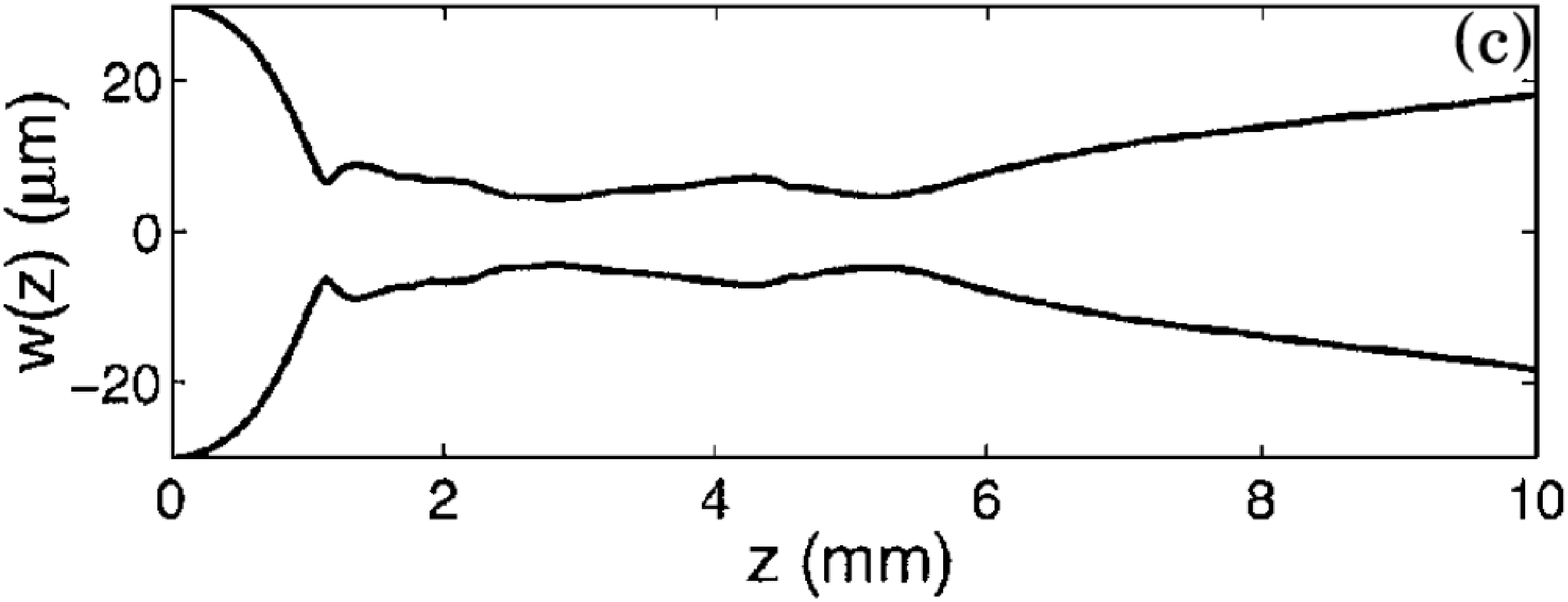}
\caption{(a) Transverse photograph of a single filament propagating 
(from left to right) in fused silica. (b) Measured
diameter of the beam comparing high (linear) and low (nonlinear) 
divergences. (c) Numerically-computed diameter of the
nonlinear filament \cite{Tzortzakis:prl:87:213902}.}
\label{fig23}
\end{figure}

Investigations mixing propagation and damage in 
dielectrics started from the early 2000's. Tzortzakis {\it et al.}
\cite{Tzortzakis:prl:87:213902} observed the 1-cm long 
self-channeling of a 160-fs focused pulse conveying 3 critical
powers at 800 nm. The pulse traveled across a fused silica
sample like a narrow waveguide with $\sim 20$ $\mu$m
waist. Local heating caused by the high repetition rate (200 kHz) and 
damage by local intensities $> 10$ TW/cm$^2$
were avoided by making the sample move in the $(x,y)$ plane. 
Auto-correlation traces revealed a two-peaked structure and spectra were asymmetrically broadened, which 
signaled a pulse splitting driven by plasma. Earlier, Brodeur 
and Chin \cite{Brodeur:josab:16:637} noticed a similar supercontinuum 
in glasses and explained it from the multiphoton excitation of electrons
into the conduction band at the focus point. The low beam divergence 
was attributed to the Kerr-lens effect. As an example, Fig.\ \ref{fig23}(a) shows a 
transverse photograph of the self-guided filament measured at input energy of 2
$\mu$J in fused silica. Fig.\ \ref{fig23}(b) details the filament waist 
in nonlinear propagation regime and Fig.\ \ref{fig23}(c) reproduces
this waist numerically computed from Eqs.\ (\ref{modeleq}).

At higher local intensities, the breakdown limit is 
approached. Because the plasma generated during breakdown remains at 
the threshold, a precise control of the interaction region can be 
reached with ultrashort pulses. This property can be used
for material processing, medical laser applications as well as 
solid-state microelectronics. The major qualitative differences 
between
damages caused by short ($< 10$ ps) and longer pulses ($> 50$ ps) 
appear in the damage morphology: Short pulses ablate the material, 
whereas long pulses produce conventional melting, boiling and 
fracture. With tightly focused fs pulses, permanent birefringent structures 
embedded in bulk fused silica have been realized over microscopic dimensions 
\cite{Sudrie:oc:171:279,Sudrie:oc:191:333}.
Arrays of parallel grooves formed 
transmission diffraction gratings with periodically-modified index changes. 
Two distinct types of damages can be created, namely, those consisting 
of local isotropic increase of the refractive index at subcritical 
powers, and those inducing local 
birefringence at powers above critical. 
The damage track consists of diffusing zones of $\sim 20$ $\mu$m 
transverse width. The total 
damage region can extend over a distance reaching 80 $\mu$m along the 
optical path. Numerical simulations accounting for photo-ionization,
avalanche, electron recombination and Kerr self-focusing reproduced experimental damage tracks 
\cite{Sudrie:prl:89:186601,Sudrie:oc:191:333}, along which electron 
densities up to $3 \times 10^{20}$ cm$^{-3}$ resulted from the combined actions of 
photo- and impact ionization. Accounting moreover for Ohmic heating and 
thermal cooling by collisional ionization, Pe{\~n}ano {\it et al.}
\cite{Penano:pre:72:036412} numerically solved the 1D electromagnetic 
wave equation to quantify the transmission, reflection and absorption
of 100-fs laser pulses by a thin plasma layer 
formed at the surface of a dielectrics. With a fluence of 3.2 J/cm$^2$, the
interaction fully ionizes the surface and heats the plasma to $\sim 
10$ eV. A significant transmission up to $30 \%$ of the laser energy
is possible even when the plasma density is above critical 
$(\omega_{\rm pe} > \omega_0)$. In the layer, the 
electron density rapidly becomes supercritical and implies a sharp rise in the
reflected pulse amplitude. Compared with MPI alone, collisional 
ionization increases the peak plasma density by $30 \%$.

\subsection{Pulse propagation in liquids and applications}

Similarly to LIB in solids, the optical breakdown in 
fluids gained considerable interest, because it finds various 
therapeutic  
applications for, e.g., plasma-mediated laser surgery and prevention 
of ocular damages. First simulations in water  
\cite{Feng:ieeejqe:33:127} emphasized the dominant role
of cascade ionization for long pulses,  
while multiphoton ionization was expected to  
prevail for shorter pulses. Further studies 
\cite{Noack:ieeejqe:35:1156} evaluated the absorption coefficients during the LIB process for  
100 ns down to 100 fs pulses.  
It was established that a critical density threshold for LIB was  
$\rho_{\rm cr} = 10^{20}$ cm$^{-3}$ for long (ns) pulses, but  
$\rho_{\rm cr} = 10^{21}$ cm$^{-3}$ for short (ps) ones. In  
water and for 100-fs durations, LIB results in bubble 
formation  
supported by thermoelastic effects at pulse powers above the  
self-focusing threshold. This process also  
causes refractive index changes in the beam path upstream over  
several hundreds of microns. Within water droplets the tight focusing 
and the nonlinearities of the  
LIB process moreover generate a nanosized plasma hot enough to emit 
in the  
visible, and preferentially in the backward direction  
\cite{Favre:prl:89:035002}.

The first experiments on fs pulse self-guiding versus the optical 
breakdown  
limit in water were performed by Liu and co-workers for different 
focusing geometries \cite{Liu:apb:75:595}. It was found that the 
shorter the  
focal length, the larger the transverse size of the optical  
breakdown plasma is. Self-focusing drives the  
initial localization of the plasma towards the beam axis. White light 
is generated along a  
short ($< 1$ mm) filament and deflected at a small constant angle 
only. The 
supercontinuum sources for the high-frequencies were identified in 
the rings formed by plasma defocusing and amplified by the back of  
the pulse, where shocklike dynamics  
blueshift the spectrum 
\cite{Kandidov:qe:34:348,Kandidov:apb:77:149}.

\begin{figure}
\includegraphics[width=0.6\columnwidth]{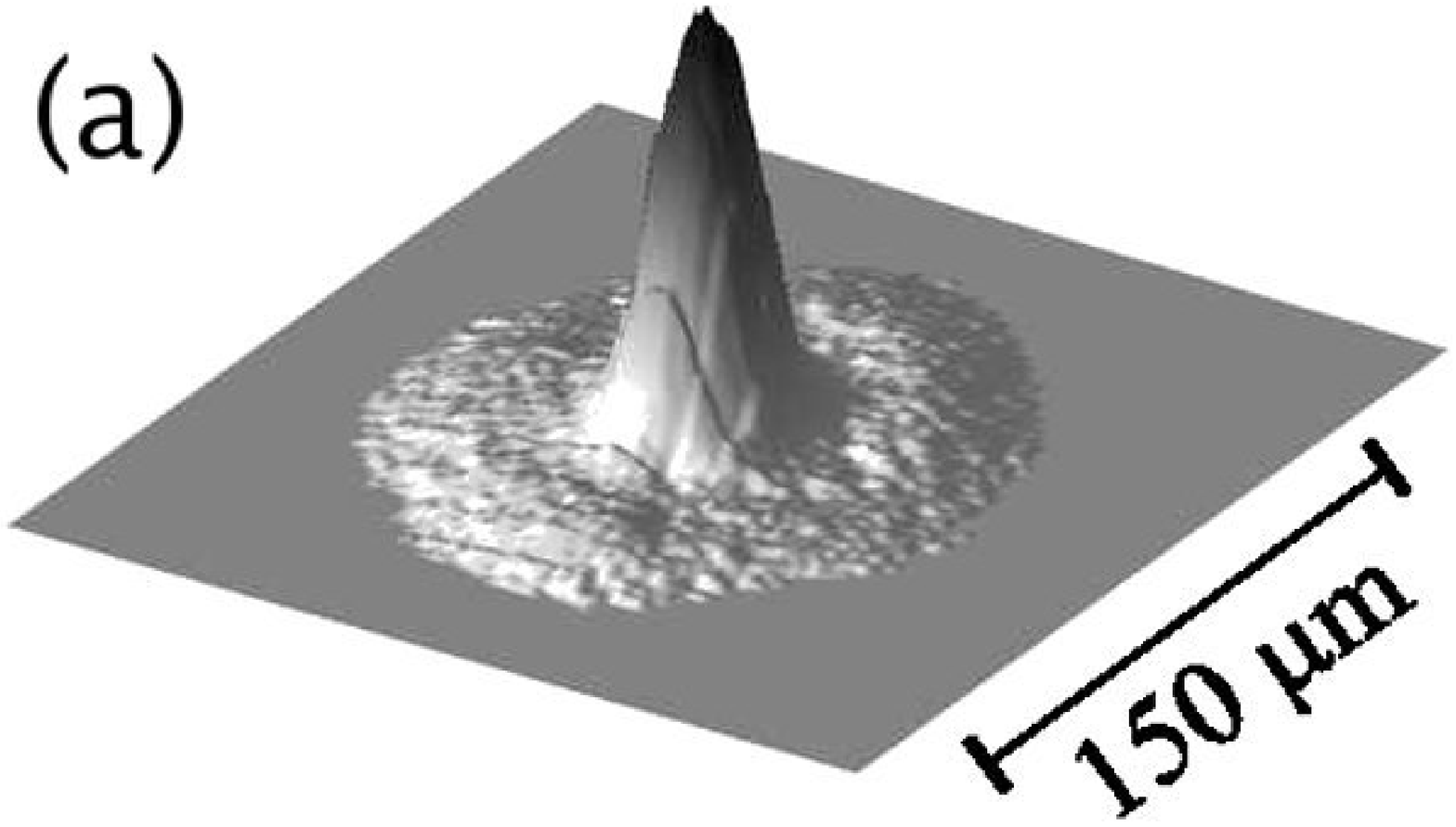}
\includegraphics[width=\columnwidth]{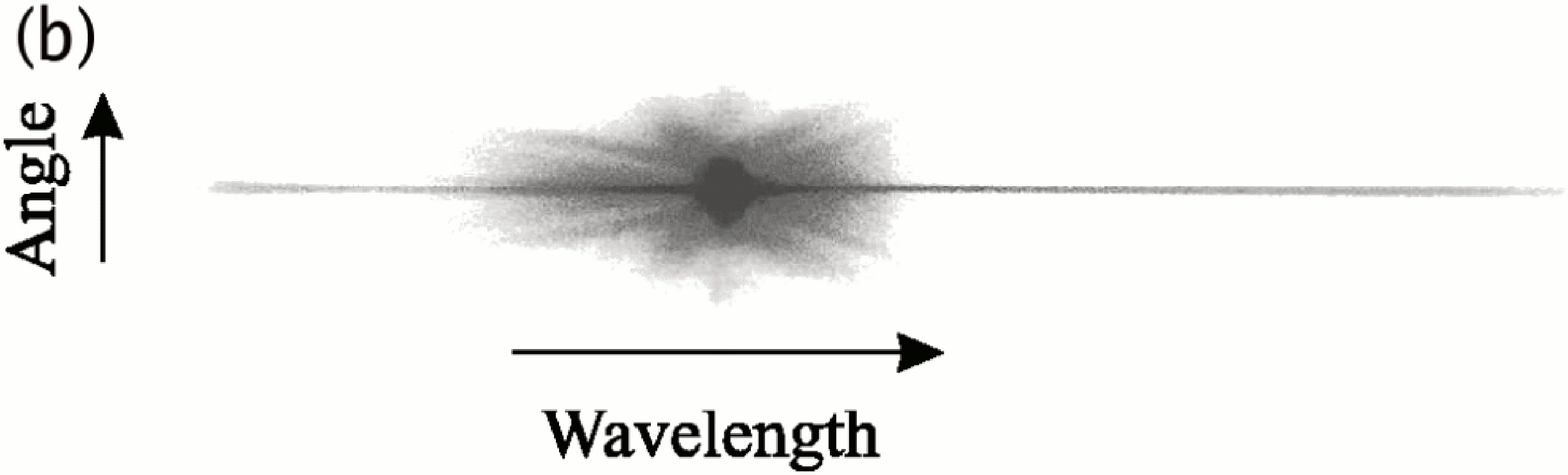}
\includegraphics[width=\columnwidth]{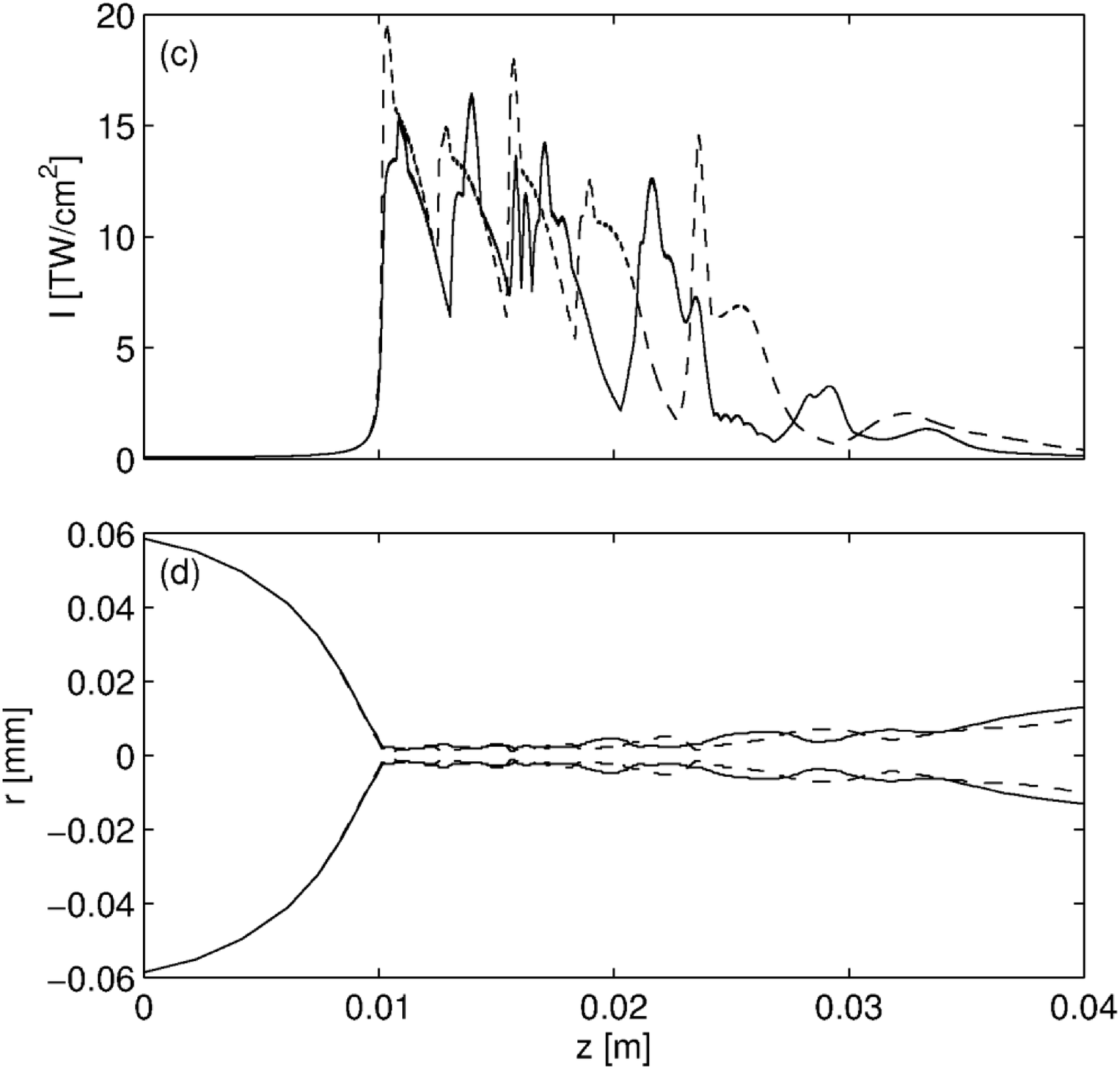}
\caption{(a) Normalized 3D beam profile at the output of the water  
cell (31 mm); (b) Spectral-angular distribution of the radiation  
\cite{Dubietis:ol:28:1269}. Courtesy of A.\ Dubietis. (c) Peak intensities computed numrically with (solid curve) 
and  
without (dashed curve) plasma gain from 170-fs, 100-$\mu$m-waisted  
pulses focused with $f = 5$ cm ($P_{\rm in} = 10 \,P_{\rm cr}$) in 
water;  
(d) Corresponding FWHM radius.}
\label{fig24}
\end{figure}

Far below LIB limits, the self-guided propagation of femtosecond filaments in water was  
thoroughly examined by Dubietis, Di Trapani and their collaborators 
a few years ago  
\cite{Dubietis:ol:28:1269,Dubietis:prl:92:253903,Porras:prl:93:153902}.  
By launching a $\sim 3\,\mu$J, 170-fs clean beam with $\sim 
100\,\mu$m  
FWHM diameter onto a water-filled cuvette in loosely focused ($f 
\geq  
5$ cm) geometry, a single filament formed at the wavelength of 527 nm 
and was capable of covering up to 4 cm along the propagation axis,  
while keeping a mean FWHM diameter of a few tens of $\mu$m. The  
experiment revealed that the filament dynamics was not sustained by 
a  
balance between Kerr-induced self-focusing and plasma-induced  
defocusing. Numerical simulations outlined, instead, the spontaneous  
reshaping of the beam into a Bessel-type $X$-wave fulfilling the  
requirements of minimum nonlinear losses, maximum stationarity and  
localization. This scenario has been discussed in Sec.\ \ref{sec4a}, where the important role  
of the laser wavelength in selecting the key player able to saturate  
the wave collapse and support the self-guiding process was underlined. 
For this purpose, Fig.\ \ref{fig24}(a) illustrates the pulse fluence 
along  
the self-guiding range measured at the exit plane of a 31-mm long 
cuvette for 
$P_{\rm in} =  6.6 \,P_{\rm cr}$. Fig.\ \ref{fig24}(b) shows the 
$X$-shaped  
spectral-angular distribution of the radiation. Plots (c,d) represent 
peak  
intensities and FWHM diameters reached with similar pulse parameters  
with and without plasma gain, which evidences the weak influence of  
plasma defocusing in this dynamics  \cite{Skupin:pra:74:043813}.

Increasing more the pulse power leads to multiple filamentation, 
which has been addressed in a few papers
\cite{Cook:apl:83:3861,Schroeder:oe:12:4768,Heck:oc:259:216}. By 
using  
a cylindrical planoconvex lens, Cook {\it et al.} produced  
horizontal arrays of stable white-light filaments in water at 800 
nm,  
allowing interference effects between neighboring cells. Similarly to 
a  
pair of Young's slits, a filament pair creates interference 
patterns,  
which is the signature for a constant phase relationship between the 
supercontinua  
generated by the filaments (the same property applies to atmospheric filaments; see Sec.\ \ref{sec7a4}). Those thus appear as coherent  
sources of white-light. Besides, Schroeder {\it et al.} 
demonstrated the possibility to arrange filaments in water into 1D 
arrays of parallel,  
non-overlapping spots by clipping the impinging laser beam by a slit  
aperture built from razorblades. Metallic wire meshes 
can also be used to generate space-controlled 2D arrays of filaments. Recently, Heck 
{\it  
et al.} demonstrated the efficiency of an adaptive control over the  
position and extent of filaments in water tanks, through a closed  
feedback loop setup employing a spatial light modulator and a 
genetic  
algorithm that allow to manipulate the amplitude and phase of the 
input  
pulse.

\begin{figure}
\includegraphics[width=\columnwidth]{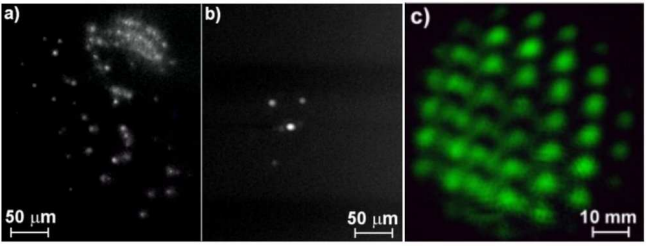}
\caption{Filamentation pattern obtained from a 1-cm long cell of (a)  
pure ethanol, (b) dilute solution of ethanol/Coumarin 153 at 4 g/l, 
(c)  
far-field fluence distribution of excited  
states of Coumarin emitting simultaneously in phase.}
\label{fig25}
\end{figure}

Apart from water, other fluids can support femtosecond  
filamentation, such as alcohols like ethanol or methanol.  
Dyes may be introduced into these liquids, in 
order to visualize the filamentary evolution through one,  
two or three-photon fluorescence and modify the multifilamentation patterns by varying the dye concentration  
\cite{Liu:oc:225:193,Schroeder:oc:234:399,Liu:oe:13:10248,Guyon:pra:73:051802}. 
Relying on the dye nonlinear absorption, the structural changes of 
the  
filamentation becomes visible to the eyes, due to the fluorescence 
from  
the dye molecules excited by multiphoton excitations at visible 
wavelengths. From a dilute solution of methanol and Rhodamine B, 
Schroeder and Chin  
figured out that femtosecond filaments propagate straight, may die off 
prematurely or fuse into new spots. Along the propagation axis, 
''mature'' filaments ending after the bright fluorescence zone were 
identified. By means of the same techniques, Liu 
{\it et al.} observed multi-focusing events by increasing the beam 
energy in  
methanol doped with $0.13 \%$ of Coumarin 440 at 800 nm. The  
photon bath surrounding the 20-$\mu$m large filament core was  
numerically examined: It was found that the near-axis region takes energy from a ring-shaped region limited to $r <  
60\,\mu$m around the filament core. Outside, the peripheral domain ($r > 
60\,\mu$m) acquires energy. The resulting energetic balance preserves 
an almost constant energy in the  
near-axis region. Nonlinear fluorescence  
techniques were recently used to discriminate between multiphoton  
absorption and conical emission by $z$-scan analysis using metal 
meshes  
\cite{Liu:oe:13:10248}. It appears that the energy loss caused by  
conversion into CE ($40 \%$) is much higher than that caused by MPA ($\sim 3 \%$) at powers $< 8 P_{\rm cr}$ (see also Sec.\ \ref{sec4a}). Finally, Guyon {\it et al.} employed Coumarin 153 in dye-doped cells of ethanol. At high dye concentration (4 g/l), the filamentation pattern was shown to self-organize into a latticelike (hexagonal) figure, by letting the  
two-photon absorption of Coumarin switch out filamentary sites in the $(x,y)$ plane.  
Pump-dump experiments furthermore revealed that, at all filamentary  
sites, excited Coumarin molecules could coherently relax by  
fluorescence and emit in-phase light stimulated by the dump pulse,  
i.e., a collection of filaments can be used as microscopic laser  
sources in dense media. Figure \ref{fig25} summarizes some of these  
observations.

\section{Filaments in the atmosphere: Conveying intense structures 
over kilometers\label{sec7}}

The final section is devoted to the medium which originally
served as the ''birthplace'' for the science of femtosecond light filaments, namely, the atmosphere. Special emphasis is laid on experimental diagnostics. From the theoretical point of view, 
filamentation in air is modeled from Eqs.\ (\ref{modeleq}), using classical dispersion relation in air \cite{Peck:josa:62:958}. 
Although of weaker percentage ($20 \%$ vs $80 \%$ in air), dioxygen 
molecules have a lower ionization potential than nitrogen and they 
provide the dominant species prevailing through ionization \cite{Couairon:josab:19:1117}.

\subsection{Long-distance propagation and white-light supercontinuum}

\subsubsection{Temporal chirping and spatial lensing}

\begin{figure}
\includegraphics[width=0.50\columnwidth]{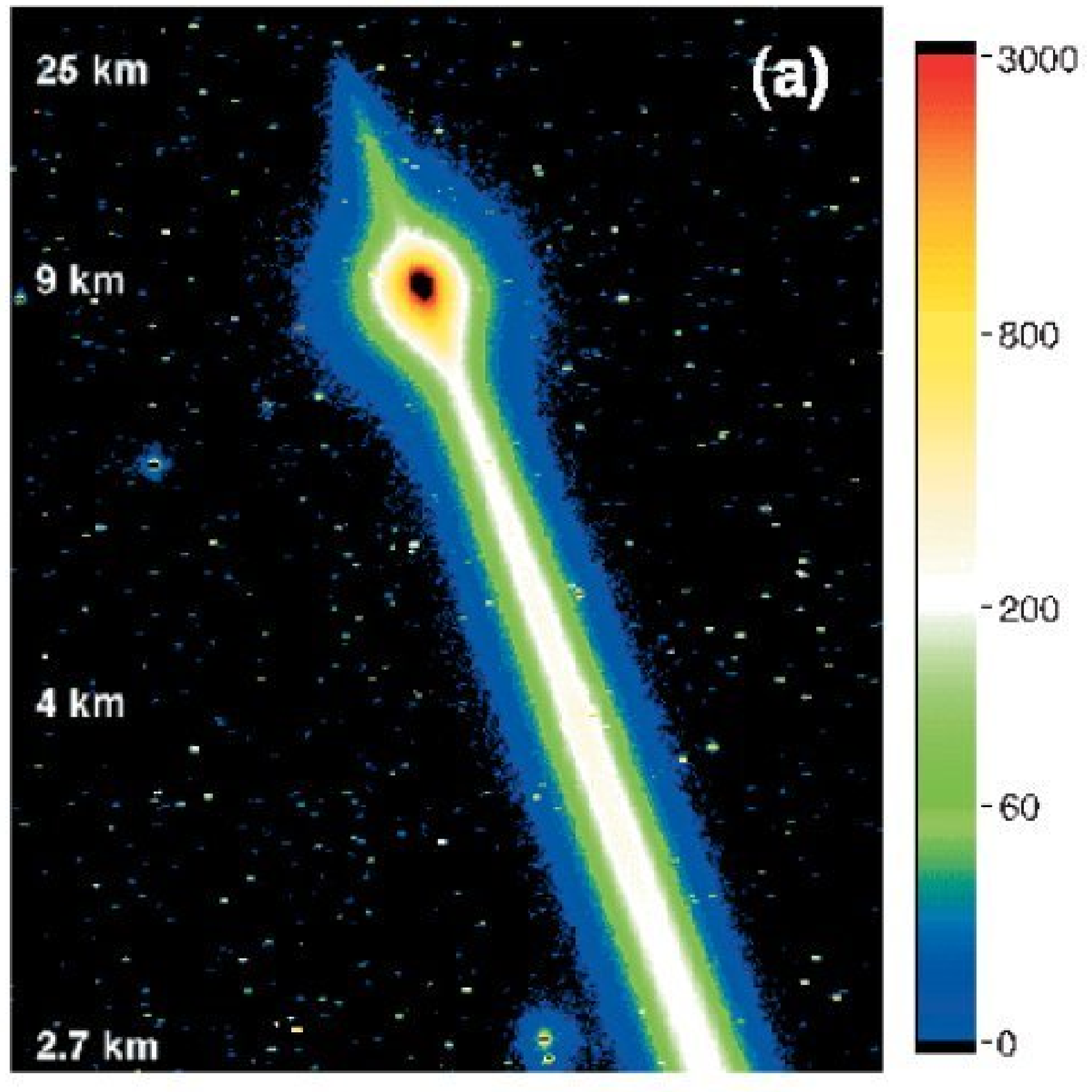}
\includegraphics[width=0.48\columnwidth]{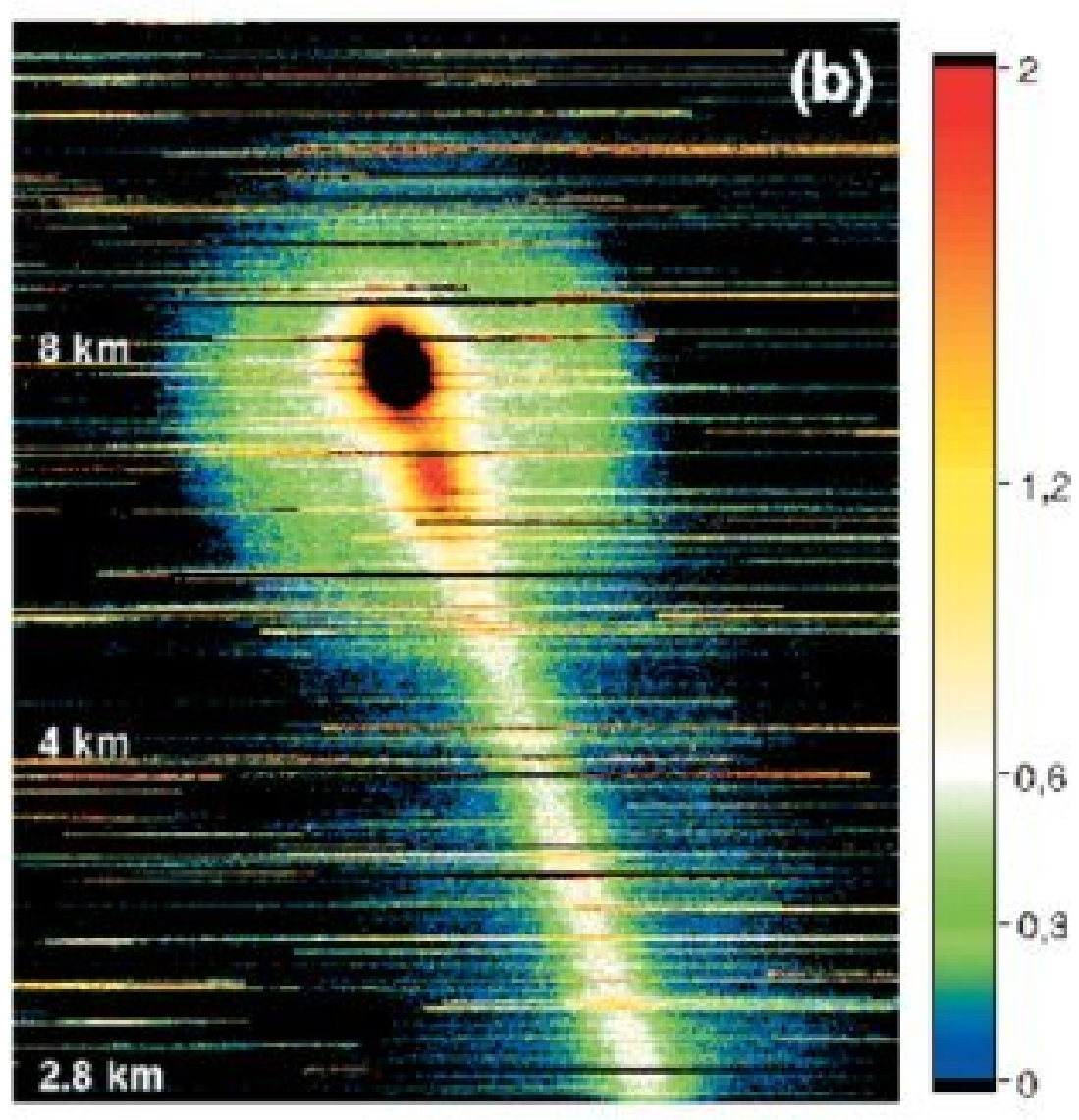}
\caption{Typical fs beam image of the Teramobile laser beam from 
Tautenburg observatory. (a) Fundamental wavelength; (b) Blue-green band of the continuum. The horizontal stripes across the pictures come from stars passing through the telescope field of view. Note the strongly nonlinear altitude scale 
due to triangulation.}
\label{fig41}
\end{figure}

The propagation of high-power (TW) femtosecond laser pulses in air 
has attracted considerable attention from the pioneering
observation of the white-light supercontinuum beyond 10 km 
\cite{Woste:lo:29:51}. A number of important practical applications 
have been suggested, including remote sensing \cite{Rairoux:apb:71:573}, 
directed energy delivery and artificial lightning
\cite{LaFontaine:ieeetps:27:688} among others. Because the filament onset and length are key parameters 
for spectroscopic measurements and for depositing high intensities on remote targets, 
monitoring these parameters are of utmost importance. To achieve high intensities at 
remote distances, a negative frequency chirp can be introduced in the laser pulse \cite{Alexeev:apl:84:4080,Wille:epjap:20:183}. In addition to transverse self-focusing, the pulse undergoes a temporal compression as it compensates the normal group-velocity dispersion along the 
propagation axis. Chirping effects can be measured by evaluating differences in the conical emission with and without pulse chirping \cite{Rodriguez:pre:69:036607}.
For instance, the vertical propagation of the Teramobile beam, presented in Appendix \ref{appC}, was examined from the ground using the 2-m astronomical telescope of the Th{\"u}ringer Landessternwarte (Thuringia State Observatory, Germany). 
Collected images were a combination of both (i) cross-section images of the 
beam impinged on the bottom of clouds or haze layers acting as 
screens, and (ii) side imaging of the Rayleigh-scattered light from 
the beam over large altitude ranges. Some of these images are shown 
in Fig.\ \ref{fig41} at both the fundamental wavelength and the 
white-light continuum. In this case, the altitude was retrieved using triangulation. Figure \ref{fig41}(b) demonstrates the efficiency of 
the supercontinuum generation, since its blue spectral 
signal has been detected from altitudes beyond 18 km for appropriate chirping. These experiments demonstrated for the first time the possibility to deliver high-intensities and generate white light upon variable km distances, by means of GVD/chirp precompensation techniques. The same 
observation assessed that the conical emission 
bears two thirds of the overall white-light energy, while the on-axis, forward-directed central component carries the remaining one third.

To modify the filamentation distance, a chirp is usually introduced by changing the distance 
between the gratings of CPA laser 
compressors. Applied to the Gaussian pulse (\ref{incond}), this 
technique modifies the second-order phase contribution $\varphi'' = d^2 \varphi/d \omega^2|_{\omega = \omega_0}$. It enhances the input pulse duration and decreases the beam power as
\begin{equation}
\label{tpc}
t_p^C = t_p^{C=0} \sqrt{1+C^2},\quad P^C =P^{C=0}/\sqrt{1 + C^2}
\end{equation}
at equal spectral content. As it linearly propagates, 
the pulse competes with normal GVD and reaches the 
minimal duration $t_{\rm min} = t_p^{C=0}$ at the distance
\begin{equation}
\label{zmin}
z_{\rm min} = \frac{|C|}{1+C^2} \frac{t_p^2}{2 k''}.
\end{equation}
Pulse chirping mixes two essential modifications.
The first one is dictated by the phase $\sim 
\mbox{e}^{-iCt^2/t_p^2}$. 
For input powers above $P_{\rm cr}$, a positive chirp $C>0$ delays 
the 
occurrence of the self-focus point, whereas this occurs earlier along 
the 
optical path for a negative chirp $(C < 0)$. The second one results 
in diminishing the effective pulse power, 
which pushes the first focus to later propagation distances and 
reduces the number of filaments in case 
of multifilamentation, whatever the sign of $C$ may be. In usual experimental conditions, this second effect prevails over the first one. Despite the strong nonlinearities driving the filament dynamics, the 
chirped-induced linear compression stage still persists with $C < 0$. 
It gives rise to additional focusing events and keeps the beam 
localized at distances close to $z_{\rm min}$ 
\cite{Nuter:ol:30:917,Golubtsov:qe:33:525}. Moreover, the generation 
efficiency of the supercontinuum may vary by several orders of magnitude compared with that of a transform-limited pulse.

\begin{figure}
\includegraphics[width=\columnwidth]{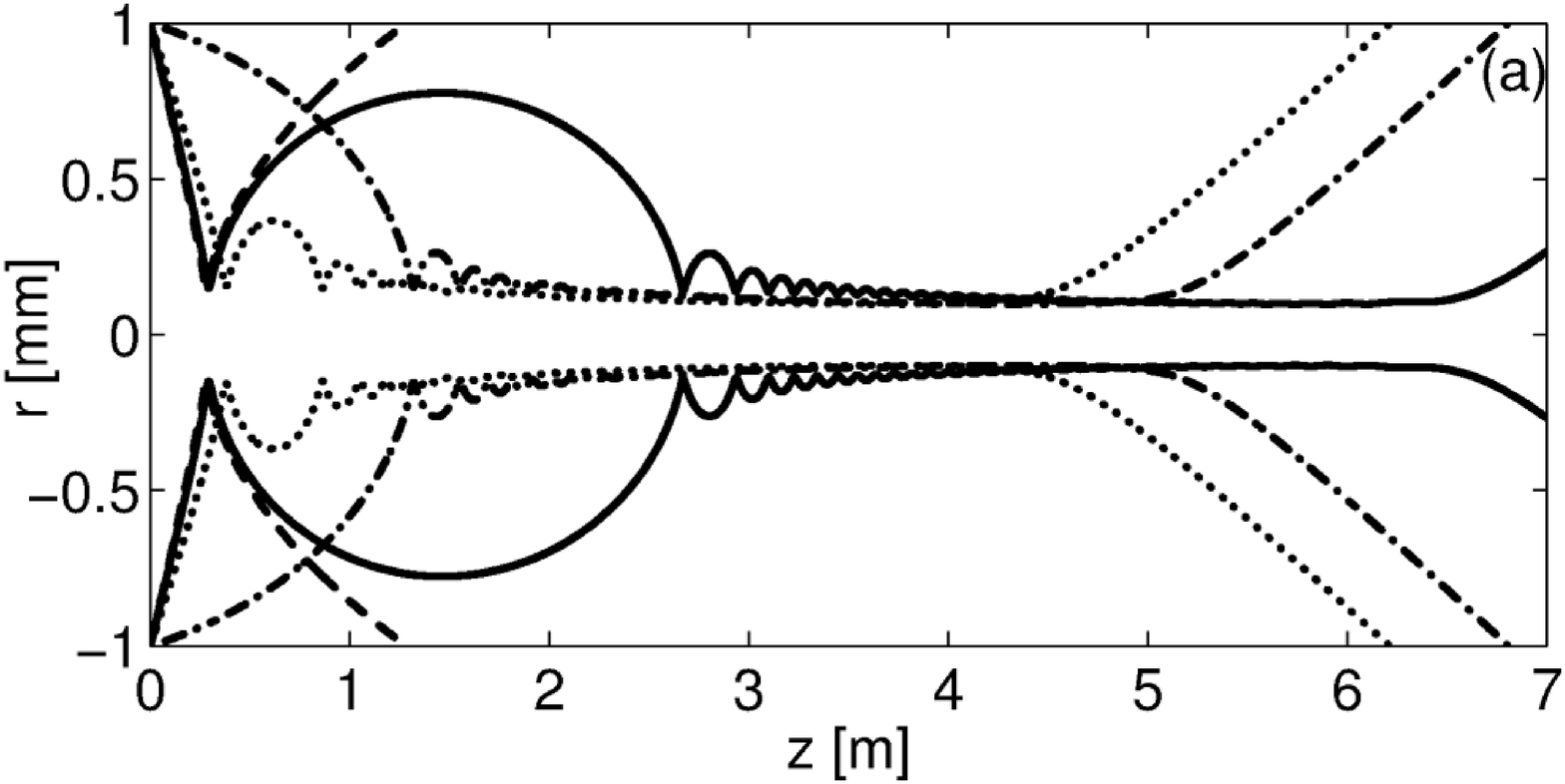}
\includegraphics[width=\columnwidth]{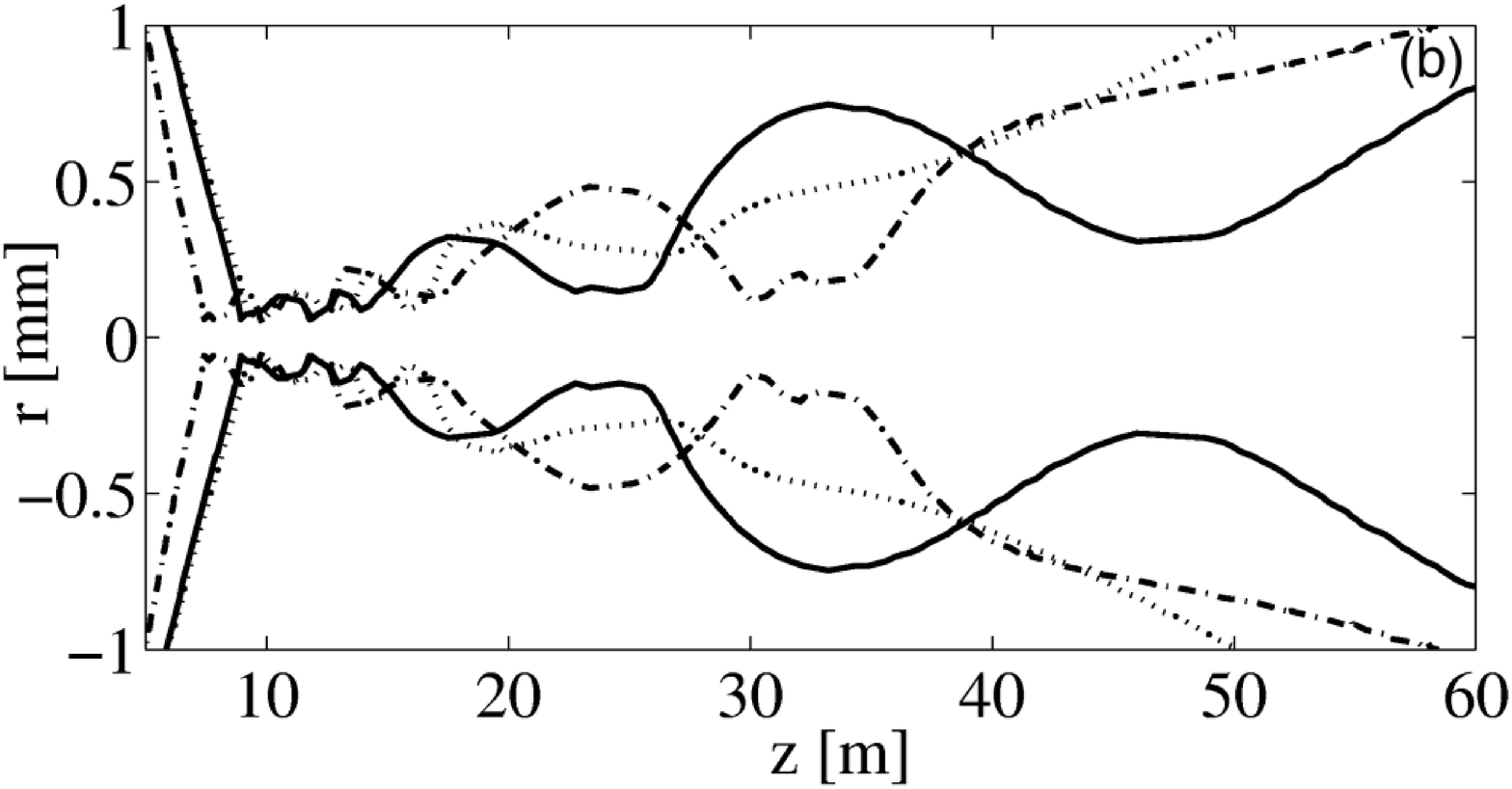}
\caption{(a) Beam diameter of pulses spatially focused with $w_0 = 
1$ mm, $f = 35$ cm (dashed curve), $f = 37$ cm (solid curve), $f = 50$ cm (dotted curve), $f = 5$ m (dash-dotted curve); (b) Beam diameter vs. 
$z \geq 5$ m for temporally-chirped, 3-mm-waisted pulses with $C = 0$ 
(dash-dotted curve), $C = 1.02$ (dotted pulse) and $C = -1.02$ (solid 
curve).}
\label{fig26}
\end{figure}

The onset distance and longitudinal extent of fs filaments are also conditioned 
by the initial spatial focusing geometry. No simple analytical rule exists on 
this point, because of the complex spatial distortions destroying the 
initial homogeneity of the beam. The filament length moreover depends 
on the energy consumed by plasma excitation and on the accessible 
peak intensity. However, most of the experiments emphasize the use of rather large ratios $f/w_0$, in order to avoid an immediate plasma
defocusing in tightly focused configuration. This property can be 
refound by integrating the dynamical equations (\ref{VA}) derived from the two-scale variational method. The variational principle indicates that at given $P_{\rm in}/P_{\rm cr}$ the ratio $f/z_0$ must be large enough to insure the self-trapping condition $z_c \leq f$. Otherwise, for $f/z_0 \ll 1$, the beam diffracts just after the focal point. Figure 
\ref{fig26}(a) shows the beam diameter of 100-fs, 1-mm-waisted 
pulses in air at $P_{\rm in} = 10\,P_{\rm cr}$ and various focal lengths 
$f$, which confirms the previous belief. Figure \ref{fig26}(b) 
represents the beam extent along the $z$ axis of temporally-chirped 
pulses with waist $w_0 = 3$ mm, input energy of 1.9 mJ and same power 
ratio, for which $t_p^C = 100$ fs while $t_p^{C=0} = 70$ fs.

\subsubsection{Plasma and optical field measurements}

\begin{figure}
\includegraphics[width=\columnwidth]{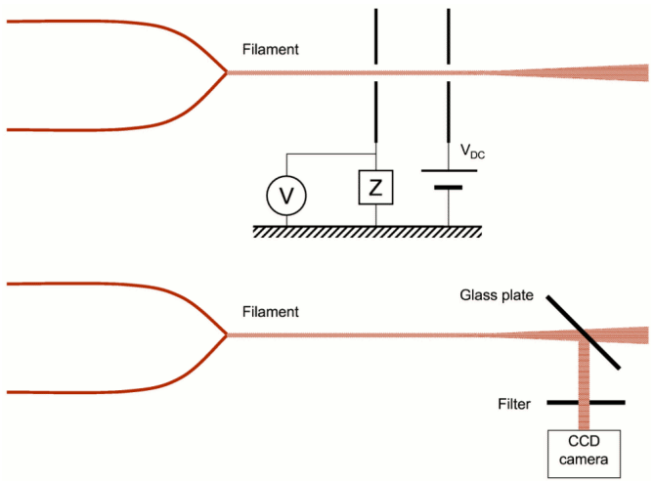}
\caption{Schemes of experimental setups for (a) conductivity 
measurements and (b) CCD imaging for optical field measurements.}
\label{fig27}
\end{figure}

Plasma detection relies on the existence of a difference of potential 
produced by the current generated by ionization
of air molecules. Working in free atmospheric medium makes it easy to measure the conductivity of the plasma channel, that causes a drastic reduction of air resistivity after the passage of a self-guided filament. This conducting
column can directly be evidenced by letting the filament pass between 
two copper electrodes drilled in their center and between which a DC 
voltage of typically 1000 V is applied (see Fig.\ \ref{fig27}). The current circulating through the plasma column is then measured by recording the voltage induced across an external load resistance \cite{Tzortzakis:pre:60:3505,Tzortzakis:pre:64:057401}. Knowing the 
current density per ion ($i = 3 \times 10^{-14}$ A/cm$^2$) and taking 
into account the volume occupied by a plasma filament, peak electron 
densities of $10^{16}-10^{17}$ cm$^{-3}$ have been reported in air at 800 nm. 
Another technique consists in resolving in time small local changes in the
atmospheric refractive index by diffractometry 
\cite{Tzortzakis:oc:181:123}. The principle is here to use a probe 
beam that crosses the filament path under a small angle. The far 
field image of the probe forms fringe patterns, that yield a
direct measurement of the accumulated phase containing the 
plasma-induced defocusing effect. Let us also mention sonographic 
methods, that take advantage of the sound signals along the plasma 
column \cite{Yu:ao:42:7117,Hosseini:apb:79:519,Hao:cpl:22:636}. This ''acoustic'' diagnostics employs a microphone placed perpendicularly to the channel and recording the sound signals by a digital oscilloscope. The 
sound emitted from a plasma string is a portion of the pressure modulation of an acoustic wave from which the absorbed
optical energy and plasma density are deduced from the electric peak 
voltage at a given distance $z$. By doing so, the variations in the 
electron density can then be plotted along the propagation axis.

Plasma lengths attained from single $\sim$ mJ femtosecond pulses currently remain of the order of the meter. The combination of twin pulses, launched collinearly in convergent geometry and separated from each other by a suitable time delay, can, however, double this length 
\cite{Tzortzakis:apb:76:609,Couairon:oc:225:177}. This process, 
called ''concatenation'' of plasma filaments, relies on locking together the 
ionized channels generated by each individual pulse. By tuning their 
focal lenses ($f_2 - f_1 \sim 1$ m) and their time separation ($ 
\simeq t_p$), the less powerful time slice ending the self-guiding 
of the first pulse coincides with the most powerful one of the second 
pulse starting a new plasma column 
\cite{Berge:pre:69:065601}.

Experiments over a horizontal path provided more information about the length over which ionized plasma channels are formed \cite{Mechain:oc:247:171}. The influence of an initial negative chirp enlarging the pulse duration from 0.2 to 
9.6 ps, corresponding to 190 and 4 $P_{\rm cr}$ at constant energy (190 mJ), respectively, was especially examined upon distances up to 2350 m. Whereas ionization is clearly observed over propagation scales $< 50$ m for short pulse durations (0.2 ps), it becomes more and more sporadic for durations above 3 ps and ceases for 9.6-ps pulses, i.e., at powers becoming close to critical. In this case, a few low-intensity ($\sim 10^{12}$ W/cm$^2$), mm-waisted spots survive inside the photon bath, still capable of covering several hundreds of meters. The bundle does not convey enough power to trigger full ionization. Instead, the beam evolves in a regime along which GVD is able to take over MPI and sustain the beam in a confined state over distances of the order of the input Rayleigh length 
\cite{Champeaux:pre:71:046604}. When the initial power is too weak, broad beams may produce a few bright spots by modulational instability, but none of these is capable of developing extensive plasma sequences (see Fig.\ \ref{fig7}).

Several diagnostics exist for optical field measurements. The first 
consists in recording intensity profiles by a thick glass plate 
placed on the propagation axis at $\sim 45^{\circ}$ angle (Fig.\ \ref{fig27}). The weak reflection from the glass is then imaged with a high-aperture lens 
onto a linear charge coupled device (CCD) camera 
\cite{Tzortzakis:prl:86:5470}. Detection is performed out of the 
highest-intensity region close to the nonlinear focus, in order to 
leave the entrance window of the glass plate undamaged and keep up 
the reflected beam undistorted. Also, the high repetition rate of the 
laser source may not avoid multi-spot measurements. For this reason, 
Bernstein {\it et al.} \cite{Bernstein:ol:28:2354} reported on 
single-shot measurements of self-focusing pulses that do not have the 
intensity required to produce ionization. Besides a CCD camera, a spectrometer 
and second-harmonic frequency-resolved optical gating (FROG) device allowed to measure the spatial, spectral and temporal distributions of the pulse, 
respectively. The data, collected from an initially collimated 
Gaussian beam, showed spatial and temporal narrowing and spectral broadening 
at discrete energy levels preceding the ionization stage. They confirmed 
a critical power value for nonlinear compression effects of 11.5 GW for 800 nm pulses, 
i.e., about $\sim 10-15$ GW. Nowadays, several diagnostics such as SPIDER and 
crossed (X)FROG traces complete standard auto-correlation pictures and 
spectra to catch the spatio-temporal structure of 
a pulse in the $(\omega,t)$ plane. To explore the 
filamentation stage along which intensities as high as $5 \times 
10^{13}$ W/cm$^2$ are attained, Ting, Gordon and co-workers 
\cite{Ting:pop:12:056705,Ting:ao:44:1474,Gordon:ieeetps:34:249} 
elaborated on a new method following which the filament is propagated 
into a helium chamber through a nozzle that creates a sharp 
air-helium interface. Because helium has lower Kerr index and higher 
ionization potential than air, nonlinear focusing and plasma 
defocusing are arrested at the transition in the chamber. The 
filament then expands due to diffraction to larger sizes and lower 
intensities. A calibrated portion of energy can safely be collected and imaged 
either directly by a CCD array or through an imaging spectrometer.

\subsubsection{Multifilamentation}

Open-air terawatt laser facilities make it possible to observe 
optical focal spots formed by a myriad of filamentary cells for input powers 
containing several thousands of critical powers in air. The 2D 
reduced model (\ref{2D_1}) using the experimental fluence as initial 
condition actually reproduces the evolution of the multifilamentation pattern over long scales, saving computational resources when one neglects the temporal dimension. As an example, Fig.\ \ref{fig29} shows the overall envelope of the 
Teramobile bundle (see Appendix \ref{appC}) launched in parallel geometry. The initial pulse contains 700 critical powers and its transverse profile is scanned at 
different propagation distances. Filaments rise from the initial beam defects, form a crown of dots growing from the diffraction ring and then excite clusters of cells. These clusters, some of whose are identified by the labels (1), (2) and (3), are faithfully reproduced by the numerics. The filament number 
remains in the order of $P_{\rm in}/P_{\rm fil}$, where $P_{\rm fil} 
\simeq 3-5 P_{\rm cr}$ is the power in one filament. At large distances, the primary brightest spots decay into secondary filaments by exchanging power
through the energy reservoir formed by the background field 
\cite{Berge:prl:92:225002}. 

\begin{figure}
\includegraphics[width=0.49\columnwidth]{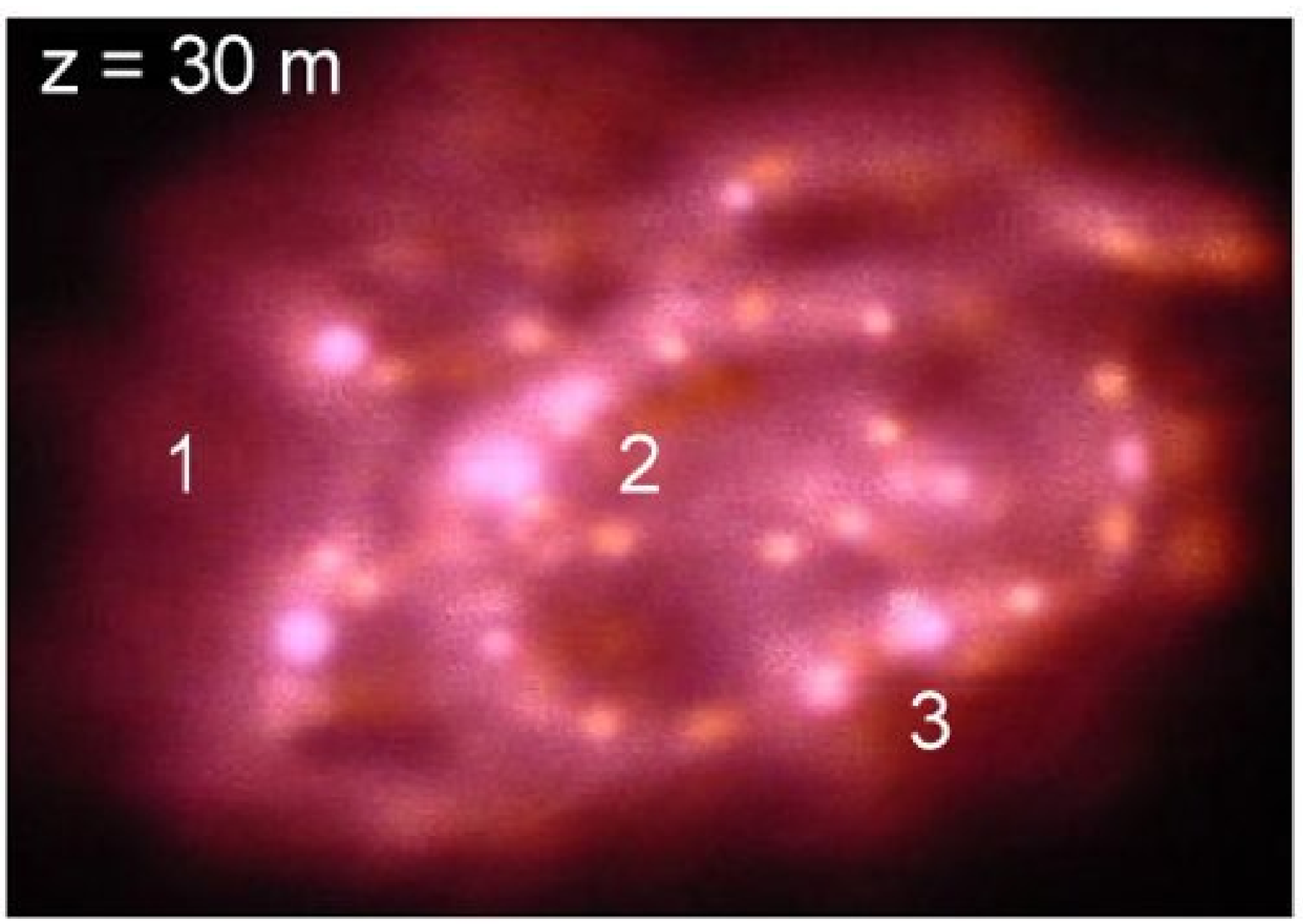}
\includegraphics[width=0.49\columnwidth]{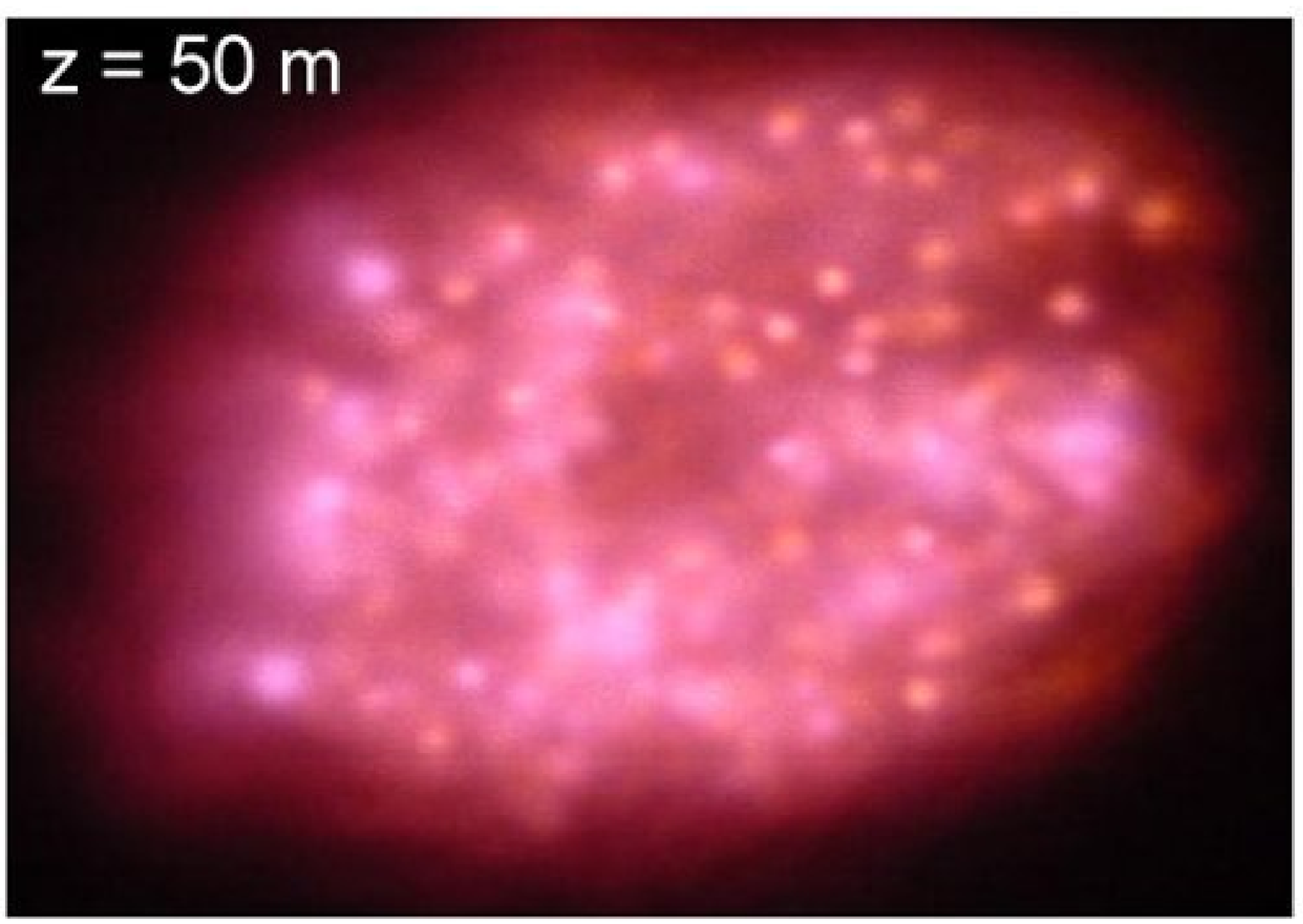}
\includegraphics[width=0.49\columnwidth]{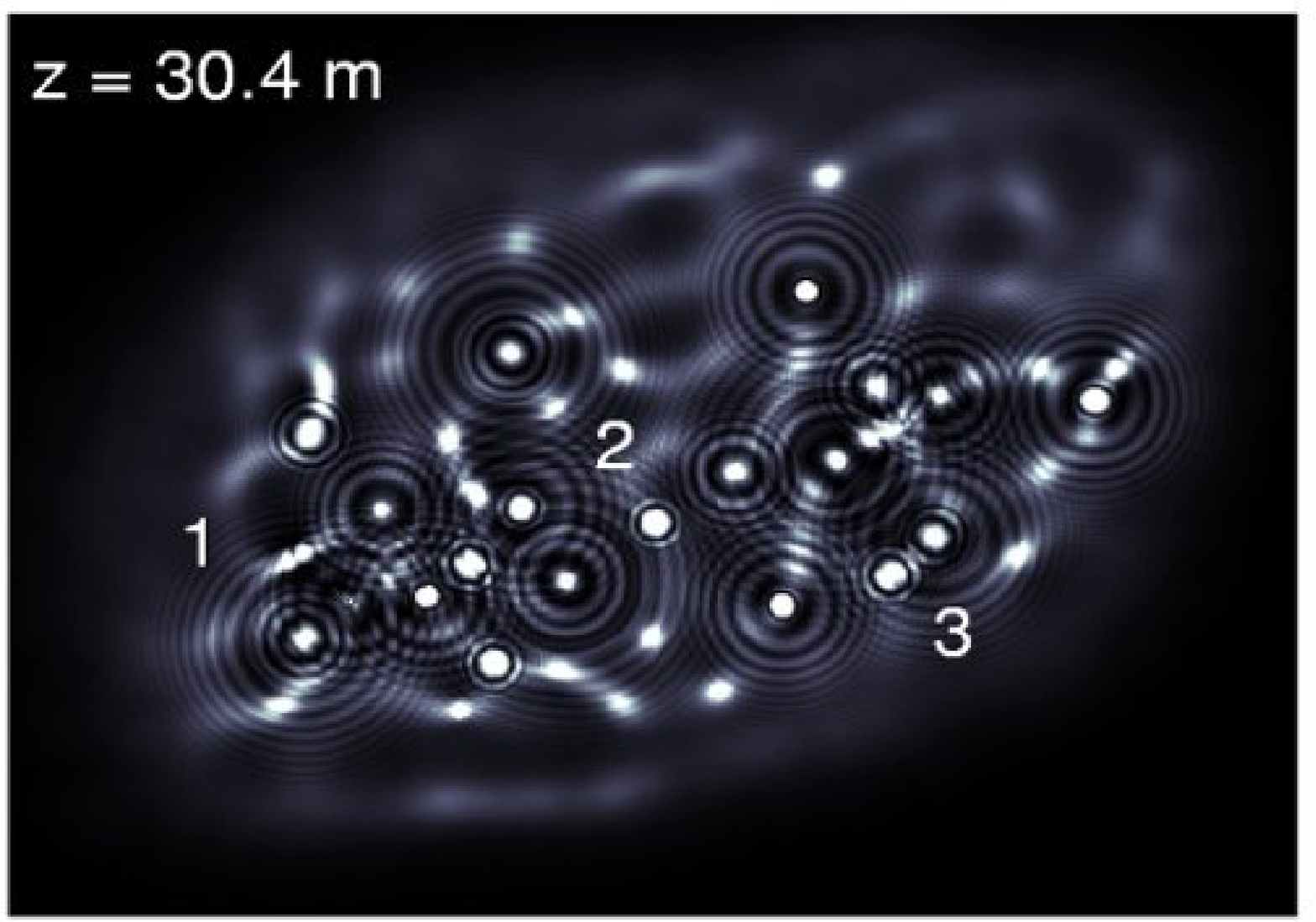}
\includegraphics[width=0.49\columnwidth]{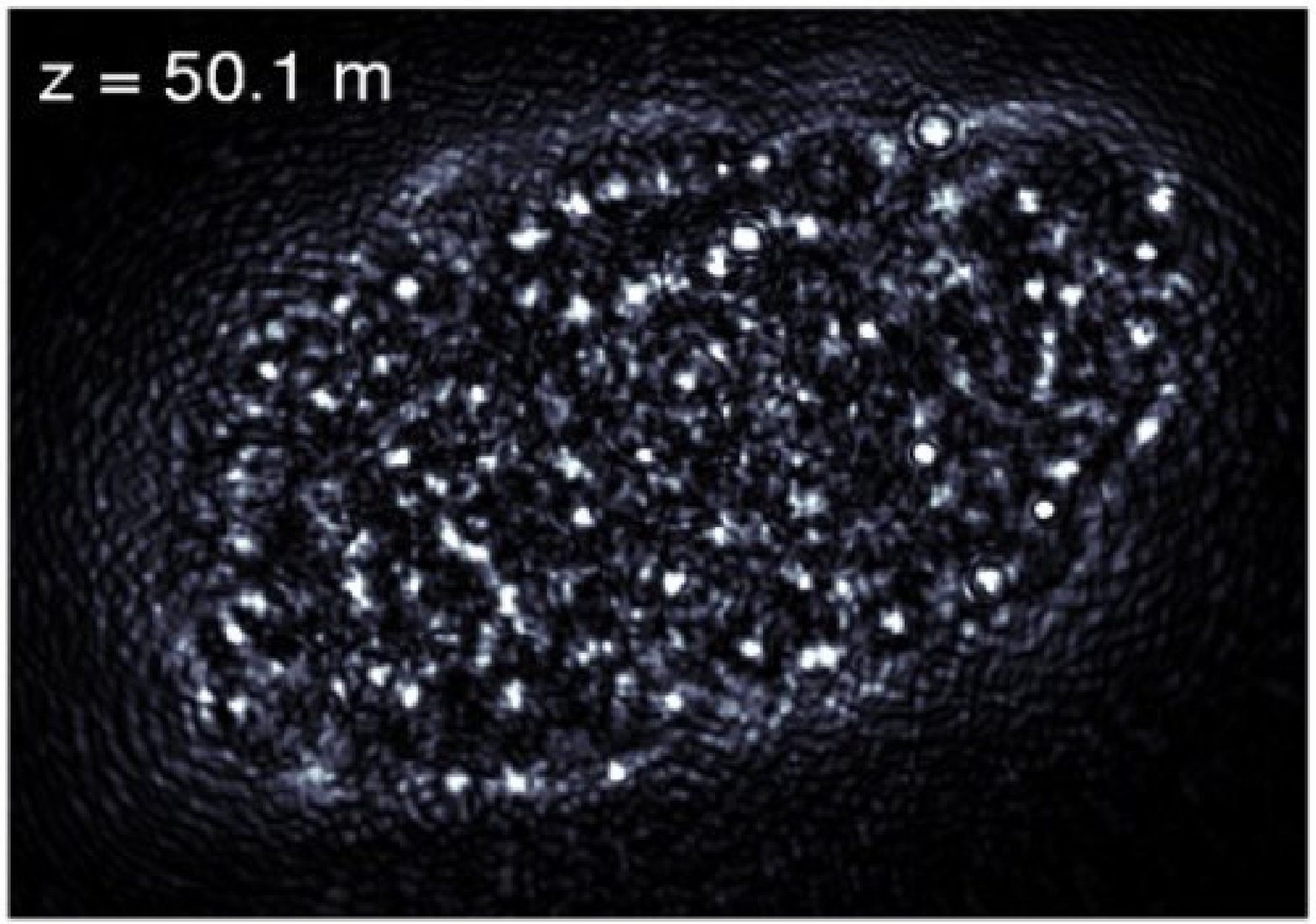}
\caption{Filamentation patterns from the Teramobile beam with $w_0 = 
2.5$ cm, FWHM duration of 100 fs and input power equal to 700 $P_{\rm 
cr}$ at different distances $z$. (top) Experiments; (bottom) 
Numerical computations.}
\label{fig29}
\end{figure}

\begin{figure}
\includegraphics[width=\columnwidth]{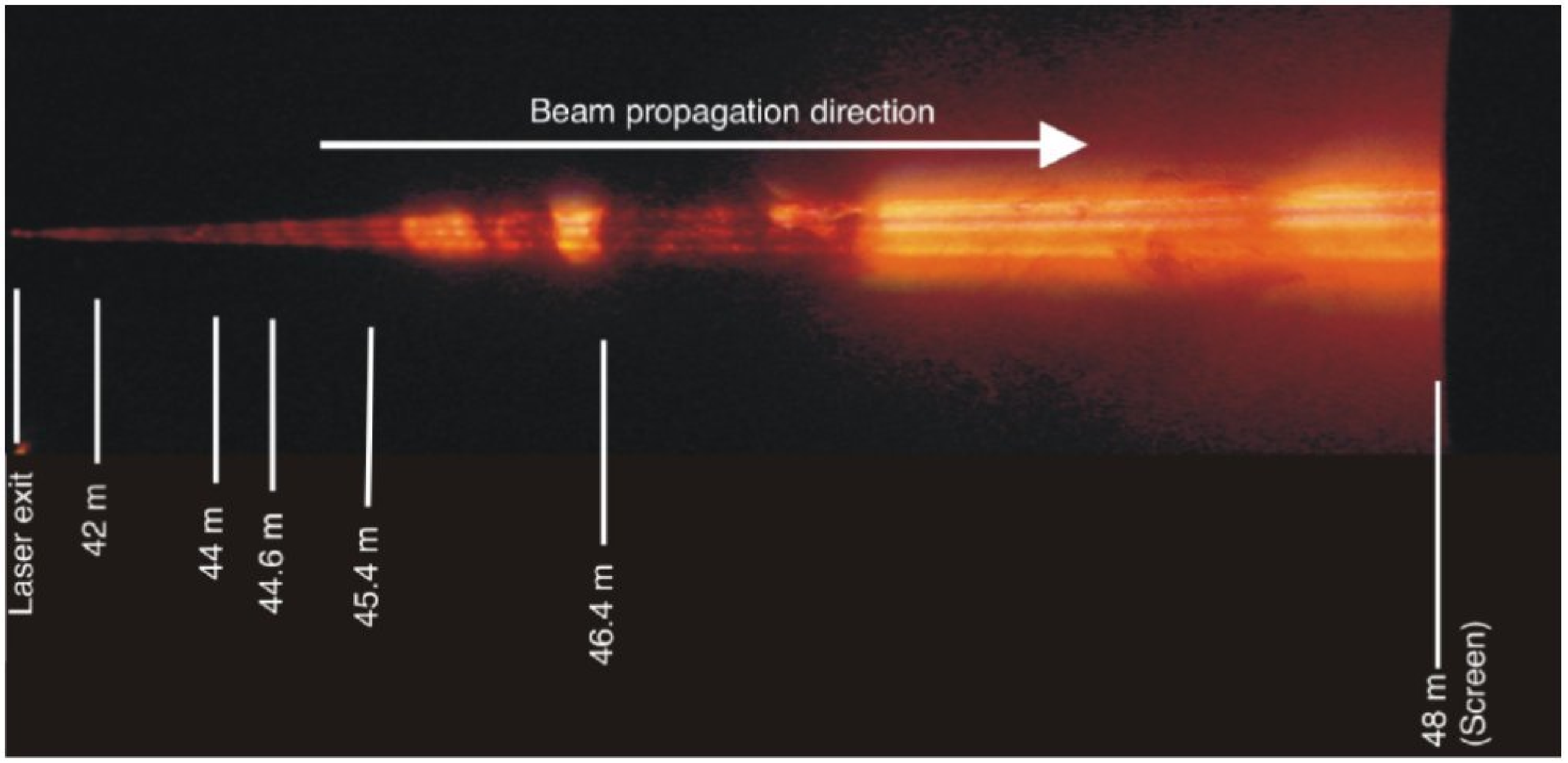}
\includegraphics[width=\columnwidth]{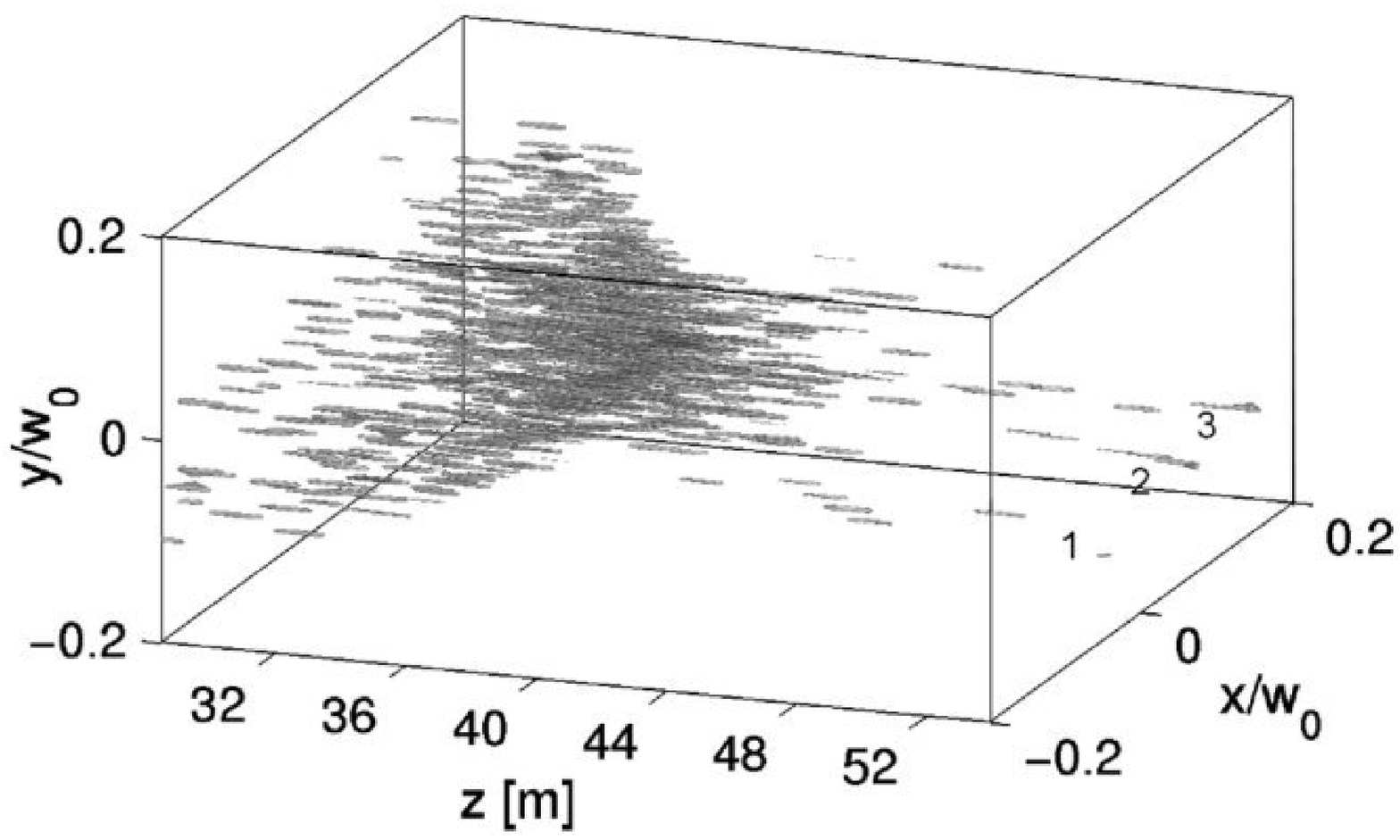}
\caption{Teramobile fluence of a focused beam $(f=40$ m) with 760 
critical powers, yielding three filamentary strings beyond the linear focal point: (top) Experiment, (bottom) numerical computation from the 2D model (\ref{2D_1}).}
\label{fig30}
\end{figure}

\begin{figure}
\includegraphics[width=\columnwidth]{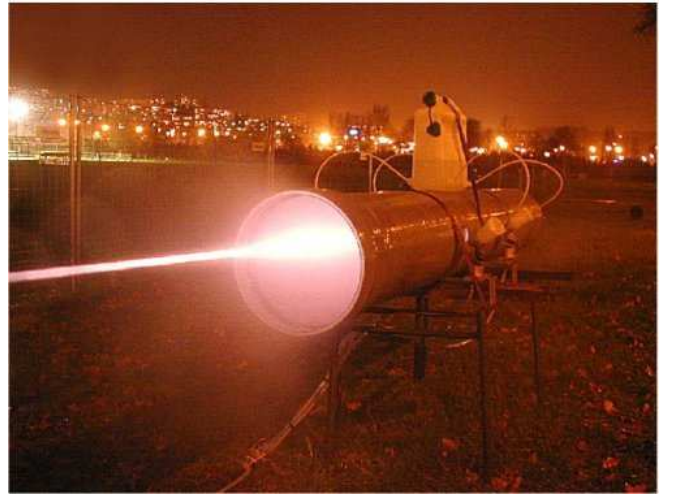}
\caption{Open cloud chamber. The cloud spans over 10 m.}
\label{fig39}
\end{figure}

\begin{figure}
\includegraphics[width=0.49\columnwidth]{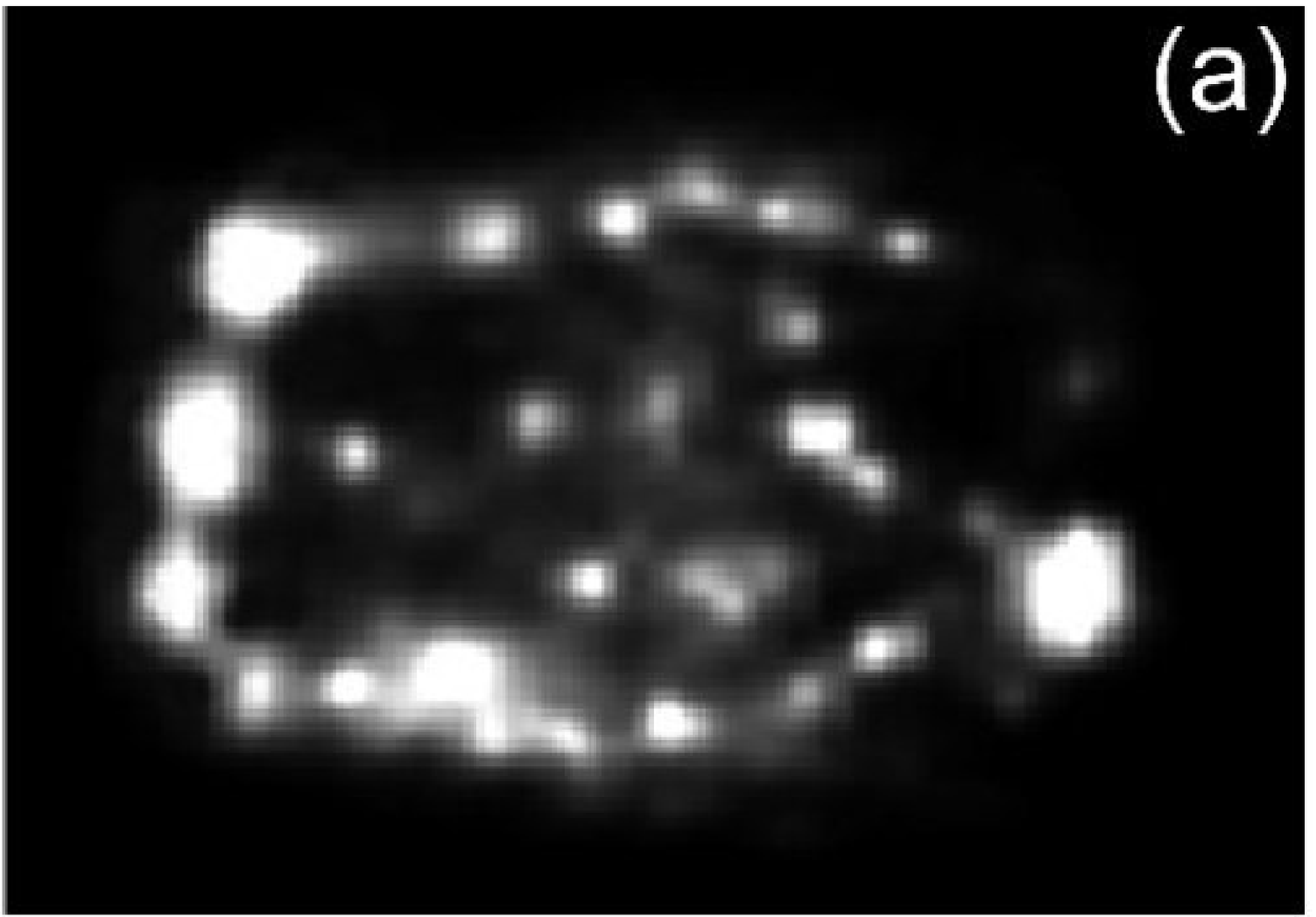}
\includegraphics[width=0.49\columnwidth]{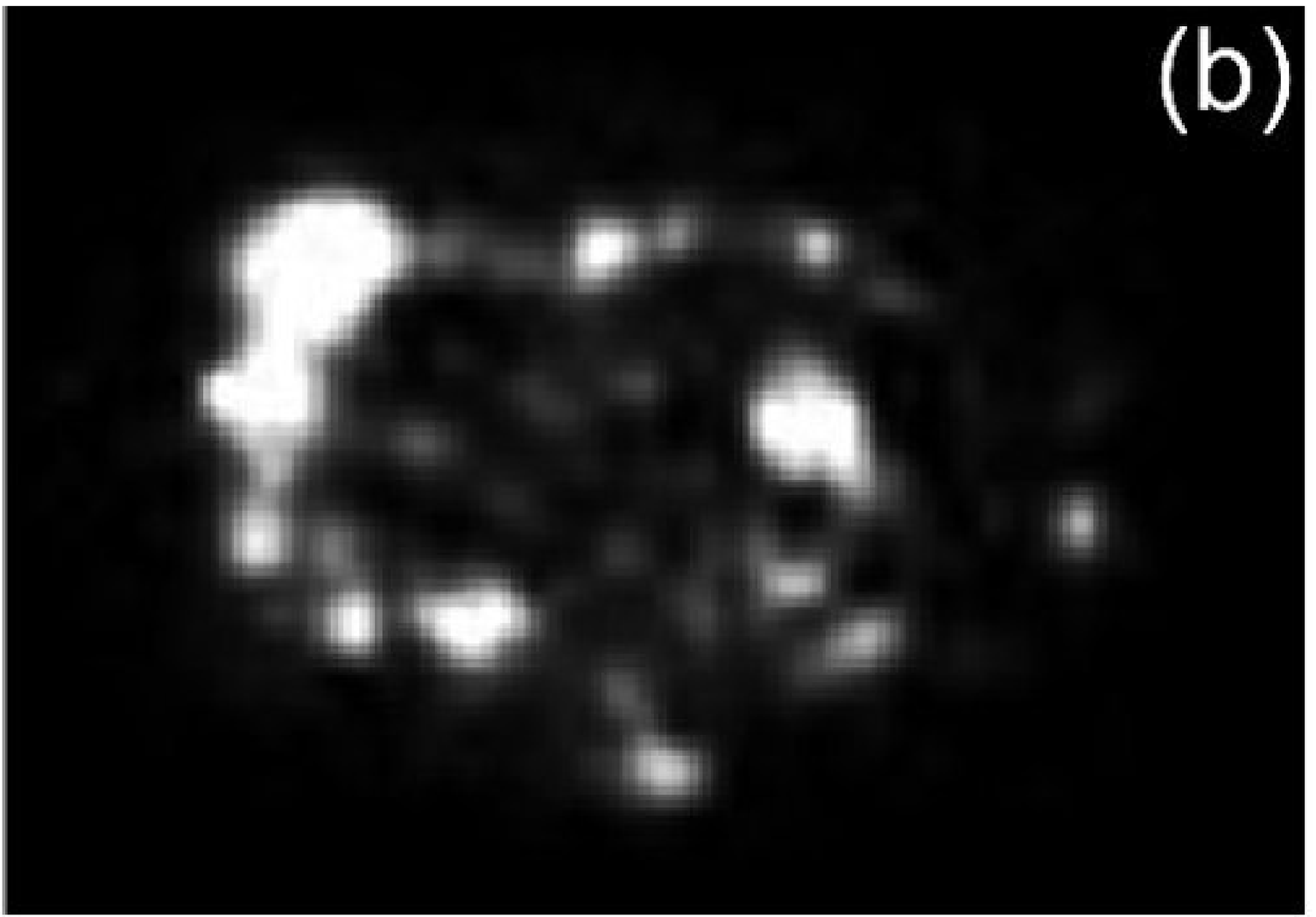}
\includegraphics[width=\columnwidth]{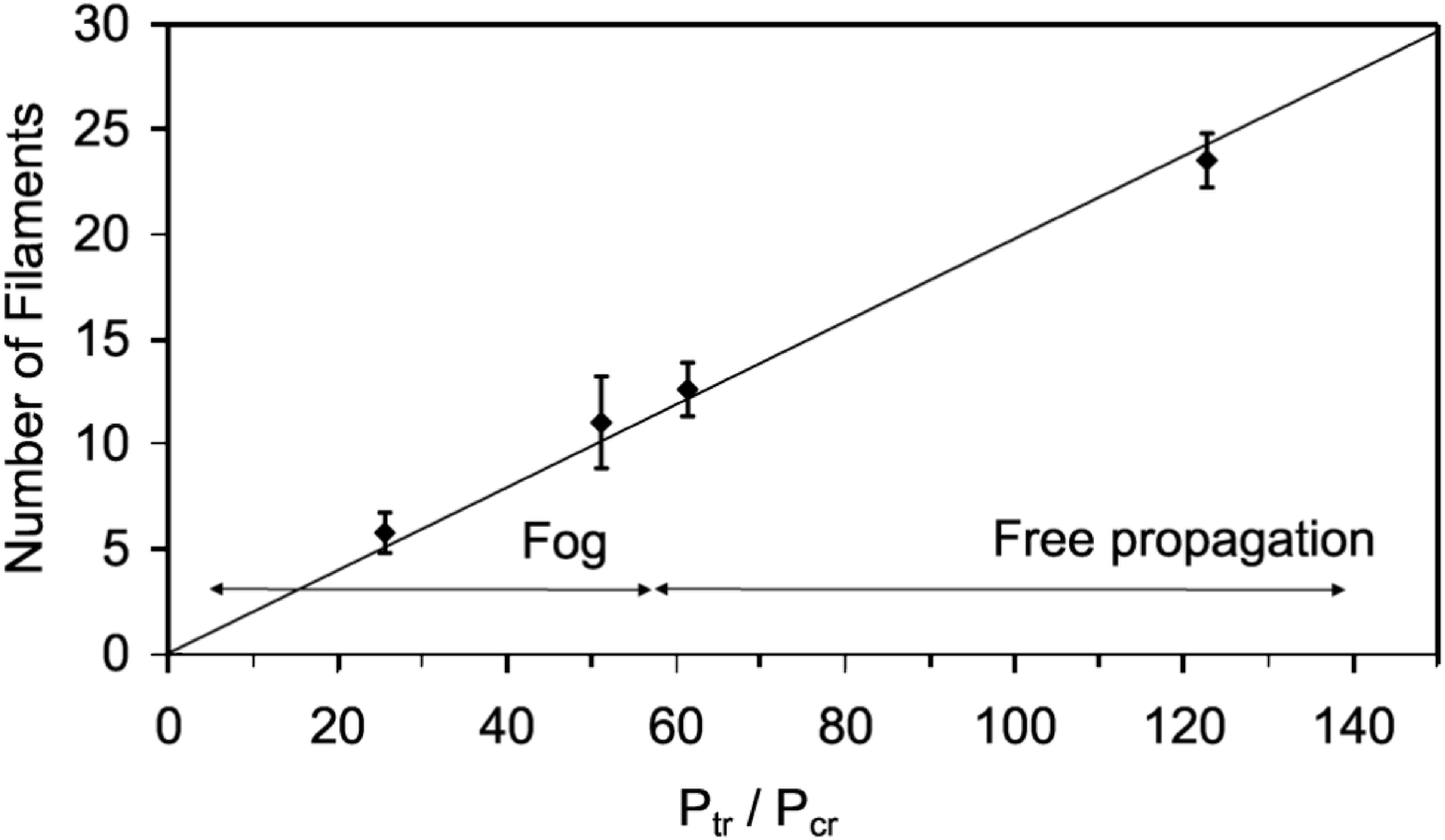}
\caption{Modification of a filamentation pattern with $\sim 120 
P_{\rm cr}$ over 50 m: (a) in dry air and (b) after traversing a 10-m long tube of fog. The bottom panel reports the filament number vs. 
the transmitted power over critical.}
\label{fig31}
\end{figure}

When the focusing geometry is  changed, the filamentation pattern 
becomes severely modified. Figure \ref{fig30} shows a longitudinal 
visualization of the filaments nucleated in a beam with 760 $P_{\rm 
cr}$ and focal length $f = 40$ m. The beam is directed towards an open cloud chamber scattering a weak-density cloud. After the focal point, only three strings of light emerge and cover about $\sim 8$ m each, while the same beam should produce more than one hundred filaments in parallel geometry. Numerical computations for this configuration clear up that many filaments are created before the 
linear focus, but they fuse into three strings of light acquiring a 
high directivity afterwards \cite{Skupin:pre:70:046602}. 
The same experimental setup (Fig.\ \ref{fig39})
put in evidence the robustness of femtosecond filaments 
through a multitude of $1 \,\mu$m large water droplets 
randomly-distributed at various densities in the 10-m long cloud chamber. 
A filament is still transmitted through a 
cloud with an optical thickness as high as 3.2. For a cloud optical density 
of 1.2 or below, corresponding to cumulus or stratocumulus, 
the filamentation does not seem affected. Filaments remain 
visible at the exit of the fog even for a droplet concentration so high ($8.6 
\times 10^4$ cm$^{-3}$) that each filament hits on the average 
2000 droplets per propagation meter. The corresponding extinction 
coefficient is 0.2 m$^{-1}$. Hence, filamentation can be transmitted 
through a fog over distances comparable with the visibility length. Energy losses due to random collisions implies an exponential decrease of power, which modifies the 
number and position of the filaments in the bundle 
\cite{Mejean:pre:72:026611}. Such experiments, whose results are 
recalled in Fig.\ \ref{fig31}, permitted to estimate the power per 
filament to about 5 $P_{\rm cr} \sim 15$ GW in air (see bottom panel).

Pulse propagation in adverse weather becomes an important topic for 
LIDAR applications. In this scope, Kandidov {\it et al.} addressed 
the point of the nucleation of filaments in a turbulent atmosphere 
supporting statistical fluctuations of the refractive index 
\cite{Kandidov:qe:29:911}. These fluctuations naturally arise as the medium becomes, e.g., locally heated. On the basis of a phase screen model, the cubic NLS equation including statistical variations of the linear optical index 
(Kolmogorov turbulence) describes random paths 
for the nucleation and position of the nonlinear focus. This focus appears on the average shorter compared with unperturbed air and the beam 
centroid moves along a random path in the $(x,y)$ plane. This induces a 
transverse deflection of the beam axis by several mm, which 
was later observed from experimental averaged data \cite{Chin:apb:74:67}. Besides this beam wandering, femtosecond filaments keep their position stable relatively to the bundle in highly turbulent regions, as the index variation caused by the nonlinearity exceeds turbulent-induced refractive-index gradients by two orders of magnitude \cite{Ackermann:ol:31:86}.

Furthermore, the influence of pressure variations on TW laser pulses was 
recently investigated. This issue is timely, as filaments propagating vertically over several km undergo an exponential decrease of the local pressure, $p(z) = p_0 \mbox{e}^{-z/L_a}$, where $L_a = 8$ km is the attenuation 
length for air density \cite{Sprangle:pre:66:046418}. Pressure 
variations induce proportional changes in the GVD coefficient, the Kerr 
refractive index and molecule density available for ionization, as recalled by 
Eq.\ (\ref{pressure}). At reduced pressure (0.7 atm), experiments 
performed at 3230-m altitude above sea level revealed 
farther filamentation onset and reduction of the filament number 
linked to the decrease of $n_2(p)$ by $30 \%$ \cite{Mechain:apb:80:785}. The filamentation process, 
especially the filament length, was shown to be qualitatively 
unaffected. Numerical simulations in this field specified that a single 
atmospheric filament does not indeed significantly change its 
self-channeling range and keeps an almost constant peak intensity. In contrast, the plasma level, the filament width [Eqs.\ (\ref{estimate13})] and the nonlinear focus (\ref{marburger}) 
evolve with $p$ and $\sqrt{p}$, respectively, due to the effective 
increase of $P_{\rm cr}(n_2)$. Multifilamentation patterns have their onset distance governed by the maximum rate for modulation 
instability $\gamma_{\rm max} \sim 1/n_2(p) I_0$ [see Eq.\ (\ref{MF_6})]. This length thus 
varies linearly with air pressure \cite{Champeaux:ol:31:1301}.

\subsubsection{White-light generation\label{sec7a4}}

\begin{figure}
\includegraphics[width=\columnwidth]{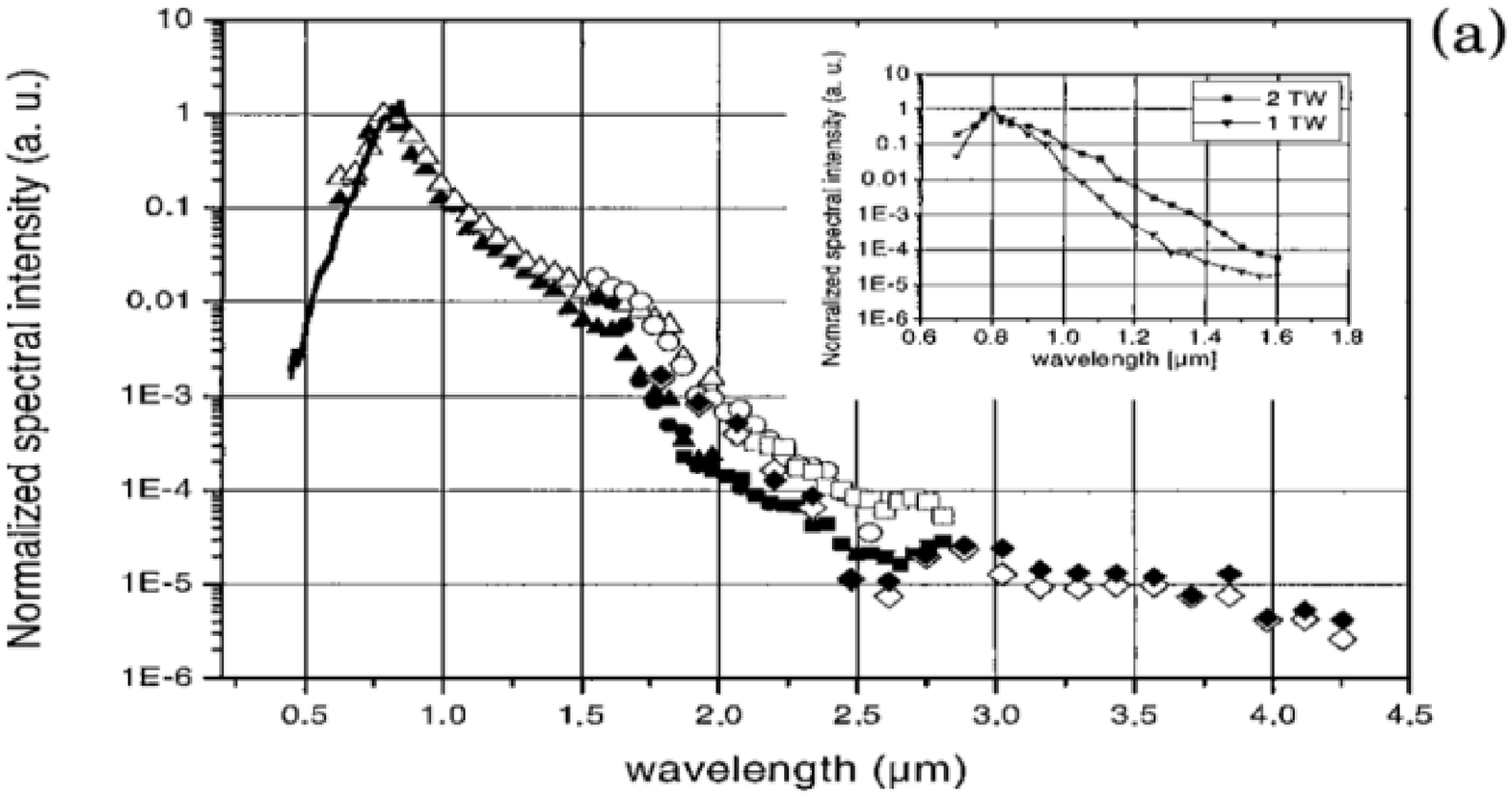}
\includegraphics[width=\columnwidth]{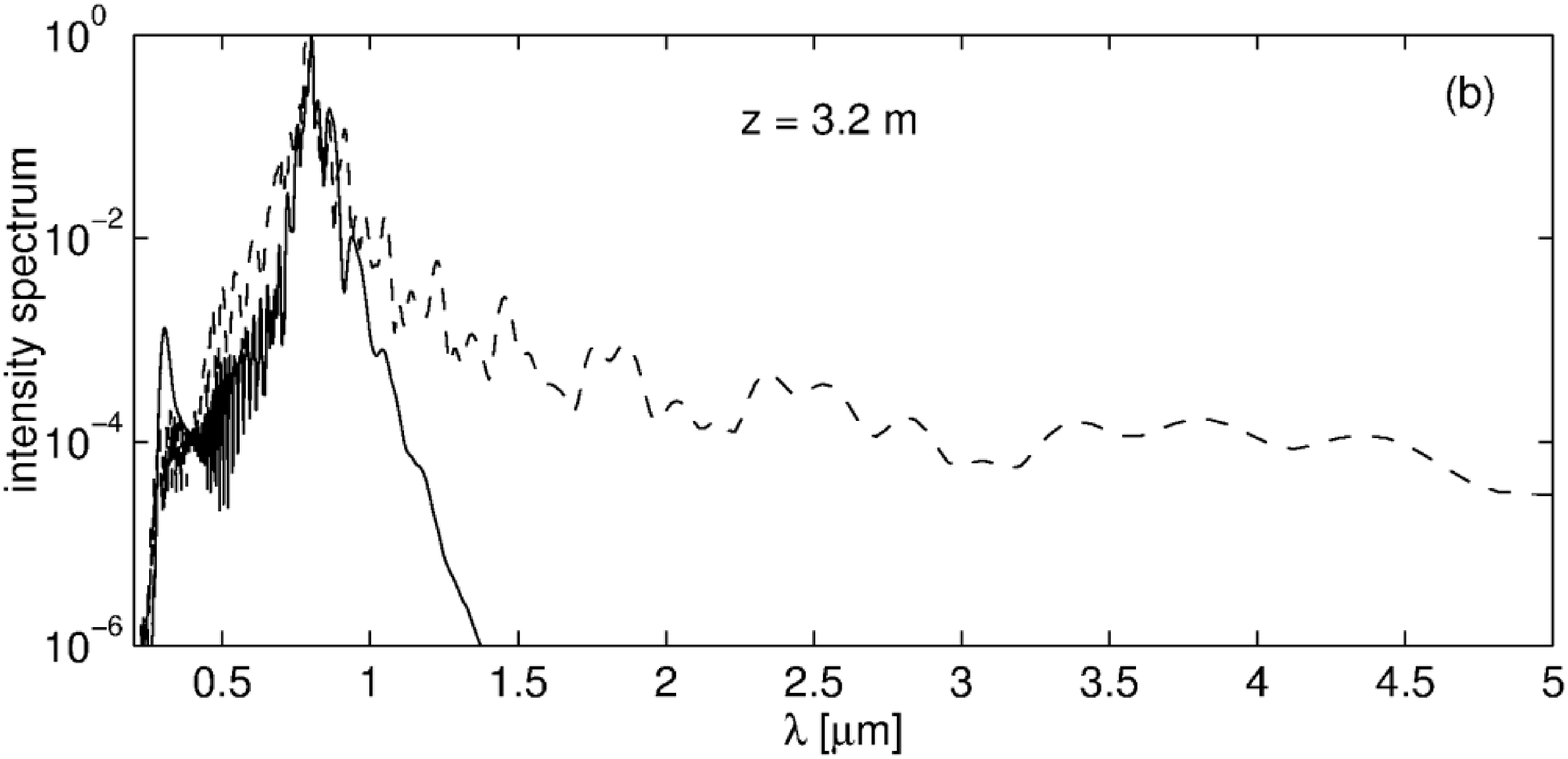}
\caption{(a) Measured white-light spectrum by 2-TW laser pulses 
(laser system A). Chirp settings correspond to 35-fs initial pulse duration 
without chirp (filled symbols) and 55-fs initial pulse duration with 
negative chirp (open symbols). Inset, spectrum measured from laser 
system B. The two curves have the same normalization factor. (b) 
Spectrum for a single filament numerically computed from Eqs.\ 
(\ref{modeleq}). The dash-dotted curve includes TH generation in the limits $T, T^{-1} \rightarrow 1$, while the solid curve represents the spectrum calculated from the complete equation set.}
\label{fig32}
\end{figure}

The white-light continuum generated in air by ultrashort laser pulses 
is essential in view of LIDAR applications, since it constitutes the light source used in 
multipollutant remote sensing. This white light was characterized over the last years, with progressively 
extending bandwidths. The supercontinuum has first been characterized 
in the visible \cite{Alfano:prl:24:584,Nishioka:ol:20:2505}. J. Kasparian {\it et al.} next investigated the infrared region \cite{Kasparian:ol:25:1397}. Two different terawatt CPA laser 
systems (A: 60-mJ energy, 35-fs minimal pulse duration, 25-mm FWHM beam 
diameter; B: 100-200 mJ, 100-fs minimal pulse duration, 35-mm FWHM 
diameters) produced spectra measured at a total distance of $\sim 30$ 
m from the lens, as the laser beam was diffracting after 20 m of 
filament propagation. Illustrated in Fig.\ 
\ref{fig32}(a), the continuum band developed from laser system A is 
very broad, 
extending at least to 4.5 $\mu$m. An almost exponential decay over 4 
orders of magnitude up to 2.5 $\mu$m is observed, followed by a slower decay of one order of magnitude only. 
As shown from the inset plotting results from laser system B, variations in the input energy makes spectral intensity 
change by only one decade in some spectral regions. The spectral shape of these different pulses remains, nevertheless, quite similar within one decade in spectral intensity.
Extension to the UV-visible domain down to 230 nm (Fig.\ \ref{fig16}) is not represented, as it was detected later.
For comparison, Fig.\ \ref{fig32}(b) shows the numerically-computed spectrum 
of a 0.5-mm waisted, $4P_{\rm cr} \sim 10$ GW, 100-fs pulse after a 
single femtosecond filament has been generated in air 
by accounting for third-harmonic generation in the limits $T, T^{-1} \rightarrow 1$ \cite{Berge:pre:71:016602} and from the full UPPE model (\ref{finalUPPE1}). The experimental and numerical spectral shapes are similar up to 1.2 $\mu$m. Differences, however, occur at larger wavelengths, as space-time focusing and self-steepening are included, which shortens the red parts of the spectrum to some extent. Recovering the experimental redshifts from the UPPE model in air is still an open issue.

The similarity of white-light spectra emitted by TW laser pulses 
(experiments) and GW pulses suggests that the multifilamentation 
produces a spectrum analogous to that generated by a single filament. 
This conclusion is consistent with Chin {\it et al.}'s experimental 
observations about the coherence properties of femtosecond filaments 
\cite{Chin:oc:210:329}. Observation of the 
interference pattern produced by two or more filaments allowed to predict that 
they emerge in phase from the background field and they possess the 
same phase relationship. This means that the laser 
spectrum around the central wavelength in the conditions of multiple 
filamentation is in principle identical to that developed by an isolated 
filament.

The previous property is important for LIDAR applications, because it 
suggests that the comprehension of the spectral dynamics of one 
filament is sufficient for understanding the spectral dynamics of multifilamented beams. In addition to the wide spectral region covered by nonlinear femtosecond pulses, the strong enhancement of the backscattered photons in filamentation regime makes ultrashort laser pulses quite promising tools for the remote identification of multipollutants in aerosols. Indeed, the supercontinuum emitted by a filament is enhanced in the backward direction [see Eq.\ (\ref{back_eq3})], i.e., towards the laser source, compared with linear 
Rayleigh-Mie scattering \cite{Yu:ol:26:533}.

\subsection{Remote sensing (LIDAR) applications}

\subsubsection{Principle of LIDAR: Towards ''Femtolidars''}

The LIDAR (LIght Detection And Ranging) technique 
\cite{Measures:LRS:84,Wolf:EAC:2000:2226,Kasparian:LRS:2005,Theopold:DR:2005} 
was demonstrated shortly after the advent of the first lasers. In 
this technique, a laser pulse is emitted into the atmosphere. The 
backscattered light is collected on a telescope and detected as a 
function of time, with a typical resolution of 1-10 nanoseconds. This temporal window yields a high spatial resolution, since the flight time 
of the detected photons is directly proportional to the 
distance where they have been backscattered. Such a spatial resolution, 
combined with the possibility of sweeping the laser beam, provides two- and three-dimensional maps of measured atmospheric species. This is the main advantage of LIDAR over other measurement methods for atmospheric trace-gases.
One of the most popular LIDAR techniques is called DIAL (DIfferential Absorbtion Lidar). It allows to selectively measure the concentration of gaseous pollutants by comparing the Lidar signals at two wavelengths nearby to one another, 
one being on an absorption line of the pollutant, and the 
other just beneath. However, this method is basically limited to 
pollutants that exhibit a narrow absorption line without interference 
from the absorption spectra of other atmospheric compounds. Moreover, 
the need to tune the laser wavelength on the absorption line forbids 
to simultaneously identify more than one pollutant within one acquisition.

\begin{figure}
\includegraphics[width=\columnwidth]{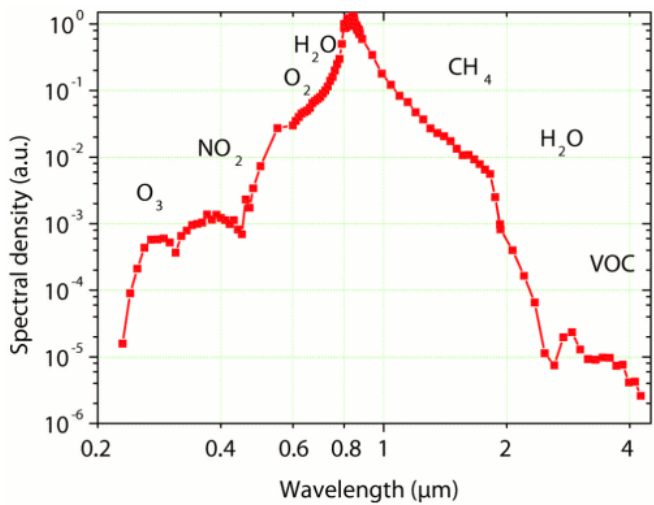}
\caption{Spectrum of the white light continuum assembled from five 
spectral regions. Main atmospheric absorption 
bands are marked above the curve.}
\label{fig33}
\end{figure}

Femtosecond Lidars (or so-called ''FemtoLidars'') overcome classical DIAL limitations. Exploiting the ''Femtosecond 
atmospheric lamp'' discovered in 1997 by W{\"o}ste {\it et al.} 
\cite{Woste:lo:29:51}, high-power
ultrashort laser pulses can be shined vertically into the sky, where they generate white-light through the filamentation mechanism. The 
backscattered
light, recorded with a telescope linked to a time-gated spectrometer, then provides a fascinating vector for atmospheric research over km ranges
\cite{Rairoux:apb:71:573,Rodriguez:pre:69:036607}. The spectral bandwidth developed by fs filaments is very broad, since it spans at 
least from 230 nm in the ultraviolet to 4.5 $\mu$m in the 
mid-infrared (Figure \ref{fig33}). On the spectrum, the absorption 
band of water between 1.8 and 2.5 $\mu$m is clearly visible, showing the 
potential of white-light for optical remote sensing in the 
atmosphere. It also covers the 
absorption band of volatile organic compounds (VOCs) between 3 and 
3.5 $\mu$m. VOCs constitute a family of organic compounds with 
strongly overlapping absorption spectra, which prevents any 
measurement by classical DIAL techniques. In addition, the flat continuum spanning from the visible down to 
230 nm through third-harmonic generation (see Fig.\ \ref{fig16}) provides a promising light 
source for the measurement of trace gases that absorb in the blue or 
the UV, such as ozone, toluene, benzene, SO$_2$ or nitrogen 
oxydes. Recently, femtosecond laser-induced filaments were also applied for 
greenhouse gaz methane (CH$_4$) in air \cite{Xu:apb:82:655}. Intense 
filaments dissociate pollutant molecules into small fragments, which 
emit characteristic fluorescence. This can be used to remotely measure 
the pollutant concentration at 
characteristic spectral lines.

\begin{figure}
\includegraphics[width=\columnwidth]{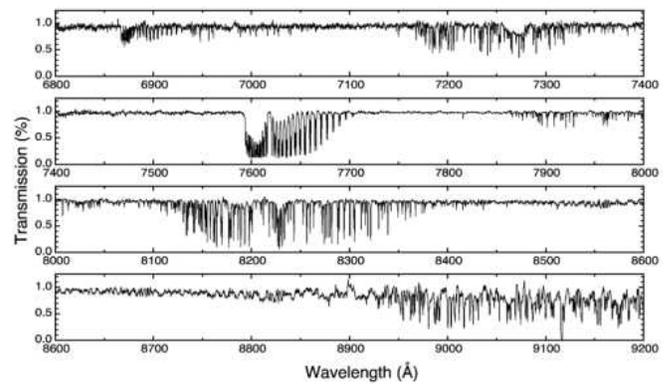}
\caption{High-resolution atmospheric absorption spectrum from an 
altitude of 4.5 km measured in a Lidar configuration. The broad 
spectrum allowed to simultaneously measure the air temperature and 
humidity.}
\label{fig34}
\end{figure}

Besides, multiparameter measurements are necessary to monitor the 
dynamics of atmospheric physico-chemistry or to remotely analyze a cocktail of species emitted during a chemical 
accident. They are also requested to characterize the nucleation and 
maturation of clouds. Such processes play an important role for 
atmospheric modeling, both on meteorological and climatological 
scales. In particular, the droplet growth and related density have a key 
influence on the forecast of both precipitations and earth 
albedo. Their characterization requires continuous measurements of 
the size distribution inside the clouds, with a temporal resolution of a 
few tens of minutes, compatible with the evaporation and growth 
time ranges. While airborne measurements are too expensive for routine 
monitoring, radiosounding does not provide the adequate repetition of 
probing. Therefore, optical techniques are promising, and they can be made optimal with fs laser pulses. As an example, 
a high-resolution absorption spectrum over a spectral interval larger 
than 200 nm is shown in Fig.\ \ref{fig34}, based on 
the white-light continuum, providing both the water vapor 
concentration and the temperature of air 
\cite{Kasparian:sc:301:61,Bourayou:josab:22:369}. Once combined, these two parameters supply the relative humidity, which is the relevant factor for atmospheric dynamics. Moreover, angular measurements of the 
multiple scattering yield the size distribution and 
concentration within the clouds, through an inversion of the laws of 
multiple Mie scattering based on a genetic algorithm. These 
measurements require the same laser source and two independent 
detectors. With fs Lidar setups, they can be implemented simultaneously so as to yield a complete characterization of the cloud 
microphysics.

\begin{figure}
\includegraphics[width=\columnwidth]{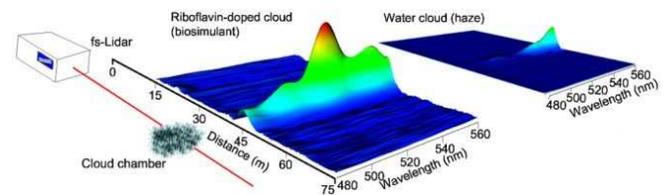}
\caption{Remote detection and identification of bioaerosols. The 
femtosecond laser illuminates a plume of riboflavin (RBF) containing 
microparticles 45 m away (left). The backward emitted 2-photon 
excited fluorescence (2-PEF), recorded as a function of distance and 
wavelength, exhibits the specific RBF fluorescence signature for the 
bioaerosols (middle) but not for haze (right).}
\label{fig35}
\end{figure}

''FemtoLidars'' are not restricted to the white-light 
Lidar. The high intensities carried in the 
filaments can generate nonlinear effects {\it in situ} on a 
target. Such nonlinear processes, which include ablation, ionization and multiphoton-excited fluorescence (M-PEF), 
constitute a supplementary information channel for remote sensing. An 
example of this technique has been provided by the remote detection 
and identification of biological aerosols through 2-PEF Lidar 
\cite{Mejean:apb:78:535}. Here, the purpose was to detect and locate 
rapidly a suspect emission, to map the emitted cloud, and to identify 
potentially pathogen agents among the multiple background atmospheric 
aerosols, including organic compounds such as soot or 
pollens. The 2-PEF allowed, for the first time, to remotely identify aerosols simulating biological agents in air by 
nonlinear Lidar (Fig.\ \ref{fig35}). Moreover, collected signals 
based on N-PEF become more 
efficient than Lidar signals based on 1-PEF for distances above a few 
kilometers. This is due to two effects: (i) the directional emission 
of the N-PEF from the aerosol particles minimizes the decrease of the collection efficiency with increasing distances \cite{Boutou:apb:75:145}, and (ii) the visible or near-infrared 
wavelengths used for exciting N-PEF, compared to the UV 
wavelength required for 1-PEF, experiences less attenuation in 
the atmosphere because of the $1/\lambda^4$ dependence of the 
Rayleigh scattering.

\subsubsection{Remote filament-induced breakdown spectroscopy}

\begin{figure}
\includegraphics[width=0.8\columnwidth]{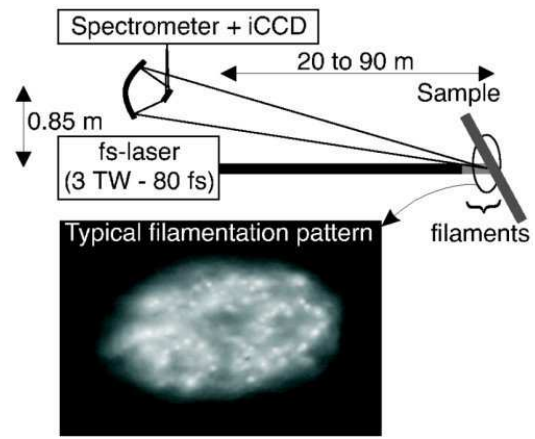}
\caption{Principle of Remote filament-induced breakdown spectroscopy 
(R-FIBS). Bottom: beam profile near to the sample showing 
multifilamentation with typically 30 filaments across the beam 
\cite{Stelmaszczyk:apb:85:3977}.}
\label{fig36}
\end{figure}

The ability of the filaments to remotely deliver intensities as high 
as $50$ TW/cm$^2$ also 
opens the way to innovative exploration techniques, such as the ''remote filament-induced breakdown spectroscopy'' (R-FIBS)
\cite{Stelmaszczyk:apb:85:3977}. This consists of a combination of Lidar and LIB spectroscopy (LIBS). LIBS \cite{Cremers:EAC:2000:9595,Cremers:LIB:06} is a 
versatile tool allowing an elemental analysis of surfaces of 
materials such as metals \cite{Angel:fjac:369:320}, plastics 
\cite{Niessner:pspie:2360:254,Jongil:as:56:852}, minerals 
\cite{Knight:as:54:331,Wiens:jgrp:107:FIDO3-1-14,Sharma:saa:59:2391}, 
aerosols, biological materials \cite{Kyuseok:asr:32:183} or liquids as well. 
It relies on the local ionization of the surface by a strongly 
focused pulsed laser, typically a Nd:YAG laser. The emission spectrum 
of the plasma generated at the surface allows a fast analysis, either 
qualitative or quantitative, with detection limits down to a few 
parts per million (ppm) for some elements. The use of subpicosecond laser pulses significantly enhances the 
reproducibility of the measurements, because the lower pulse energy 
limits the heating of the sample 
\cite{Angel:fjac:369:320,Rohwetter:jaas:19:437,Albert:apa:76:319,Dou:apa:76:303}. The use of broadband detection systems makes LIBS a flexible 
technique which requires neither preparation of the sample, nor {\it a priori} knowledge of the elements to be found.
Applications such as the identification of highly radioactive nuclear 
waste or real-time monitoring of melted alloys in industrial 
processes require a remote analysis technique. LIBS, which is 
suitable for raw samples, is a good candidate in this regard, since 
it only requires a direct view to the sample. However, due to the limited size 
of the optical components, diffraction intrinsically limits the 
intensity that can be focused on a remote target. On the contrary, self-guided filaments can deliver much higher intensities than the ablation 
threshold of many species, at distances of hundreds of meters or even 
kilometers. So, in R-FIBS techniques, the 
laser-generated filaments are launched on a remote target, and the 
light emitted by the excited plasma plume is collected through a detection 
setup comparable to a Lidar-based one (Fig.\ \ref{fig36}). Although issued from 180 m distances \cite{Stelmaszczyk:apb:85:3977}, the data suggest that 
measurements can be dimensioned up to the kilometer range 
\cite{Rohwetter:jaas:19:437}. This technique is currently 
developed for various applications, e.g., the monitoring of heritage 
\cite{Tzortzakis:ol:31:1139} or bacteria 
\cite{Baudelet:jap:99:084701,Xu:ol:31:1540}. 

\subsection{Towards a laser lightning rod}

\begin{figure}
\includegraphics[width=\columnwidth]{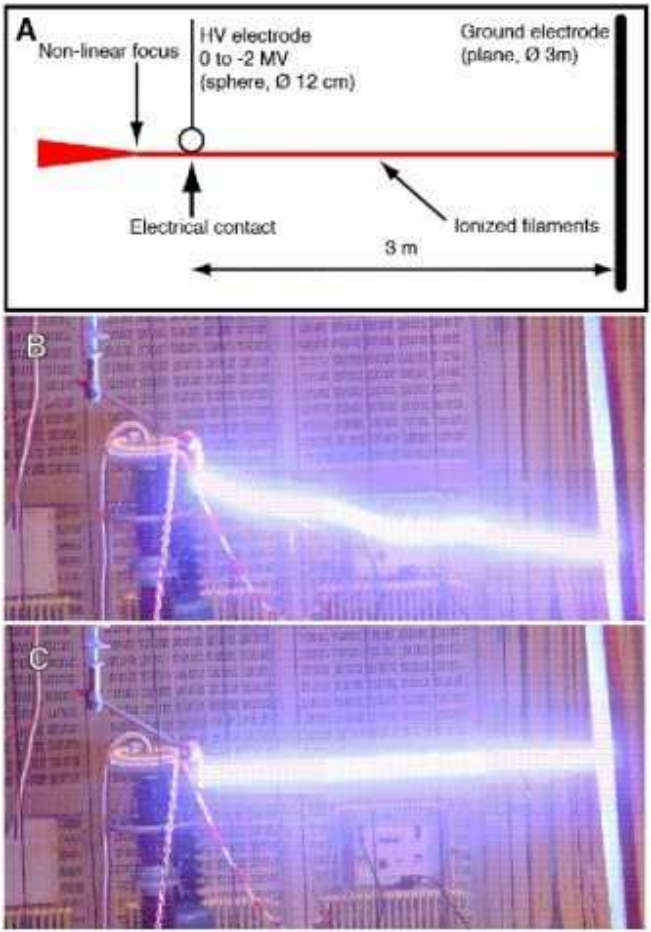}
\caption{Laser control of high-voltage discharges. (A) Experimental 
setup. (B) Free discharge over 3 m, without laser filaments. Note the 
erratic path. (C) Straight discharge guided along laser filaments 
\cite{Kasparian:sc:301:61}.}
\label{fig37}
\end{figure}

Besides remote sensing applications, ultrashort filaments may give access 
to the 
control of lightning strikes. Classical techniques to trigger lightning have employed rocket-pulled wires since the 1970's. However, the number of rockets available per storm is necessarily limited and rockets must be launched synchronized with the raise of the ambient electric field. Moreover, the wire falling down may pollute the 
measurement as well as the environment. Therefore, the idea emerged 
to apply lasers to control lightning by ionizing the ambient air along the 
beam and forming a conducting plasma "wire". First attempts used nanosecond pulses 
\cite{Koopman:jap:42:1883,Ball:ao:13:2292,Miki:jpd:26:1244}. They 
were unsuccessful, because such lasers could not produce continuously-ionized plasma channels. More recently, this field was 
renewed by the advent of CPA lasers providing higher intensities in 
shorter pulses, therefore avoiding laser absorption by inverse 
bremsstrahlung. Encouraging results have been obtained using focused ultrashort 
laser pulses to trigger and guide high-voltage discharges over several 
meters in the laboratory \cite{Pepin:pop:8:2532,Comtois:apl:76:819,Ting:pop:12:056705}. 
Others have been obtained on smaller scales using UV 
ultrashort lasers \cite{Rambo:joapao:3:146}. Since the filaments 
provide a conducting path over several meters or even longer, they are particularly suitable for the atmosphere. Spectacular experiments with the Teramobile laser installed in a high-voltage facility showed 
that ultrashort filaments can guide discharges 
over up to 4.5 m \cite{Ackermann:apb:82:561}. Instead of their usual erratic path, discharges are guided along the triggering laser beam (Fig.\ \ref{fig37}). 
Moreover, the breakdown voltage is typically reduced by $30 \%$ 
\cite{Rodriguez:ol:27:772}. Partly guided discharges also occur in 
sphere-plane gaps, providing valuable information about the 
mechanism of the initiation and propagation of laser-triggered 
streamers, with, e.g., plasma lifetimes of about $1 \mu$s \cite{Ackermann:apb:82:561}. Furthermore, an 
artificial rain does not prevent the laser filaments from triggering such 
discharges \cite{Ackermann:apl:85:5781}. Current research now 
focuses on the possibility to extend the plasma lifetime, in order to increase the guiding length and 
improve scalability to the atmosphere. This approach relies 
on re-heating and photodetaching electrons of the plasma channel by 
subsequent pulses, either in the nanosecond \cite{Rambo:joapao:3:146} 
or in the femtosecond regime \cite{Hao:apb:80:627}. Although high 
laser powers are usually believed to efficiently detach electrons from O$_2^-$ ions in the plasma, 
it was recently demonstrated that a subsequent Nd:YAG laser pulse of 
moderate energy (sub-Joule) at 532 nm efficiently supports the 
triggering of discharges by an infrared, femtosecond laser 
\cite{Mejean:apl:88:021101}. This effect was interpreted as resulting 
from a positive retroaction loop where Joule heating of the plasma channel 
enhances photodetachment, while the resulting higher 
electron density boosts in turn the Joule effect. 

\section{Outlook\label{sec8}}

Femtosecond lasers opened up beam parameter ranges from which universal, robust ultrashort filamentary structures conveying high intensities can propagate over long distances in a rich variety of transparent media. These ''femtosecond filaments'' are characterized by sizes of a few tens up to the hundred of microns and can reach ultrashort FWHM durations down to the optical cycle limit. They design new objects, that can be exploited for delivering few-cycle pulses at very high power levels in the future. Ionization processes do not break this property. They even amplify it by cutting the back time slices of the pulse. An accurate control of plasma generation should push the frontiers in nonlinear optics and allow for a precise monitoring of high-order harmonics. The basic scenario of a dynamical balance between Kerr focusing and plasma defocusing explains the existence and the long life of femtosecond filaments in most of propagation configurations. This scenario must, however, be revised at visible wavelengths in condensed materials, in which chromatic dispersion and nonlinear absorption (at high enough power) seem to drive the self-guiding mechanism through an important conical emission managed by $X$-shaped waveforms in normally dispersive regimes. This aspect requires more investigations. Also, propagation regimes of anomalous dispersion, which promote temporal compression and may further improve self-compression techniques, should be deepened.

In the last years, impressive progresses have been made in the 
understanding of the nonlinear propagation of high-power laser 
pulses over long distances in the atmosphere. For broad beams, multiple filamentation becomes more and more controllable by the use of deformable mirrors and genetic algorithms. Besides classical techniques of linear focusing and phase chirping, this will soon enable experimentalists to optimize the propagation range of very powerful pulses. 
Filaments can therefore be used to remotely deliver high intensities 
generating {\it in-situ} nonlinear effects. The spectrum of 
the white-light continuum is much broader than that expected a few years 
ago, especially in the ultraviolet where the mixing between the 
fundamental and third harmonics wavelengths  together with the spectral amplification caused by pulse steepening give rise to a spectral plateau 
extending down to 230 nm.
These progresses open the way to applications such as remote sensing 
of gaseous pollutants and aerosols by Lidar, LIBS analysis of solid 
samples, or the control of high-voltage electric 
discharges and lightning strikes. Due to its mobility, the 
Teramobile allowed to demonstrate the feasibility of many of these 
techniques. These applications will probably be pushed forward in the 
future by technological improvements involving more compact systems, diode pumping, as 
well as spatial and temporal pulse shaping. Novel active media, such as 
Ytterbium doping or OPCPA, also open the way to new spectral regions, 
especially the infrared. This spectral region, where eye-safety 
constraints are easier to fulfill, is nearer the absorption region of 
many pollutants, such as volatile organic compounds. These more 
flexible systems will be easier to operate, allowing routine uses for future industrial or atmospheric applications.

\begin{acknowledgments}
The Teramobile project is funded by the Centre National de la 
Recherche Scientifique (CNRS, France) and the Deutsche 
Forschungsgemeinschaft (DFG, Germany), with contribution by the 
French and German ministries of Foreign Affairs and the Agence Nationale de la Recherche (ANR, France). The authors 
gratefully acknowledge its team members, formed by 
the groups of L.\ W{\"o}ste at the Freie Universit{\"a}t Berlin, 
R.\ Sauerbrey at the Friedrich Schiller Universit{\"a}t Jena, A.\ Mysyrowicz at Laboratory for Applied Optics, Palaiseau (France), and the LASIM team led by J.-P.\ Wolf.
\end{acknowledgments}

\appendix

\section{Ionization rates for atoms and molecules\label{appA}}

This appendix details different photo-ionization theories for gases and
dense transparent materials. 

\subsection{Ionization in gases}

To model the generation of free electrons by an intense light field interacting with a gaseous medium, 
photo-ionization theories were first developed for atomic systems by 
Keldysh \cite{Keldysh:spjetp:20:1307}, later by Perelomov, Popov and 
Terent'ev (PPT) 
\cite{Perelomov:spjetp:23:924,Perelomov:spjetp:24:207}, 
before being refined by Ammosov, Delone and Krainov (ADK) twenty years 
after \cite{Ammosov:spjetp:64:1191}. More recently, Tong {\it et al.} 
proposed a theory for molecular species in tunnel ionization regime 
\cite{Tong:pra:66:033402}. 

\subsubsection{The Keldysh theory}

Keldysh's theory is limited to hydrogenoid atoms in 
their fundamental electronic
state and does not consider the Coulomb interaction between the 
leaving 
electron and the residual ion. The ionization rate $W$ of an atom 
irradiated by a laser field $\vec{E}(t) = \vec{E}_p \cos(\omega t)$ is evaluated by
\begin{equation}
\label{eq:rate}
\begin{split}
W & = \frac{q_e^2}{\hbar^{2}}
\lim_{t \rightarrow \infty} \int \frac{d \vec p}{(2\pi\hbar)^{3}} 
\frac{d}{dt}
\left| \int_{0}^{t} dt' \cos(\omega t')\right. \\
& \quad \left. \times \left<\Psi_{\vec p}(\vec r, t') |\vec r \cdot \vec{E}_p | 
\Psi_{g}(\vec r, t')
\right> \right|^{2},
\end{split}
\end{equation}
where $q_e$ is the electron charge and the ground state of the 
hydrogenlike atom, 
characterized by the energy $E_{g} = -U_i$ (ionization potential), is
\begin{equation}
\Psi_{g}(\vec r, t) = \frac{1}{\sqrt{\pi a^{3}}} e^{-\frac{r}{a}} 
e^{-\frac{i}{\hbar} E_{g} t }.
\label{eq:fo_fondamental}
\end{equation}
Here, $a = a_B/Z$ includes the Bohr radius $a_B$ of hydrogen. The 
continuum electronic states are described with the Volkov functions 
\begin{equation}
\Psi_{\vec p}(\vec r, t) = e^{\frac{i}{\hbar} [ (\vec p +q_e\vec 
A(t))\cdot \vec r -
\frac{1}{2m_e} \int_{0}^{t}dt'[\vec p + q_e \vec A(t')]^{2}]},
\label{eq:fo_continu}
\end{equation}
where $\vec A(t)$ is the vector potential of the laser field. 
Insertion of (\ref{eq:fo_fondamental}) 
and (\ref{eq:fo_continu}) into Eq.\ (\ref{eq:rate}) and integration 
over the spatial coordinates result in the expression
\begin{equation}
\label{eq:rate2}
\begin{split}
W  & =  \frac{2}{\hbar^{2}} \lim_{t \rightarrow \infty} \textrm{Re}
\left[\int \frac{d\vec p}{(2\pi\hbar)^{3}} \right.\\
 & \quad \left. \times \int_{0}^{t}dt' \cos(\omega t)
 \cos(\omega t') L(\vec p, t') L^{*}(\vec p, t) \right], 
\end{split}
\end{equation}
where $^*$ means complex conjugate and
\begin{subequations}
\begin{align}
\label{eq:L1}
\begin{split}
L(\vec p, t) & = V_{0}\left[\vec p - \frac{q_e}{\omega}\vec{E}_p \sin(\omega 
t)\right] \\
& \quad \times e^{\frac{i}{\hbar} \left\{U_it + \frac{1}{2m_e} \int_{0}^{t}dt'
\left[\vec p - \frac{q_e}{\omega} \vec{E}_p \sin(\omega t')\right]^{2}\right\}},
\end{split}\\
V_{0}(\vec p) & = -i 8 q_e  \sqrt{\pi a^{3}}\hbar \vec{E}_p \cdot \vec 
\nabla_{\vec p} 
\left[\frac{1}{(1+\frac{a^{2}p^{2}}{\hbar^{2}})^{2}}\right].
\end{align}
\end{subequations}
The function $L(\vec p, t)$ is periodic, with period equal to $T=2\pi 
/\omega$. It can then be decomposed into Fourier series. Plugging this 
series into Eq.\ (\ref{eq:rate2}) leads to
\begin{equation}
\label{eq:etape1}
\begin{split}
W & = \frac{2\pi}{\hbar} \int \frac{d\vec p}{(2\pi\hbar)^{3}} |L(\vec 
p)|^{2} \\
 & \quad \times \sum_{n=-\infty}^{+\infty}\delta\left({U_i} + \frac{p^{2}}{2m_e}+ 
\frac{q_e^{2}E_p^{2}}
{4m_e\omega^{2}}
-n\hbar \omega\right),
\end{split}
\end{equation}
with 
\begin{equation}
\label{eq:L}
\begin{split}
L(\vec p) & =   16 i q_e \frac{\sqrt{\pi a^{7}}}{\hbar \pi} U_i^{3} 
\int_{-1}^{1}du \frac{\vec{E}_p \cdot \left(\vec p - \frac{q_e}{\omega}\vec{E}_p u \right)}
{\left[U_i+\frac{1}{2m_e}\left(\vec p - \frac{q_e}{\omega} \vec{E}_p u\right)^{2}\right]^{3}} \\ 
 & \quad \times
e^{\frac{i}{\hbar\omega} \int_{0}^{u} \frac{dv}{\sqrt{1-v^{2}}}
\left[U_i+\frac{1}{2m_e}\left(\vec p - \frac{q_e}{\omega} \vec{E}_p v\right)^{2}\right]}.
\end{split}
\end{equation}
Assuming that the electron leaves the atom with a small kinetic 
energy ($\frac{p^{2}}{2m_e} \ll U_i$), the poles in the denominator of 
$L({\vec p})$ are written as
\begin{equation}
u_{s}^{\pm} = i\gamma \left[ \pm 1 + i 
\frac{p\cos(\theta)}{\sqrt{2m_eU_i}}
\pm p^{2}\frac{\sin^{2}(\theta)}{2m_eU_i}\right],
\end{equation}
where $\theta$ is the angle between the impulsion vector $\vec p$ and 
the electric field $\vec{E}(t)$. $\gamma = \omega \sqrt{2 m_e U_i} 
/ (|q_e|E_p)$
is the adiabaticity {\it Keldysh parameter}. It involves the ratio of the ionization potential over the ponderomotive energy $U_p \equiv q_e^2 E_p^2/4m_e \omega^2$. By means of the Saddle 
point
method and the residue theorem, Eq.\ (\ref{eq:L}) is next integrated 
to yield
\begin{equation}
\label{eq:keldysh1}
\begin{split}
W & =  4 \sqrt{2}\omega \sqrt{\frac{U_i}{\hbar 
\omega}}\left[\frac{\gamma}
{\sqrt{1+\gamma^{2}}}\right]^{\frac{3}
{2}} \\
& \quad \times e^{-\frac{2\tilde{U_i}}{\hbar\omega}
\left[\sinh^{-1}(\gamma)-\frac{\gamma 
\sqrt{1+\gamma^{2}}}{1+2\gamma^{2}}\right]} 
S(\gamma, \frac{\tilde{U_i}}{\hbar\omega}),
\end{split}
\end{equation}
where $\tilde{U_i} \equiv U_i + U_p$ and $S(\gamma, x)$ is defined by
\begin{multline}
\label{eq:s_keldysh1}
S(\gamma, x) = 
\sum_{n=0}^{+\infty}e^{-2\left[\sinh^{-1}(\gamma)-
\frac{\gamma}{\sqrt{1+\gamma^{2}}}\right](<x+1> -x+n)} \\
 \times \int_{0}^{\sqrt{\frac{2\gamma}{\sqrt{1+\gamma^{2}}}
(<x+1> - x+n)}} e^{y^{2}-\frac{2\gamma}{\sqrt{1+\gamma^{2}}}
(<x+1> - x+n)}dy.
\end{multline} 
To take electron-ion correlation into account, 
Keldysh eventually multiplies the ionization rate by the factor 
$(U_i/\hbar\omega) \gamma/\sqrt{1+\gamma^{2}}$. Expressed in 
atomic units (a.u.) $m_e=|q_e|=\hbar=a_B=1$ \cite{Bransden:PAM:03}, the resulting 
ionization rate expresses
\begin{equation}
\label{eq:keldysh}
\begin{split}
W & = 2\sqrt{2}\left(\frac{2E_{0}}{E_p \sqrt{1+\gamma^{2}}} 
\right)^{\frac{1}{2}}
  e^{-2\nu\left[\sinh^{-1}(\gamma)-\frac{\gamma \sqrt{1+\gamma^{2}}}
{1+2\gamma^{2}}\right]} \\
& \quad \times U_i \frac{\gamma^{2}}{1+\gamma^{2}}\sum_{\kappa \geq \nu_{0}}^{+\infty}e^{-\alpha (\kappa-\nu)}
\Phi_{0}(\sqrt{\beta(\kappa-\nu)}),
\end{split}
\end{equation}
where $E_{0} = (2U_i)^{3/2}$, 
$\gamma = \omega\sqrt{2U_i}/E_p$,
$\nu = \tilde{U_i}/[\hbar \omega]_{\rm a.u.}$,
$\beta = 2\gamma/\sqrt{1+\gamma^{2}}$,
$\alpha = 2 \left[\sinh^{-1}(\gamma) - 
\gamma/\sqrt{1+\gamma^{2}}\right]$,
$\nu_{0} = <\nu +1 >$
and $\Phi_{m}(x) = e^{-x^{2}} 
\int_{0}^{x} (x^{2}-y^{2})^{|m|}e^{y^{2}} dy$.
Equation (\ref{eq:keldysh}) differs from the original Keldysh formulation by a factor 4,
originating from a corrected version of the residue theorem.

This theory was the first one able to describe atom 
ionization by an alternating field in low intensity regime ($\gamma \gg 1$) as well as in high 
intensity regime ($\gamma \ll 1$). The former regime refers to 
multiphoton ionization (MPI), through which the electron 
is freed as the atom absorbs 
$K = <U_i/(\hbar \omega) +1 >$ photons. The latter one 
corresponds to the tunnel regime, for which the electron leaves
the ion by passing through the Coulomb barrier.
In MPI regime, the ionization rate is 
obtained by taking the limit $\gamma \rightarrow +\infty$ in Eq.\ 
(\ref{eq:keldysh}), which reduces to
\begin{equation}
W = \sigma^{(K)} \times I^K,
\end{equation}
where $I$ is the laser intensity and $\sigma^{(K)}$ is the 
photo-ionization cross-section
\begin{equation}
\begin{split}
\sigma^{(K)} &= 4\sqrt{2} \omega \left(\frac{U_i}{[\hbar \omega]_{\rm a.u.}} \right)^{2K+3/2} \\
&\quad\times\frac{e^{2K-U_i/[\hbar \omega]_{a.u.}}}{E_0^{2K}} 
\Phi_{0}(\sqrt{2K-\frac{2U_i}{[\hbar \omega]_{\rm a.u.}}}),
\end{split}
\end{equation}
Expressed in s$^{-1}$cm$^{2 
K}$/W$^K$, the above ''cross-section'' parameter must be converted as 
$\sigma^{(K)} \rightarrow \sigma^{(K)}[\mbox{a.u.}]/[2.42 \times 
10^{-17} \times (3.51 \times 10^{16})^K]$.

\subsubsection{The PPT theory}

Later, Perelomov, Popov and Terent'ev 
\cite{Perelomov:spjetp:23:924,Perelomov:spjetp:24:207} developed a 
more accurate model. First, they included the Coulomb 
interaction between the ion and the electron, when the latter leaves 
the atomic core. Second, they considered any atomic bound states as 
initial. The resulting rate is then
\begin{equation}
\label{eq:PPT_ADK}
\begin{split}
W & = \frac{4\sqrt{2}}{\pi} |C_{n^{*},l^{*}}|^{2}
\left(\frac{2E_{0}}{E_p \sqrt{1+\gamma^{2}}} 
\right)^{2n^{*}-\frac{3}{2}-|m|}\\
&\quad\times  \frac{f(l,m)}{|m|!}
e^{-2\nu\left[\sinh^{-1}(\gamma)-\frac{\gamma \sqrt{1+\gamma^{2}}}
{1+2\gamma^{2}}\right]}\\ 
&\quad\times U_i \frac{\gamma^{2}}{1+\gamma^{2}}\sum_{\kappa \geq \nu_{0}}^{+\infty}e^{-\alpha (\kappa-\nu)}
\Phi_{m}(\sqrt{\beta(\kappa-\nu)}),
\end{split}
\end{equation}
where $n^{*} = Z/\sqrt{2U_i}$ is the effective quantum number, $Z$ is the residual ion charge, $l^{*}=n^{*}-1$ and $n,l,m$ are the principal quantum number, the orbital momentum and the magnetic quantum number, respectively.
The factors $|C_{n^{*},l^{*}}|$ and $f(l,m)$ are
\begin{subequations}
\begin{align}
|C_{n^{*},l^{*}}|^{2} & = 
\frac{2^{2n^{*}}}{n^{*}\Gamma(n^{*}+l^{*}+1)
\Gamma(n^{*}-l^{*})}, \\
f(l,m) &= \frac{(2l+1)(l+|m|)!}{2^{|m|}|m|!\ (l-|m|)!}.
\end{align}
\end{subequations}
Even if Eq.\ (\ref{eq:PPT_ADK}) is usually presented as the PPT 
formula, the coefficients $|C_{n^{*},l^{*}}|$ are in fact extracted 
from the tunneling theory derived by Ammosov, Delone and Krainov 
\cite{Ammosov:spjetp:64:1191}. Differences between PPT and ADK 
coefficients essentially lie in the fact that ADK theory employs 
electron wavefunctions in a Coulomb potential (Volkov states), which 
are connected by continuity with the continuum states at large distances $(r \gg 1/\sqrt{2 U_i})$.

\subsubsection{The ADK molecular theory}

The PPT rate [Eq.\ (\ref{eq:PPT_ADK})] holds to describe photo-ionization of atoms. It can lead to some discrepancy when 
it is applied to molecular systems, because the coefficients 
$|C_{n^{*},l^{*}}|$, originally evaluated from atomic wavefunctions, 
cannot reproduce molecular peculiarities, such as, for example, the 
suppression of ionization observed from the molecule O$_2$ 
\cite{Dewitt:prl:87:153001}. To overcome such limitations,
we may extend the molecular tunneling theory by Tong {\it et al.} 
\cite{Tong:pra:66:033402} by plugging molecular coefficients
into the tunnel limit of the PPT formula and prolonging the latter to 
low intensity MPI regimes analytically. By doing so, Eq.\ 
(\ref{eq:PPT_ADK}) is able to describe molecule ionization after the 
substitution
\begin{multline}
|C_{n^{*},l^{*}}|^{2} f(l,m) \rightarrow \\
\left[\sum_{l} \frac{C_{l}}{(2U_i)^{(n^*/2)+(1/4)}}
\sqrt{\frac{(2l+1)(l+|m|)!}{2^{|m|} |m|!(l-|m|)!}}\right]^{2},
\end{multline}
where the coefficients $C_l$ have been established for different 
molecules. For dioxygen,  $C_{2}$ = 0.683 and $C_{4}$ = 
0.033 while $C_1$=$C_3$=0, and $m=1$ \cite{Tong:pra:66:033402}.

\subsection{Ionization in dense media}

Plasma generation in dense media is described 
with the ionization rate for crystals developed by Keldysh 
\cite{Keldysh:spjetp:20:1307}. Its analytical
evaluation is identical to that applying to atoms, except that the 
initial states are now modeled by Bloch wave functions. Following 
similar procedure, the ionization rate for crystals with energy gap 
$E_g$ irradiated
by an electromagnetic field $E_p \cos(\omega t)$ is then
\begin{equation}
\label{eq:crystal}
\begin{split}
W & = \frac{2\omega}{9\pi} \left(\frac{\sqrt{1+\gamma^2}}{\gamma}
\frac{m^*\omega}{\hbar}\right)^{\frac{3}{2}}
Q(\gamma, \frac{\tilde{\Delta}}{\hbar \omega}) \\
& \quad \times e^{-\pi <\frac{\tilde{\Delta}}{\hbar \omega}+1> \times \left[\frac{{\bar {\cal K}} (\frac{\gamma^2}{1+\gamma^2}) -
{\bar {\cal E}}(\frac{\gamma^2}{1+\gamma^2})}{{\bar {\cal E}} (\frac{1}{1+\gamma^2})}\right]}
\end{split}
\end{equation}
with $\gamma = \omega \sqrt{m^* E_g}/(|q_e|E_p)$,
${m^*}^{-1} = {m_e}^{-1} + {m_h}^{-1}$,
\begin{subequations}
\begin{align}
\tilde{E_g}& = \frac{2}{\pi} E_g \frac{\sqrt{1+\gamma^2}}{\gamma} 
{\bar {\cal E}}(\frac{1}{1+\gamma^2}), \\
\begin{split}
Q(\gamma, x) &= \sqrt{\frac{\pi}{2 {\bar {\cal K}}(\frac{1}{1+\gamma^2})}} 
\times
\sum_{n=0}^{+\infty}e^{-\pi n 
\left[\frac{{\bar {\cal K}}(\frac{\gamma^2}{1+\gamma^2})
-{\bar {\cal E}}(\frac{\gamma^2}{1+\gamma^2})} 
{{\bar {\cal E}}(\frac{1}{1+\gamma^2})}\right]} \\
& \quad \times \Phi_0(\sqrt{\frac{\pi^2}{4}\frac{2<x+1>-2x+n}
{{\bar {\cal K}} (\frac{1}{1+\gamma^2}) {\bar {\cal E}} (\frac{1}{1+\gamma^2})}}).
\end{split}
\end{align}
\end{subequations}
Here, the functions ${\bar {\cal K}}(x) \equiv \int_0^{\pi/2} (1 - x 
\sin^2{\theta})^{-1/2} d\theta$ and ${\bar {\cal E}}(x) \equiv \int_0^{\pi/2} (1 - 
x \sin^2\theta)^{1/2}d\theta$ are the complete elliptic integrals of 
the first and
second kind \cite{Abramovitz:HMF:72} and $m^*$ is the reduced mass 
for the electron/hole pair. The above equation corrects a slip of pen 
occurring in the original Keldysh's formula. 
Note that whereas $W$ for gas [Eq.\ (\ref{eq:keldysh})] is expressed 
per time unit, $W$ for condensed media [Eq.\ (\ref{eq:crystal})] is 
expressed per time unit and per cubic meter. This ionization rate 
reduces at low intensities to its multiphoton limit ($\gamma 
\rightarrow +\infty$) taking the form
\begin{equation}
W = \sigma^{(K)}\times I^K
\end{equation}
where 
\begin{equation}
\begin{split}
\sigma^{(K)} & = \frac{2\omega}{9\pi} \left(\frac{m^* 
\omega}{\hbar}\right)^{\frac{3}{2}}
\Phi_0(\sqrt{2(K-\frac{E_g}{\hbar \omega})})  \\
&\quad\times e^{2K}
\left(\frac{q_e^2}{8m^* \omega^2 E_g \epsilon_0 c n_0}\right)^K
\end{split}
\end{equation}
and $n_0 = n(\omega=\omega_0)$ is the linear refractive index.

\section{Atomic dipole for High-Harmonic Generation\label{appB}}

The nonlinear polarization vector $\vec{P}_{\rm NL}(\vec{r} ,t)$ used 
to describe HHG in gases is expressed as
\begin{equation}
\vec{P}_{\rm NL}(\vec{r} ,t) = \rho_{\rm nt} \, \vec{d}(\vec{r},t),
\end{equation}
where $\vec{d}(\vec{r},t)$ is the time-dependent atomic dipole 
calculated by 
Lewenstein {\it et al.} \cite{Lewenstein:pra:49:2117}. For a 
single electron interacting with
a linearly-polarized laser field $\vec{E}(t) = E_p \hat{x} \cos(\omega t)$, the 
atomic dipole is defined by
\begin{equation}
\vec{d}(\vec{r},t) = q_e \langle \psi(\vec{r},t) | \vec{r} | 
\psi(\vec{r},t) \rangle.
\label{eq:TDdipole}
\end{equation}
Solving the time-dependent Schr{\"o}dinger equation, $\psi(\vec{r},t)$ 
denotes the electronic wavefunction
\begin{equation}
|\psi(\vec{r},t) \rangle  = e^{i\frac{U_i t}{\hbar}}\times 
\left[|0\rangle + \int d^{3}v\ b(\vec{v},t)\ |v\rangle \right],
\label{eq:wavefunction}
\end{equation}
where $U_i$ is the ionization potential of the atom, $|0>$ is the ground
state and $b(\vec{v},t)$ 
is the amplitude of the continuum states $|v>$, 
respectively.
In Eq.\ (\ref{eq:wavefunction}), several assumptions have been made, 
which limits the validity domain of this theory: 
\begin{itemize}
\item  All excited electronic states are ignored, reducing the 
model to the harmonic production of orders $2K+1 \geq \frac{U_i}{\hbar 
\omega}$.
\item  The depletion of the ground state is neglected such that the 
ground state amplitude is equal to 1.
\item  Electrons in the continuum are treated as free particles 
moving in an oscillating laser field,
with no Coulomb potential. This assumption is 
valid when the ponderomotive energy
$U_p$ is higher than the ionization potential.
\end{itemize}
Inserting (\ref{eq:wavefunction}) into Eq.\ (\ref{eq:TDdipole}), we 
obtain
\begin{equation}
\begin{split}
\vec{d}(t) & =   i \frac{q_e^2 E_p \hat{x}}{\hbar \omega}\int_{0}^{\omega t} 
d(\omega t') \cos(\omega t')
\int d^3p\ d_x^{*}[\vec{p}+q_e\vec{A}(t)] \\
& \quad \times e^{-i\frac{S(\vec{p},t,t')}{\hbar \omega}} \times 
d_x[\vec{p}+q_e\vec{A}(t')] + c.c.,
\end{split}
\label{eq:TDdipole2}
\end{equation}
where $\vec{d}(\vec{p}+q_e\vec{A}(t))$ is the field-free dipole 
transition matrix element
between the ground state and the continuum state characterized by the 
momentum $\vec{v}=\vec{p}+q_e\vec{A}(t)$. Here, $\vec{p}$ is the 
canonical momentum and $\vec{A}(t)$ is the vector potential of the 
laser field
[$\vec{E}(t)=- \partial \vec{A}(t)/\partial t$].
Its formulation for transition from state $| 1s \rangle$ is
\begin{equation}
\vec{d}(\vec{p})= - i \hbar 
\frac{2^{7/2}}{\pi}(2m_eU_i)^{5/4}\frac{\vec{p}}{(\vec{p}^2 + 2m_e 
U_i)^3},
\end{equation}
where $m_e$ is the electron mass. $S(\vec{p},t,t')$ is the 
quasi-classical action describing the motion of an electron freely 
moving in the
laser field with constant momentum $\vec{p}$ as
\begin{equation}
S(\vec{p},t,t') = \int_{\omega t'}^{\omega t}\left\{U_i + 
\frac{[\vec{p}+q_e\vec{A}(t'')]^2}{2m_e}\right\} d(\omega t'').
\end{equation}
The harmonic amplitude $\vec{d}_{2K+1}$ follows from Fourier 
transforming the time-dependent
dipole moment $\vec{d}(t)$ 
\begin{equation}
\vec{d}_{2K+1} = \frac{1}{T}\int_{0}^{T}\ \vec{d}(t)\ 
e^{i(2K+1)\omega t} dt.
\end{equation}
For isotropic media, only odd harmonics are produced. The dominant 
contributions to $\vec{d}_{2K+1}$ comes from the stationary points of 
the Legendre-transformed quasi-classical action, for which the 
derivatives
of $S(\vec{p},t,t') - (2K+1) \hbar \omega \times \omega t$ with 
respect to $\vec{p}$, $t$ and $t'$ vanish. Introducing
the returning time $\tau = t - t'$, the Saddle point equations read
\begin{subequations}
\begin{align}
\vec{\nabla}_{\vec{p}}S & = \vec{p}_{st} \omega \tau 
+ q_e \int_{\omega(t-\tau)}^{\omega t}
\vec{A}(t'')\ d(\omega t'') = 0 
\label{eq:pst} \\
\frac{\partial S}{\partial(\omega \tau)} & = U_i + 
\frac{[\vec{p}_{st}+q_e\vec{A}(t-\tau)]^2}{2m_e}=0 
\label{eq:SP1}\\
\label{eq:SP2}
\begin{split}
\frac{\partial S}{\partial(\omega t)}& = 
\frac{[\vec{p}_{st}+q_e\vec{A}(t)]^2}{2m_e}- 
\frac{[\vec{p}_{st}+q_e\vec{A}(t-\tau)]^2}{2m_e}\\
& =  (2K+1) \hbar \omega
\end{split}
\end{align}
\end{subequations}
Equation (\ref{eq:pst}) reduces to $\vec{\nabla}_{\vec{p}}S(\vec{p},t,t') = \vec{x}(t) - \vec{x}(t') = 
\vec{0}$ where $\vec{x}(t) \equiv \int_0^t v(t') dt'$. This relation imposes that the electron trajectories return at time $t$ 
to the same point they left at the time $t'$ of ionization. Using Eqs.\ (\ref{eq:SP1}) and 
(\ref{eq:SP2}) together with the Fourier expansion of $\big[(\vec{p}_{st}+q_e\vec{A}(t-\tau))^2/(m_e\hbar\omega) + 2 
U_i/(\hbar\omega) \big]^{-3}$ yields the final expression for the Fourier component 
of the dipole, i.e.,
\begin{widetext}
\begin{equation}
\begin{split}
\vec{d}_{2K+1} & = -i \frac{q_e \left(\frac{U_p}{\hbar \omega}\right)^{\frac{3}{2}} 
\frac{32 \hbar}{\pi^2} 
\left(\frac{2U_i}{\hbar \omega} \right)^{\frac{5}{2}} 
}{(2K+1)^3\sqrt{m_e\hbar\omega}} \sum_{M=-\infty}^{\infty} \int_{0}^{+\infty} d\Phi 
\left[\frac{\pi}{\epsilon + i\frac{\Phi}{2}}\right]^{\frac{3}{2}} b_{M}(\Phi) e^{-i\frac{F_{K}(\Phi)}{\hbar\omega}} i^{K-M} \\
& \quad \times \left\{ -B(\Phi) J_{K-M+2}\left[\frac{U_p C(\Phi)}{\hbar\omega}\right]
-i B(\Phi) e^{i\Phi} J_{K-M-1}(\frac{U_p  C(\Phi)}{\hbar\omega})  \right. \\
& \quad \left. + i \left[ B(\Phi) e^{i\Phi} + D(\Phi) \right] J_{K-M+1}(\frac{U_p 
C(\Phi)}{\hbar\omega})+
\left[ B(\Phi) + D(\Phi)e^{i\Phi}  \right] J_{K-M}(\frac{U_p 
C(\Phi)}{\hbar\omega}) \right\},
\end{split}
\end{equation}
\end{widetext}
where $F_K(x) = ( U_i + U_p - K\hbar\omega ) x - 2 U_p 
[1-\cos(x)]/x$,
$B(x) = -\frac{1}{2}+ \sin(x)/x -2 \sin^2(x/2)/x^2$ and $D(x) = -2 
B(x) -1 + \cos(x)$. The function $b_M(\Phi)$ has a cumbersome 
expression, whose analytical methods to compute it can be found in 
\cite{Antoine:pra:53:1725}.

\section{The Teramobile laser\label{appC}}

\begin{figure}
\includegraphics[width=\columnwidth]{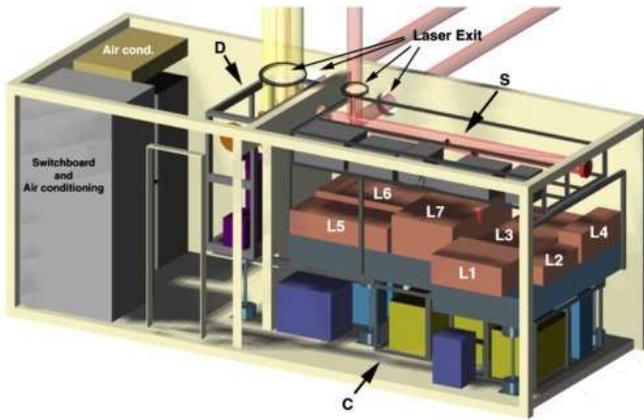}
\caption{Three-dimensional view of the Teramobile. (L): Laser system: 
Ti:Sa oscillator and its Nd:YAG pump laser (L1), stretcher (L2), 
regenerative amplifier, multipass preamplifier (L3) and their Nd:YAG 
pump laser (L4); Multipass main amplifier (L5) pumped by two Nd:YAG 
units (L6); Compressor (L7). (S) Beam expanding system; (C) Power 
supplies; (D) Lidar detection system.}
\label{fig38}
\end{figure}

Field experiments are required to characterize the filamentation over 
long distances as well as to develop atmospheric applications in real 
scale. Such experiments demand mobility to perform investigations at 
adequate locations. Studies of high-power fs-laser beam propagation 
over km-range distances can only be performed outdoors, where relevant aerosol pollutants take place, e.g., in urban areas or at industrial 
sites. Laser-induced lightning investigations require spots where the lightning 
probability is high, as well as test experiments at high voltage 
facilities. These considerations clearly define the need for a mobile 
fs-TW laser system, embedded in a standard freight 
container-integrated laboratory equipped with the necessary Lidar 
detection, power and cooling supplies, temperature stabilization, 
vibration control, and an additional standard Lidar system to assure 
eye safety. These specifications were first achieved by the Teramobile system 
\cite{Wille:epjap:20:183}. The laser itself is based on a 
Ti:Sapphire CPA oscillator and a Nd:YAG pumped Ti:Sa amplification 
chain. It provides 350 mJ pulses with 70 fs duration resulting in a 
peak power of 5 TW at around 800 nm and with a 
repetition rate of 10 Hz. Its integration in the reduced space of the 
mobile laboratory required a particularly compact design (Fig.\ 
\ref{fig38}). The classical compressor setup has been improved into 
a chirp generator to pre-compensate the group velocity dispersion 
in air. Combined with an adjustable focus, this permits to 
control the location of the onset of filamentation and its length. 
Mechanical and 
thermal stabilities of the mobile laboratory are kept under control, so that the Teramobile 
can be transported to any place in the world and operated 
even under adverse weather conditions.

The Teramobile container also includes a Lidar detection chain based 
on a 40 cm receiving telescope, a high-resolution spectrometer 
equipped with a set of gratings and detectors allowing simultaneous 
temporal and spectral analysis of the return signal in a wavelength 
range comprised between 190 nm and 2.5 $\mu$m.

\bibliography{/cea/dsku/trekking/hal1/home/ppe/berge/BIBTEX/references}

\begin{thebibliography}{379}
\expandafter\ifx\csname natexlab\endcsname\relax\def\natexlab#1{#1}\fi
\expandafter\ifx\csname bibnamefont\endcsname\relax
  \def\bibnamefont#1{#1}\fi
\expandafter\ifx\csname bibfnamefont\endcsname\relax
  \def\bibfnamefont#1{#1}\fi
\expandafter\ifx\csname citenamefont\endcsname\relax
  \def\citenamefont#1{#1}\fi
\expandafter\ifx\csname url\endcsname\relax
  \def\url#1{\texttt{#1}}\fi
\expandafter\ifx\csname urlprefix\endcsname\relax\def\urlprefix{URL }\fi
\providecommand{\bibinfo}[2]{#2}
\providecommand{\eprint}[2][]{\url{#2}}

\bibitem[{\citenamefont{Abramovitz and Stegun}(1972)}]{Abramovitz:HMF:72}
\bibinfo{author}{\bibnamefont{Abramovitz}, \bibfnamefont{M.}}, and
  \bibinfo{author}{\bibfnamefont{I.~A.} \bibnamefont{Stegun}},
  \bibinfo{year}{1972}, \emph{\bibinfo{title}{Handbook of Mathematical
  Functions}} (\bibinfo{publisher}{Dover}, \bibinfo{address}{New York}).

\bibitem[{\citenamefont{Ackermann}
  \emph{et~al.}(2006{\natexlab{a}})\citenamefont{Ackermann, M\'echain,
  M\'ejean, Bourayou, Rodriguez, Stelmaszczyk, Kasparian, Yu, Salmon,
  Tzortzakis, Andr\'e, Bourrillon} \emph{et~al.}}]{Ackermann:apb:82:561}
\bibinfo{author}{\bibnamefont{Ackermann}, \bibfnamefont{R.}},
  \bibinfo{author}{\bibfnamefont{G.}~\bibnamefont{M\'echain}},
  \bibinfo{author}{\bibfnamefont{G.}~\bibnamefont{M\'ejean}},
  \bibinfo{author}{\bibfnamefont{R.}~\bibnamefont{Bourayou}},
  \bibinfo{author}{\bibfnamefont{M.}~\bibnamefont{Rodriguez}},
  \bibinfo{author}{\bibfnamefont{K.}~\bibnamefont{Stelmaszczyk}},
  \bibinfo{author}{\bibfnamefont{J.}~\bibnamefont{Kasparian}},
  \bibinfo{author}{\bibfnamefont{J.}~\bibnamefont{Yu}},
  \bibinfo{author}{\bibfnamefont{E.}~\bibnamefont{Salmon}},
  \bibinfo{author}{\bibfnamefont{S.}~\bibnamefont{Tzortzakis}},
  \bibinfo{author}{\bibfnamefont{Y.-B.} \bibnamefont{Andr\'e}},
  \bibinfo{author}{\bibfnamefont{J.-F.} \bibnamefont{Bourrillon}},
  \emph{et~al.}, \bibinfo{year}{2006}{\natexlab{a}}, \bibinfo{journal}{Appl.\
  Phys.\ B: Lasers \& Optics} \textbf{\bibinfo{volume}{82}},
  \bibinfo{pages}{561}.

\bibitem[{\citenamefont{Ackermann}
  \emph{et~al.}(2006{\natexlab{b}})\citenamefont{Ackermann, M{\'e}jean,
  Kasparian, Yu, Salmon, and Wolf}}]{Ackermann:ol:31:86}
\bibinfo{author}{\bibnamefont{Ackermann}, \bibfnamefont{R.}},
  \bibinfo{author}{\bibfnamefont{G.}~\bibnamefont{M{\'e}jean}},
  \bibinfo{author}{\bibfnamefont{J.}~\bibnamefont{Kasparian}},
  \bibinfo{author}{\bibfnamefont{J.}~\bibnamefont{Yu}},
  \bibinfo{author}{\bibfnamefont{E.}~\bibnamefont{Salmon}}, and
  \bibinfo{author}{\bibfnamefont{J.-P.} \bibnamefont{Wolf}},
  \bibinfo{year}{2006}{\natexlab{b}}, \bibinfo{journal}{Opt.\ Lett.}
  \textbf{\bibinfo{volume}{31}}, \bibinfo{pages}{86}.

\bibitem[{\citenamefont{Ackermann} \emph{et~al.}(2004)\citenamefont{Ackermann,
  Stelmaszczyk, Rohwetter, M{\'e}jean, Salmon, Yu, Kasparian, M{\'e}chain,
  Bergmann, Schaper, Weise, Kumm} \emph{et~al.}}]{Ackermann:apl:85:5781}
\bibinfo{author}{\bibnamefont{Ackermann}, \bibfnamefont{R.}},
  \bibinfo{author}{\bibfnamefont{K.}~\bibnamefont{Stelmaszczyk}},
  \bibinfo{author}{\bibfnamefont{P.}~\bibnamefont{Rohwetter}},
  \bibinfo{author}{\bibfnamefont{G.}~\bibnamefont{M{\'e}jean}},
  \bibinfo{author}{\bibfnamefont{E.}~\bibnamefont{Salmon}},
  \bibinfo{author}{\bibfnamefont{J.}~\bibnamefont{Yu}},
  \bibinfo{author}{\bibfnamefont{J.}~\bibnamefont{Kasparian}},
  \bibinfo{author}{\bibfnamefont{G.}~\bibnamefont{M{\'e}chain}},
  \bibinfo{author}{\bibfnamefont{V.}~\bibnamefont{Bergmann}},
  \bibinfo{author}{\bibfnamefont{S.}~\bibnamefont{Schaper}},
  \bibinfo{author}{\bibfnamefont{B.}~\bibnamefont{Weise}},
  \bibinfo{author}{\bibfnamefont{T.}~\bibnamefont{Kumm}}, \emph{et~al.},
  \bibinfo{year}{2004}, \bibinfo{journal}{Appl.\ Phys.\ Lett.}
  \textbf{\bibinfo{volume}{85}}, \bibinfo{pages}{5781}.

\bibitem[{\citenamefont{Agostini} \emph{et~al.}(1979)\citenamefont{Agostini,
  Fabre, Mainfray, Petite, and Rahman}}]{Agostini:prl:42:1127}
\bibinfo{author}{\bibnamefont{Agostini}, \bibfnamefont{P.}},
  \bibinfo{author}{\bibfnamefont{F.}~\bibnamefont{Fabre}},
  \bibinfo{author}{\bibfnamefont{G.}~\bibnamefont{Mainfray}},
  \bibinfo{author}{\bibfnamefont{G.}~\bibnamefont{Petite}}, and
  \bibinfo{author}{\bibfnamefont{N.~K.} \bibnamefont{Rahman}},
  \bibinfo{year}{1979}, \bibinfo{journal}{Phys.\ Rev.\ Lett.}
  \textbf{\bibinfo{volume}{42}}, \bibinfo{pages}{1127}.

\bibitem[{\citenamefont{Agrawal}(2001)}]{Agrawal:NFO:01}
\bibinfo{author}{\bibnamefont{Agrawal}, \bibfnamefont{G.~P.}},
  \bibinfo{year}{2001}, \emph{\bibinfo{title}{Nonlinear Fiber Optics}}
  (\bibinfo{publisher}{Academic Press}, \bibinfo{address}{San Diego}),
  \bibinfo{edition}{third} edition.

\bibitem[{\citenamefont{Akhmediev} \emph{et~al.}(1992)\citenamefont{Akhmediev,
  Korneev, and Nabiev}}]{Akhmediev:ol:17:393}
\bibinfo{author}{\bibnamefont{Akhmediev}, \bibfnamefont{N.~N.}},
  \bibinfo{author}{\bibfnamefont{V.~I.} \bibnamefont{Korneev}}, and
  \bibinfo{author}{\bibfnamefont{R.~F.} \bibnamefont{Nabiev}},
  \bibinfo{year}{1992}, \bibinfo{journal}{Opt.\ Lett.}
  \textbf{\bibinfo{volume}{17}}, \bibinfo{pages}{393}.

\bibitem[{\citenamefont{Ak{\"o}zbek}
  \emph{et~al.}(2003)\citenamefont{Ak{\"o}zbek, Becker, Scalora, Chin, and
  Bowden}}]{Akozbek:apb:77:177}
\bibinfo{author}{\bibnamefont{Ak{\"o}zbek}, \bibfnamefont{N.}},
  \bibinfo{author}{\bibfnamefont{A.}~\bibnamefont{Becker}},
  \bibinfo{author}{\bibfnamefont{M.}~\bibnamefont{Scalora}},
  \bibinfo{author}{\bibfnamefont{S.~L.} \bibnamefont{Chin}}, and
  \bibinfo{author}{\bibfnamefont{C.~M.} \bibnamefont{Bowden}},
  \bibinfo{year}{2003}, \bibinfo{journal}{Appl.\ Phys.\ B: Lasers \& Optics}
  \textbf{\bibinfo{volume}{77}}, \bibinfo{pages}{177}.

\bibitem[{\citenamefont{Ak{\"o}zbek}
  \emph{et~al.}(2002)\citenamefont{Ak{\"o}zbek, Iwasaki, Becker, Scalora, Chin,
  and Bowden}}]{Akozbek:prl:89:143901}
\bibinfo{author}{\bibnamefont{Ak{\"o}zbek}, \bibfnamefont{N.}},
  \bibinfo{author}{\bibfnamefont{A.}~\bibnamefont{Iwasaki}},
  \bibinfo{author}{\bibfnamefont{A.}~\bibnamefont{Becker}},
  \bibinfo{author}{\bibfnamefont{M.}~\bibnamefont{Scalora}},
  \bibinfo{author}{\bibfnamefont{S.~L.} \bibnamefont{Chin}}, and
  \bibinfo{author}{\bibfnamefont{C.~M.} \bibnamefont{Bowden}},
  \bibinfo{year}{2002}, \bibinfo{journal}{Phys.\ Rev.\ Lett.}
  \textbf{\bibinfo{volume}{89}}, \bibinfo{pages}{143901}.

\bibitem[{\citenamefont{Ak{\"o}zbek}
  \emph{et~al.}(2001)\citenamefont{Ak{\"o}zbek, Scalora, Bowden, , and
  Chin}}]{Akozbek:oc:191:353}
\bibinfo{author}{\bibnamefont{Ak{\"o}zbek}, \bibfnamefont{N.}},
  \bibinfo{author}{\bibfnamefont{M.}~\bibnamefont{Scalora}},
  \bibinfo{author}{\bibfnamefont{C.~M.} \bibnamefont{Bowden}}, , and
  \bibinfo{author}{\bibfnamefont{S.~L.} \bibnamefont{Chin}},
  \bibinfo{year}{2001}, \bibinfo{journal}{Opt.\ Commun.}
  \textbf{\bibinfo{volume}{191}}, \bibinfo{pages}{353}.

\bibitem[{\citenamefont{Ak{\"o}zbek}
  \emph{et~al.}(2006)\citenamefont{Ak{\"o}zbek, Trushin, Baltu\v{s}ka, Fuss,
  Goulielmakis, Kosma, Krausz, Panja, Uiberacker, Schmid, Becker, Scalora}
  \emph{et~al.}}]{Akozbek:njp:8:177}
\bibinfo{author}{\bibnamefont{Ak{\"o}zbek}, \bibfnamefont{N.}},
  \bibinfo{author}{\bibfnamefont{S.~A.} \bibnamefont{Trushin}},
  \bibinfo{author}{\bibfnamefont{A.}~\bibnamefont{Baltu\v{s}ka}},
  \bibinfo{author}{\bibfnamefont{W.}~\bibnamefont{Fuss}},
  \bibinfo{author}{\bibfnamefont{E.}~\bibnamefont{Goulielmakis}},
  \bibinfo{author}{\bibfnamefont{K.}~\bibnamefont{Kosma}},
  \bibinfo{author}{\bibfnamefont{F.}~\bibnamefont{Krausz}},
  \bibinfo{author}{\bibfnamefont{S.}~\bibnamefont{Panja}},
  \bibinfo{author}{\bibfnamefont{M.}~\bibnamefont{Uiberacker}},
  \bibinfo{author}{\bibfnamefont{W.~E.} \bibnamefont{Schmid}},
  \bibinfo{author}{\bibfnamefont{A.}~\bibnamefont{Becker}},
  \bibinfo{author}{\bibfnamefont{M.}~\bibnamefont{Scalora}}, \emph{et~al.},
  \bibinfo{year}{2006}, \bibinfo{journal}{New J.\ Phys.}
  \textbf{\bibinfo{volume}{8}}, \bibinfo{pages}{177}.

\bibitem[{\citenamefont{Albert} \emph{et~al.}(2003)\citenamefont{Albert, Roger,
  Glinec, Loulergue, Etchepare, Boulmer-Leborgne, Peeri{\`e}re, and
  Millon}}]{Albert:apa:76:319}
\bibinfo{author}{\bibnamefont{Albert}, \bibfnamefont{O.}},
  \bibinfo{author}{\bibfnamefont{S.}~\bibnamefont{Roger}},
  \bibinfo{author}{\bibfnamefont{Y.}~\bibnamefont{Glinec}},
  \bibinfo{author}{\bibfnamefont{J.~C.} \bibnamefont{Loulergue}},
  \bibinfo{author}{\bibfnamefont{J.}~\bibnamefont{Etchepare}},
  \bibinfo{author}{\bibfnamefont{C.}~\bibnamefont{Boulmer-Leborgne}},
  \bibinfo{author}{\bibfnamefont{J.}~\bibnamefont{Peeri{\`e}re}}, and
  \bibinfo{author}{\bibfnamefont{E.}~\bibnamefont{Millon}},
  \bibinfo{year}{2003}, \bibinfo{journal}{Appl.\ Phys.\ A: Materials Science \&
  Processing} \textbf{\bibinfo{volume}{76}}, \bibinfo{pages}{319}.

\bibitem[{\citenamefont{Alexeev} \emph{et~al.}(2004)\citenamefont{Alexeev,
  Ting, Gordon, Briscoe, Pe{\~n}ano, Hubbard, and
  Sprangle}}]{Alexeev:apl:84:4080}
\bibinfo{author}{\bibnamefont{Alexeev}, \bibfnamefont{I.}},
  \bibinfo{author}{\bibfnamefont{A.}~\bibnamefont{Ting}},
  \bibinfo{author}{\bibfnamefont{D.~F.} \bibnamefont{Gordon}},
  \bibinfo{author}{\bibfnamefont{E.}~\bibnamefont{Briscoe}},
  \bibinfo{author}{\bibfnamefont{J.~R.} \bibnamefont{Pe{\~n}ano}},
  \bibinfo{author}{\bibfnamefont{R.~F.} \bibnamefont{Hubbard}}, and
  \bibinfo{author}{\bibfnamefont{P.}~\bibnamefont{Sprangle}},
  \bibinfo{year}{2004}, \bibinfo{journal}{Appl.\ Phys.\ Lett.}
  \textbf{\bibinfo{volume}{84}}, \bibinfo{pages}{4080}.

\bibitem[{\citenamefont{Alexeev} \emph{et~al.}(2005)\citenamefont{Alexeev,
  Ting, Gordon, Briscoe, Hafizi, and Sprangle}}]{Alexeev:ol:30:1503}
\bibinfo{author}{\bibnamefont{Alexeev}, \bibfnamefont{I.}},
  \bibinfo{author}{\bibfnamefont{A.~C.} \bibnamefont{Ting}},
  \bibinfo{author}{\bibfnamefont{D.~F.} \bibnamefont{Gordon}},
  \bibinfo{author}{\bibfnamefont{E.}~\bibnamefont{Briscoe}},
  \bibinfo{author}{\bibfnamefont{B.}~\bibnamefont{Hafizi}}, and
  \bibinfo{author}{\bibfnamefont{P.}~\bibnamefont{Sprangle}},
  \bibinfo{year}{2005}, \bibinfo{journal}{Opt.\ Lett.}
  \textbf{\bibinfo{volume}{30}}, \bibinfo{pages}{1503}.

\bibitem[{\citenamefont{Alfano and Shapiro}(1970)}]{Alfano:prl:24:584}
\bibinfo{author}{\bibnamefont{Alfano}, \bibfnamefont{R.~R.}}, and
  \bibinfo{author}{\bibfnamefont{S.~L.} \bibnamefont{Shapiro}},
  \bibinfo{year}{1970}, \bibinfo{journal}{Phys.\ Rev.\ Lett.}
  \textbf{\bibinfo{volume}{24}}, \bibinfo{pages}{584}.

\bibitem[{\citenamefont{Ammosov} \emph{et~al.}(1986)\citenamefont{Ammosov,
  Delone, and Kra\v{i}nov}}]{Ammosov:spjetp:64:1191}
\bibinfo{author}{\bibnamefont{Ammosov}, \bibfnamefont{M.~V.}},
  \bibinfo{author}{\bibfnamefont{N.~B.} \bibnamefont{Delone}}, and
  \bibinfo{author}{\bibfnamefont{V.~P.} \bibnamefont{Kra\v{i}nov}},
  \bibinfo{year}{1986}, \bibinfo{journal}{Sov.\ Phys.\ JETP}
  \textbf{\bibinfo{volume}{64}}, \bibinfo{pages}{1191}.

\bibitem[{\citenamefont{Anderson}(1983)}]{Anderson:pra:27:3135}
\bibinfo{author}{\bibnamefont{Anderson}, \bibfnamefont{D.}},
  \bibinfo{year}{1983}, \bibinfo{journal}{Phys.\ Rev.\ A}
  \textbf{\bibinfo{volume}{27}}, \bibinfo{pages}{3135}.

\bibitem[{\citenamefont{Anderson and Bonnedal}(1979)}]{Anderson:pf:22:105}
\bibinfo{author}{\bibnamefont{Anderson}, \bibfnamefont{D.}}, and
  \bibinfo{author}{\bibfnamefont{M.}~\bibnamefont{Bonnedal}},
  \bibinfo{year}{1979}, \bibinfo{journal}{Phys.\ Fluids}
  \textbf{\bibinfo{volume}{22}}, \bibinfo{pages}{105}.

\bibitem[{\citenamefont{Anderson} \emph{et~al.}(1979)\citenamefont{Anderson,
  Bonnedal, and Lisak}}]{Anderson:pf:22:1838}
\bibinfo{author}{\bibnamefont{Anderson}, \bibfnamefont{D.}},
  \bibinfo{author}{\bibfnamefont{M.}~\bibnamefont{Bonnedal}}, and
  \bibinfo{author}{\bibfnamefont{M.}~\bibnamefont{Lisak}},
  \bibinfo{year}{1979}, \bibinfo{journal}{Phys.\ Fluids}
  \textbf{\bibinfo{volume}{22}}, \bibinfo{pages}{1838}.

\bibitem[{\citenamefont{Anderson and Lisak}(1983)}]{Anderson:pra:27:1393}
\bibinfo{author}{\bibnamefont{Anderson}, \bibfnamefont{D.}}, and
  \bibinfo{author}{\bibfnamefont{M.}~\bibnamefont{Lisak}},
  \bibinfo{year}{1983}, \bibinfo{journal}{Phys.\ Rev.\ A}
  \textbf{\bibinfo{volume}{27}}, \bibinfo{pages}{1393}.

\bibitem[{\citenamefont{Angel} \emph{et~al.}(2001)\citenamefont{Angel, Stratis,
  Eland, Lai, Berg, and Gold}}]{Angel:fjac:369:320}
\bibinfo{author}{\bibnamefont{Angel}, \bibfnamefont{S.~M.}},
  \bibinfo{author}{\bibfnamefont{D.~N.} \bibnamefont{Stratis}},
  \bibinfo{author}{\bibfnamefont{K.~L.} \bibnamefont{Eland}},
  \bibinfo{author}{\bibfnamefont{T.}~\bibnamefont{Lai}},
  \bibinfo{author}{\bibfnamefont{M.~A.} \bibnamefont{Berg}}, and
  \bibinfo{author}{\bibfnamefont{D.~M.} \bibnamefont{Gold}},
  \bibinfo{year}{2001}, \bibinfo{journal}{Fresenius J. \ Anal. \ Chem.}
  \textbf{\bibinfo{volume}{369}}, \bibinfo{pages}{320}.

\bibitem[{\citenamefont{Antoine} \emph{et~al.}(1996)\citenamefont{Antoine,
  L'Huillier, Lewenstein, Sali\`eres, and Carr\'e}}]{Antoine:pra:53:1725}
\bibinfo{author}{\bibnamefont{Antoine}, \bibfnamefont{P.}},
  \bibinfo{author}{\bibfnamefont{A.}~\bibnamefont{L'Huillier}},
  \bibinfo{author}{\bibfnamefont{M.}~\bibnamefont{Lewenstein}},
  \bibinfo{author}{\bibfnamefont{P.}~\bibnamefont{Sali\`eres}}, and
  \bibinfo{author}{\bibfnamefont{B.}~\bibnamefont{Carr\'e}},
  \bibinfo{year}{1996}, \bibinfo{journal}{Phys.\ Rev.\ A}
  \textbf{\bibinfo{volume}{53}}, \bibinfo{pages}{1725}.

\bibitem[{\citenamefont{Ar{\'e}valo and Becker}(2005)}]{Arevalo:pre:72:026605}
\bibinfo{author}{\bibnamefont{Ar{\'e}valo}, \bibfnamefont{E.}}, and
  \bibinfo{author}{\bibfnamefont{A.}~\bibnamefont{Becker}},
  \bibinfo{year}{2005}, \bibinfo{journal}{Phys.\ Rev.\ E}
  \textbf{\bibinfo{volume}{72}}, \bibinfo{pages}{026605}.

\bibitem[{\citenamefont{Askar'yan}(1962)}]{Askaryan:spjetp:15:1088}
\bibinfo{author}{\bibnamefont{Askar'yan}, \bibfnamefont{G.~A.}},
  \bibinfo{year}{1962}, \bibinfo{journal}{Sov.\ Phys.\ JETP}
  \textbf{\bibinfo{volume}{15}}, \bibinfo{pages}{1088}.

\bibitem[{\citenamefont{Atai} \emph{et~al.}(1994)\citenamefont{Atai, Chen, and
  Soto-Crespo}}]{Atai:pra:49:3170}
\bibinfo{author}{\bibnamefont{Atai}, \bibfnamefont{J.}},
  \bibinfo{author}{\bibfnamefont{Y.}~\bibnamefont{Chen}}, and
  \bibinfo{author}{\bibfnamefont{J.~M.} \bibnamefont{Soto-Crespo}},
  \bibinfo{year}{1994}, \bibinfo{journal}{Phys.\ Rev.\ A}
  \textbf{\bibinfo{volume}{49}}, \bibinfo{pages}{R3170}.

\bibitem[{\citenamefont{Audebert} \emph{et~al.}(1994)\citenamefont{Audebert,
  Daguzan, {Dos Santos}, Gauthier, Geindre, Guizard, Hamoniaux, Krastev,
  Martin, Petite, and Antonetti}}]{Audebert:prl:73:1990}
\bibinfo{author}{\bibnamefont{Audebert}, \bibfnamefont{P.}},
  \bibinfo{author}{\bibfnamefont{P.}~\bibnamefont{Daguzan}},
  \bibinfo{author}{\bibfnamefont{A.}~\bibnamefont{{Dos Santos}}},
  \bibinfo{author}{\bibfnamefont{J.-C.} \bibnamefont{Gauthier}},
  \bibinfo{author}{\bibfnamefont{J.-P.} \bibnamefont{Geindre}},
  \bibinfo{author}{\bibfnamefont{S.}~\bibnamefont{Guizard}},
  \bibinfo{author}{\bibfnamefont{G.}~\bibnamefont{Hamoniaux}},
  \bibinfo{author}{\bibfnamefont{K.}~\bibnamefont{Krastev}},
  \bibinfo{author}{\bibfnamefont{P.}~\bibnamefont{Martin}},
  \bibinfo{author}{\bibfnamefont{G.}~\bibnamefont{Petite}}, and
  \bibinfo{author}{\bibfnamefont{A.}~\bibnamefont{Antonetti}},
  \bibinfo{year}{1994}, \bibinfo{journal}{Phys.\ Rev.\ Lett.}
  \textbf{\bibinfo{volume}{73}}, \bibinfo{pages}{1990}.

\bibitem[{\citenamefont{Augst} \emph{et~al.}(1991)\citenamefont{Augst,
  Meyerhofer, Strickland, and Chin}}]{Augst:josab:8:858}
\bibinfo{author}{\bibnamefont{Augst}, \bibfnamefont{S.}},
  \bibinfo{author}{\bibfnamefont{D.~D.} \bibnamefont{Meyerhofer}},
  \bibinfo{author}{\bibfnamefont{D.}~\bibnamefont{Strickland}}, and
  \bibinfo{author}{\bibfnamefont{S.~L.} \bibnamefont{Chin}},
  \bibinfo{year}{1991}, \bibinfo{journal}{J.\ Opt.\ Soc.\ Am.\ B}
  \textbf{\bibinfo{volume}{8}}, \bibinfo{pages}{858}.

\bibitem[{\citenamefont{Backus} \emph{et~al.}(1996)\citenamefont{Backus,
  Peatross, Zeek, Rundquist, Taft, Murnane, and Kapteyn}}]{Backus:ol:21:665}
\bibinfo{author}{\bibnamefont{Backus}, \bibfnamefont{S.}},
  \bibinfo{author}{\bibfnamefont{J.}~\bibnamefont{Peatross}},
  \bibinfo{author}{\bibfnamefont{Z.}~\bibnamefont{Zeek}},
  \bibinfo{author}{\bibfnamefont{A.}~\bibnamefont{Rundquist}},
  \bibinfo{author}{\bibfnamefont{G.}~\bibnamefont{Taft}},
  \bibinfo{author}{\bibfnamefont{M.~M.} \bibnamefont{Murnane}}, and
  \bibinfo{author}{\bibfnamefont{H.~C.} \bibnamefont{Kapteyn}},
  \bibinfo{year}{1996}, \bibinfo{journal}{Opt.\ Lett.}
  \textbf{\bibinfo{volume}{21}}, \bibinfo{pages}{665}.

\bibitem[{\citenamefont{Ball}(1974)}]{Ball:ao:13:2292}
\bibinfo{author}{\bibnamefont{Ball}, \bibfnamefont{L.~M.}},
  \bibinfo{year}{1974}, \bibinfo{journal}{Appl.\ Opt.}
  \textbf{\bibinfo{volume}{13}}, \bibinfo{pages}{2292}.

\bibitem[{\citenamefont{Baudelet} \emph{et~al.}(2006)\citenamefont{Baudelet,
  Guyon, Yu, Wolf, Fr{\'e}jafon, and Laloi}}]{Baudelet:jap:99:084701}
\bibinfo{author}{\bibnamefont{Baudelet}, \bibfnamefont{M.}},
  \bibinfo{author}{\bibfnamefont{L.}~\bibnamefont{Guyon}},
  \bibinfo{author}{\bibfnamefont{J.}~\bibnamefont{Yu}},
  \bibinfo{author}{\bibfnamefont{J.-P.} \bibnamefont{Wolf}},
  \bibinfo{author}{\bibfnamefont{T.~A.~E.} \bibnamefont{Fr{\'e}jafon}}, and
  \bibinfo{author}{\bibfnamefont{P.}~\bibnamefont{Laloi}},
  \bibinfo{year}{2006}, \bibinfo{journal}{J.\ Appl.\ Phys.}
  \textbf{\bibinfo{volume}{99}}, \bibinfo{pages}{084701}.

\bibitem[{\citenamefont{Becker} \emph{et~al.}(2001)\citenamefont{Becker,
  Ak{\"o}zbek, Vijayalakshmi, Oral, Bowden, and Chin}}]{Becker:apb:73:287}
\bibinfo{author}{\bibnamefont{Becker}, \bibfnamefont{A.}},
  \bibinfo{author}{\bibfnamefont{N.}~\bibnamefont{Ak{\"o}zbek}},
  \bibinfo{author}{\bibfnamefont{K.}~\bibnamefont{Vijayalakshmi}},
  \bibinfo{author}{\bibfnamefont{E.}~\bibnamefont{Oral}},
  \bibinfo{author}{\bibfnamefont{C.~M.} \bibnamefont{Bowden}}, and
  \bibinfo{author}{\bibfnamefont{S.~L.} \bibnamefont{Chin}},
  \bibinfo{year}{2001}, \bibinfo{journal}{Appl.\ Phys.\ B: Lasers \& Optics}
  \textbf{\bibinfo{volume}{73}}, \bibinfo{pages}{287}.

\bibitem[{\citenamefont{Berezhiani}
  \emph{et~al.}(2001)\citenamefont{Berezhiani, Skarka, and
  Aleksi\'c}}]{Berezhiani:pre:64:057601}
\bibinfo{author}{\bibnamefont{Berezhiani}, \bibfnamefont{V.~I.}},
  \bibinfo{author}{\bibfnamefont{V.}~\bibnamefont{Skarka}}, and
  \bibinfo{author}{\bibfnamefont{N.~B.} \bibnamefont{Aleksi\'c}},
  \bibinfo{year}{2001}, \bibinfo{journal}{Phys.\ Rev.\ E}
  \textbf{\bibinfo{volume}{64}}, \bibinfo{pages}{057601}.

\bibitem[{\citenamefont{Berg\'e}(1998)}]{Berge:pr:303:259}
\bibinfo{author}{\bibnamefont{Berg\'e}, \bibfnamefont{L.}},
  \bibinfo{year}{1998}, \bibinfo{journal}{Phys.\ Rep.}
  \textbf{\bibinfo{volume}{303}}, \bibinfo{pages}{259}.

\bibitem[{\citenamefont{Berg\'e}(2004)}]{Berge:pre:69:065601}
\bibinfo{author}{\bibnamefont{Berg\'e}, \bibfnamefont{L.}},
  \bibinfo{year}{2004}, \bibinfo{journal}{Phys.\ Rev.\ E}
  \textbf{\bibinfo{volume}{69}}, \bibinfo{pages}{065601(R)}.

\bibitem[{\citenamefont{Berg\'e and Couairon}(2000)}]{Berge:pop:7:210}
\bibinfo{author}{\bibnamefont{Berg\'e}, \bibfnamefont{L.}}, and
  \bibinfo{author}{\bibfnamefont{A.}~\bibnamefont{Couairon}},
  \bibinfo{year}{2000}, \bibinfo{journal}{Phys.\ Plasmas}
  \textbf{\bibinfo{volume}{7}}, \bibinfo{pages}{210}.

\bibitem[{\citenamefont{Berg\'e and
  Couairon}(2001{\natexlab{a}})}]{Berge:prl:86:1003}
\bibinfo{author}{\bibnamefont{Berg\'e}, \bibfnamefont{L.}}, and
  \bibinfo{author}{\bibfnamefont{A.}~\bibnamefont{Couairon}},
  \bibinfo{year}{2001}{\natexlab{a}}, \bibinfo{journal}{Phys.\ Rev.\ Lett.}
  \textbf{\bibinfo{volume}{86}}, \bibinfo{pages}{1003}.

\bibitem[{\citenamefont{Berg\'e and
  Couairon}(2001{\natexlab{b}})}]{Berge:pd:152:752}
\bibinfo{author}{\bibnamefont{Berg\'e}, \bibfnamefont{L.}}, and
  \bibinfo{author}{\bibfnamefont{A.}~\bibnamefont{Couairon}},
  \bibinfo{year}{2001}{\natexlab{b}}, \bibinfo{journal}{Physica D}
  \textbf{\bibinfo{volume}{152-153}}, \bibinfo{pages}{752}.

\bibitem[{\citenamefont{Berg\'e} \emph{et~al.}(2002)\citenamefont{Berg\'e,
  Germaschewski, Grauer, and Rasmussen}}]{Berge:prl:89:153902}
\bibinfo{author}{\bibnamefont{Berg\'e}, \bibfnamefont{L.}},
  \bibinfo{author}{\bibfnamefont{K.}~\bibnamefont{Germaschewski}},
  \bibinfo{author}{\bibfnamefont{R.}~\bibnamefont{Grauer}}, and
  \bibinfo{author}{\bibfnamefont{J.~J.} \bibnamefont{Rasmussen}},
  \bibinfo{year}{2002}, \bibinfo{journal}{Phys.\ Rev.\ Lett.}
  \textbf{\bibinfo{volume}{89}}, \bibinfo{pages}{153902}.

\bibitem[{\citenamefont{Berg\'e} \emph{et~al.}(2003)\citenamefont{Berg\'e,
  Gou{\'e}dard, Schj{\o}dt-Eriksen, and Ward}}]{Berge:pd:176:181}
\bibinfo{author}{\bibnamefont{Berg\'e}, \bibfnamefont{L.}},
  \bibinfo{author}{\bibfnamefont{C.}~\bibnamefont{Gou{\'e}dard}},
  \bibinfo{author}{\bibfnamefont{J.}~\bibnamefont{Schj{\o}dt-Eriksen}}, and
  \bibinfo{author}{\bibfnamefont{H.}~\bibnamefont{Ward}}, \bibinfo{year}{2003},
  \bibinfo{journal}{Physica D} \textbf{\bibinfo{volume}{176}},
  \bibinfo{pages}{181}.

\bibitem[{\citenamefont{Berg\'e} \emph{et~al.}(1997)\citenamefont{Berg\'e,
  Schmidt, Rasmussen, Christiansen, and Rasmussen}}]{Berge:josab:14:2550}
\bibinfo{author}{\bibnamefont{Berg\'e}, \bibfnamefont{L.}},
  \bibinfo{author}{\bibfnamefont{M.~R.} \bibnamefont{Schmidt}},
  \bibinfo{author}{\bibfnamefont{J.~J.} \bibnamefont{Rasmussen}},
  \bibinfo{author}{\bibfnamefont{P.~L.} \bibnamefont{Christiansen}}, and
  \bibinfo{author}{\bibfnamefont{K.~{\O}.} \bibnamefont{Rasmussen}},
  \bibinfo{year}{1997}, \bibinfo{journal}{J.\ Opt.\ Soc.\ Am.\ B}
  \textbf{\bibinfo{volume}{14}}, \bibinfo{pages}{2550}.

\bibitem[{\citenamefont{Berg\'e and Skupin}(2005)}]{Berge:pre:71:065601}
\bibinfo{author}{\bibnamefont{Berg\'e}, \bibfnamefont{L.}}, and
  \bibinfo{author}{\bibfnamefont{S.}~\bibnamefont{Skupin}},
  \bibinfo{year}{2005}, \bibinfo{journal}{Phys.\ Rev.\ E}
  \textbf{\bibinfo{volume}{71}}, \bibinfo{pages}{065601(R)}.

\bibitem[{\citenamefont{Berg\'e} \emph{et~al.}(2004)\citenamefont{Berg\'e,
  Skupin, Lederer, M{\'e}jean, Yu, Kasparian, Salmon, Wolf, Rodriguez,
  W{\"o}ste, Bourayou, and Sauerbrey}}]{Berge:prl:92:225002}
\bibinfo{author}{\bibnamefont{Berg\'e}, \bibfnamefont{L.}},
  \bibinfo{author}{\bibfnamefont{S.}~\bibnamefont{Skupin}},
  \bibinfo{author}{\bibfnamefont{F.}~\bibnamefont{Lederer}},
  \bibinfo{author}{\bibfnamefont{G.}~\bibnamefont{M{\'e}jean}},
  \bibinfo{author}{\bibfnamefont{J.}~\bibnamefont{Yu}},
  \bibinfo{author}{\bibfnamefont{J.}~\bibnamefont{Kasparian}},
  \bibinfo{author}{\bibfnamefont{E.}~\bibnamefont{Salmon}},
  \bibinfo{author}{\bibfnamefont{J.~P.} \bibnamefont{Wolf}},
  \bibinfo{author}{\bibfnamefont{M.}~\bibnamefont{Rodriguez}},
  \bibinfo{author}{\bibfnamefont{L.}~\bibnamefont{W{\"o}ste}},
  \bibinfo{author}{\bibfnamefont{R.}~\bibnamefont{Bourayou}}, and
  \bibinfo{author}{\bibfnamefont{R.}~\bibnamefont{Sauerbrey}},
  \bibinfo{year}{2004}, \bibinfo{journal}{Phys.\ Rev.\ Lett.}
  \textbf{\bibinfo{volume}{92}}, \bibinfo{pages}{225002}.

\bibitem[{\citenamefont{Berg\'e} \emph{et~al.}(2005)\citenamefont{Berg\'e,
  Skupin, M{\'e}jean, Kasparian, Yu, Frey, Salmon, and
  Wolf}}]{Berge:pre:71:016602}
\bibinfo{author}{\bibnamefont{Berg\'e}, \bibfnamefont{L.}},
  \bibinfo{author}{\bibfnamefont{S.}~\bibnamefont{Skupin}},
  \bibinfo{author}{\bibfnamefont{G.}~\bibnamefont{M{\'e}jean}},
  \bibinfo{author}{\bibfnamefont{J.}~\bibnamefont{Kasparian}},
  \bibinfo{author}{\bibfnamefont{J.}~\bibnamefont{Yu}},
  \bibinfo{author}{\bibfnamefont{S.}~\bibnamefont{Frey}},
  \bibinfo{author}{\bibfnamefont{E.}~\bibnamefont{Salmon}}, and
  \bibinfo{author}{\bibfnamefont{J.~P.} \bibnamefont{Wolf}},
  \bibinfo{year}{2005}, \bibinfo{journal}{Phys.\ Rev.\ E}
  \textbf{\bibinfo{volume}{71}}, \bibinfo{pages}{016602}.

\bibitem[{\citenamefont{Berkovsky} \emph{et~al.}(2005)\citenamefont{Berkovsky,
  Kozlov, and Shpolyanskiy}}]{Berkovsky:pra:72:043821}
\bibinfo{author}{\bibnamefont{Berkovsky}, \bibfnamefont{A.~N.}},
  \bibinfo{author}{\bibfnamefont{S.~A.} \bibnamefont{Kozlov}}, and
  \bibinfo{author}{\bibfnamefont{Y.~A.} \bibnamefont{Shpolyanskiy}},
  \bibinfo{year}{2005}, \bibinfo{journal}{Phys.\ Rev.\ A}
  \textbf{\bibinfo{volume}{72}}, \bibinfo{pages}{043821}.

\bibitem[{\citenamefont{Bernstein} \emph{et~al.}(2003)\citenamefont{Bernstein,
  Diels, Luk, Nelson, McPherson, and Cameron}}]{Bernstein:ol:28:2354}
\bibinfo{author}{\bibnamefont{Bernstein}, \bibfnamefont{A.~C.}},
  \bibinfo{author}{\bibfnamefont{J.-C.} \bibnamefont{Diels}},
  \bibinfo{author}{\bibfnamefont{T.~S.} \bibnamefont{Luk}},
  \bibinfo{author}{\bibfnamefont{T.~R.} \bibnamefont{Nelson}},
  \bibinfo{author}{\bibfnamefont{A.}~\bibnamefont{McPherson}}, and
  \bibinfo{author}{\bibfnamefont{S.~M.} \bibnamefont{Cameron}},
  \bibinfo{year}{2003}, \bibinfo{journal}{Opt.\ Lett.}
  \textbf{\bibinfo{volume}{28}}, \bibinfo{pages}{2354}.

\bibitem[{\citenamefont{Bespalov and Talanov}(1966)}]{Bespalov:jetp:3:307}
\bibinfo{author}{\bibnamefont{Bespalov}, \bibfnamefont{V.~I.}}, and
  \bibinfo{author}{\bibfnamefont{V.~I.} \bibnamefont{Talanov}},
  \bibinfo{year}{1966}, \bibinfo{journal}{JETP Lett.}
  \textbf{\bibinfo{volume}{3}}, \bibinfo{pages}{307}.

\bibitem[{\citenamefont{Bondeson} \emph{et~al.}(1979)\citenamefont{Bondeson,
  Lisak, and Anderson}}]{Bondeson:ps:20:479}
\bibinfo{author}{\bibnamefont{Bondeson}, \bibfnamefont{A.}},
  \bibinfo{author}{\bibfnamefont{M.}~\bibnamefont{Lisak}}, and
  \bibinfo{author}{\bibfnamefont{D.}~\bibnamefont{Anderson}},
  \bibinfo{year}{1979}, \bibinfo{journal}{Phys.\ Scr.}
  \textbf{\bibinfo{volume}{20}}, \bibinfo{pages}{479}.

\bibitem[{\citenamefont{Bourayou} \emph{et~al.}(2005)\citenamefont{Bourayou,
  M{\'e}jean, Kasparian, Rodriguez, Salmon, Yu, Lehmann, Stecklum, Laux,
  Eisl{\"o}ffel, Scholz, Hatzes} \emph{et~al.}}]{Bourayou:josab:22:369}
\bibinfo{author}{\bibnamefont{Bourayou}, \bibfnamefont{R.}},
  \bibinfo{author}{\bibfnamefont{G.}~\bibnamefont{M{\'e}jean}},
  \bibinfo{author}{\bibfnamefont{J.}~\bibnamefont{Kasparian}},
  \bibinfo{author}{\bibfnamefont{M.}~\bibnamefont{Rodriguez}},
  \bibinfo{author}{\bibfnamefont{E.}~\bibnamefont{Salmon}},
  \bibinfo{author}{\bibfnamefont{J.}~\bibnamefont{Yu}},
  \bibinfo{author}{\bibfnamefont{H.}~\bibnamefont{Lehmann}},
  \bibinfo{author}{\bibfnamefont{B.}~\bibnamefont{Stecklum}},
  \bibinfo{author}{\bibfnamefont{U.}~\bibnamefont{Laux}},
  \bibinfo{author}{\bibfnamefont{J.}~\bibnamefont{Eisl{\"o}ffel}},
  \bibinfo{author}{\bibfnamefont{A.}~\bibnamefont{Scholz}},
  \bibinfo{author}{\bibfnamefont{A.~P.} \bibnamefont{Hatzes}}, \emph{et~al.},
  \bibinfo{year}{2005}, \bibinfo{journal}{J.\ Opt.\ Soc.\ Am.\ B}
  \textbf{\bibinfo{volume}{22}}, \bibinfo{pages}{369}.

\bibitem[{\citenamefont{Boutou} \emph{et~al.}(2002)\citenamefont{Boutou, Favre,
  Hill, Pan, Chang, and Wolf}}]{Boutou:apb:75:145}
\bibinfo{author}{\bibnamefont{Boutou}, \bibfnamefont{V.}},
  \bibinfo{author}{\bibfnamefont{C.}~\bibnamefont{Favre}},
  \bibinfo{author}{\bibfnamefont{S.~C.} \bibnamefont{Hill}},
  \bibinfo{author}{\bibfnamefont{Y.~L.} \bibnamefont{Pan}},
  \bibinfo{author}{\bibfnamefont{R.~K.} \bibnamefont{Chang}}, and
  \bibinfo{author}{\bibfnamefont{J.-P.} \bibnamefont{Wolf}},
  \bibinfo{year}{2002}, \bibinfo{journal}{Appl.\ Phys.\ B: Lasers \& Optics}
  \textbf{\bibinfo{volume}{75}}, \bibinfo{pages}{145}.

\bibitem[{\citenamefont{Boyd}(1992)}]{Boyd:NO:92}
\bibinfo{editor}{\bibnamefont{Boyd}, \bibfnamefont{R.~W.}} (ed.),
  \bibinfo{year}{1992}, \emph{\bibinfo{title}{Nonlinear Optics}}
  (\bibinfo{publisher}{Academic Press}, \bibinfo{address}{San Diego}).

\bibitem[{\citenamefont{Brabec and Krausz}(1997)}]{Brabec:prl:78:3282}
\bibinfo{author}{\bibnamefont{Brabec}, \bibfnamefont{T.}}, and
  \bibinfo{author}{\bibfnamefont{F.}~\bibnamefont{Krausz}},
  \bibinfo{year}{1997}, \bibinfo{journal}{Phys.\ Rev.\ Lett.}
  \textbf{\bibinfo{volume}{78}}, \bibinfo{pages}{3282}.

\bibitem[{\citenamefont{Brabec and Krausz}(2000)}]{Brabec:rmp:72:545}
\bibinfo{author}{\bibnamefont{Brabec}, \bibfnamefont{T.}}, and
  \bibinfo{author}{\bibfnamefont{F.}~\bibnamefont{Krausz}},
  \bibinfo{year}{2000}, \bibinfo{journal}{Rev.\ Mod.\ Phys.}
  \textbf{\bibinfo{volume}{72}}, \bibinfo{pages}{545}.

\bibitem[{\citenamefont{Bransden and Joachain}(2003)}]{Bransden:PAM:03}
\bibinfo{author}{\bibnamefont{Bransden}, \bibfnamefont{B.~H.}}, and
  \bibinfo{author}{\bibfnamefont{C.~J.} \bibnamefont{Joachain}},
  \bibinfo{year}{2003}, \emph{\bibinfo{title}{Physics of Atoms and Molecules}}
  (\bibinfo{publisher}{Pearson Education Limited, Prentice Hall}).

\bibitem[{\citenamefont{Braun} \emph{et~al.}(1995)\citenamefont{Braun, Korn,
  Liu, Du, Squier, and Mourou}}]{Braun:ol:20:73}
\bibinfo{author}{\bibnamefont{Braun}, \bibfnamefont{A.}},
  \bibinfo{author}{\bibfnamefont{G.}~\bibnamefont{Korn}},
  \bibinfo{author}{\bibfnamefont{X.}~\bibnamefont{Liu}},
  \bibinfo{author}{\bibfnamefont{D.}~\bibnamefont{Du}},
  \bibinfo{author}{\bibfnamefont{J.}~\bibnamefont{Squier}}, and
  \bibinfo{author}{\bibfnamefont{G.}~\bibnamefont{Mourou}},
  \bibinfo{year}{1995}, \bibinfo{journal}{Opt.\ Lett.}
  \textbf{\bibinfo{volume}{20}}, \bibinfo{pages}{73}.

\bibitem[{\citenamefont{Brodeur} \emph{et~al.}(1997)\citenamefont{Brodeur,
  Chien, Ilkov, Chin, Kosareva, and Kandidov}}]{Brodeur:ol:22:304}
\bibinfo{author}{\bibnamefont{Brodeur}, \bibfnamefont{A.}},
  \bibinfo{author}{\bibfnamefont{C.~Y.} \bibnamefont{Chien}},
  \bibinfo{author}{\bibfnamefont{F.~A.} \bibnamefont{Ilkov}},
  \bibinfo{author}{\bibfnamefont{S.~L.} \bibnamefont{Chin}},
  \bibinfo{author}{\bibfnamefont{O.~G.} \bibnamefont{Kosareva}}, and
  \bibinfo{author}{\bibfnamefont{V.~P.} \bibnamefont{Kandidov}},
  \bibinfo{year}{1997}, \bibinfo{journal}{Opt.\ Lett.}
  \textbf{\bibinfo{volume}{22}}, \bibinfo{pages}{304}.

\bibitem[{\citenamefont{Brodeur and Chin}(1998)}]{Brodeur:prl:80:4406}
\bibinfo{author}{\bibnamefont{Brodeur}, \bibfnamefont{A.}}, and
  \bibinfo{author}{\bibfnamefont{S.~L.} \bibnamefont{Chin}},
  \bibinfo{year}{1998}, \bibinfo{journal}{Phys.\ Rev.\ Lett.}
  \textbf{\bibinfo{volume}{80}}, \bibinfo{pages}{4406}.

\bibitem[{\citenamefont{Brodeur and Chin}(1999)}]{Brodeur:josab:16:637}
\bibinfo{author}{\bibnamefont{Brodeur}, \bibfnamefont{A.}}, and
  \bibinfo{author}{\bibfnamefont{S.~L.} \bibnamefont{Chin}},
  \bibinfo{year}{1999}, \bibinfo{journal}{J.\ Opt.\ Soc.\ Am.\ B}
  \textbf{\bibinfo{volume}{16}}, \bibinfo{pages}{637}.

\bibitem[{\citenamefont{Buryak} \emph{et~al.}(1999)\citenamefont{Buryak,
  Steblina, and Sammut}}]{Buryak:ol:24:1859}
\bibinfo{author}{\bibnamefont{Buryak}, \bibfnamefont{A.~V.}},
  \bibinfo{author}{\bibfnamefont{V.~V.} \bibnamefont{Steblina}}, and
  \bibinfo{author}{\bibfnamefont{R.~A.} \bibnamefont{Sammut}},
  \bibinfo{year}{1999}, \bibinfo{journal}{Opt.\ Lett.}
  \textbf{\bibinfo{volume}{24}}, \bibinfo{pages}{1859}.

\bibitem[{\citenamefont{Campillo} \emph{et~al.}(1973)\citenamefont{Campillo,
  Shapiro, and Suydam}}]{Campillo:apl:23:628}
\bibinfo{author}{\bibnamefont{Campillo}, \bibfnamefont{A.~J.}},
  \bibinfo{author}{\bibfnamefont{S.~L.} \bibnamefont{Shapiro}}, and
  \bibinfo{author}{\bibfnamefont{B.~R.} \bibnamefont{Suydam}},
  \bibinfo{year}{1973}, \bibinfo{journal}{Appl.\ Phys.\ Lett.}
  \textbf{\bibinfo{volume}{23}}, \bibinfo{pages}{628}.

\bibitem[{\citenamefont{Campillo} \emph{et~al.}(1974)\citenamefont{Campillo,
  Shapiro, and Suydam}}]{Campillo:apl:24:178}
\bibinfo{author}{\bibnamefont{Campillo}, \bibfnamefont{A.~J.}},
  \bibinfo{author}{\bibfnamefont{S.~L.} \bibnamefont{Shapiro}}, and
  \bibinfo{author}{\bibfnamefont{B.~R.} \bibnamefont{Suydam}},
  \bibinfo{year}{1974}, \bibinfo{journal}{Appl.\ Phys.\ Lett.}
  \textbf{\bibinfo{volume}{24}}, \bibinfo{pages}{178}.

\bibitem[{\citenamefont{Cerullo} \emph{et~al.}(1996)\citenamefont{Cerullo,
  Dienes, and Magni}}]{Cerullo:ol:21:65}
\bibinfo{author}{\bibnamefont{Cerullo}, \bibfnamefont{G.}},
  \bibinfo{author}{\bibfnamefont{A.}~\bibnamefont{Dienes}}, and
  \bibinfo{author}{\bibfnamefont{V.}~\bibnamefont{Magni}},
  \bibinfo{year}{1996}, \bibinfo{journal}{Opt.\ Lett.}
  \textbf{\bibinfo{volume}{21}}, \bibinfo{pages}{65}.

\bibitem[{\citenamefont{Champeaux and
  Berg{\'e}}(2003)}]{Champeaux:pre:68:066603}
\bibinfo{author}{\bibnamefont{Champeaux}, \bibfnamefont{S.}}, and
  \bibinfo{author}{\bibfnamefont{L.}~\bibnamefont{Berg{\'e}}},
  \bibinfo{year}{2003}, \bibinfo{journal}{Phys.\ Rev.\ E}
  \textbf{\bibinfo{volume}{68}}, \bibinfo{pages}{066603}.

\bibitem[{\citenamefont{Champeaux and
  Berg{\'e}}(2005)}]{Champeaux:pre:71:046604}
\bibinfo{author}{\bibnamefont{Champeaux}, \bibfnamefont{S.}}, and
  \bibinfo{author}{\bibfnamefont{L.}~\bibnamefont{Berg{\'e}}},
  \bibinfo{year}{2005}, \bibinfo{journal}{Phys.\ Rev.\ E}
  \textbf{\bibinfo{volume}{71}}, \bibinfo{pages}{046604}.

\bibitem[{\citenamefont{Champeaux and Berg{\'e}}(2006)}]{Champeaux:ol:31:1301}
\bibinfo{author}{\bibnamefont{Champeaux}, \bibfnamefont{S.}}, and
  \bibinfo{author}{\bibfnamefont{L.}~\bibnamefont{Berg{\'e}}},
  \bibinfo{year}{2006}, \bibinfo{journal}{Opt.\ Lett.}
  \textbf{\bibinfo{volume}{31}}, \bibinfo{pages}{1301}.

\bibitem[{\citenamefont{Chen} \emph{et~al.}(2006)\citenamefont{Chen, Leng, Liu,
  Zhu, Li, and Xu}}]{Chen:oc:259:331}
\bibinfo{author}{\bibnamefont{Chen}, \bibfnamefont{X.}},
  \bibinfo{author}{\bibfnamefont{Y.}~\bibnamefont{Leng}},
  \bibinfo{author}{\bibfnamefont{J.}~\bibnamefont{Liu}},
  \bibinfo{author}{\bibfnamefont{Y.}~\bibnamefont{Zhu}},
  \bibinfo{author}{\bibfnamefont{R.}~\bibnamefont{Li}}, and
  \bibinfo{author}{\bibfnamefont{Z.}~\bibnamefont{Xu}}, \bibinfo{year}{2006},
  \bibinfo{journal}{Opt.\ Commun.} \textbf{\bibinfo{volume}{259}},
  \bibinfo{pages}{331}.

\bibitem[{\citenamefont{Cheng} \emph{et~al.}(2001)\citenamefont{Cheng, Wright,
  and Moloney}}]{Cheng:prl:87:213001}
\bibinfo{author}{\bibnamefont{Cheng}, \bibfnamefont{C.-C.}},
  \bibinfo{author}{\bibfnamefont{E.~M.} \bibnamefont{Wright}}, and
  \bibinfo{author}{\bibfnamefont{J.~V.} \bibnamefont{Moloney}},
  \bibinfo{year}{2001}, \bibinfo{journal}{Phys.\ Rev.\ Lett.}
  \textbf{\bibinfo{volume}{87}}, \bibinfo{pages}{213001}.

\bibitem[{\citenamefont{Cheng} \emph{et~al.}(2002)\citenamefont{Cheng, Wright,
  and Moloney}}]{Cheng:prl:89:139302}
\bibinfo{author}{\bibnamefont{Cheng}, \bibfnamefont{C.-C.}},
  \bibinfo{author}{\bibfnamefont{E.~M.} \bibnamefont{Wright}}, and
  \bibinfo{author}{\bibfnamefont{J.~V.} \bibnamefont{Moloney}},
  \bibinfo{year}{2002}, \bibinfo{journal}{Phys.\ Rev.\ Lett.}
  \textbf{\bibinfo{volume}{89}}, \bibinfo{pages}{139302}.

\bibitem[{\citenamefont{Chernev and
  Petrov}(1992{\natexlab{a}})}]{Chernev:ol:17:172}
\bibinfo{author}{\bibnamefont{Chernev}, \bibfnamefont{P.}}, and
  \bibinfo{author}{\bibfnamefont{V.}~\bibnamefont{Petrov}},
  \bibinfo{year}{1992}{\natexlab{a}}, \bibinfo{journal}{Opt.\ Lett.}
  \textbf{\bibinfo{volume}{17}}, \bibinfo{pages}{172}.

\bibitem[{\citenamefont{Chernev and
  Petrov}(1992{\natexlab{b}})}]{Chernev:oc:87:28}
\bibinfo{author}{\bibnamefont{Chernev}, \bibfnamefont{P.}}, and
  \bibinfo{author}{\bibfnamefont{V.}~\bibnamefont{Petrov}},
  \bibinfo{year}{1992}{\natexlab{b}}, \bibinfo{journal}{Opt.\ Commun.}
  \textbf{\bibinfo{volume}{87}}, \bibinfo{pages}{28}.

\bibitem[{\citenamefont{Chiao} \emph{et~al.}(1964)\citenamefont{Chiao, Garmire,
  and Townes}}]{Chiao:prl:13:479}
\bibinfo{author}{\bibnamefont{Chiao}, \bibfnamefont{R.~Y.}},
  \bibinfo{author}{\bibfnamefont{E.}~\bibnamefont{Garmire}}, and
  \bibinfo{author}{\bibfnamefont{C.~H.} \bibnamefont{Townes}},
  \bibinfo{year}{1964}, \bibinfo{journal}{Phys.\ Rev.\ Lett.}
  \textbf{\bibinfo{volume}{13}}, \bibinfo{pages}{479}.

\bibitem[{\citenamefont{Chin}
  \emph{et~al.}(1999{\natexlab{a}})\citenamefont{Chin, Brodeur, Petit,
  Kosareva, and Kandidov}}]{Chin:jnop:8:121}
\bibinfo{author}{\bibnamefont{Chin}, \bibfnamefont{S.~L.}},
  \bibinfo{author}{\bibfnamefont{A.}~\bibnamefont{Brodeur}},
  \bibinfo{author}{\bibfnamefont{S.}~\bibnamefont{Petit}},
  \bibinfo{author}{\bibfnamefont{O.~G.} \bibnamefont{Kosareva}}, and
  \bibinfo{author}{\bibfnamefont{V.~P.} \bibnamefont{Kandidov}},
  \bibinfo{year}{1999}{\natexlab{a}}, \bibinfo{journal}{J.\ Nonlinear Opt.\
  Phys.\ Mater.} \textbf{\bibinfo{volume}{8}}, \bibinfo{pages}{121}.

\bibitem[{\citenamefont{Chin}
  \emph{et~al.}(1999{\natexlab{b}})\citenamefont{Chin, Petit, Borne, and
  Miyazaki}}]{Chin:jjap:38:L126}
\bibinfo{author}{\bibnamefont{Chin}, \bibfnamefont{S.~L.}},
  \bibinfo{author}{\bibfnamefont{S.}~\bibnamefont{Petit}},
  \bibinfo{author}{\bibfnamefont{F.}~\bibnamefont{Borne}}, and
  \bibinfo{author}{\bibfnamefont{K.}~\bibnamefont{Miyazaki}},
  \bibinfo{year}{1999}{\natexlab{b}}, \bibinfo{journal}{Jpn.\ J.\ Appl.\ Phys.}
  \textbf{\bibinfo{volume}{38}}, \bibinfo{pages}{L126}.

\bibitem[{\citenamefont{Chin}
  \emph{et~al.}(2002{\natexlab{a}})\citenamefont{Chin, Petit, Liu, Iwasaki,
  Nadeau, Kandidov, Kosareva, and Andrianov}}]{Chin:oc:210:329}
\bibinfo{author}{\bibnamefont{Chin}, \bibfnamefont{S.~L.}},
  \bibinfo{author}{\bibfnamefont{S.}~\bibnamefont{Petit}},
  \bibinfo{author}{\bibfnamefont{W.}~\bibnamefont{Liu}},
  \bibinfo{author}{\bibfnamefont{A.}~\bibnamefont{Iwasaki}},
  \bibinfo{author}{\bibfnamefont{M.-C.} \bibnamefont{Nadeau}},
  \bibinfo{author}{\bibfnamefont{V.~P.} \bibnamefont{Kandidov}},
  \bibinfo{author}{\bibfnamefont{O.~G.} \bibnamefont{Kosareva}}, and
  \bibinfo{author}{\bibfnamefont{K.~Y.} \bibnamefont{Andrianov}},
  \bibinfo{year}{2002}{\natexlab{a}}, \bibinfo{journal}{Opt.\ Commun.}
  \textbf{\bibinfo{volume}{210}}, \bibinfo{pages}{329}.

\bibitem[{\citenamefont{Chin}
  \emph{et~al.}(2002{\natexlab{b}})\citenamefont{Chin, Talebpour, Yang, Petit,
  Kandidov, Kosareva, and Tamarov}}]{Chin:apb:74:67}
\bibinfo{author}{\bibnamefont{Chin}, \bibfnamefont{S.~L.}},
  \bibinfo{author}{\bibfnamefont{A.}~\bibnamefont{Talebpour}},
  \bibinfo{author}{\bibfnamefont{J.}~\bibnamefont{Yang}},
  \bibinfo{author}{\bibfnamefont{S.}~\bibnamefont{Petit}},
  \bibinfo{author}{\bibfnamefont{V.~P.} \bibnamefont{Kandidov}},
  \bibinfo{author}{\bibfnamefont{O.~G.} \bibnamefont{Kosareva}}, and
  \bibinfo{author}{\bibfnamefont{M.~P.} \bibnamefont{Tamarov}},
  \bibinfo{year}{2002}{\natexlab{b}}, \bibinfo{journal}{Appl.\ Phys.\ B: Lasers
  \& Optics} \textbf{\bibinfo{volume}{74}}, \bibinfo{pages}{67}.

\bibitem[{\citenamefont{Chiron} \emph{et~al.}(1999)\citenamefont{Chiron,
  Lamouroux, Lange, Ripoche, Franco, Prade, Bonnaud, Riazuelo, and
  Mysyrowicz}}]{Chiron:epjd:6:383}
\bibinfo{author}{\bibnamefont{Chiron}, \bibfnamefont{A.}},
  \bibinfo{author}{\bibfnamefont{B.}~\bibnamefont{Lamouroux}},
  \bibinfo{author}{\bibfnamefont{R.}~\bibnamefont{Lange}},
  \bibinfo{author}{\bibfnamefont{J.-F.} \bibnamefont{Ripoche}},
  \bibinfo{author}{\bibfnamefont{M.}~\bibnamefont{Franco}},
  \bibinfo{author}{\bibfnamefont{B.}~\bibnamefont{Prade}},
  \bibinfo{author}{\bibfnamefont{G.}~\bibnamefont{Bonnaud}},
  \bibinfo{author}{\bibfnamefont{G.}~\bibnamefont{Riazuelo}}, and
  \bibinfo{author}{\bibfnamefont{A.}~\bibnamefont{Mysyrowicz}},
  \bibinfo{year}{1999}, \bibinfo{journal}{Eur.\ Phys.\ J.\ D}
  \textbf{\bibinfo{volume}{6}}, \bibinfo{pages}{383}.

\bibitem[{\citenamefont{Christo\-doulides}
  \emph{et~al.}(2004)\citenamefont{Christo\-doulides, Efremidis, {Di Trapani},
  and Malomed}}]{Christodoulides:ol:29:1446}
\bibinfo{author}{\bibnamefont{Christo\-doulides}, \bibfnamefont{D.~N.}},
  \bibinfo{author}{\bibfnamefont{N.~K.} \bibnamefont{Efremidis}},
  \bibinfo{author}{\bibfnamefont{P.}~\bibnamefont{{Di Trapani}}}, and
  \bibinfo{author}{\bibfnamefont{B.~A.} \bibnamefont{Malomed}},
  \bibinfo{year}{2004}, \bibinfo{journal}{Opt.\ Lett.}
  \textbf{\bibinfo{volume}{29}}, \bibinfo{pages}{1446}.

\bibitem[{\citenamefont{Comtois} \emph{et~al.}(2000)\citenamefont{Comtois,
  Chien, Desparois, Gu{\'e}rin, Jarry, Johnston, Kieffer, LaFontaine, Martin,
  Potvin, Bondiou-Clergerie, and Gallimberti}}]{Comtois:apl:76:819}
\bibinfo{author}{\bibnamefont{Comtois}, \bibfnamefont{D.}},
  \bibinfo{author}{\bibfnamefont{C.~Y.} \bibnamefont{Chien}},
  \bibinfo{author}{\bibfnamefont{A.}~\bibnamefont{Desparois}},
  \bibinfo{author}{\bibfnamefont{F.}~\bibnamefont{Gu{\'e}rin}},
  \bibinfo{author}{\bibfnamefont{G.}~\bibnamefont{Jarry}},
  \bibinfo{author}{\bibfnamefont{T.~W.} \bibnamefont{Johnston}},
  \bibinfo{author}{\bibfnamefont{J.-C.} \bibnamefont{Kieffer}},
  \bibinfo{author}{\bibfnamefont{B.}~\bibnamefont{LaFontaine}},
  \bibinfo{author}{\bibfnamefont{F.}~\bibnamefont{Martin}},
  \bibinfo{author}{\bibfnamefont{C.}~\bibnamefont{Potvin}},
  \bibinfo{author}{\bibfnamefont{A.}~\bibnamefont{Bondiou-Clergerie}}, and
  \bibinfo{author}{\bibfnamefont{I.}~\bibnamefont{Gallimberti}},
  \bibinfo{year}{2000}, \bibinfo{journal}{Appl.\ Phys.\ Lett.}
  \textbf{\bibinfo{volume}{76}}, \bibinfo{pages}{819}.

\bibitem[{\citenamefont{Conti} \emph{et~al.}(2003)\citenamefont{Conti, Trillo,
  {Di Trapani}, Valiulis, Piskarskas, Jedrkiewicz, and
  Trull}}]{Conti:prl:90:170406}
\bibinfo{author}{\bibnamefont{Conti}, \bibfnamefont{C.}},
  \bibinfo{author}{\bibfnamefont{S.}~\bibnamefont{Trillo}},
  \bibinfo{author}{\bibfnamefont{P.}~\bibnamefont{{Di Trapani}}},
  \bibinfo{author}{\bibfnamefont{G.}~\bibnamefont{Valiulis}},
  \bibinfo{author}{\bibfnamefont{A.}~\bibnamefont{Piskarskas}},
  \bibinfo{author}{\bibfnamefont{O.}~\bibnamefont{Jedrkiewicz}}, and
  \bibinfo{author}{\bibfnamefont{J.}~\bibnamefont{Trull}},
  \bibinfo{year}{2003}, \bibinfo{journal}{Phys.\ Rev.\ Lett.}
  \textbf{\bibinfo{volume}{90}}, \bibinfo{pages}{170406}.

\bibitem[{\citenamefont{Cook} \emph{et~al.}(2003)\citenamefont{Cook, Kar, and
  Lamb}}]{Cook:apl:83:3861}
\bibinfo{author}{\bibnamefont{Cook}, \bibfnamefont{K.}},
  \bibinfo{author}{\bibfnamefont{A.~K.} \bibnamefont{Kar}}, and
  \bibinfo{author}{\bibfnamefont{R.~A.} \bibnamefont{Lamb}},
  \bibinfo{year}{2003}, \bibinfo{journal}{Appl.\ Phys.\ Lett.}
  \textbf{\bibinfo{volume}{83}}, \bibinfo{pages}{3861}.

\bibitem[{\citenamefont{Cook} \emph{et~al.}(2005)\citenamefont{Cook, McGeorge,
  Kar, Taghizadeh, and Lamb}}]{Cook:apl:86:021105}
\bibinfo{author}{\bibnamefont{Cook}, \bibfnamefont{K.}},
  \bibinfo{author}{\bibfnamefont{R.}~\bibnamefont{McGeorge}},
  \bibinfo{author}{\bibfnamefont{A.~K.} \bibnamefont{Kar}},
  \bibinfo{author}{\bibfnamefont{M.~R.} \bibnamefont{Taghizadeh}}, and
  \bibinfo{author}{\bibfnamefont{R.~A.} \bibnamefont{Lamb}},
  \bibinfo{year}{2005}, \bibinfo{journal}{Appl.\ Phys.\ Lett.}
  \textbf{\bibinfo{volume}{86}}, \bibinfo{pages}{021105}.

\bibitem[{\citenamefont{Corkum}(1993)}]{Corkum:prl:71:1994}
\bibinfo{author}{\bibnamefont{Corkum}, \bibfnamefont{P.~B.}},
  \bibinfo{year}{1993}, \bibinfo{journal}{Phys.\ Rev.\ Lett.}
  \textbf{\bibinfo{volume}{71}}, \bibinfo{pages}{1994}.

\bibitem[{\citenamefont{Corkum} \emph{et~al.}(1989)\citenamefont{Corkum,
  Burnett, and Brunel}}]{Corkum:prl:62:1259}
\bibinfo{author}{\bibnamefont{Corkum}, \bibfnamefont{P.~B.}},
  \bibinfo{author}{\bibfnamefont{N.~H.} \bibnamefont{Burnett}}, and
  \bibinfo{author}{\bibfnamefont{F.}~\bibnamefont{Brunel}},
  \bibinfo{year}{1989}, \bibinfo{journal}{Phys.\ Rev.\ Lett.}
  \textbf{\bibinfo{volume}{62}}, \bibinfo{pages}{1259}.

\bibitem[{\citenamefont{Cornaggia and Hering}(2000)}]{Cornaggia:pra:62:023403}
\bibinfo{author}{\bibnamefont{Cornaggia}, \bibfnamefont{C.}}, and
  \bibinfo{author}{\bibfnamefont{P.}~\bibnamefont{Hering}},
  \bibinfo{year}{2000}, \bibinfo{journal}{Phys.\ Rev.\ A}
  \textbf{\bibinfo{volume}{62}}, \bibinfo{pages}{023403}.

\bibitem[{\citenamefont{Couairon}(2003{\natexlab{a}})}]{Couairon:pra:68:015801}
\bibinfo{author}{\bibnamefont{Couairon}, \bibfnamefont{A.}},
  \bibinfo{year}{2003}{\natexlab{a}}, \bibinfo{journal}{Phys.\ Rev.\ A}
  \textbf{\bibinfo{volume}{68}}, \bibinfo{pages}{015801}.

\bibitem[{\citenamefont{Couairon}(2003{\natexlab{b}})}]{Couairon:epjd:27:159}
\bibinfo{author}{\bibnamefont{Couairon}, \bibfnamefont{A.}},
  \bibinfo{year}{2003}{\natexlab{b}}, \bibinfo{journal}{Eur.\ Phys.\ J.\ D}
  \textbf{\bibinfo{volume}{27}}, \bibinfo{pages}{159}.

\bibitem[{\citenamefont{Couairon and Berg{\'e}}(2002)}]{Couairon:prl:88:135003}
\bibinfo{author}{\bibnamefont{Couairon}, \bibfnamefont{A.}}, and
  \bibinfo{author}{\bibfnamefont{L.}~\bibnamefont{Berg{\'e}}},
  \bibinfo{year}{2002}, \bibinfo{journal}{Phys.\ Rev.\ Lett.}
  \textbf{\bibinfo{volume}{88}}, \bibinfo{pages}{135003}.

\bibitem[{\citenamefont{Couairon} \emph{et~al.}(2006)\citenamefont{Couairon,
  Biegert, Hauri, Kornelis, Helbing, Keller, and
  Mysyrowicz}}]{Couairon:jmo:53:75}
\bibinfo{author}{\bibnamefont{Couairon}, \bibfnamefont{A.}},
  \bibinfo{author}{\bibfnamefont{J.}~\bibnamefont{Biegert}},
  \bibinfo{author}{\bibfnamefont{C.~P.} \bibnamefont{Hauri}},
  \bibinfo{author}{\bibfnamefont{W.}~\bibnamefont{Kornelis}},
  \bibinfo{author}{\bibfnamefont{F.~W.} \bibnamefont{Helbing}},
  \bibinfo{author}{\bibfnamefont{U.}~\bibnamefont{Keller}}, and
  \bibinfo{author}{\bibfnamefont{A.}~\bibnamefont{Mysyrowicz}},
  \bibinfo{year}{2006}, \bibinfo{journal}{J.\ Mod.\ Opt.}
  \textbf{\bibinfo{volume}{53}}, \bibinfo{pages}{75}.

\bibitem[{\citenamefont{Couairon} \emph{et~al.}(2005)\citenamefont{Couairon,
  Franco, Mysyrowicz, Biegert, and Keller}}]{Couairon:ol:30:2657}
\bibinfo{author}{\bibnamefont{Couairon}, \bibfnamefont{A.}},
  \bibinfo{author}{\bibfnamefont{M.}~\bibnamefont{Franco}},
  \bibinfo{author}{\bibfnamefont{A.}~\bibnamefont{Mysyrowicz}},
  \bibinfo{author}{\bibfnamefont{J.}~\bibnamefont{Biegert}}, and
  \bibinfo{author}{\bibfnamefont{U.}~\bibnamefont{Keller}},
  \bibinfo{year}{2005}, \bibinfo{journal}{Opt.\ Lett.}
  \textbf{\bibinfo{volume}{30}}, \bibinfo{pages}{2657}.

\bibitem[{\citenamefont{Couairon} \emph{et~al.}(2003)\citenamefont{Couairon,
  M\'echain, Tzortzakis, Franco, Lamouroux, Prade, and
  Mysyrowicz}}]{Couairon:oc:225:177}
\bibinfo{author}{\bibnamefont{Couairon}, \bibfnamefont{A.}},
  \bibinfo{author}{\bibfnamefont{G.}~\bibnamefont{M\'echain}},
  \bibinfo{author}{\bibfnamefont{S.}~\bibnamefont{Tzortzakis}},
  \bibinfo{author}{\bibfnamefont{M.}~\bibnamefont{Franco}},
  \bibinfo{author}{\bibfnamefont{B.}~\bibnamefont{Lamouroux}},
  \bibinfo{author}{\bibfnamefont{B.}~\bibnamefont{Prade}}, and
  \bibinfo{author}{\bibfnamefont{A.}~\bibnamefont{Mysyrowicz}},
  \bibinfo{year}{2003}, \bibinfo{journal}{Opt.\ Commun.}
  \textbf{\bibinfo{volume}{225}}, \bibinfo{pages}{177}.

\bibitem[{\citenamefont{Couairon} \emph{et~al.}(2002)\citenamefont{Couairon,
  Tzortzakis, Berg{\'e}, Franco, Prade, and
  Mysyrowicz}}]{Couairon:josab:19:1117}
\bibinfo{author}{\bibnamefont{Couairon}, \bibfnamefont{A.}},
  \bibinfo{author}{\bibfnamefont{S.}~\bibnamefont{Tzortzakis}},
  \bibinfo{author}{\bibfnamefont{L.}~\bibnamefont{Berg{\'e}}},
  \bibinfo{author}{\bibfnamefont{M.}~\bibnamefont{Franco}},
  \bibinfo{author}{\bibfnamefont{B.}~\bibnamefont{Prade}}, and
  \bibinfo{author}{\bibfnamefont{A.}~\bibnamefont{Mysyrowicz}},
  \bibinfo{year}{2002}, \bibinfo{journal}{J.\ Opt.\ Soc.\ Am.\ B}
  \textbf{\bibinfo{volume}{19}}, \bibinfo{pages}{1117}.

\bibitem[{\citenamefont{Courvoisier}
  \emph{et~al.}(2003)\citenamefont{Courvoisier, Boutou, Kasparian, Salmon,
  M{\'e}jean, Yu, and Wolf}}]{Courvoisier:apl:83:213}
\bibinfo{author}{\bibnamefont{Courvoisier}, \bibfnamefont{F.}},
  \bibinfo{author}{\bibfnamefont{V.}~\bibnamefont{Boutou}},
  \bibinfo{author}{\bibfnamefont{J.}~\bibnamefont{Kasparian}},
  \bibinfo{author}{\bibfnamefont{E.}~\bibnamefont{Salmon}},
  \bibinfo{author}{\bibfnamefont{G.}~\bibnamefont{M{\'e}jean}},
  \bibinfo{author}{\bibfnamefont{J.}~\bibnamefont{Yu}}, and
  \bibinfo{author}{\bibfnamefont{J.~P.} \bibnamefont{Wolf}},
  \bibinfo{year}{2003}, \bibinfo{journal}{Appl.\ Phys.\ Lett.}
  \textbf{\bibinfo{volume}{83}}, \bibinfo{pages}{213}.

\bibitem[{\citenamefont{Cremers and Knight}(2000)}]{Cremers:EAC:2000:9595}
\bibinfo{author}{\bibnamefont{Cremers}, \bibfnamefont{D.~A.}}, and
  \bibinfo{author}{\bibfnamefont{A.~K.} \bibnamefont{Knight}},
  \bibinfo{year}{2000}, \emph{\bibinfo{title}{Encyclopedia of Analytical
  Chemistry}}, in  \cite{Meyers:EAC:2000}, p. \bibinfo{pages}{9595}.

\bibitem[{\citenamefont{Cremers and Radziemski}(2006)}]{Cremers:LIB:06}
\bibinfo{author}{\bibnamefont{Cremers}, \bibfnamefont{D.~A.}}, and
  \bibinfo{author}{\bibfnamefont{L.~J.} \bibnamefont{Radziemski}},
  \bibinfo{year}{2006}, \emph{\bibinfo{title}{Handbook of Laser-Induced
  Breakdown measurements}} (\bibinfo{publisher}{John Wiley \& Sons},
  \bibinfo{address}{Chichester}).

\bibitem[{\citenamefont{Desaix} \emph{et~al.}(1991)\citenamefont{Desaix,
  Anderson, and Lisak}}]{Desaix:josab:8:2082}
\bibinfo{author}{\bibnamefont{Desaix}, \bibfnamefont{M.}},
  \bibinfo{author}{\bibfnamefont{D.}~\bibnamefont{Anderson}}, and
  \bibinfo{author}{\bibfnamefont{M.}~\bibnamefont{Lisak}},
  \bibinfo{year}{1991}, \bibinfo{journal}{J.\ Opt.\ Soc.\ Am.\ B}
  \textbf{\bibinfo{volume}{8}}, \bibinfo{pages}{2082}.

\bibitem[{\citenamefont{Desyatnikov}
  \emph{et~al.}(2000)\citenamefont{Desyatnikov, Maimistov, and
  Malomed}}]{Desyatnikov:pre:61:3107}
\bibinfo{author}{\bibnamefont{Desyatnikov}, \bibfnamefont{A.}},
  \bibinfo{author}{\bibfnamefont{A.}~\bibnamefont{Maimistov}}, and
  \bibinfo{author}{\bibfnamefont{B.}~\bibnamefont{Malomed}},
  \bibinfo{year}{2000}, \bibinfo{journal}{Phys.\ Rev.\ E}
  \textbf{\bibinfo{volume}{61}}, \bibinfo{pages}{3107}.

\bibitem[{\citenamefont{Desyatnikov}
  \emph{et~al.}(2005)\citenamefont{Desyatnikov, Torner, and
  Kivshar}}]{Desyatnikov:po:47:291}
\bibinfo{author}{\bibnamefont{Desyatnikov}, \bibfnamefont{A.}},
  \bibinfo{author}{\bibfnamefont{L.}~\bibnamefont{Torner}}, and
  \bibinfo{author}{\bibfnamefont{Y.~S.} \bibnamefont{Kivshar}},
  \bibinfo{year}{2005}, \bibinfo{journal}{Prog.\ in Opt.}
  \textbf{\bibinfo{volume}{47}}, \bibinfo{pages}{291}.

\bibitem[{\citenamefont{DeWitt} \emph{et~al.}(2001)\citenamefont{DeWitt, Wells,
  and Jones}}]{Dewitt:prl:87:153001}
\bibinfo{author}{\bibnamefont{DeWitt}, \bibfnamefont{M.~J.}},
  \bibinfo{author}{\bibfnamefont{E.}~\bibnamefont{Wells}}, and
  \bibinfo{author}{\bibfnamefont{R.~R.} \bibnamefont{Jones}},
  \bibinfo{year}{2001}, \bibinfo{journal}{Phys.\ Rev.\ Lett.}
  \textbf{\bibinfo{volume}{87}}, \bibinfo{pages}{153001}.

\bibitem[{\citenamefont{{Di Trapani}} \emph{et~al.}(2003)\citenamefont{{Di
  Trapani}, Valiulis, Piskarskas, Jedrkiewicz, Trull, Conti, and
  Trillo}}]{Ditrapani:prl:91:093904}
\bibinfo{author}{\bibnamefont{{Di Trapani}}, \bibfnamefont{P.}},
  \bibinfo{author}{\bibfnamefont{G.}~\bibnamefont{Valiulis}},
  \bibinfo{author}{\bibfnamefont{A.}~\bibnamefont{Piskarskas}},
  \bibinfo{author}{\bibfnamefont{O.}~\bibnamefont{Jedrkiewicz}},
  \bibinfo{author}{\bibfnamefont{J.}~\bibnamefont{Trull}},
  \bibinfo{author}{\bibfnamefont{C.}~\bibnamefont{Conti}}, and
  \bibinfo{author}{\bibfnamefont{S.}~\bibnamefont{Trillo}},
  \bibinfo{year}{2003}, \bibinfo{journal}{Phys.\ Rev.\ Lett.}
  \textbf{\bibinfo{volume}{91}}, \bibinfo{pages}{093904}.

\bibitem[{\citenamefont{Donnat} \emph{et~al.}(1992)\citenamefont{Donnat,
  Gouedard, Veron, Bonville, Sauteret, and Migus}}]{Donnat:ol:17:331}
\bibinfo{author}{\bibnamefont{Donnat}, \bibfnamefont{P.}},
  \bibinfo{author}{\bibfnamefont{C.}~\bibnamefont{Gouedard}},
  \bibinfo{author}{\bibfnamefont{D.}~\bibnamefont{Veron}},
  \bibinfo{author}{\bibfnamefont{O.}~\bibnamefont{Bonville}},
  \bibinfo{author}{\bibfnamefont{C.}~\bibnamefont{Sauteret}}, and
  \bibinfo{author}{\bibfnamefont{A.}~\bibnamefont{Migus}},
  \bibinfo{year}{1992}, \bibinfo{journal}{Opt.\ Lett.}
  \textbf{\bibinfo{volume}{17}}, \bibinfo{pages}{331}.

\bibitem[{\citenamefont{Dou} \emph{et~al.}(2003)\citenamefont{Dou, Knobbe,
  Parkhill, Irwin, Matthews, and Church}}]{Dou:apa:76:303}
\bibinfo{author}{\bibnamefont{Dou}, \bibfnamefont{K.}},
  \bibinfo{author}{\bibfnamefont{E.~T.} \bibnamefont{Knobbe}},
  \bibinfo{author}{\bibfnamefont{R.~L.} \bibnamefont{Parkhill}},
  \bibinfo{author}{\bibfnamefont{B.}~\bibnamefont{Irwin}},
  \bibinfo{author}{\bibfnamefont{L.}~\bibnamefont{Matthews}}, and
  \bibinfo{author}{\bibfnamefont{K.~H.} \bibnamefont{Church}},
  \bibinfo{year}{2003}, \bibinfo{journal}{Appl.\ Phys.\ A: Materials Science \&
  Processing} \textbf{\bibinfo{volume}{76}}, \bibinfo{pages}{303}.

\bibitem[{\citenamefont{Drescher} \emph{et~al.}(2001)\citenamefont{Drescher,
  Hentschel, Kienberger, Tempea, Spielmann, Reider, Corkum, and
  Krausz}}]{Drescher:sc:291:1923}
\bibinfo{author}{\bibnamefont{Drescher}, \bibfnamefont{M.}},
  \bibinfo{author}{\bibfnamefont{M.}~\bibnamefont{Hentschel}},
  \bibinfo{author}{\bibfnamefont{R.}~\bibnamefont{Kienberger}},
  \bibinfo{author}{\bibfnamefont{G.}~\bibnamefont{Tempea}},
  \bibinfo{author}{\bibfnamefont{C.}~\bibnamefont{Spielmann}},
  \bibinfo{author}{\bibfnamefont{G.~A.} \bibnamefont{Reider}},
  \bibinfo{author}{\bibfnamefont{P.~B.} \bibnamefont{Corkum}}, and
  \bibinfo{author}{\bibfnamefont{F.}~\bibnamefont{Krausz}},
  \bibinfo{year}{2001}, \bibinfo{journal}{Science}
  \textbf{\bibinfo{volume}{291}}, \bibinfo{pages}{1923}.

\bibitem[{\citenamefont{Du} \emph{et~al.}(1994)\citenamefont{Du, Liu, Korn,
  Squier, and Mourou}}]{Du:apl:64:3071}
\bibinfo{author}{\bibnamefont{Du}, \bibfnamefont{D.}},
  \bibinfo{author}{\bibfnamefont{X.}~\bibnamefont{Liu}},
  \bibinfo{author}{\bibfnamefont{G.}~\bibnamefont{Korn}},
  \bibinfo{author}{\bibfnamefont{J.}~\bibnamefont{Squier}}, and
  \bibinfo{author}{\bibfnamefont{G.}~\bibnamefont{Mourou}},
  \bibinfo{year}{1994}, \bibinfo{journal}{Appl.\ Phys.\ Lett.}
  \textbf{\bibinfo{volume}{64}}, \bibinfo{pages}{3071}.

\bibitem[{\citenamefont{Dubietis}
  \emph{et~al.}(2004{\natexlab{a}})\citenamefont{Dubietis, Gai\v{z}auskas,
  Tamo\v{s}auskas, and {Di Trapani}}}]{Dubietis:prl:92:253903}
\bibinfo{author}{\bibnamefont{Dubietis}, \bibfnamefont{A.}},
  \bibinfo{author}{\bibfnamefont{E.}~\bibnamefont{Gai\v{z}auskas}},
  \bibinfo{author}{\bibfnamefont{G.}~\bibnamefont{Tamo\v{s}auskas}}, and
  \bibinfo{author}{\bibfnamefont{P.}~\bibnamefont{{Di Trapani}}},
  \bibinfo{year}{2004}{\natexlab{a}}, \bibinfo{journal}{Phys.\ Rev.\ Lett.}
  \textbf{\bibinfo{volume}{92}}, \bibinfo{pages}{253903}.

\bibitem[{\citenamefont{Dubietis}
  \emph{et~al.}(2004{\natexlab{b}})\citenamefont{Dubietis, Ku\v{c}inska,
  Tamo\v{s}auskas, Gai\v{z}auskas, Porras, and {Di
  Trapani}}}]{Dubietis:ol:29:2893}
\bibinfo{author}{\bibnamefont{Dubietis}, \bibfnamefont{A.}},
  \bibinfo{author}{\bibfnamefont{E.}~\bibnamefont{Ku\v{c}inska}},
  \bibinfo{author}{\bibfnamefont{G.}~\bibnamefont{Tamo\v{s}auskas}},
  \bibinfo{author}{\bibfnamefont{E.}~\bibnamefont{Gai\v{z}auskas}},
  \bibinfo{author}{\bibfnamefont{M.~A.} \bibnamefont{Porras}}, and
  \bibinfo{author}{\bibfnamefont{P.}~\bibnamefont{{Di Trapani}}},
  \bibinfo{year}{2004}{\natexlab{b}}, \bibinfo{journal}{Opt.\ Lett.}
  \textbf{\bibinfo{volume}{29}}, \bibinfo{pages}{2893}.

\bibitem[{\citenamefont{Dubietis} \emph{et~al.}(2003)\citenamefont{Dubietis,
  Tamo\v{s}auskas, Diomin, and Varavavi\v{c}ius}}]{Dubietis:ol:28:1269}
\bibinfo{author}{\bibnamefont{Dubietis}, \bibfnamefont{A.}},
  \bibinfo{author}{\bibfnamefont{G.}~\bibnamefont{Tamo\v{s}auskas}},
  \bibinfo{author}{\bibfnamefont{I.}~\bibnamefont{Diomin}}, and
  \bibinfo{author}{\bibfnamefont{A.}~\bibnamefont{Varavavi\v{c}ius}},
  \bibinfo{year}{2003}, \bibinfo{journal}{Opt.\ Lett.}
  \textbf{\bibinfo{volume}{28}}, \bibinfo{pages}{1269}.

\bibitem[{\citenamefont{Dubietis}
  \emph{et~al.}(2004{\natexlab{c}})\citenamefont{Dubietis, Tamo\v{s}auskas,
  Fibich, and Ilan}}]{Dubietis:ol:29:1126}
\bibinfo{author}{\bibnamefont{Dubietis}, \bibfnamefont{A.}},
  \bibinfo{author}{\bibfnamefont{G.}~\bibnamefont{Tamo\v{s}auskas}},
  \bibinfo{author}{\bibfnamefont{G.}~\bibnamefont{Fibich}}, and
  \bibinfo{author}{\bibfnamefont{B.}~\bibnamefont{Ilan}},
  \bibinfo{year}{2004}{\natexlab{c}}, \bibinfo{journal}{Opt.\ Lett.}
  \textbf{\bibinfo{volume}{29}}, \bibinfo{pages}{1126}.

\bibitem[{\citenamefont{Esarey} \emph{et~al.}(1997)\citenamefont{Esarey,
  Sprangle, Krall, and Ting}}]{Esarey:ieeejqe:33:1879}
\bibinfo{author}{\bibnamefont{Esarey}, \bibfnamefont{E.}},
  \bibinfo{author}{\bibfnamefont{P.}~\bibnamefont{Sprangle}},
  \bibinfo{author}{\bibfnamefont{J.}~\bibnamefont{Krall}}, and
  \bibinfo{author}{\bibfnamefont{A.}~\bibnamefont{Ting}}, \bibinfo{year}{1997},
  \bibinfo{journal}{IEEE J.\ Quant.\ Electron.} \textbf{\bibinfo{volume}{33}},
  \bibinfo{pages}{1879}.

\bibitem[{\citenamefont{Favre} \emph{et~al.}(2002)\citenamefont{Favre, Boutou,
  Hill, Zimmer, Krenz, Lambrecht, Yu, Chang, Woeste, and
  Wolf}}]{Favre:prl:89:035002}
\bibinfo{author}{\bibnamefont{Favre}, \bibfnamefont{C.}},
  \bibinfo{author}{\bibfnamefont{V.}~\bibnamefont{Boutou}},
  \bibinfo{author}{\bibfnamefont{S.~C.} \bibnamefont{Hill}},
  \bibinfo{author}{\bibfnamefont{W.}~\bibnamefont{Zimmer}},
  \bibinfo{author}{\bibfnamefont{M.}~\bibnamefont{Krenz}},
  \bibinfo{author}{\bibfnamefont{H.}~\bibnamefont{Lambrecht}},
  \bibinfo{author}{\bibfnamefont{J.}~\bibnamefont{Yu}},
  \bibinfo{author}{\bibfnamefont{R.~K.} \bibnamefont{Chang}},
  \bibinfo{author}{\bibfnamefont{L.}~\bibnamefont{Woeste}}, and
  \bibinfo{author}{\bibfnamefont{J.-P.} \bibnamefont{Wolf}},
  \bibinfo{year}{2002}, \bibinfo{journal}{Phys.\ Rev.\ Lett.}
  \textbf{\bibinfo{volume}{89}}, \bibinfo{pages}{035002}.

\bibitem[{\citenamefont{Feit and Fleck}(1974)}]{Feit:apl:24:169}
\bibinfo{author}{\bibnamefont{Feit}, \bibfnamefont{M.~D.}}, and
  \bibinfo{author}{\bibfnamefont{J.~A.} \bibnamefont{Fleck}},
  \bibinfo{year}{1974}, \bibinfo{journal}{Appl.\ Phys.\ Lett.}
  \textbf{\bibinfo{volume}{24}}, \bibinfo{pages}{169}.

\bibitem[{\citenamefont{Feit and Fleck}(1988)}]{Feit:josab:5:633}
\bibinfo{author}{\bibnamefont{Feit}, \bibfnamefont{M.~D.}}, and
  \bibinfo{author}{\bibfnamefont{J.~A.} \bibnamefont{Fleck}},
  \bibinfo{year}{1988}, \bibinfo{journal}{J.\ Opt.\ Soc.\ Am.\ B}
  \textbf{\bibinfo{volume}{5}}, \bibinfo{pages}{633}.

\bibitem[{\citenamefont{Feng} \emph{et~al.}(1995)\citenamefont{Feng, Moloney,
  Newell, and Wright}}]{Feng:ol:20:1958}
\bibinfo{author}{\bibnamefont{Feng}, \bibfnamefont{Q.}},
  \bibinfo{author}{\bibfnamefont{J.~V.} \bibnamefont{Moloney}},
  \bibinfo{author}{\bibfnamefont{A.~C.} \bibnamefont{Newell}}, and
  \bibinfo{author}{\bibfnamefont{E.~M.} \bibnamefont{Wright}},
  \bibinfo{year}{1995}, \bibinfo{journal}{Opt.\ Lett.}
  \textbf{\bibinfo{volume}{20}}, \bibinfo{pages}{1958}.

\bibitem[{\citenamefont{Feng} \emph{et~al.}(1997)\citenamefont{Feng, Moloney,
  Newell, Wright, Cook, Kennedy, Hammer, Rockwell, and
  Thompson}}]{Feng:ieeejqe:33:127}
\bibinfo{author}{\bibnamefont{Feng}, \bibfnamefont{Q.}},
  \bibinfo{author}{\bibfnamefont{J.~V.} \bibnamefont{Moloney}},
  \bibinfo{author}{\bibfnamefont{A.~C.} \bibnamefont{Newell}},
  \bibinfo{author}{\bibfnamefont{E.~M.} \bibnamefont{Wright}},
  \bibinfo{author}{\bibfnamefont{K.}~\bibnamefont{Cook}},
  \bibinfo{author}{\bibfnamefont{P.~K.} \bibnamefont{Kennedy}},
  \bibinfo{author}{\bibfnamefont{D.~X.} \bibnamefont{Hammer}},
  \bibinfo{author}{\bibfnamefont{B.~A.} \bibnamefont{Rockwell}}, and
  \bibinfo{author}{\bibfnamefont{C.~R.} \bibnamefont{Thompson}},
  \bibinfo{year}{1997}, \bibinfo{journal}{IEEE J.\ Quant.\ Electron.}
  \textbf{\bibinfo{volume}{33}}, \bibinfo{pages}{127}.

\bibitem[{\citenamefont{Fibich}(1996)}]{Fibich:prl:76:4356}
\bibinfo{author}{\bibnamefont{Fibich}, \bibfnamefont{G.}},
  \bibinfo{year}{1996}, \bibinfo{journal}{Phys.\ Rev.\ Lett.}
  \textbf{\bibinfo{volume}{76}}, \bibinfo{pages}{4356}.

\bibitem[{\citenamefont{Fibich} \emph{et~al.}(2005)\citenamefont{Fibich,
  Eisenmann, Ilan, Erlich, Fraenkel, Henis, Gaeta, and
  Zigler}}]{Fibich:oe:13:5897}
\bibinfo{author}{\bibnamefont{Fibich}, \bibfnamefont{G.}},
  \bibinfo{author}{\bibfnamefont{S.}~\bibnamefont{Eisenmann}},
  \bibinfo{author}{\bibfnamefont{B.}~\bibnamefont{Ilan}},
  \bibinfo{author}{\bibfnamefont{Y.}~\bibnamefont{Erlich}},
  \bibinfo{author}{\bibfnamefont{M.}~\bibnamefont{Fraenkel}},
  \bibinfo{author}{\bibfnamefont{Z.}~\bibnamefont{Henis}},
  \bibinfo{author}{\bibfnamefont{A.~L.} \bibnamefont{Gaeta}}, and
  \bibinfo{author}{\bibfnamefont{A.}~\bibnamefont{Zigler}},
  \bibinfo{year}{2005}, \bibinfo{journal}{Opt.\ Express}
  \textbf{\bibinfo{volume}{13}}, \bibinfo{pages}{5897}.

\bibitem[{\citenamefont{Fibich} \emph{et~al.}(2004)\citenamefont{Fibich,
  Eisenmann, Ilan, and Zigler}}]{Fibich:ol:29:1772}
\bibinfo{author}{\bibnamefont{Fibich}, \bibfnamefont{G.}},
  \bibinfo{author}{\bibfnamefont{S.}~\bibnamefont{Eisenmann}},
  \bibinfo{author}{\bibfnamefont{B.}~\bibnamefont{Ilan}}, and
  \bibinfo{author}{\bibfnamefont{A.}~\bibnamefont{Zigler}},
  \bibinfo{year}{2004}, \bibinfo{journal}{Opt.\ Lett.}
  \textbf{\bibinfo{volume}{29}}, \bibinfo{pages}{1772}.

\bibitem[{\citenamefont{Fibich and
  Ilan}(2001{\natexlab{a}})}]{Fibich:ol:26:840}
\bibinfo{author}{\bibnamefont{Fibich}, \bibfnamefont{G.}}, and
  \bibinfo{author}{\bibfnamefont{B.}~\bibnamefont{Ilan}},
  \bibinfo{year}{2001}{\natexlab{a}}, \bibinfo{journal}{Opt.\ Lett.}
  \textbf{\bibinfo{volume}{26}}, \bibinfo{pages}{840}.

\bibitem[{\citenamefont{Fibich and
  Ilan}(2001{\natexlab{b}})}]{Fibich:pd:157:112}
\bibinfo{author}{\bibnamefont{Fibich}, \bibfnamefont{G.}}, and
  \bibinfo{author}{\bibfnamefont{B.}~\bibnamefont{Ilan}},
  \bibinfo{year}{2001}{\natexlab{b}}, \bibinfo{journal}{Physica D}
  \textbf{\bibinfo{volume}{157}}, \bibinfo{pages}{112}.

\bibitem[{\citenamefont{Fibich and Ilan}(2004)}]{Fibich:ol:29:887}
\bibinfo{author}{\bibnamefont{Fibich}, \bibfnamefont{G.}}, and
  \bibinfo{author}{\bibfnamefont{B.}~\bibnamefont{Ilan}}, \bibinfo{year}{2004},
  \bibinfo{journal}{Opt.\ Lett.} \textbf{\bibinfo{volume}{29}},
  \bibinfo{pages}{887}.

\bibitem[{\citenamefont{Fibich} \emph{et~al.}(2002)\citenamefont{Fibich, Ilan,
  and Tsynkov}}]{Fibich:jsc:17:351}
\bibinfo{author}{\bibnamefont{Fibich}, \bibfnamefont{G.}},
  \bibinfo{author}{\bibfnamefont{B.}~\bibnamefont{Ilan}}, and
  \bibinfo{author}{\bibfnamefont{S.}~\bibnamefont{Tsynkov}},
  \bibinfo{year}{2002}, \bibinfo{journal}{J.\ Scien.\ Comput.}
  \textbf{\bibinfo{volume}{17}}, \bibinfo{pages}{351}.

\bibitem[{\citenamefont{Fibich and Papanicolaou}(1999)}]{Fibich:siamjam:60:183}
\bibinfo{author}{\bibnamefont{Fibich}, \bibfnamefont{G.}}, and
  \bibinfo{author}{\bibfnamefont{G.}~\bibnamefont{Papanicolaou}},
  \bibinfo{year}{1999}, \bibinfo{journal}{SIAM J.\ Appl.\ Math.}
  \textbf{\bibinfo{volume}{60}}, \bibinfo{pages}{183}.

\bibitem[{\citenamefont{Fibich and Papanicolaou}(1997)}]{Fibich:ol:22:1379}
\bibinfo{author}{\bibnamefont{Fibich}, \bibfnamefont{G.}}, and
  \bibinfo{author}{\bibfnamefont{G.~C.} \bibnamefont{Papanicolaou}},
  \bibinfo{year}{1997}, \bibinfo{journal}{Opt.\ Lett.}
  \textbf{\bibinfo{volume}{22}}, \bibinfo{pages}{1379}.

\bibitem[{\citenamefont{Fibich} \emph{et~al.}(2003)\citenamefont{Fibich, Ren,
  and Wang}}]{Fibich:pre:67:056603}
\bibinfo{author}{\bibnamefont{Fibich}, \bibfnamefont{G.}},
  \bibinfo{author}{\bibfnamefont{W.}~\bibnamefont{Ren}}, and
  \bibinfo{author}{\bibfnamefont{X.-P.} \bibnamefont{Wang}},
  \bibinfo{year}{2003}, \bibinfo{journal}{Phys.\ Rev.\ E}
  \textbf{\bibinfo{volume}{67}}, \bibinfo{pages}{056603}.

\bibitem[{\citenamefont{Fibich} \emph{et~al.}(2006)\citenamefont{Fibich, Sivan,
  Ehrlich, Louzon, Fraenkel, Eisenmann, Katzir, and
  Ziegler}}]{Fibich:oe:14:4946}
\bibinfo{author}{\bibnamefont{Fibich}, \bibfnamefont{G.}},
  \bibinfo{author}{\bibfnamefont{Y.}~\bibnamefont{Sivan}},
  \bibinfo{author}{\bibfnamefont{Y.}~\bibnamefont{Ehrlich}},
  \bibinfo{author}{\bibfnamefont{E.}~\bibnamefont{Louzon}},
  \bibinfo{author}{\bibfnamefont{M.}~\bibnamefont{Fraenkel}},
  \bibinfo{author}{\bibfnamefont{S.}~\bibnamefont{Eisenmann}},
  \bibinfo{author}{\bibfnamefont{Y.}~\bibnamefont{Katzir}}, and
  \bibinfo{author}{\bibfnamefont{A.}~\bibnamefont{Ziegler}},
  \bibinfo{year}{2006}, \bibinfo{journal}{Opt.\ Express}
  \textbf{\bibinfo{volume}{14}}, \bibinfo{pages}{4946}.

\bibitem[{\citenamefont{Firth and Skryabin}(1997)}]{Firth:prl:79:2450}
\bibinfo{author}{\bibnamefont{Firth}, \bibfnamefont{W.~J.}}, and
  \bibinfo{author}{\bibfnamefont{D.~V.} \bibnamefont{Skryabin}},
  \bibinfo{year}{1997}, \bibinfo{journal}{Phys.\ Rev.\ Lett.}
  \textbf{\bibinfo{volume}{79}}, \bibinfo{pages}{2450}.

\bibitem[{\citenamefont{Fra\v{i}man}(1985)}]{Fraiman:spjetp:61:228}
\bibinfo{author}{\bibnamefont{Fra\v{i}man}, \bibfnamefont{G.~M.}},
  \bibinfo{year}{1985}, \bibinfo{journal}{Sov.\ Phys.\ JETP}
  \textbf{\bibinfo{volume}{61}}, \bibinfo{pages}{228}.

\bibitem[{\citenamefont{Fujii and Fukuchi}(2005)}]{Kasparian:LRS:2005}
\bibinfo{editor}{\bibnamefont{Fujii}, \bibfnamefont{T.}}, and
  \bibinfo{editor}{\bibfnamefont{T.}~\bibnamefont{Fukuchi}} (eds.),
  \bibinfo{year}{2005}, \emph{\bibinfo{title}{Femtosecond white-light lidar}}
  (\bibinfo{publisher}{Marcel Dekker, Inc.}, \bibinfo{address}{New York}).

\bibitem[{\citenamefont{Gaeta}(2000)}]{Gaeta:prl:84:3582}
\bibinfo{author}{\bibnamefont{Gaeta}, \bibfnamefont{A.~L.}},
  \bibinfo{year}{2000}, \bibinfo{journal}{Phys.\ Rev.\ Lett.}
  \textbf{\bibinfo{volume}{84}}, \bibinfo{pages}{3582}.

\bibitem[{\citenamefont{Gaeta and Wise}(2001)}]{Gaeta:prl:87:229401}
\bibinfo{author}{\bibnamefont{Gaeta}, \bibfnamefont{A.~L.}}, and
  \bibinfo{author}{\bibfnamefont{F.}~\bibnamefont{Wise}}, \bibinfo{year}{2001},
  \bibinfo{journal}{Phys.\ Rev.\ Lett.} \textbf{\bibinfo{volume}{87}},
  \bibinfo{pages}{229401}.

\bibitem[{\citenamefont{Gatz and Herrmann}(1997)}]{Gatz:josab:14:1795}
\bibinfo{author}{\bibnamefont{Gatz}, \bibfnamefont{S.}}, and
  \bibinfo{author}{\bibfnamefont{J.}~\bibnamefont{Herrmann}},
  \bibinfo{year}{1997}, \bibinfo{journal}{J.\ Opt.\ Soc.\ Am.\ B}
  \textbf{\bibinfo{volume}{14}}, \bibinfo{pages}{1795}.

\bibitem[{\citenamefont{Geissler} \emph{et~al.}(1999)\citenamefont{Geissler,
  Tempea, Scrinzi, Schn{\"u}rer, Krausz, and Brabec}}]{Geissler:prl:83:2930}
\bibinfo{author}{\bibnamefont{Geissler}, \bibfnamefont{M.}},
  \bibinfo{author}{\bibfnamefont{G.}~\bibnamefont{Tempea}},
  \bibinfo{author}{\bibfnamefont{A.}~\bibnamefont{Scrinzi}},
  \bibinfo{author}{\bibfnamefont{M.}~\bibnamefont{Schn{\"u}rer}},
  \bibinfo{author}{\bibfnamefont{F.}~\bibnamefont{Krausz}}, and
  \bibinfo{author}{\bibfnamefont{T.}~\bibnamefont{Brabec}},
  \bibinfo{year}{1999}, \bibinfo{journal}{Phys.\ Rev.\ Lett.}
  \textbf{\bibinfo{volume}{83}}, \bibinfo{pages}{2930}.

\bibitem[{\citenamefont{Germaschewski}
  \emph{et~al.}(2001)\citenamefont{Germaschewski, Grauer, Berg\'e, Mezentsev,
  and Rasmussen}}]{Germaschewski:pd:151:175}
\bibinfo{author}{\bibnamefont{Germaschewski}, \bibfnamefont{K.}},
  \bibinfo{author}{\bibfnamefont{R.}~\bibnamefont{Grauer}},
  \bibinfo{author}{\bibfnamefont{L.}~\bibnamefont{Berg\'e}},
  \bibinfo{author}{\bibfnamefont{V.~K.} \bibnamefont{Mezentsev}}, and
  \bibinfo{author}{\bibfnamefont{J.~J.} \bibnamefont{Rasmussen}},
  \bibinfo{year}{2001}, \bibinfo{journal}{Physica D}
  \textbf{\bibinfo{volume}{151}}, \bibinfo{pages}{175}.

\bibitem[{\citenamefont{Gil'denburg}
  \emph{et~al.}(1995)\citenamefont{Gil'denburg, Pozdnyakova, and
  Shereshevskii}}]{Gildenburg:pla:203:214}
\bibinfo{author}{\bibnamefont{Gil'denburg}, \bibfnamefont{V.~B.}},
  \bibinfo{author}{\bibfnamefont{V.~I.} \bibnamefont{Pozdnyakova}}, and
  \bibinfo{author}{\bibfnamefont{I.~A.} \bibnamefont{Shereshevskii}},
  \bibinfo{year}{1995}, \bibinfo{journal}{Phys.\ Lett.\ A}
  \textbf{\bibinfo{volume}{203}}, \bibinfo{pages}{214}.

\bibitem[{\citenamefont{Glassey}(1977)}]{Glassey:jmp:18:1794}
\bibinfo{author}{\bibnamefont{Glassey}, \bibfnamefont{R.~T.}},
  \bibinfo{year}{1977}, \bibinfo{journal}{J.\ Math.\ Phys.}
  \textbf{\bibinfo{volume}{18}}, \bibinfo{pages}{1794}.

\bibitem[{\citenamefont{Golubtsov} \emph{et~al.}(2003)\citenamefont{Golubtsov,
  Kandidov, and Kosareva}}]{Golubtsov:qe:33:525}
\bibinfo{author}{\bibnamefont{Golubtsov}, \bibfnamefont{I.~S.}},
  \bibinfo{author}{\bibfnamefont{V.~P.} \bibnamefont{Kandidov}}, and
  \bibinfo{author}{\bibfnamefont{O.~G.} \bibnamefont{Kosareva}},
  \bibinfo{year}{2003}, \bibinfo{journal}{Quant.\ Electron.}
  \textbf{\bibinfo{volume}{33}}, \bibinfo{pages}{525}.

\bibitem[{\citenamefont{Gong} \emph{et~al.}(1998)\citenamefont{Gong, Li, Zhang,
  and Yang}}]{Gong:cpl:15:30}
\bibinfo{author}{\bibnamefont{Gong}, \bibfnamefont{Q.-H.}},
  \bibinfo{author}{\bibfnamefont{J.-L.} \bibnamefont{Li}},
  \bibinfo{author}{\bibfnamefont{T.-Q.} \bibnamefont{Zhang}}, and
  \bibinfo{author}{\bibfnamefont{H.}~\bibnamefont{Yang}}, \bibinfo{year}{1998},
  \bibinfo{journal}{Chin.\ Phys.\ Lett.} \textbf{\bibinfo{volume}{15}},
  \bibinfo{pages}{30}.

\bibitem[{\citenamefont{Gordon} \emph{et~al.}(2006)\citenamefont{Gordon, Ting,
  Alexeev, Fischer, and Sprangle}}]{Gordon:ieeetps:34:249}
\bibinfo{author}{\bibnamefont{Gordon}, \bibfnamefont{D.~F.}},
  \bibinfo{author}{\bibfnamefont{A.~C.} \bibnamefont{Ting}},
  \bibinfo{author}{\bibfnamefont{I.}~\bibnamefont{Alexeev}},
  \bibinfo{author}{\bibfnamefont{R.~P.} \bibnamefont{Fischer}}, and
  \bibinfo{author}{\bibfnamefont{P.}~\bibnamefont{Sprangle}},
  \bibinfo{year}{2006}, \bibinfo{journal}{IEEE Trans.\ Plasma Sc.}
  \textbf{\bibinfo{volume}{34}}, \bibinfo{pages}{249}.

\bibitem[{\citenamefont{Grow and Gaeta}(2005)}]{Grow:oe:13:4594}
\bibinfo{author}{\bibnamefont{Grow}, \bibfnamefont{T.~D.}}, and
  \bibinfo{author}{\bibfnamefont{A.~L.} \bibnamefont{Gaeta}},
  \bibinfo{year}{2005}, \bibinfo{journal}{Opt.\ Express}
  \textbf{\bibinfo{volume}{13}}, \bibinfo{pages}{4594}.

\bibitem[{\citenamefont{Guyon} \emph{et~al.}(2006)\citenamefont{Guyon,
  Courvoisier, Boutou, Nuter, Vin{\c{c}}otte, Champeaux, Berg\'e, Glorieux, and
  Wolf}}]{Guyon:pra:73:051802}
\bibinfo{author}{\bibnamefont{Guyon}, \bibfnamefont{L.}},
  \bibinfo{author}{\bibfnamefont{F.}~\bibnamefont{Courvoisier}},
  \bibinfo{author}{\bibfnamefont{V.}~\bibnamefont{Boutou}},
  \bibinfo{author}{\bibfnamefont{R.}~\bibnamefont{Nuter}},
  \bibinfo{author}{\bibfnamefont{A.}~\bibnamefont{Vin{\c{c}}otte}},
  \bibinfo{author}{\bibfnamefont{S.}~\bibnamefont{Champeaux}},
  \bibinfo{author}{\bibfnamefont{L.}~\bibnamefont{Berg\'e}},
  \bibinfo{author}{\bibfnamefont{P.}~\bibnamefont{Glorieux}}, and
  \bibinfo{author}{\bibfnamefont{J.-P.} \bibnamefont{Wolf}},
  \bibinfo{year}{2006}, \bibinfo{journal}{Phys.\ Rev.\ A}
  \textbf{\bibinfo{volume}{73}}, \bibinfo{pages}{051802(R)}.

\bibitem[{\citenamefont{Hao}
  \emph{et~al.}(2005{\natexlab{a}})\citenamefont{Hao, Zhang, Li, Lu, Yuan,
  Zheng, Wang, Ling, and Wei}}]{Hao:apb:80:627}
\bibinfo{author}{\bibnamefont{Hao}, \bibfnamefont{Z.}},
  \bibinfo{author}{\bibfnamefont{J.}~\bibnamefont{Zhang}},
  \bibinfo{author}{\bibfnamefont{Y.~T.} \bibnamefont{Li}},
  \bibinfo{author}{\bibfnamefont{X.}~\bibnamefont{Lu}},
  \bibinfo{author}{\bibfnamefont{X.~H.} \bibnamefont{Yuan}},
  \bibinfo{author}{\bibfnamefont{Z.~Y.} \bibnamefont{Zheng}},
  \bibinfo{author}{\bibfnamefont{Z.~H.} \bibnamefont{Wang}},
  \bibinfo{author}{\bibfnamefont{W.~J.} \bibnamefont{Ling}}, and
  \bibinfo{author}{\bibfnamefont{Z.~Y.} \bibnamefont{Wei}},
  \bibinfo{year}{2005}{\natexlab{a}}, \bibinfo{journal}{Appl.\ Phys.\ B: Lasers
  \& Optics} \textbf{\bibinfo{volume}{80}}, \bibinfo{pages}{627}.

\bibitem[{\citenamefont{Hao}
  \emph{et~al.}(2005{\natexlab{b}})\citenamefont{Hao, Yu, Zhang, Li, Yuan,
  Zheng, Wang, Wang, Ling, and Wei}}]{Hao:cpl:22:636}
\bibinfo{author}{\bibnamefont{Hao}, \bibfnamefont{Z.-Q.}},
  \bibinfo{author}{\bibfnamefont{J.}~\bibnamefont{Yu}},
  \bibinfo{author}{\bibfnamefont{J.}~\bibnamefont{Zhang}},
  \bibinfo{author}{\bibfnamefont{Y.-T.} \bibnamefont{Li}},
  \bibinfo{author}{\bibfnamefont{X.-H.} \bibnamefont{Yuan}},
  \bibinfo{author}{\bibfnamefont{Z.-Y.} \bibnamefont{Zheng}},
  \bibinfo{author}{\bibfnamefont{P.}~\bibnamefont{Wang}},
  \bibinfo{author}{\bibfnamefont{Z.-H.} \bibnamefont{Wang}},
  \bibinfo{author}{\bibfnamefont{W.-J.} \bibnamefont{Ling}}, and
  \bibinfo{author}{\bibfnamefont{Z.-Y.} \bibnamefont{Wei}},
  \bibinfo{year}{2005}{\natexlab{b}}, \bibinfo{journal}{Chin.\ Phys.\ Lett.}
  \textbf{\bibinfo{volume}{22}}, \bibinfo{pages}{636}.

\bibitem[{\citenamefont{Hauri} \emph{et~al.}(2004)\citenamefont{Hauri,
  Kornelis, Helbing, Heinrich, Couairon, Mysyrowicz, Biegert, and
  Keller}}]{Hauri:apb:79:673}
\bibinfo{author}{\bibnamefont{Hauri}, \bibfnamefont{C.~P.}},
  \bibinfo{author}{\bibfnamefont{W.}~\bibnamefont{Kornelis}},
  \bibinfo{author}{\bibfnamefont{F.~W.} \bibnamefont{Helbing}},
  \bibinfo{author}{\bibfnamefont{A.}~\bibnamefont{Heinrich}},
  \bibinfo{author}{\bibfnamefont{A.}~\bibnamefont{Couairon}},
  \bibinfo{author}{\bibfnamefont{A.}~\bibnamefont{Mysyrowicz}},
  \bibinfo{author}{\bibfnamefont{J.}~\bibnamefont{Biegert}}, and
  \bibinfo{author}{\bibfnamefont{U.}~\bibnamefont{Keller}},
  \bibinfo{year}{2004}, \bibinfo{journal}{Appl.\ Phys.\ B: Lasers \& Optics}
  \textbf{\bibinfo{volume}{79}}, \bibinfo{pages}{673}.

\bibitem[{\citenamefont{He and Liu}(1999)}]{He:PNO:99}
\bibinfo{author}{\bibnamefont{He}, \bibfnamefont{G.~S.}}, and
  \bibinfo{author}{\bibfnamefont{S.~H.} \bibnamefont{Liu}},
  \bibinfo{year}{1999}, \emph{\bibinfo{title}{Physics of Nonlinear Optics}}
  (\bibinfo{publisher}{World Scientific}, \bibinfo{address}{Singapore}).

\bibitem[{\citenamefont{Heck} \emph{et~al.}(2006)\citenamefont{Heck, Sloss, and
  Levis}}]{Heck:oc:259:216}
\bibinfo{author}{\bibnamefont{Heck}, \bibfnamefont{G.}},
  \bibinfo{author}{\bibfnamefont{J.}~\bibnamefont{Sloss}}, and
  \bibinfo{author}{\bibfnamefont{R.~J.} \bibnamefont{Levis}},
  \bibinfo{year}{2006}, \bibinfo{journal}{Opt.\ Commun.}
  \textbf{\bibinfo{volume}{259}}, \bibinfo{pages}{216}.

\bibitem[{\citenamefont{Hellwarth} \emph{et~al.}(1990)\citenamefont{Hellwarth,
  Pennington, and Henesian}}]{Hellwarth:pra:41:2766}
\bibinfo{author}{\bibnamefont{Hellwarth}, \bibfnamefont{R.~W.}},
  \bibinfo{author}{\bibfnamefont{D.~M.} \bibnamefont{Pennington}}, and
  \bibinfo{author}{\bibfnamefont{M.~A.} \bibnamefont{Henesian}},
  \bibinfo{year}{1990}, \bibinfo{journal}{Phys.\ Rev.\ A}
  \textbf{\bibinfo{volume}{41}}, \bibinfo{pages}{2766}.

\bibitem[{\citenamefont{Henz and Herrmann}(1999)}]{Henz:pra:59:2528}
\bibinfo{author}{\bibnamefont{Henz}, \bibfnamefont{S.}}, and
  \bibinfo{author}{\bibfnamefont{J.}~\bibnamefont{Herrmann}},
  \bibinfo{year}{1999}, \bibinfo{journal}{Phys.\ Rev.\ A}
  \textbf{\bibinfo{volume}{59}}, \bibinfo{pages}{2528}.

\bibitem[{\citenamefont{Homoelle and Gaeta}(2000)}]{Homoelle:ol:25:761}
\bibinfo{author}{\bibnamefont{Homoelle}, \bibfnamefont{D.}}, and
  \bibinfo{author}{\bibfnamefont{A.~L.} \bibnamefont{Gaeta}},
  \bibinfo{year}{2000}, \bibinfo{journal}{Opt.\ Lett.}
  \textbf{\bibinfo{volume}{25}}, \bibinfo{pages}{761}.

\bibitem[{\citenamefont{Hosseini}
  \emph{et~al.}(2004{\natexlab{a}})\citenamefont{Hosseini, Luo, Ferland, Liu,
  Chin, Kosareva, Panov, Ak{\"o}zbek, and Kandidov}}]{Hosseini:pra:70:033802}
\bibinfo{author}{\bibnamefont{Hosseini}, \bibfnamefont{S.~A.}},
  \bibinfo{author}{\bibfnamefont{Q.}~\bibnamefont{Luo}},
  \bibinfo{author}{\bibfnamefont{B.}~\bibnamefont{Ferland}},
  \bibinfo{author}{\bibfnamefont{W.}~\bibnamefont{Liu}},
  \bibinfo{author}{\bibfnamefont{S.~L.} \bibnamefont{Chin}},
  \bibinfo{author}{\bibfnamefont{O.~G.} \bibnamefont{Kosareva}},
  \bibinfo{author}{\bibfnamefont{N.~A.} \bibnamefont{Panov}},
  \bibinfo{author}{\bibfnamefont{N.}~\bibnamefont{Ak{\"o}zbek}}, and
  \bibinfo{author}{\bibfnamefont{V.~P.} \bibnamefont{Kandidov}},
  \bibinfo{year}{2004}{\natexlab{a}}, \bibinfo{journal}{Phys.\ Rev.\ A}
  \textbf{\bibinfo{volume}{70}}, \bibinfo{pages}{033802}.

\bibitem[{\citenamefont{Hosseini}
  \emph{et~al.}(2004{\natexlab{b}})\citenamefont{Hosseini, Yu, Luo, and
  Chin}}]{Hosseini:apb:79:519}
\bibinfo{author}{\bibnamefont{Hosseini}, \bibfnamefont{S.~A.}},
  \bibinfo{author}{\bibfnamefont{J.}~\bibnamefont{Yu}},
  \bibinfo{author}{\bibfnamefont{Q.}~\bibnamefont{Luo}}, and
  \bibinfo{author}{\bibfnamefont{S.~L.} \bibnamefont{Chin}},
  \bibinfo{year}{2004}{\natexlab{b}}, \bibinfo{journal}{Appl.\ Phys.\ B: Lasers
  \& Optics} \textbf{\bibinfo{volume}{79}}, \bibinfo{pages}{519}.

\bibitem[{\citenamefont{Husakou and Herrmann}(2001)}]{Husakou:prl:87:203901}
\bibinfo{author}{\bibnamefont{Husakou}, \bibfnamefont{A.~V.}}, and
  \bibinfo{author}{\bibfnamefont{J.}~\bibnamefont{Herrmann}},
  \bibinfo{year}{2001}, \bibinfo{journal}{Phys.\ Rev.\ Lett.}
  \textbf{\bibinfo{volume}{87}}, \bibinfo{pages}{203901}.

\bibitem[{\citenamefont{Jin} \emph{et~al.}(2005)\citenamefont{Jin, Zhang, Xu,
  Lu, Li, Wang, Wei, Yuan, and Yu}}]{Jin:oe:13:10424}
\bibinfo{author}{\bibnamefont{Jin}, \bibfnamefont{Z.}},
  \bibinfo{author}{\bibfnamefont{J.}~\bibnamefont{Zhang}},
  \bibinfo{author}{\bibfnamefont{M.~H.} \bibnamefont{Xu}},
  \bibinfo{author}{\bibfnamefont{X.}~\bibnamefont{Lu}},
  \bibinfo{author}{\bibfnamefont{Y.~T.} \bibnamefont{Li}},
  \bibinfo{author}{\bibfnamefont{Z.~H.} \bibnamefont{Wang}},
  \bibinfo{author}{\bibfnamefont{Z.~Y.} \bibnamefont{Wei}},
  \bibinfo{author}{\bibfnamefont{X.~H.} \bibnamefont{Yuan}}, and
  \bibinfo{author}{\bibfnamefont{W.}~\bibnamefont{Yu}}, \bibinfo{year}{2005},
  \bibinfo{journal}{Opt.\ Express} \textbf{\bibinfo{volume}{13}},
  \bibinfo{pages}{10424}.

\bibitem[{\citenamefont{Jong-Il} \emph{et~al.}(2002)\citenamefont{Jong-Il,
  Klenze, and Kim}}]{Jongil:as:56:852}
\bibinfo{author}{\bibnamefont{Jong-Il}, \bibfnamefont{Y.}},
  \bibinfo{author}{\bibfnamefont{R.}~\bibnamefont{Klenze}}, and
  \bibinfo{author}{\bibfnamefont{J.~I.} \bibnamefont{Kim}},
  \bibinfo{year}{2002}, \bibinfo{journal}{Appl. \ Spectroscopy}
  \textbf{\bibinfo{volume}{56}}, \bibinfo{pages}{852}.

\bibitem[{\citenamefont{Kandidov} \emph{et~al.}(2005)\citenamefont{Kandidov,
  Ak{\"o}zbek, Scalora, Kosareva, Nyakk, Luo, Hosseini, and
  Chin}}]{Kandidov:apb:80:267}
\bibinfo{author}{\bibnamefont{Kandidov}, \bibfnamefont{V.~P.}},
  \bibinfo{author}{\bibfnamefont{N.}~\bibnamefont{Ak{\"o}zbek}},
  \bibinfo{author}{\bibfnamefont{M.}~\bibnamefont{Scalora}},
  \bibinfo{author}{\bibfnamefont{O.~G.} \bibnamefont{Kosareva}},
  \bibinfo{author}{\bibfnamefont{A.~V.} \bibnamefont{Nyakk}},
  \bibinfo{author}{\bibfnamefont{Q.}~\bibnamefont{Luo}},
  \bibinfo{author}{\bibfnamefont{S.~A.} \bibnamefont{Hosseini}}, and
  \bibinfo{author}{\bibfnamefont{S.~L.} \bibnamefont{Chin}},
  \bibinfo{year}{2005}, \bibinfo{journal}{Appl.\ Phys.\ B: Lasers \& Optics}
  \textbf{\bibinfo{volume}{80}}, \bibinfo{pages}{267}.

\bibitem[{\citenamefont{Kandidov} \emph{et~al.}(2004)\citenamefont{Kandidov,
  Golubtsov, and Kosareva}}]{Kandidov:qe:34:348}
\bibinfo{author}{\bibnamefont{Kandidov}, \bibfnamefont{V.~P.}},
  \bibinfo{author}{\bibfnamefont{I.~S.} \bibnamefont{Golubtsov}}, and
  \bibinfo{author}{\bibfnamefont{O.~G.} \bibnamefont{Kosareva}},
  \bibinfo{year}{2004}, \bibinfo{journal}{Quant.\ Electron.}
  \textbf{\bibinfo{volume}{34}}, \bibinfo{pages}{348}.

\bibitem[{\citenamefont{Kandidov} \emph{et~al.}(2003)\citenamefont{Kandidov,
  Kosareva, Golubtsov, Liu, Becker, Ak{\"o}zbeck, Bowden, and
  Chin}}]{Kandidov:apb:77:149}
\bibinfo{author}{\bibnamefont{Kandidov}, \bibfnamefont{V.~P.}},
  \bibinfo{author}{\bibfnamefont{O.~G.} \bibnamefont{Kosareva}},
  \bibinfo{author}{\bibfnamefont{I.~S.} \bibnamefont{Golubtsov}},
  \bibinfo{author}{\bibfnamefont{W.}~\bibnamefont{Liu}},
  \bibinfo{author}{\bibfnamefont{A.}~\bibnamefont{Becker}},
  \bibinfo{author}{\bibfnamefont{N.}~\bibnamefont{Ak{\"o}zbeck}},
  \bibinfo{author}{\bibfnamefont{C.~M.} \bibnamefont{Bowden}}, and
  \bibinfo{author}{\bibfnamefont{S.~L.} \bibnamefont{Chin}},
  \bibinfo{year}{2003}, \bibinfo{journal}{Appl.\ Phys.\ B: Lasers \& Optics}
  \textbf{\bibinfo{volume}{77}}, \bibinfo{pages}{149}.

\bibitem[{\citenamefont{Kandidov} \emph{et~al.}(1994)\citenamefont{Kandidov,
  Kosareva, and Shlenov}}]{Kandidov:qe:24:905}
\bibinfo{author}{\bibnamefont{Kandidov}, \bibfnamefont{V.~P.}},
  \bibinfo{author}{\bibfnamefont{O.~G.} \bibnamefont{Kosareva}}, and
  \bibinfo{author}{\bibfnamefont{S.~A.} \bibnamefont{Shlenov}},
  \bibinfo{year}{1994}, \bibinfo{journal}{Quant.\ Electron.}
  \textbf{\bibinfo{volume}{24}}, \bibinfo{pages}{905}.

\bibitem[{\citenamefont{Kandidov} \emph{et~al.}(1999)\citenamefont{Kandidov,
  Kosareva, Tamarov, Brodeur, and Chin}}]{Kandidov:qe:29:911}
\bibinfo{author}{\bibnamefont{Kandidov}, \bibfnamefont{V.~P.}},
  \bibinfo{author}{\bibfnamefont{O.~G.} \bibnamefont{Kosareva}},
  \bibinfo{author}{\bibfnamefont{M.~P.} \bibnamefont{Tamarov}},
  \bibinfo{author}{\bibfnamefont{A.}~\bibnamefont{Brodeur}}, and
  \bibinfo{author}{\bibfnamefont{S.~L.} \bibnamefont{Chin}},
  \bibinfo{year}{1999}, \bibinfo{journal}{Quant.\ Electron.}
  \textbf{\bibinfo{volume}{29}}, \bibinfo{pages}{911}.

\bibitem[{\citenamefont{Karlsson} \emph{et~al.}(1992)\citenamefont{Karlsson,
  Anderson, and Desaix}}]{Karlsson:ol:17:22}
\bibinfo{author}{\bibnamefont{Karlsson}, \bibfnamefont{M.}},
  \bibinfo{author}{\bibfnamefont{D.}~\bibnamefont{Anderson}}, and
  \bibinfo{author}{\bibfnamefont{M.}~\bibnamefont{Desaix}},
  \bibinfo{year}{1992}, \bibinfo{journal}{Opt.\ Lett.}
  \textbf{\bibinfo{volume}{17}}, \bibinfo{pages}{22}.

\bibitem[{\citenamefont{Kasparian} \emph{et~al.}(2003)\citenamefont{Kasparian,
  Rodriguez, M{\'e}jean, Yu, Salmon, Wille, {Bou\-rayou}, Frey, Andr{\'e},
  Mysyrowicz, Sauerbrey, Wolf} \emph{et~al.}}]{Kasparian:sc:301:61}
\bibinfo{author}{\bibnamefont{Kasparian}, \bibfnamefont{J.}},
  \bibinfo{author}{\bibfnamefont{M.}~\bibnamefont{Rodriguez}},
  \bibinfo{author}{\bibfnamefont{G.}~\bibnamefont{M{\'e}jean}},
  \bibinfo{author}{\bibfnamefont{J.}~\bibnamefont{Yu}},
  \bibinfo{author}{\bibfnamefont{E.}~\bibnamefont{Salmon}},
  \bibinfo{author}{\bibfnamefont{H.}~\bibnamefont{Wille}},
  \bibinfo{author}{\bibfnamefont{R.}~\bibnamefont{{Bou\-rayou}}},
  \bibinfo{author}{\bibfnamefont{S.}~\bibnamefont{Frey}},
  \bibinfo{author}{\bibfnamefont{Y.~B.} \bibnamefont{Andr{\'e}}},
  \bibinfo{author}{\bibfnamefont{A.}~\bibnamefont{Mysyrowicz}},
  \bibinfo{author}{\bibfnamefont{R.}~\bibnamefont{Sauerbrey}},
  \bibinfo{author}{\bibfnamefont{J.~P.} \bibnamefont{Wolf}}, \emph{et~al.},
  \bibinfo{year}{2003}, \bibinfo{journal}{Science}
  \textbf{\bibinfo{volume}{301}}, \bibinfo{pages}{61}.

\bibitem[{\citenamefont{Kasparian}
  \emph{et~al.}(2000{\natexlab{a}})\citenamefont{Kasparian, Sauerbrey, and
  Chin}}]{Kasparian:apb:71:877}
\bibinfo{author}{\bibnamefont{Kasparian}, \bibfnamefont{J.}},
  \bibinfo{author}{\bibfnamefont{R.}~\bibnamefont{Sauerbrey}}, and
  \bibinfo{author}{\bibfnamefont{S.~L.} \bibnamefont{Chin}},
  \bibinfo{year}{2000}{\natexlab{a}}, \bibinfo{journal}{Appl.\ Phys.\ B: Lasers
  \& Optics} \textbf{\bibinfo{volume}{71}}, \bibinfo{pages}{877}.

\bibitem[{\citenamefont{Kasparian}
  \emph{et~al.}(2000{\natexlab{b}})\citenamefont{Kasparian, Sauerbrey,
  Mondelain, Niedermeier, Yu, Wolf, Andr{\'e}, Franco, Prade, Tzortzakis,
  Mysyrowicz, Rodriguez} \emph{et~al.}}]{Kasparian:ol:25:1397}
\bibinfo{author}{\bibnamefont{Kasparian}, \bibfnamefont{J.}},
  \bibinfo{author}{\bibfnamefont{R.}~\bibnamefont{Sauerbrey}},
  \bibinfo{author}{\bibfnamefont{D.}~\bibnamefont{Mondelain}},
  \bibinfo{author}{\bibfnamefont{S.}~\bibnamefont{Niedermeier}},
  \bibinfo{author}{\bibfnamefont{J.}~\bibnamefont{Yu}},
  \bibinfo{author}{\bibfnamefont{J.~P.} \bibnamefont{Wolf}},
  \bibinfo{author}{\bibfnamefont{Y.~B.} \bibnamefont{Andr{\'e}}},
  \bibinfo{author}{\bibfnamefont{M.}~\bibnamefont{Franco}},
  \bibinfo{author}{\bibfnamefont{B.}~\bibnamefont{Prade}},
  \bibinfo{author}{\bibfnamefont{S.}~\bibnamefont{Tzortzakis}},
  \bibinfo{author}{\bibfnamefont{A.}~\bibnamefont{Mysyrowicz}},
  \bibinfo{author}{\bibfnamefont{M.}~\bibnamefont{Rodriguez}}, \emph{et~al.},
  \bibinfo{year}{2000}{\natexlab{b}}, \bibinfo{journal}{Opt.\ Lett.}
  \textbf{\bibinfo{volume}{25}}, \bibinfo{pages}{1397}.

\bibitem[{\citenamefont{Kath and Smyth}(1995)}]{Kath:pre:51:1484}
\bibinfo{author}{\bibnamefont{Kath}, \bibfnamefont{W.~L.}}, and
  \bibinfo{author}{\bibfnamefont{N.~F.} \bibnamefont{Smyth}},
  \bibinfo{year}{1995}, \bibinfo{journal}{Phys.\ Rev.\ E}
  \textbf{\bibinfo{volume}{51}}, \bibinfo{pages}{1484}.

\bibitem[{\citenamefont{Keldysh}(1965)}]{Keldysh:spjetp:20:1307}
\bibinfo{author}{\bibnamefont{Keldysh}, \bibfnamefont{L.~V.}},
  \bibinfo{year}{1965}, \bibinfo{journal}{Sov.\ Phys.\ JETP}
  \textbf{\bibinfo{volume}{20}}, \bibinfo{pages}{1307}.

\bibitem[{\citenamefont{Kelley}(1965)}]{Kelley:prl:15:1005}
\bibinfo{author}{\bibnamefont{Kelley}, \bibfnamefont{P.~L.}},
  \bibinfo{year}{1965}, \bibinfo{journal}{Phys.\ Rev.\ Lett.}
  \textbf{\bibinfo{volume}{15}}, \bibinfo{pages}{1005}.

\bibitem[{\citenamefont{Kennedy}(1995)}]{Kennedy:ieeejqe:31:2241}
\bibinfo{author}{\bibnamefont{Kennedy}, \bibfnamefont{P.~K.}},
  \bibinfo{year}{1995}, \bibinfo{journal}{IEEE J.\ Quant.\ Electron.}
  \textbf{\bibinfo{volume}{31}}, \bibinfo{pages}{2241}.

\bibitem[{\citenamefont{Knight} \emph{et~al.}(2000)\citenamefont{Knight,
  Scherbarth, Cremers, and Ferris}}]{Knight:as:54:331}
\bibinfo{author}{\bibnamefont{Knight}, \bibfnamefont{A.~K.}},
  \bibinfo{author}{\bibfnamefont{N.~L.} \bibnamefont{Scherbarth}},
  \bibinfo{author}{\bibfnamefont{D.~A.} \bibnamefont{Cremers}}, and
  \bibinfo{author}{\bibfnamefont{M.~J.} \bibnamefont{Ferris}},
  \bibinfo{year}{2000}, \bibinfo{journal}{Appl. \ Spectroscopy}
  \textbf{\bibinfo{volume}{54}}, \bibinfo{pages}{331}.

\bibitem[{\citenamefont{Kolesik}
  \emph{et~al.}(2003{\natexlab{a}})\citenamefont{Kolesik, Katona, Moloney, and
  Wright}}]{Kolesik:prl:91:043905}
\bibinfo{author}{\bibnamefont{Kolesik}, \bibfnamefont{M.}},
  \bibinfo{author}{\bibfnamefont{G.}~\bibnamefont{Katona}},
  \bibinfo{author}{\bibfnamefont{J.~V.} \bibnamefont{Moloney}}, and
  \bibinfo{author}{\bibfnamefont{E.~M.} \bibnamefont{Wright}},
  \bibinfo{year}{2003}{\natexlab{a}}, \bibinfo{journal}{Phys.\ Rev.\ Lett.}
  \textbf{\bibinfo{volume}{91}}, \bibinfo{pages}{043905}.

\bibitem[{\citenamefont{Kolesik}
  \emph{et~al.}(2003{\natexlab{b}})\citenamefont{Kolesik, Katona, Moloney, and
  Wright}}]{Kolesik:apb:77:185}
\bibinfo{author}{\bibnamefont{Kolesik}, \bibfnamefont{M.}},
  \bibinfo{author}{\bibfnamefont{G.}~\bibnamefont{Katona}},
  \bibinfo{author}{\bibfnamefont{J.~V.} \bibnamefont{Moloney}}, and
  \bibinfo{author}{\bibfnamefont{E.~M.} \bibnamefont{Wright}},
  \bibinfo{year}{2003}{\natexlab{b}}, \bibinfo{journal}{Appl.\ Phys.\ B: Lasers
  \& Optics} \textbf{\bibinfo{volume}{77}}, \bibinfo{pages}{185}.

\bibitem[{\citenamefont{Kolesik and
  Moloney}(2004{\natexlab{a}})}]{Kolesik:pre:70:036604}
\bibinfo{author}{\bibnamefont{Kolesik}, \bibfnamefont{M.}}, and
  \bibinfo{author}{\bibfnamefont{J.~V.} \bibnamefont{Moloney}},
  \bibinfo{year}{2004}{\natexlab{a}}, \bibinfo{journal}{Phys.\ Rev.\ E}
  \textbf{\bibinfo{volume}{70}}, \bibinfo{pages}{036604}.

\bibitem[{\citenamefont{Kolesik and
  Moloney}(2004{\natexlab{b}})}]{Kolesik:ol:29:590}
\bibinfo{author}{\bibnamefont{Kolesik}, \bibfnamefont{M.}}, and
  \bibinfo{author}{\bibfnamefont{J.~V.} \bibnamefont{Moloney}},
  \bibinfo{year}{2004}{\natexlab{b}}, \bibinfo{journal}{Opt.\ Lett.}
  \textbf{\bibinfo{volume}{29}}, \bibinfo{pages}{590}.

\bibitem[{\citenamefont{Kolesik} \emph{et~al.}(2002)\citenamefont{Kolesik,
  Moloney, and Mlejnek}}]{Kolesik:prl:89:283902}
\bibinfo{author}{\bibnamefont{Kolesik}, \bibfnamefont{M.}},
  \bibinfo{author}{\bibfnamefont{J.~V.} \bibnamefont{Moloney}}, and
  \bibinfo{author}{\bibfnamefont{M.}~\bibnamefont{Mlejnek}},
  \bibinfo{year}{2002}, \bibinfo{journal}{Phys.\ Rev.\ Lett.}
  \textbf{\bibinfo{volume}{89}}, \bibinfo{pages}{283902}.

\bibitem[{\citenamefont{Kolesik} \emph{et~al.}(2001)\citenamefont{Kolesik,
  Moloney, and Wright}}]{Kolesik:pre:64:046607}
\bibinfo{author}{\bibnamefont{Kolesik}, \bibfnamefont{M.}},
  \bibinfo{author}{\bibfnamefont{J.~V.} \bibnamefont{Moloney}}, and
  \bibinfo{author}{\bibfnamefont{E.~M.} \bibnamefont{Wright}},
  \bibinfo{year}{2001}, \bibinfo{journal}{Phys.\ Rev.\ E}
  \textbf{\bibinfo{volume}{64}}, \bibinfo{pages}{046607}.

\bibitem[{\citenamefont{Kolokolov}(1976)}]{Kolokolov:rqe:17:1016}
\bibinfo{author}{\bibnamefont{Kolokolov}, \bibfnamefont{A.~A.}},
  \bibinfo{year}{1976}, \bibinfo{journal}{Radiophys.\ Quant.\ Electron.}
  \textbf{\bibinfo{volume}{17}}, \bibinfo{pages}{1016}.

\bibitem[{\citenamefont{Konno and Suzuki}(1979)}]{Konno:ps:20:382}
\bibinfo{author}{\bibnamefont{Konno}, \bibfnamefont{K.}}, and
  \bibinfo{author}{\bibfnamefont{H.}~\bibnamefont{Suzuki}},
  \bibinfo{year}{1979}, \bibinfo{journal}{Phys.\ Scr.}
  \textbf{\bibinfo{volume}{20}}, \bibinfo{pages}{382}.

\bibitem[{\citenamefont{Koopman and Wilkerson}(1971)}]{Koopman:jap:42:1883}
\bibinfo{author}{\bibnamefont{Koopman}, \bibfnamefont{D.~W.}}, and
  \bibinfo{author}{\bibfnamefont{T.~D.} \bibnamefont{Wilkerson}},
  \bibinfo{year}{1971}, \bibinfo{journal}{J.\ Appl.\ Phys.}
  \textbf{\bibinfo{volume}{42}}, \bibinfo{pages}{1883}.

\bibitem[{\citenamefont{Koprinkov} \emph{et~al.}(2000)\citenamefont{Koprinkov,
  Suda, Wang, and Midorikawa}}]{Koprinkov:prl:84:3847}
\bibinfo{author}{\bibnamefont{Koprinkov}, \bibfnamefont{I.~G.}},
  \bibinfo{author}{\bibfnamefont{A.}~\bibnamefont{Suda}},
  \bibinfo{author}{\bibfnamefont{P.}~\bibnamefont{Wang}}, and
  \bibinfo{author}{\bibfnamefont{K.}~\bibnamefont{Midorikawa}},
  \bibinfo{year}{2000}, \bibinfo{journal}{Phys.\ Rev.\ Lett.}
  \textbf{\bibinfo{volume}{84}}, \bibinfo{pages}{3847}.

\bibitem[{\citenamefont{Koprinkov} \emph{et~al.}(2001)\citenamefont{Koprinkov,
  Suda, Wang, and Midorikawa}}]{Koprinkov:prl:87:229401}
\bibinfo{author}{\bibnamefont{Koprinkov}, \bibfnamefont{I.~G.}},
  \bibinfo{author}{\bibfnamefont{A.}~\bibnamefont{Suda}},
  \bibinfo{author}{\bibfnamefont{P.}~\bibnamefont{Wang}}, and
  \bibinfo{author}{\bibfnamefont{K.}~\bibnamefont{Midorikawa}},
  \bibinfo{year}{2001}, \bibinfo{journal}{Phys.\ Rev.\ Lett.}
  \textbf{\bibinfo{volume}{87}}, \bibinfo{pages}{229402}.

\bibitem[{\citenamefont{Kosareva} \emph{et~al.}(1997)\citenamefont{Kosareva,
  Kandidov, Brodeur, Chien, and Chin}}]{Kosareva:ol:22:1332}
\bibinfo{author}{\bibnamefont{Kosareva}, \bibfnamefont{O.}},
  \bibinfo{author}{\bibfnamefont{V.~P.} \bibnamefont{Kandidov}},
  \bibinfo{author}{\bibfnamefont{A.}~\bibnamefont{Brodeur}},
  \bibinfo{author}{\bibfnamefont{C.~Y.} \bibnamefont{Chien}}, and
  \bibinfo{author}{\bibfnamefont{S.~L.} \bibnamefont{Chin}},
  \bibinfo{year}{1997}, \bibinfo{journal}{Opt.\ Lett.}
  \textbf{\bibinfo{volume}{22}}, \bibinfo{pages}{1332}.

\bibitem[{\citenamefont{Kosmatov} \emph{et~al.}(1991)\citenamefont{Kosmatov,
  Shvets, and Zakharov}}]{Kosmatov:pd:52:16}
\bibinfo{author}{\bibnamefont{Kosmatov}, \bibfnamefont{N.~E.}},
  \bibinfo{author}{\bibfnamefont{V.~F.} \bibnamefont{Shvets}}, and
  \bibinfo{author}{\bibfnamefont{V.~E.} \bibnamefont{Zakharov}},
  \bibinfo{year}{1991}, \bibinfo{journal}{Physica D}
  \textbf{\bibinfo{volume}{52}}, \bibinfo{pages}{16}.

\bibitem[{\citenamefont{Krainov}(1997)}]{Krainov:josab:14:425}
\bibinfo{author}{\bibnamefont{Krainov}, \bibfnamefont{V.~P.}},
  \bibinfo{year}{1997}, \bibinfo{journal}{J.\ Opt.\ Soc.\ Am.\ B}
  \textbf{\bibinfo{volume}{14}}, \bibinfo{pages}{425}.

\bibitem[{\citenamefont{Kruglov} \emph{et~al.}(1992)\citenamefont{Kruglov,
  Logvin, and Volkov}}]{Kruglov:jmo:39:2277}
\bibinfo{author}{\bibnamefont{Kruglov}, \bibfnamefont{V.~I.}},
  \bibinfo{author}{\bibfnamefont{Y.~A.} \bibnamefont{Logvin}}, and
  \bibinfo{author}{\bibfnamefont{V.~M.} \bibnamefont{Volkov}},
  \bibinfo{year}{1992}, \bibinfo{journal}{J.\ Mod.\ Opt.}
  \textbf{\bibinfo{volume}{39}}, \bibinfo{pages}{2277}.

\bibitem[{\citenamefont{Kruglov and Vlasov}(1985)}]{Kruglov:pl:111A:401}
\bibinfo{author}{\bibnamefont{Kruglov}, \bibfnamefont{V.~I.}}, and
  \bibinfo{author}{\bibfnamefont{R.~A.} \bibnamefont{Vlasov}},
  \bibinfo{year}{1985}, \bibinfo{journal}{Phys.\ Lett.}
  \textbf{\bibinfo{volume}{111A}}, \bibinfo{pages}{401}.

\bibitem[{\citenamefont{Kuznetsov}(1996)}]{Kuznetsov:chaos:6:381}
\bibinfo{author}{\bibnamefont{Kuznetsov}, \bibfnamefont{E.~A.}},
  \bibinfo{year}{1996}, \bibinfo{journal}{CHAOS} \textbf{\bibinfo{volume}{6}},
  \bibinfo{pages}{381}.

\bibitem[{\citenamefont{Kuznetsov} \emph{et~al.}(1995)\citenamefont{Kuznetsov,
  Rasmussen, Rypdal, and Turitsyn}}]{Kuznetsov:pd:87:273}
\bibinfo{author}{\bibnamefont{Kuznetsov}, \bibfnamefont{E.~A.}},
  \bibinfo{author}{\bibfnamefont{J.~J.} \bibnamefont{Rasmussen}},
  \bibinfo{author}{\bibfnamefont{K.}~\bibnamefont{Rypdal}}, and
  \bibinfo{author}{\bibfnamefont{S.~K.} \bibnamefont{Turitsyn}},
  \bibinfo{year}{1995}, \bibinfo{journal}{Physica D}
  \textbf{\bibinfo{volume}{87}}, \bibinfo{pages}{273}.

\bibitem[{\citenamefont{Kuznetsov} \emph{et~al.}(1986)\citenamefont{Kuznetsov,
  Rubenchik, and Zakharov}}]{Kuznetsov:pr:142:103}
\bibinfo{author}{\bibnamefont{Kuznetsov}, \bibfnamefont{E.~A.}},
  \bibinfo{author}{\bibfnamefont{A.~M.} \bibnamefont{Rubenchik}}, and
  \bibinfo{author}{\bibfnamefont{V.~E.} \bibnamefont{Zakharov}},
  \bibinfo{year}{1986}, \bibinfo{journal}{Phys.\ Rep.}
  \textbf{\bibinfo{volume}{142}}, \bibinfo{pages}{103}.

\bibitem[{\citenamefont{Kyuseok} \emph{et~al.}(1997)\citenamefont{Kyuseok,
  Yong-Ill, and Sneddon}}]{Kyuseok:asr:32:183}
\bibinfo{author}{\bibnamefont{Kyuseok}, \bibfnamefont{S.}},
  \bibinfo{author}{\bibfnamefont{L.}~\bibnamefont{Yong-Ill}}, and
  \bibinfo{author}{\bibfnamefont{J.}~\bibnamefont{Sneddon}},
  \bibinfo{year}{1997}, \bibinfo{journal}{Appl. \ Spectroscopy Reviews}
  \textbf{\bibinfo{volume}{32}}, \bibinfo{pages}{183}.

\bibitem[{\citenamefont{Labaune} \emph{et~al.}(1992)\citenamefont{Labaune,
  Baton, Jalinaud, Baldis, and Pesme}}]{Labaune:pfb:4:2224}
\bibinfo{author}{\bibnamefont{Labaune}, \bibfnamefont{C.}},
  \bibinfo{author}{\bibfnamefont{S.}~\bibnamefont{Baton}},
  \bibinfo{author}{\bibfnamefont{T.}~\bibnamefont{Jalinaud}},
  \bibinfo{author}{\bibfnamefont{H.~A.} \bibnamefont{Baldis}}, and
  \bibinfo{author}{\bibfnamefont{D.}~\bibnamefont{Pesme}},
  \bibinfo{year}{1992}, \bibinfo{journal}{Phys.\ Fluids B}
  \textbf{\bibinfo{volume}{4}}, \bibinfo{pages}{2224}.

\bibitem[{\citenamefont{{LaFontaine}}
  \emph{et~al.}(1999{\natexlab{a}})\citenamefont{{LaFontaine}, Vidal, Comtois,
  Chien, Desparois, Johnston, Kieffer, Mercure, P\'epin, and
  Rizk}}]{LaFontaine:ieeetps:27:688}
\bibinfo{author}{\bibnamefont{{LaFontaine}}, \bibfnamefont{B.}},
  \bibinfo{author}{\bibfnamefont{F.}~\bibnamefont{Vidal}},
  \bibinfo{author}{\bibfnamefont{D.}~\bibnamefont{Comtois}},
  \bibinfo{author}{\bibfnamefont{C.-Y.} \bibnamefont{Chien}},
  \bibinfo{author}{\bibfnamefont{A.}~\bibnamefont{Desparois}},
  \bibinfo{author}{\bibfnamefont{T.-W.} \bibnamefont{Johnston}},
  \bibinfo{author}{\bibfnamefont{J.-C.} \bibnamefont{Kieffer}},
  \bibinfo{author}{\bibfnamefont{H.-P.} \bibnamefont{Mercure}},
  \bibinfo{author}{\bibfnamefont{H.}~\bibnamefont{P\'epin}}, and
  \bibinfo{author}{\bibfnamefont{F.~A.~M.} \bibnamefont{Rizk}},
  \bibinfo{year}{1999}{\natexlab{a}}, \bibinfo{journal}{IEEE Trans.\ Plasma
  Sc.} \textbf{\bibinfo{volume}{27}}, \bibinfo{pages}{688}.

\bibitem[{\citenamefont{{LaFontaine}}
  \emph{et~al.}(1999{\natexlab{b}})\citenamefont{{LaFontaine}, Vidal, Jiang,
  Chien, Comtois, Desparois, Johnston, Kieffer, P{\'e}pin, and
  Mercure}}]{LaFontaine:pop:6:1615}
\bibinfo{author}{\bibnamefont{{LaFontaine}}, \bibfnamefont{B.}},
  \bibinfo{author}{\bibfnamefont{F.}~\bibnamefont{Vidal}},
  \bibinfo{author}{\bibfnamefont{Z.}~\bibnamefont{Jiang}},
  \bibinfo{author}{\bibfnamefont{C.~Y.} \bibnamefont{Chien}},
  \bibinfo{author}{\bibfnamefont{D.}~\bibnamefont{Comtois}},
  \bibinfo{author}{\bibfnamefont{A.}~\bibnamefont{Desparois}},
  \bibinfo{author}{\bibfnamefont{T.~W.} \bibnamefont{Johnston}},
  \bibinfo{author}{\bibfnamefont{J.~C.} \bibnamefont{Kieffer}},
  \bibinfo{author}{\bibfnamefont{H.}~\bibnamefont{P{\'e}pin}}, and
  \bibinfo{author}{\bibfnamefont{H.~P.} \bibnamefont{Mercure}},
  \bibinfo{year}{1999}{\natexlab{b}}, \bibinfo{journal}{Phys.\ Plasmas}
  \textbf{\bibinfo{volume}{6}}, \bibinfo{pages}{1615}.

\bibitem[{\citenamefont{Landman} \emph{et~al.}(1988)\citenamefont{Landman,
  Papanicolaou, Sulem, and Sulem}}]{Landman:pra:38:3837}
\bibinfo{author}{\bibnamefont{Landman}, \bibfnamefont{M.~J.}},
  \bibinfo{author}{\bibfnamefont{G.~C.} \bibnamefont{Papanicolaou}},
  \bibinfo{author}{\bibfnamefont{C.}~\bibnamefont{Sulem}}, and
  \bibinfo{author}{\bibfnamefont{P.~L.} \bibnamefont{Sulem}},
  \bibinfo{year}{1988}, \bibinfo{journal}{Phys.\ Rev.\ A}
  \textbf{\bibinfo{volume}{38}}, \bibinfo{pages}{3837}.

\bibitem[{\citenamefont{Lange}
  \emph{et~al.}(1998{\natexlab{a}})\citenamefont{Lange, Chiron, Ripoche,
  Mysyrowicz, Breger, and Agostini}}]{Lange:prl:81:1611}
\bibinfo{author}{\bibnamefont{Lange}, \bibfnamefont{H.~R.}},
  \bibinfo{author}{\bibfnamefont{A.}~\bibnamefont{Chiron}},
  \bibinfo{author}{\bibfnamefont{J.-F.} \bibnamefont{Ripoche}},
  \bibinfo{author}{\bibfnamefont{A.}~\bibnamefont{Mysyrowicz}},
  \bibinfo{author}{\bibfnamefont{P.}~\bibnamefont{Breger}}, and
  \bibinfo{author}{\bibfnamefont{P.}~\bibnamefont{Agostini}},
  \bibinfo{year}{1998}{\natexlab{a}}, \bibinfo{journal}{Phys.\ Rev.\ Lett.}
  \textbf{\bibinfo{volume}{81}}, \bibinfo{pages}{1611}.

\bibitem[{\citenamefont{Lange}
  \emph{et~al.}(1998{\natexlab{b}})\citenamefont{Lange, Grillon, Ripoche,
  Franco, Lamouroux, Prade, Mysyrowicz, Nibbering, and
  Chiron}}]{Lange:ol:23:120}
\bibinfo{author}{\bibnamefont{Lange}, \bibfnamefont{H.~R.}},
  \bibinfo{author}{\bibfnamefont{G.}~\bibnamefont{Grillon}},
  \bibinfo{author}{\bibfnamefont{J.-F.} \bibnamefont{Ripoche}},
  \bibinfo{author}{\bibfnamefont{M.~A.} \bibnamefont{Franco}},
  \bibinfo{author}{\bibfnamefont{B.}~\bibnamefont{Lamouroux}},
  \bibinfo{author}{\bibfnamefont{B.~S.} \bibnamefont{Prade}},
  \bibinfo{author}{\bibfnamefont{A.}~\bibnamefont{Mysyrowicz}},
  \bibinfo{author}{\bibfnamefont{E.~T.~J.} \bibnamefont{Nibbering}}, and
  \bibinfo{author}{\bibfnamefont{A.}~\bibnamefont{Chiron}},
  \bibinfo{year}{1998}{\natexlab{b}}, \bibinfo{journal}{Opt.\ Lett.}
  \textbf{\bibinfo{volume}{23}}, \bibinfo{pages}{120}.

\bibitem[{\citenamefont{Lehmeier} \emph{et~al.}(1985)\citenamefont{Lehmeier,
  Leupacher, and Penzkofer}}]{Lehmeier:oc:56:67}
\bibinfo{author}{\bibnamefont{Lehmeier}, \bibfnamefont{H.~J.}},
  \bibinfo{author}{\bibfnamefont{W.}~\bibnamefont{Leupacher}}, and
  \bibinfo{author}{\bibfnamefont{A.}~\bibnamefont{Penzkofer}},
  \bibinfo{year}{1985}, \bibinfo{journal}{Opt.\ Commun.}
  \textbf{\bibinfo{volume}{56}}, \bibinfo{pages}{67}.

\bibitem[{\citenamefont{Lehner and Auby}(2000)}]{Lehner:pre:61:1996}
\bibinfo{author}{\bibnamefont{Lehner}, \bibfnamefont{T.}}, and
  \bibinfo{author}{\bibfnamefont{N.}~\bibnamefont{Auby}}, \bibinfo{year}{2000},
  \bibinfo{journal}{Phys.\ Rev.\ E} \textbf{\bibinfo{volume}{61}},
  \bibinfo{pages}{1996}.

\bibitem[{\citenamefont{{leMesurier}}(2000)}]{Lemesurier:pd:138:334}
\bibinfo{author}{\bibnamefont{{leMesurier}}, \bibfnamefont{B.~J.}},
  \bibinfo{year}{2000}, \bibinfo{journal}{Physica D}
  \textbf{\bibinfo{volume}{138}}, \bibinfo{pages}{334}.

\bibitem[{\citenamefont{{leMesurier}}
  \emph{et~al.}(2004)\citenamefont{{leMesurier}, Christiansen, Gaididei, and
  Rasmussen}}]{Lemesurier:pre:70:046614}
\bibinfo{author}{\bibnamefont{{leMesurier}}, \bibfnamefont{B.~J.}},
  \bibinfo{author}{\bibfnamefont{P.~L.} \bibnamefont{Christiansen}},
  \bibinfo{author}{\bibfnamefont{Y.~B.} \bibnamefont{Gaididei}}, and
  \bibinfo{author}{\bibfnamefont{J.~J.} \bibnamefont{Rasmussen}},
  \bibinfo{year}{2004}, \bibinfo{journal}{Phys.\ Rev.\ E}
  \textbf{\bibinfo{volume}{70}}, \bibinfo{pages}{046614}.

\bibitem[{\citenamefont{Lenzner} \emph{et~al.}(1998)\citenamefont{Lenzner,
  Kr{\"u}ger, Sartania, Cheng, Spielmann, Mourou, Kautek, and
  Krausz}}]{Lenzner:prl:80:4076}
\bibinfo{author}{\bibnamefont{Lenzner}, \bibfnamefont{M.}},
  \bibinfo{author}{\bibfnamefont{J.}~\bibnamefont{Kr{\"u}ger}},
  \bibinfo{author}{\bibfnamefont{S.}~\bibnamefont{Sartania}},
  \bibinfo{author}{\bibfnamefont{Z.}~\bibnamefont{Cheng}},
  \bibinfo{author}{\bibfnamefont{C.}~\bibnamefont{Spielmann}},
  \bibinfo{author}{\bibfnamefont{G.}~\bibnamefont{Mourou}},
  \bibinfo{author}{\bibfnamefont{W.}~\bibnamefont{Kautek}}, and
  \bibinfo{author}{\bibfnamefont{F.}~\bibnamefont{Krausz}},
  \bibinfo{year}{1998}, \bibinfo{journal}{Phys.\ Rev.\ Lett.}
  \textbf{\bibinfo{volume}{80}}, \bibinfo{pages}{4076}.

\bibitem[{\citenamefont{Lewenstein}
  \emph{et~al.}(1994)\citenamefont{Lewenstein, Balcou, Ivanov, L'Huillier, and
  Corkum}}]{Lewenstein:pra:49:2117}
\bibinfo{author}{\bibnamefont{Lewenstein}, \bibfnamefont{M.}},
  \bibinfo{author}{\bibfnamefont{P.}~\bibnamefont{Balcou}},
  \bibinfo{author}{\bibfnamefont{M.~Y.} \bibnamefont{Ivanov}},
  \bibinfo{author}{\bibfnamefont{A.}~\bibnamefont{L'Huillier}}, and
  \bibinfo{author}{\bibfnamefont{P.~B.} \bibnamefont{Corkum}},
  \bibinfo{year}{1994}, \bibinfo{journal}{Phys.\ Rev.\ A}
  \textbf{\bibinfo{volume}{49}}, \bibinfo{pages}{2117}.

\bibitem[{\citenamefont{Li} \emph{et~al.}(1999)\citenamefont{Li, Menon,
  Nibarger, and Gibson}}]{Li:prl:82:2394}
\bibinfo{author}{\bibnamefont{Li}, \bibfnamefont{M.}},
  \bibinfo{author}{\bibfnamefont{S.}~\bibnamefont{Menon}},
  \bibinfo{author}{\bibfnamefont{J.~P.} \bibnamefont{Nibarger}}, and
  \bibinfo{author}{\bibfnamefont{G.~N.} \bibnamefont{Gibson}},
  \bibinfo{year}{1999}, \bibinfo{journal}{Phys.\ Rev.\ Lett.}
  \textbf{\bibinfo{volume}{82}}, \bibinfo{pages}{2394}.

\bibitem[{\citenamefont{Litvak}
  \emph{et~al.}(2000{\natexlab{a}})\citenamefont{Litvak, Mironov, and
  Sher}}]{Litvak:jetp:91:1268}
\bibinfo{author}{\bibnamefont{Litvak}, \bibfnamefont{A.~G.}},
  \bibinfo{author}{\bibfnamefont{V.~A.} \bibnamefont{Mironov}}, and
  \bibinfo{author}{\bibfnamefont{E.~M.} \bibnamefont{Sher}},
  \bibinfo{year}{2000}{\natexlab{a}}, \bibinfo{journal}{J.\ Exp.\ Theor.\
  Phys.} \textbf{\bibinfo{volume}{91}}, \bibinfo{pages}{1268}.

\bibitem[{\citenamefont{Litvak}
  \emph{et~al.}(2000{\natexlab{b}})\citenamefont{Litvak, Mironov, and
  Sher}}]{Litvak:pre:61:891}
\bibinfo{author}{\bibnamefont{Litvak}, \bibfnamefont{A.~G.}},
  \bibinfo{author}{\bibfnamefont{V.~A.} \bibnamefont{Mironov}}, and
  \bibinfo{author}{\bibfnamefont{E.~M.} \bibnamefont{Sher}},
  \bibinfo{year}{2000}{\natexlab{b}}, \bibinfo{journal}{Phys.\ Rev.\ E}
  \textbf{\bibinfo{volume}{61}}, \bibinfo{pages}{891}.

\bibitem[{\citenamefont{Liu}
  \emph{et~al.}(2005{\natexlab{a}})\citenamefont{Liu, Schroeder, Chin, Li, and
  Xu}}]{Liu:oe:13:10248}
\bibinfo{author}{\bibnamefont{Liu}, \bibfnamefont{J.}},
  \bibinfo{author}{\bibfnamefont{H.}~\bibnamefont{Schroeder}},
  \bibinfo{author}{\bibfnamefont{S.~L.} \bibnamefont{Chin}},
  \bibinfo{author}{\bibfnamefont{R.}~\bibnamefont{Li}}, and
  \bibinfo{author}{\bibfnamefont{Z.}~\bibnamefont{Xu}},
  \bibinfo{year}{2005}{\natexlab{a}}, \bibinfo{journal}{Opt.\ Express}
  \textbf{\bibinfo{volume}{13}}, \bibinfo{pages}{10248}.

\bibitem[{\citenamefont{Liu}
  \emph{et~al.}(2005{\natexlab{b}})\citenamefont{Liu, Schroeder, Chin, Li, Yu,
  and Xu}}]{Liu:pra:72:053817}
\bibinfo{author}{\bibnamefont{Liu}, \bibfnamefont{J.}},
  \bibinfo{author}{\bibfnamefont{H.}~\bibnamefont{Schroeder}},
  \bibinfo{author}{\bibfnamefont{S.~L.} \bibnamefont{Chin}},
  \bibinfo{author}{\bibfnamefont{R.}~\bibnamefont{Li}},
  \bibinfo{author}{\bibfnamefont{W.}~\bibnamefont{Yu}}, and
  \bibinfo{author}{\bibfnamefont{Z.}~\bibnamefont{Xu}},
  \bibinfo{year}{2005}{\natexlab{b}}, \bibinfo{journal}{Phys.\ Rev.\ A}
  \textbf{\bibinfo{volume}{72}}, \bibinfo{pages}{053817}.

\bibitem[{\citenamefont{Liu and Chin}(2005)}]{Liu:oe:13:5750}
\bibinfo{author}{\bibnamefont{Liu}, \bibfnamefont{W.}}, and
  \bibinfo{author}{\bibfnamefont{S.~L.} \bibnamefont{Chin}},
  \bibinfo{year}{2005}, \bibinfo{journal}{Opt.\ Express}
  \textbf{\bibinfo{volume}{13}}, \bibinfo{pages}{5750}.

\bibitem[{\citenamefont{Liu} \emph{et~al.}(2003)\citenamefont{Liu, Chin,
  Kosareva, Golubtsov, and Kandidov}}]{Liu:oc:225:193}
\bibinfo{author}{\bibnamefont{Liu}, \bibfnamefont{W.}},
  \bibinfo{author}{\bibfnamefont{S.~L.} \bibnamefont{Chin}},
  \bibinfo{author}{\bibfnamefont{O.}~\bibnamefont{Kosareva}},
  \bibinfo{author}{\bibfnamefont{I.~S.} \bibnamefont{Golubtsov}}, and
  \bibinfo{author}{\bibfnamefont{V.~P.} \bibnamefont{Kandidov}},
  \bibinfo{year}{2003}, \bibinfo{journal}{Opt.\ Commun.}
  \textbf{\bibinfo{volume}{225}}, \bibinfo{pages}{193}.

\bibitem[{\citenamefont{Liu} \emph{et~al.}(2004)\citenamefont{Liu, Hosseini,
  Luo, Ferland, Chin, Kosareva, Panov, and Kandidov}}]{Liu:njp:6:1}
\bibinfo{author}{\bibnamefont{Liu}, \bibfnamefont{W.}},
  \bibinfo{author}{\bibfnamefont{S.~A.} \bibnamefont{Hosseini}},
  \bibinfo{author}{\bibfnamefont{Q.}~\bibnamefont{Luo}},
  \bibinfo{author}{\bibfnamefont{B.}~\bibnamefont{Ferland}},
  \bibinfo{author}{\bibfnamefont{S.~L.} \bibnamefont{Chin}},
  \bibinfo{author}{\bibfnamefont{O.~G.} \bibnamefont{Kosareva}},
  \bibinfo{author}{\bibfnamefont{N.~A.} \bibnamefont{Panov}}, and
  \bibinfo{author}{\bibfnamefont{V.~P.} \bibnamefont{Kandidov}},
  \bibinfo{year}{2004}, \bibinfo{journal}{New J.\ Phys.}
  \textbf{\bibinfo{volume}{6}}, \bibinfo{pages}{1}.

\bibitem[{\citenamefont{Liu} \emph{et~al.}(2002)\citenamefont{Liu, Kosareva,
  Golubtsov, Iwasaki, Becker, Kandidov, and Chin}}]{Liu:apb:75:595}
\bibinfo{author}{\bibnamefont{Liu}, \bibfnamefont{W.}},
  \bibinfo{author}{\bibfnamefont{O.}~\bibnamefont{Kosareva}},
  \bibinfo{author}{\bibfnamefont{I.~S.} \bibnamefont{Golubtsov}},
  \bibinfo{author}{\bibfnamefont{A.}~\bibnamefont{Iwasaki}},
  \bibinfo{author}{\bibfnamefont{A.}~\bibnamefont{Becker}},
  \bibinfo{author}{\bibfnamefont{V.~P.} \bibnamefont{Kandidov}}, and
  \bibinfo{author}{\bibfnamefont{S.~L.} \bibnamefont{Chin}},
  \bibinfo{year}{2002}, \bibinfo{journal}{Appl.\ Phys.\ B: Lasers \& Optics}
  \textbf{\bibinfo{volume}{75}}, \bibinfo{pages}{595}.

\bibitem[{\citenamefont{Liu}
  \emph{et~al.}(2005{\natexlab{c}})\citenamefont{Liu, Th\'eberge, Arevalo,
  Gravel, Becker, and Chin}}]{Liu:ol:30:2602}
\bibinfo{author}{\bibnamefont{Liu}, \bibfnamefont{W.}},
  \bibinfo{author}{\bibfnamefont{F.}~\bibnamefont{Th\'eberge}},
  \bibinfo{author}{\bibfnamefont{E.}~\bibnamefont{Arevalo}},
  \bibinfo{author}{\bibfnamefont{J.-F.} \bibnamefont{Gravel}},
  \bibinfo{author}{\bibfnamefont{A.}~\bibnamefont{Becker}}, and
  \bibinfo{author}{\bibfnamefont{S.~L.} \bibnamefont{Chin}},
  \bibinfo{year}{2005}{\natexlab{c}}, \bibinfo{journal}{Opt.\ Lett.}
  \textbf{\bibinfo{volume}{30}}, \bibinfo{pages}{2602}.

\bibitem[{\citenamefont{Liu} \emph{et~al.}(2006)\citenamefont{Liu, Th\'eberge,
  Daigle, Simard, Sarifi, Kamali, Xu, and Chin}}]{Liu:apb:85:55}
\bibinfo{author}{\bibnamefont{Liu}, \bibfnamefont{W.}},
  \bibinfo{author}{\bibfnamefont{F.}~\bibnamefont{Th\'eberge}},
  \bibinfo{author}{\bibfnamefont{J.-F.} \bibnamefont{Daigle}},
  \bibinfo{author}{\bibfnamefont{P.~T.} \bibnamefont{Simard}},
  \bibinfo{author}{\bibfnamefont{S.~M.} \bibnamefont{Sarifi}},
  \bibinfo{author}{\bibfnamefont{Y.}~\bibnamefont{Kamali}},
  \bibinfo{author}{\bibfnamefont{H.~L.} \bibnamefont{Xu}}, and
  \bibinfo{author}{\bibfnamefont{S.~L.} \bibnamefont{Chin}},
  \bibinfo{year}{2006}, \bibinfo{journal}{Appl.\ Phys.\ B: Lasers \& Optics}
  \textbf{\bibinfo{volume}{85}}, \bibinfo{pages}{55}.

\bibitem[{\citenamefont{Lotz}(1967{\natexlab{a}})}]{Lotz:zp:206:205}
\bibinfo{author}{\bibnamefont{Lotz}, \bibfnamefont{W.}},
  \bibinfo{year}{1967}{\natexlab{a}}, \bibinfo{journal}{Zeitschrift Phys.}
  \textbf{\bibinfo{volume}{206}}, \bibinfo{pages}{205}.

\bibitem[{\citenamefont{Lotz}(1967{\natexlab{b}})}]{Lotz:josab:57:873}
\bibinfo{author}{\bibnamefont{Lotz}, \bibfnamefont{W.}},
  \bibinfo{year}{1967}{\natexlab{b}}, \bibinfo{journal}{J.\ Opt.\ Soc.\ Am.\ B}
  \textbf{\bibinfo{volume}{57}}, \bibinfo{pages}{873}.

\bibitem[{\citenamefont{Lugovoi and Prokhorov}(1974)}]{Lugovoi:spu:16:658}
\bibinfo{author}{\bibnamefont{Lugovoi}, \bibfnamefont{V.~N.}}, and
  \bibinfo{author}{\bibfnamefont{A.~M.} \bibnamefont{Prokhorov}},
  \bibinfo{year}{1974}, \bibinfo{journal}{Sov.\ Phys.\ Usp.}
  \textbf{\bibinfo{volume}{16}}, \bibinfo{pages}{658}.

\bibitem[{\citenamefont{Luo} \emph{et~al.}(1995)\citenamefont{Luo, {\AA}gren,
  Minaev, and J{\o}rgensen}}]{Luo:jms:336:61}
\bibinfo{author}{\bibnamefont{Luo}, \bibfnamefont{Y.}},
  \bibinfo{author}{\bibfnamefont{H.}~\bibnamefont{{\AA}gren}},
  \bibinfo{author}{\bibfnamefont{B.}~\bibnamefont{Minaev}}, and
  \bibinfo{author}{\bibfnamefont{P.}~\bibnamefont{J{\o}rgensen}},
  \bibinfo{year}{1995}, \bibinfo{journal}{J.\ Mol.\ Struct.}
  \textbf{\bibinfo{volume}{336}}, \bibinfo{pages}{61}.

\bibitem[{\citenamefont{Luther}
  \emph{et~al.}(1994{\natexlab{a}})\citenamefont{Luther, Moloney, Newell, and
  Wright}}]{Luther:ol:19:862}
\bibinfo{author}{\bibnamefont{Luther}, \bibfnamefont{G.~G.}},
  \bibinfo{author}{\bibfnamefont{J.~V.} \bibnamefont{Moloney}},
  \bibinfo{author}{\bibfnamefont{A.~C.} \bibnamefont{Newell}}, and
  \bibinfo{author}{\bibfnamefont{E.~M.} \bibnamefont{Wright}},
  \bibinfo{year}{1994}{\natexlab{a}}, \bibinfo{journal}{Opt.\ Lett.}
  \textbf{\bibinfo{volume}{19}}, \bibinfo{pages}{862}.

\bibitem[{\citenamefont{Luther}
  \emph{et~al.}(1994{\natexlab{b}})\citenamefont{Luther, Newell, and
  Moloney}}]{Luther:pd:74:59}
\bibinfo{author}{\bibnamefont{Luther}, \bibfnamefont{G.~G.}},
  \bibinfo{author}{\bibfnamefont{A.~C.} \bibnamefont{Newell}}, and
  \bibinfo{author}{\bibfnamefont{J.~V.} \bibnamefont{Moloney}},
  \bibinfo{year}{1994}{\natexlab{b}}, \bibinfo{journal}{Physica D}
  \textbf{\bibinfo{volume}{74}}, \bibinfo{pages}{59}.

\bibitem[{\citenamefont{Mairesse} \emph{et~al.}(2003)\citenamefont{Mairesse,
  de~Bohan, Frasinski, Merdji, Dinu, Monchicourt, Breger, Kova\v{c}ev,
  Ta{\"i}eb, Carr\'e, Muller, Agostini} \emph{et~al.}}]{Mairesse:sc:302:1540}
\bibinfo{author}{\bibnamefont{Mairesse}, \bibfnamefont{Y.}},
  \bibinfo{author}{\bibfnamefont{A.}~\bibnamefont{de~Bohan}},
  \bibinfo{author}{\bibfnamefont{L.~J.} \bibnamefont{Frasinski}},
  \bibinfo{author}{\bibfnamefont{H.}~\bibnamefont{Merdji}},
  \bibinfo{author}{\bibfnamefont{L.~C.} \bibnamefont{Dinu}},
  \bibinfo{author}{\bibfnamefont{P.}~\bibnamefont{Monchicourt}},
  \bibinfo{author}{\bibfnamefont{P.}~\bibnamefont{Breger}},
  \bibinfo{author}{\bibfnamefont{M.}~\bibnamefont{Kova\v{c}ev}},
  \bibinfo{author}{\bibfnamefont{R.}~\bibnamefont{Ta{\"i}eb}},
  \bibinfo{author}{\bibfnamefont{B.}~\bibnamefont{Carr\'e}},
  \bibinfo{author}{\bibfnamefont{H.~G.} \bibnamefont{Muller}},
  \bibinfo{author}{\bibfnamefont{P.}~\bibnamefont{Agostini}}, \emph{et~al.},
  \bibinfo{year}{2003}, \bibinfo{journal}{Science}
  \textbf{\bibinfo{volume}{302}}, \bibinfo{pages}{1540}.

\bibitem[{\citenamefont{Malkin}(1990)}]{Malkin:pla:151:285}
\bibinfo{author}{\bibnamefont{Malkin}, \bibfnamefont{V.~M.}},
  \bibinfo{year}{1990}, \bibinfo{journal}{Phys.\ Lett.\ A}
  \textbf{\bibinfo{volume}{151}}, \bibinfo{pages}{285}.

\bibitem[{\citenamefont{Malkin}(1993)}]{Malkin:pd:64:251}
\bibinfo{author}{\bibnamefont{Malkin}, \bibfnamefont{V.~M.}},
  \bibinfo{year}{1993}, \bibinfo{journal}{Physica D}
  \textbf{\bibinfo{volume}{64}}, \bibinfo{pages}{251}.

\bibitem[{\citenamefont{Manassah}(1992)}]{Manassah:ol:17:1259}
\bibinfo{author}{\bibnamefont{Manassah}, \bibfnamefont{J.~T.}},
  \bibinfo{year}{1992}, \bibinfo{journal}{Opt.\ Lett.}
  \textbf{\bibinfo{volume}{17}}, \bibinfo{pages}{1259}.

\bibitem[{\citenamefont{Manassah} \emph{et~al.}(1988)\citenamefont{Manassah,
  Baldeck, and Alfano}}]{Manassah:ol:13:589}
\bibinfo{author}{\bibnamefont{Manassah}, \bibfnamefont{J.~T.}},
  \bibinfo{author}{\bibfnamefont{P.~L.} \bibnamefont{Baldeck}}, and
  \bibinfo{author}{\bibfnamefont{R.~R.} \bibnamefont{Alfano}},
  \bibinfo{year}{1988}, \bibinfo{journal}{Opt.\ Lett.}
  \textbf{\bibinfo{volume}{13}}, \bibinfo{pages}{589}.

\bibitem[{\citenamefont{Marburger}(1975)}]{Marburger:pqe:4:35}
\bibinfo{author}{\bibnamefont{Marburger}, \bibfnamefont{J.~H.}},
  \bibinfo{year}{1975}, \bibinfo{journal}{Prog.\ Quantum Electron.}
  \textbf{\bibinfo{volume}{4}}, \bibinfo{pages}{35}.

\bibitem[{\citenamefont{Mariyenko} \emph{et~al.}(2005)\citenamefont{Mariyenko,
  Strohaber, and Uiterwaal}}]{Mariyenko:oe:13:7599}
\bibinfo{author}{\bibnamefont{Mariyenko}, \bibfnamefont{I.~G.}},
  \bibinfo{author}{\bibfnamefont{J.}~\bibnamefont{Strohaber}}, and
  \bibinfo{author}{\bibfnamefont{C.~J. G.~J.} \bibnamefont{Uiterwaal}},
  \bibinfo{year}{2005}, \bibinfo{journal}{Opt.\ Express}
  \textbf{\bibinfo{volume}{13}}, \bibinfo{pages}{7599}.

\bibitem[{\citenamefont{Marklund and Shukla}(2006)}]{Marklund:ol:31:1884}
\bibinfo{author}{\bibnamefont{Marklund}, \bibfnamefont{M.}}, and
  \bibinfo{author}{\bibfnamefont{P.~K.} \bibnamefont{Shukla}},
  \bibinfo{year}{2006}, \bibinfo{journal}{Opt.\ Lett.}
  \textbf{\bibinfo{volume}{31}}, \bibinfo{pages}{1884}.

\bibitem[{\citenamefont{Martin} \emph{et~al.}(1997)\citenamefont{Martin,
  Guizard, Daguzan, Petite, D'Oliveira, Meynadier, and
  Perdrix}}]{Martin:prb:55:5799}
\bibinfo{author}{\bibnamefont{Martin}, \bibfnamefont{P.}},
  \bibinfo{author}{\bibfnamefont{S.}~\bibnamefont{Guizard}},
  \bibinfo{author}{\bibfnamefont{P.}~\bibnamefont{Daguzan}},
  \bibinfo{author}{\bibfnamefont{G.}~\bibnamefont{Petite}},
  \bibinfo{author}{\bibfnamefont{P.}~\bibnamefont{D'Oliveira}},
  \bibinfo{author}{\bibfnamefont{P.}~\bibnamefont{Meynadier}}, and
  \bibinfo{author}{\bibfnamefont{M.}~\bibnamefont{Perdrix}},
  \bibinfo{year}{1997}, \bibinfo{journal}{Phys.\ Rev.\ B}
  \textbf{\bibinfo{volume}{55}}, \bibinfo{pages}{5799}.

\bibitem[{\citenamefont{{McKinstrie} and
  Russell}(1988)}]{Mckinstrie:prl:61:2929}
\bibinfo{author}{\bibnamefont{{McKinstrie}}, \bibfnamefont{C.~J.}}, and
  \bibinfo{author}{\bibfnamefont{D.~A.} \bibnamefont{Russell}},
  \bibinfo{year}{1988}, \bibinfo{journal}{Phys.\ Rev.\ Lett.}
  \textbf{\bibinfo{volume}{61}}, \bibinfo{pages}{2929}.

\bibitem[{\citenamefont{Measures}(1984)}]{Measures:LRS:84}
\bibinfo{author}{\bibnamefont{Measures}, \bibfnamefont{R.~M.}},
  \bibinfo{year}{1984}, \emph{\bibinfo{title}{Laser remote sensing -
  Fundamentals and applications}} (\bibinfo{publisher}{Wiley Interscience},
  \bibinfo{address}{New York}).

\bibitem[{\citenamefont{M\'echain} \emph{et~al.}(2004)\citenamefont{M\'echain,
  Couairon, Franco, Prade, and Mysyrowicz}}]{Mechain:prl:93:035003}
\bibinfo{author}{\bibnamefont{M\'echain}, \bibfnamefont{G.}},
  \bibinfo{author}{\bibfnamefont{A.}~\bibnamefont{Couairon}},
  \bibinfo{author}{\bibfnamefont{M.}~\bibnamefont{Franco}},
  \bibinfo{author}{\bibfnamefont{B.}~\bibnamefont{Prade}}, and
  \bibinfo{author}{\bibfnamefont{A.}~\bibnamefont{Mysyrowicz}},
  \bibinfo{year}{2004}, \bibinfo{journal}{Phys.\ Rev.\ Lett.}
  \textbf{\bibinfo{volume}{93}}, \bibinfo{pages}{035003}.

\bibitem[{\citenamefont{M\'echain}
  \emph{et~al.}(2005{\natexlab{a}})\citenamefont{M\'echain, D'Amico, Andr\'e,
  Tzortzakis, Franco, Prade, Mysyrowicz, Couairon, Salmon, and
  Sauerbrey}}]{Mechain:oc:247:171}
\bibinfo{author}{\bibnamefont{M\'echain}, \bibfnamefont{G.}},
  \bibinfo{author}{\bibfnamefont{C.}~\bibnamefont{D'Amico}},
  \bibinfo{author}{\bibfnamefont{Y.-B.} \bibnamefont{Andr\'e}},
  \bibinfo{author}{\bibfnamefont{S.}~\bibnamefont{Tzortzakis}},
  \bibinfo{author}{\bibfnamefont{M.}~\bibnamefont{Franco}},
  \bibinfo{author}{\bibfnamefont{B.}~\bibnamefont{Prade}},
  \bibinfo{author}{\bibfnamefont{A.}~\bibnamefont{Mysyrowicz}},
  \bibinfo{author}{\bibfnamefont{A.}~\bibnamefont{Couairon}},
  \bibinfo{author}{\bibfnamefont{E.}~\bibnamefont{Salmon}}, and
  \bibinfo{author}{\bibfnamefont{R.}~\bibnamefont{Sauerbrey}},
  \bibinfo{year}{2005}{\natexlab{a}}, \bibinfo{journal}{Opt.\ Commun.}
  \textbf{\bibinfo{volume}{247}}, \bibinfo{pages}{171}.

\bibitem[{\citenamefont{M\'echain}
  \emph{et~al.}(2005{\natexlab{b}})\citenamefont{M\'echain, M\'ejean,
  Ackermann, Rohwetter, Andr\'e, Kasparian, Prade, Stelmaszczyk, Yu, Salmon,
  Winn, Schlie} \emph{et~al.}}]{Mechain:apb:80:785}
\bibinfo{author}{\bibnamefont{M\'echain}, \bibfnamefont{G.}},
  \bibinfo{author}{\bibfnamefont{G.}~\bibnamefont{M\'ejean}},
  \bibinfo{author}{\bibfnamefont{R.}~\bibnamefont{Ackermann}},
  \bibinfo{author}{\bibfnamefont{P.}~\bibnamefont{Rohwetter}},
  \bibinfo{author}{\bibfnamefont{Y.-B.} \bibnamefont{Andr\'e}},
  \bibinfo{author}{\bibfnamefont{J.}~\bibnamefont{Kasparian}},
  \bibinfo{author}{\bibfnamefont{B.}~\bibnamefont{Prade}},
  \bibinfo{author}{\bibfnamefont{K.}~\bibnamefont{Stelmaszczyk}},
  \bibinfo{author}{\bibfnamefont{J.}~\bibnamefont{Yu}},
  \bibinfo{author}{\bibfnamefont{E.}~\bibnamefont{Salmon}},
  \bibinfo{author}{\bibfnamefont{W.}~\bibnamefont{Winn}},
  \bibinfo{author}{\bibfnamefont{L.~A.~V.} \bibnamefont{Schlie}},
  \emph{et~al.}, \bibinfo{year}{2005}{\natexlab{b}}, \bibinfo{journal}{Appl.\
  Phys.\ B: Lasers \& Optics} \textbf{\bibinfo{volume}{80}},
  \bibinfo{pages}{785}.

\bibitem[{\citenamefont{M\'ejean}
  \emph{et~al.}(2006{\natexlab{a}})\citenamefont{M\'ejean, Ackermann,
  Kasparian, Salmon, Yu, Wolf, Rethmeier, Kalkner, Rohwetter, Stelmaszczyk, and
  W{\"o}ste}}]{Mejean:apl:88:021101}
\bibinfo{author}{\bibnamefont{M\'ejean}, \bibfnamefont{G.}},
  \bibinfo{author}{\bibfnamefont{R.}~\bibnamefont{Ackermann}},
  \bibinfo{author}{\bibfnamefont{J.}~\bibnamefont{Kasparian}},
  \bibinfo{author}{\bibfnamefont{E.}~\bibnamefont{Salmon}},
  \bibinfo{author}{\bibfnamefont{J.}~\bibnamefont{Yu}},
  \bibinfo{author}{\bibfnamefont{J.-P.} \bibnamefont{Wolf}},
  \bibinfo{author}{\bibfnamefont{K.}~\bibnamefont{Rethmeier}},
  \bibinfo{author}{\bibfnamefont{W.}~\bibnamefont{Kalkner}},
  \bibinfo{author}{\bibfnamefont{P.}~\bibnamefont{Rohwetter}},
  \bibinfo{author}{\bibfnamefont{K.}~\bibnamefont{Stelmaszczyk}}, and
  \bibinfo{author}{\bibfnamefont{L.}~\bibnamefont{W{\"o}ste}},
  \bibinfo{year}{2006}{\natexlab{a}}, \bibinfo{journal}{Appl.\ Phys.\ Lett.}
  \textbf{\bibinfo{volume}{88}}, \bibinfo{pages}{021101}.

\bibitem[{\citenamefont{M\'ejean}
  \emph{et~al.}(2006{\natexlab{b}})\citenamefont{M\'ejean, Kasparian, Yu, Frey,
  Salmon, Ackermann, Wolf, Berg\'e, and Skupin}}]{Mejean:apb:82:341}
\bibinfo{author}{\bibnamefont{M\'ejean}, \bibfnamefont{G.}},
  \bibinfo{author}{\bibfnamefont{J.}~\bibnamefont{Kasparian}},
  \bibinfo{author}{\bibfnamefont{J.}~\bibnamefont{Yu}},
  \bibinfo{author}{\bibfnamefont{S.}~\bibnamefont{Frey}},
  \bibinfo{author}{\bibfnamefont{E.}~\bibnamefont{Salmon}},
  \bibinfo{author}{\bibfnamefont{R.}~\bibnamefont{Ackermann}},
  \bibinfo{author}{\bibfnamefont{J.-P.} \bibnamefont{Wolf}},
  \bibinfo{author}{\bibfnamefont{L.}~\bibnamefont{Berg\'e}}, and
  \bibinfo{author}{\bibfnamefont{S.}~\bibnamefont{Skupin}},
  \bibinfo{year}{2006}{\natexlab{b}}, \bibinfo{journal}{Appl.\ Phys.\ B: Lasers
  \& Optics} \textbf{\bibinfo{volume}{82}}, \bibinfo{pages}{341}.

\bibitem[{\citenamefont{M\'ejean} \emph{et~al.}(2004)\citenamefont{M\'ejean,
  Kasparian, Yu, Frey, Salmon, and Wolf}}]{Mejean:apb:78:535}
\bibinfo{author}{\bibnamefont{M\'ejean}, \bibfnamefont{G.}},
  \bibinfo{author}{\bibfnamefont{J.}~\bibnamefont{Kasparian}},
  \bibinfo{author}{\bibfnamefont{J.}~\bibnamefont{Yu}},
  \bibinfo{author}{\bibfnamefont{S.}~\bibnamefont{Frey}},
  \bibinfo{author}{\bibfnamefont{E.}~\bibnamefont{Salmon}}, and
  \bibinfo{author}{\bibfnamefont{J.~P.} \bibnamefont{Wolf}},
  \bibinfo{year}{2004}, \bibinfo{journal}{Appl.\ Phys.\ B: Lasers \& Optics}
  \textbf{\bibinfo{volume}{78}}, \bibinfo{pages}{535}.

\bibitem[{\citenamefont{M\'ejean} \emph{et~al.}(2005)\citenamefont{M\'ejean,
  Kasparian, Yu, Salmon, Frey, Wolf, Skupin, Vin{\c{c}}otte, Nuter, Champeaux,
  and Berg\'e}}]{Mejean:pre:72:026611}
\bibinfo{author}{\bibnamefont{M\'ejean}, \bibfnamefont{G.}},
  \bibinfo{author}{\bibfnamefont{J.}~\bibnamefont{Kasparian}},
  \bibinfo{author}{\bibfnamefont{J.}~\bibnamefont{Yu}},
  \bibinfo{author}{\bibfnamefont{E.}~\bibnamefont{Salmon}},
  \bibinfo{author}{\bibfnamefont{S.}~\bibnamefont{Frey}},
  \bibinfo{author}{\bibfnamefont{J.-P.} \bibnamefont{Wolf}},
  \bibinfo{author}{\bibfnamefont{S.}~\bibnamefont{Skupin}},
  \bibinfo{author}{\bibfnamefont{A.}~\bibnamefont{Vin{\c{c}}otte}},
  \bibinfo{author}{\bibfnamefont{R.}~\bibnamefont{Nuter}},
  \bibinfo{author}{\bibfnamefont{S.}~\bibnamefont{Champeaux}}, and
  \bibinfo{author}{\bibfnamefont{L.}~\bibnamefont{Berg\'e}},
  \bibinfo{year}{2005}, \bibinfo{journal}{Phys.\ Rev.\ E}
  \textbf{\bibinfo{volume}{72}}, \bibinfo{pages}{026611}.

\bibitem[{\citenamefont{M\'evel} \emph{et~al.}(2003)\citenamefont{M\'evel,
  Tcherbakoff, Salin, and Constant}}]{Mevel:josab:20:105}
\bibinfo{author}{\bibnamefont{M\'evel}, \bibfnamefont{E.}},
  \bibinfo{author}{\bibfnamefont{O.}~\bibnamefont{Tcherbakoff}},
  \bibinfo{author}{\bibfnamefont{F.}~\bibnamefont{Salin}}, and
  \bibinfo{author}{\bibfnamefont{E.}~\bibnamefont{Constant}},
  \bibinfo{year}{2003}, \bibinfo{journal}{J.\ Opt.\ Soc.\ Am.\ B}
  \textbf{\bibinfo{volume}{20}}, \bibinfo{pages}{105}.

\bibitem[{\citenamefont{Meyers}(2000)}]{Meyers:EAC:2000}
\bibinfo{author}{\bibnamefont{Meyers}, \bibfnamefont{R.~A.}},
  \bibinfo{year}{2000}, \emph{\bibinfo{title}{Encyclopedia of Analytical
  Chemistry}} (\bibinfo{publisher}{John Wiley \& Sons},
  \bibinfo{address}{Chichester}).

\bibitem[{\citenamefont{Michinel} \emph{et~al.}(2001)\citenamefont{Michinel,
  Campo-T\'aboas, Quiroga-Teixeiro, Salgueiro, and
  Garc\'ia-Fern\'andez}}]{Michinel:jobqso:3:314}
\bibinfo{author}{\bibnamefont{Michinel}, \bibfnamefont{H.}},
  \bibinfo{author}{\bibfnamefont{J.}~\bibnamefont{Campo-T\'aboas}},
  \bibinfo{author}{\bibfnamefont{M.~L.} \bibnamefont{Quiroga-Teixeiro}},
  \bibinfo{author}{\bibfnamefont{J.~R.} \bibnamefont{Salgueiro}}, and
  \bibinfo{author}{\bibfnamefont{R.}~\bibnamefont{Garc\'ia-Fern\'andez}},
  \bibinfo{year}{2001}, \bibinfo{journal}{J.\ Opt.\ B: Quantum Semiclass.\
  Opt.} \textbf{\bibinfo{volume}{3}}, \bibinfo{pages}{314}.

\bibitem[{\citenamefont{Miki} \emph{et~al.}(1993)\citenamefont{Miki, Aihara,
  and Shindo}}]{Miki:jpd:26:1244}
\bibinfo{author}{\bibnamefont{Miki}, \bibfnamefont{M.}},
  \bibinfo{author}{\bibfnamefont{Y.}~\bibnamefont{Aihara}}, and
  \bibinfo{author}{\bibfnamefont{T.}~\bibnamefont{Shindo}},
  \bibinfo{year}{1993}, \bibinfo{journal}{J.\ Phys.\ D: Appl. \ Phys.}
  \textbf{\bibinfo{volume}{26}}, \bibinfo{pages}{1244}.

\bibitem[{\citenamefont{{Milsted Jr.} and
  Cantrell}(1996)}]{Milsted:pra:53:3536}
\bibinfo{author}{\bibnamefont{{Milsted Jr.}}, \bibfnamefont{C.~S.}}, and
  \bibinfo{author}{\bibfnamefont{C.~D.} \bibnamefont{Cantrell}},
  \bibinfo{year}{1996}, \bibinfo{journal}{Phys.\ Rev.\ A}
  \textbf{\bibinfo{volume}{53}}, \bibinfo{pages}{3536}.

\bibitem[{\citenamefont{Mlejnek} \emph{et~al.}(1999)\citenamefont{Mlejnek,
  Kolesik, Moloney, and Wright}}]{Mlejnek:prl:83:2938}
\bibinfo{author}{\bibnamefont{Mlejnek}, \bibfnamefont{M.}},
  \bibinfo{author}{\bibfnamefont{M.}~\bibnamefont{Kolesik}},
  \bibinfo{author}{\bibfnamefont{J.~V.} \bibnamefont{Moloney}}, and
  \bibinfo{author}{\bibfnamefont{E.~M.} \bibnamefont{Wright}},
  \bibinfo{year}{1999}, \bibinfo{journal}{Phys.\ Rev.\ Lett.}
  \textbf{\bibinfo{volume}{83}}, \bibinfo{pages}{2938}.

\bibitem[{\citenamefont{Mlejnek}
  \emph{et~al.}(1998{\natexlab{a}})\citenamefont{Mlejnek, Wright, and
  Moloney}}]{Mlejnek:ol:23:382}
\bibinfo{author}{\bibnamefont{Mlejnek}, \bibfnamefont{M.}},
  \bibinfo{author}{\bibfnamefont{E.~M.} \bibnamefont{Wright}}, and
  \bibinfo{author}{\bibfnamefont{J.~V.} \bibnamefont{Moloney}},
  \bibinfo{year}{1998}{\natexlab{a}}, \bibinfo{journal}{Opt.\ Lett.}
  \textbf{\bibinfo{volume}{23}}, \bibinfo{pages}{382}.

\bibitem[{\citenamefont{Mlejnek}
  \emph{et~al.}(1998{\natexlab{b}})\citenamefont{Mlejnek, Wright, and
  Moloney}}]{Mlejnek:pre:58:4903}
\bibinfo{author}{\bibnamefont{Mlejnek}, \bibfnamefont{M.}},
  \bibinfo{author}{\bibfnamefont{E.~M.} \bibnamefont{Wright}}, and
  \bibinfo{author}{\bibfnamefont{J.~V.} \bibnamefont{Moloney}},
  \bibinfo{year}{1998}{\natexlab{b}}, \bibinfo{journal}{Phys.\ Rev.\ E}
  \textbf{\bibinfo{volume}{58}}, \bibinfo{pages}{4903}.

\bibitem[{\citenamefont{Moll and Gaeta}(2004)}]{Moll:ol:29:995}
\bibinfo{author}{\bibnamefont{Moll}, \bibfnamefont{K.~D.}}, and
  \bibinfo{author}{\bibfnamefont{A.~L.} \bibnamefont{Gaeta}},
  \bibinfo{year}{2004}, \bibinfo{journal}{Opt.\ Lett.}
  \textbf{\bibinfo{volume}{29}}, \bibinfo{pages}{995}.

\bibitem[{\citenamefont{Moll} \emph{et~al.}(2003)\citenamefont{Moll, Gaeta, and
  Fibich}}]{Moll:prl:90:203902}
\bibinfo{author}{\bibnamefont{Moll}, \bibfnamefont{K.~D.}},
  \bibinfo{author}{\bibfnamefont{A.~L.} \bibnamefont{Gaeta}}, and
  \bibinfo{author}{\bibfnamefont{G.}~\bibnamefont{Fibich}},
  \bibinfo{year}{2003}, \bibinfo{journal}{Phys.\ Rev.\ Lett.}
  \textbf{\bibinfo{volume}{90}}, \bibinfo{pages}{203902}.

\bibitem[{\citenamefont{Mourou} \emph{et~al.}(2006)\citenamefont{Mourou,
  Tajima, and Bulanov}}]{Mourou:rmp:78:309}
\bibinfo{author}{\bibnamefont{Mourou}, \bibfnamefont{G.}},
  \bibinfo{author}{\bibfnamefont{T.}~\bibnamefont{Tajima}}, and
  \bibinfo{author}{\bibfnamefont{S.~V.} \bibnamefont{Bulanov}},
  \bibinfo{year}{2006}, \bibinfo{journal}{Rev.\ Mod.\ Phys.}
  \textbf{\bibinfo{volume}{78}}, \bibinfo{pages}{309}.

\bibitem[{\citenamefont{Nibbering} \emph{et~al.}(1996)\citenamefont{Nibbering,
  Curley, Grillon, Prade, Franco, Salin, and Mysyrowicz}}]{Nibbering:ol:21:62}
\bibinfo{author}{\bibnamefont{Nibbering}, \bibfnamefont{E.~T.~J.}},
  \bibinfo{author}{\bibfnamefont{P.~F.} \bibnamefont{Curley}},
  \bibinfo{author}{\bibfnamefont{G.}~\bibnamefont{Grillon}},
  \bibinfo{author}{\bibfnamefont{B.~S.} \bibnamefont{Prade}},
  \bibinfo{author}{\bibfnamefont{M.~A.} \bibnamefont{Franco}},
  \bibinfo{author}{\bibfnamefont{F.}~\bibnamefont{Salin}}, and
  \bibinfo{author}{\bibfnamefont{A.}~\bibnamefont{Mysyrowicz}},
  \bibinfo{year}{1996}, \bibinfo{journal}{Opt.\ Lett.}
  \textbf{\bibinfo{volume}{21}}, \bibinfo{pages}{62}.

\bibitem[{\citenamefont{Nibbering} \emph{et~al.}(1997)\citenamefont{Nibbering,
  Grillon, Franco, Prade, and Mysyrowicz}}]{Nibbering:josab:14:650}
\bibinfo{author}{\bibnamefont{Nibbering}, \bibfnamefont{E.~T.~J.}},
  \bibinfo{author}{\bibfnamefont{G.}~\bibnamefont{Grillon}},
  \bibinfo{author}{\bibfnamefont{M.~A.} \bibnamefont{Franco}},
  \bibinfo{author}{\bibfnamefont{B.~S.} \bibnamefont{Prade}}, and
  \bibinfo{author}{\bibfnamefont{A.}~\bibnamefont{Mysyrowicz}},
  \bibinfo{year}{1997}, \bibinfo{journal}{J.\ Opt.\ Soc.\ Am.\ B}
  \textbf{\bibinfo{volume}{14}}, \bibinfo{pages}{650}.

\bibitem[{\citenamefont{Niessner}(1994)}]{Niessner:pspie:2360:254}
\bibinfo{author}{\bibnamefont{Niessner}, \bibfnamefont{R.}},
  \bibinfo{year}{1994}, \bibinfo{journal}{Proceedings of the SPIE}
  \textbf{\bibinfo{volume}{2360}}, \bibinfo{pages}{254}.

\bibitem[{\citenamefont{Nishioka} \emph{et~al.}(1995)\citenamefont{Nishioka,
  Odajima, Ueda, and Takuma}}]{Nishioka:ol:20:2505}
\bibinfo{author}{\bibnamefont{Nishioka}, \bibfnamefont{H.}},
  \bibinfo{author}{\bibfnamefont{W.}~\bibnamefont{Odajima}},
  \bibinfo{author}{\bibfnamefont{K.}~\bibnamefont{Ueda}}, and
  \bibinfo{author}{\bibfnamefont{H.}~\bibnamefont{Takuma}},
  \bibinfo{year}{1995}, \bibinfo{journal}{Opt.\ Lett.}
  \textbf{\bibinfo{volume}{20}}, \bibinfo{pages}{2505}.

\bibitem[{\citenamefont{Nisoli}
  \emph{et~al.}(1997{\natexlab{a}})\citenamefont{Nisoli, {De Silvestri},
  Svelto, Szip{\"o}cs, Ferencz, Spielmann, Sartania, and
  Krausz}}]{Nisoli:ol:22:522}
\bibinfo{author}{\bibnamefont{Nisoli}, \bibfnamefont{M.}},
  \bibinfo{author}{\bibfnamefont{S.}~\bibnamefont{{De Silvestri}}},
  \bibinfo{author}{\bibfnamefont{O.}~\bibnamefont{Svelto}},
  \bibinfo{author}{\bibfnamefont{R.}~\bibnamefont{Szip{\"o}cs}},
  \bibinfo{author}{\bibfnamefont{K.}~\bibnamefont{Ferencz}},
  \bibinfo{author}{\bibfnamefont{C.}~\bibnamefont{Spielmann}},
  \bibinfo{author}{\bibfnamefont{S.}~\bibnamefont{Sartania}}, and
  \bibinfo{author}{\bibfnamefont{F.}~\bibnamefont{Krausz}},
  \bibinfo{year}{1997}{\natexlab{a}}, \bibinfo{journal}{Opt.\ Lett.}
  \textbf{\bibinfo{volume}{22}}, \bibinfo{pages}{522}.

\bibitem[{\citenamefont{Nisoli}
  \emph{et~al.}(1997{\natexlab{b}})\citenamefont{Nisoli, Stagira, Silvestri,
  Svelto, Sartania, Cheng, Lenzner, Spielmann, and Krausz}}]{Nisoli:apb:65:189}
\bibinfo{author}{\bibnamefont{Nisoli}, \bibfnamefont{M.}},
  \bibinfo{author}{\bibfnamefont{S.}~\bibnamefont{Stagira}},
  \bibinfo{author}{\bibfnamefont{S.~D.} \bibnamefont{Silvestri}},
  \bibinfo{author}{\bibfnamefont{O.}~\bibnamefont{Svelto}},
  \bibinfo{author}{\bibfnamefont{S.}~\bibnamefont{Sartania}},
  \bibinfo{author}{\bibfnamefont{Z.}~\bibnamefont{Cheng}},
  \bibinfo{author}{\bibfnamefont{M.}~\bibnamefont{Lenzner}},
  \bibinfo{author}{\bibfnamefont{C.}~\bibnamefont{Spielmann}}, and
  \bibinfo{author}{\bibfnamefont{F.}~\bibnamefont{Krausz}},
  \bibinfo{year}{1997}{\natexlab{b}}, \bibinfo{journal}{Appl.\ Phys.\ B: Lasers
  \& Optics} \textbf{\bibinfo{volume}{65}}, \bibinfo{pages}{189}.

\bibitem[{\citenamefont{Nisoli} \emph{et~al.}(1998)\citenamefont{Nisoli,
  Stagira, Silvestri, Svelto, Sartania, Cheng, Tempea, Spielmann, and
  Krausz}}]{Nisoli:ieeejstqe:4:414}
\bibinfo{author}{\bibnamefont{Nisoli}, \bibfnamefont{M.}},
  \bibinfo{author}{\bibfnamefont{S.}~\bibnamefont{Stagira}},
  \bibinfo{author}{\bibfnamefont{S.~D.} \bibnamefont{Silvestri}},
  \bibinfo{author}{\bibfnamefont{O.}~\bibnamefont{Svelto}},
  \bibinfo{author}{\bibfnamefont{S.}~\bibnamefont{Sartania}},
  \bibinfo{author}{\bibfnamefont{Z.}~\bibnamefont{Cheng}},
  \bibinfo{author}{\bibfnamefont{G.}~\bibnamefont{Tempea}},
  \bibinfo{author}{\bibfnamefont{C.}~\bibnamefont{Spielmann}}, and
  \bibinfo{author}{\bibfnamefont{F.}~\bibnamefont{Krausz}},
  \bibinfo{year}{1998}, \bibinfo{journal}{IEEE J.\ Selec.\ Top.\ Quant.\
  Electron.} \textbf{\bibinfo{volume}{4}}, \bibinfo{pages}{414}.

\bibitem[{\citenamefont{Noack and Vogel}(1999)}]{Noack:ieeejqe:35:1156}
\bibinfo{author}{\bibnamefont{Noack}, \bibfnamefont{J.}}, and
  \bibinfo{author}{\bibfnamefont{A.}~\bibnamefont{Vogel}},
  \bibinfo{year}{1999}, \bibinfo{journal}{IEEE J.\ Quant.\ Electron.}
  \textbf{\bibinfo{volume}{35}}, \bibinfo{pages}{1156}.

\bibitem[{\citenamefont{Nurhuda}
  \emph{et~al.}(2002{\natexlab{a}})\citenamefont{Nurhuda, Suda, Hatayama,
  Nagasaka, and Midorikawa}}]{Nurhuda:pra:66:023811}
\bibinfo{author}{\bibnamefont{Nurhuda}, \bibfnamefont{M.}},
  \bibinfo{author}{\bibfnamefont{A.}~\bibnamefont{Suda}},
  \bibinfo{author}{\bibfnamefont{M.}~\bibnamefont{Hatayama}},
  \bibinfo{author}{\bibfnamefont{K.}~\bibnamefont{Nagasaka}}, and
  \bibinfo{author}{\bibfnamefont{K.}~\bibnamefont{Midorikawa}},
  \bibinfo{year}{2002}{\natexlab{a}}, \bibinfo{journal}{Phys.\ Rev.\ A}
  \textbf{\bibinfo{volume}{66}}, \bibinfo{pages}{023811}.

\bibitem[{\citenamefont{Nurhuda}
  \emph{et~al.}(2002{\natexlab{b}})\citenamefont{Nurhuda, Suda, and
  Midorikawa}}]{Nurhuda:rr:48:40}
\bibinfo{author}{\bibnamefont{Nurhuda}, \bibfnamefont{M.}},
  \bibinfo{author}{\bibfnamefont{A.}~\bibnamefont{Suda}}, and
  \bibinfo{author}{\bibfnamefont{K.}~\bibnamefont{Midorikawa}},
  \bibinfo{year}{2002}{\natexlab{b}}, \bibinfo{journal}{RIKEN Rev.}
  \textbf{\bibinfo{volume}{48}}, \bibinfo{pages}{40}.

\bibitem[{\citenamefont{Nuter and Berg\'e}(2006)}]{Nuter:josab:23:874}
\bibinfo{author}{\bibnamefont{Nuter}, \bibfnamefont{R.}}, and
  \bibinfo{author}{\bibfnamefont{L.}~\bibnamefont{Berg\'e}},
  \bibinfo{year}{2006}, \bibinfo{journal}{J.\ Opt.\ Soc.\ Am.\ B}
  \textbf{\bibinfo{volume}{23}}, \bibinfo{pages}{874}.

\bibitem[{\citenamefont{Nuter} \emph{et~al.}(2005)\citenamefont{Nuter, Skupin,
  and Berg\'e}}]{Nuter:ol:30:917}
\bibinfo{author}{\bibnamefont{Nuter}, \bibfnamefont{R.}},
  \bibinfo{author}{\bibfnamefont{S.}~\bibnamefont{Skupin}}, and
  \bibinfo{author}{\bibfnamefont{L.}~\bibnamefont{Berg\'e}},
  \bibinfo{year}{2005}, \bibinfo{journal}{Opt.\ Lett.}
  \textbf{\bibinfo{volume}{30}}, \bibinfo{pages}{917}.

\bibitem[{\citenamefont{Pan} \emph{et~al.}(1990)\citenamefont{Pan, Taylor, and
  Clark}}]{Pan:josab:7:509}
\bibinfo{author}{\bibnamefont{Pan}, \bibfnamefont{L.}},
  \bibinfo{author}{\bibfnamefont{K.~T.} \bibnamefont{Taylor}}, and
  \bibinfo{author}{\bibfnamefont{C.~W.} \bibnamefont{Clark}},
  \bibinfo{year}{1990}, \bibinfo{journal}{J.\ Opt.\ Soc.\ Am.\ B}
  \textbf{\bibinfo{volume}{7}}, \bibinfo{pages}{509}.

\bibitem[{\citenamefont{Peck and Reeder}(1972)}]{Peck:josa:62:958}
\bibinfo{author}{\bibnamefont{Peck}, \bibfnamefont{E.~R.}}, and
  \bibinfo{author}{\bibfnamefont{K.}~\bibnamefont{Reeder}},
  \bibinfo{year}{1972}, \bibinfo{journal}{J.\ Opt.\ Soc.\ Am.}
  \textbf{\bibinfo{volume}{62}}, \bibinfo{pages}{958}.

\bibitem[{\citenamefont{Pe{\~n}ano}
  \emph{et~al.}(2005)\citenamefont{Pe{\~n}ano, Sprangle, Hafizi, Manheimer, and
  Zigler}}]{Penano:pre:72:036412}
\bibinfo{author}{\bibnamefont{Pe{\~n}ano}, \bibfnamefont{J.~R.}},
  \bibinfo{author}{\bibfnamefont{P.}~\bibnamefont{Sprangle}},
  \bibinfo{author}{\bibfnamefont{B.}~\bibnamefont{Hafizi}},
  \bibinfo{author}{\bibfnamefont{W.}~\bibnamefont{Manheimer}}, and
  \bibinfo{author}{\bibfnamefont{A.}~\bibnamefont{Zigler}},
  \bibinfo{year}{2005}, \bibinfo{journal}{Phys.\ Rev.\ E}
  \textbf{\bibinfo{volume}{72}}, \bibinfo{pages}{036412}.

\bibitem[{\citenamefont{Pe{\~n}ano}
  \emph{et~al.}(2004)\citenamefont{Pe{\~n}ano, Sprangle, Hafizi, Ting, Gordon,
  and Kapetanakos}}]{Penano:pop:11:2865}
\bibinfo{author}{\bibnamefont{Pe{\~n}ano}, \bibfnamefont{J.~R.}},
  \bibinfo{author}{\bibfnamefont{P.}~\bibnamefont{Sprangle}},
  \bibinfo{author}{\bibfnamefont{B.}~\bibnamefont{Hafizi}},
  \bibinfo{author}{\bibfnamefont{A.}~\bibnamefont{Ting}},
  \bibinfo{author}{\bibfnamefont{D.~F.} \bibnamefont{Gordon}}, and
  \bibinfo{author}{\bibfnamefont{C.~A.} \bibnamefont{Kapetanakos}},
  \bibinfo{year}{2004}, \bibinfo{journal}{Phys.\ Plasmas}
  \textbf{\bibinfo{volume}{11}}, \bibinfo{pages}{2865}.

\bibitem[{\citenamefont{Pe{\~n}ano}
  \emph{et~al.}(2003)\citenamefont{Pe{\~n}ano, Sprangle, Serafim, Hafizi, and
  Ting}}]{Penano:pre:68:056502}
\bibinfo{author}{\bibnamefont{Pe{\~n}ano}, \bibfnamefont{J.~R.}},
  \bibinfo{author}{\bibfnamefont{P.}~\bibnamefont{Sprangle}},
  \bibinfo{author}{\bibfnamefont{P.}~\bibnamefont{Serafim}},
  \bibinfo{author}{\bibfnamefont{B.}~\bibnamefont{Hafizi}}, and
  \bibinfo{author}{\bibfnamefont{A.}~\bibnamefont{Ting}}, \bibinfo{year}{2003},
  \bibinfo{journal}{Phys.\ Rev.\ E} \textbf{\bibinfo{volume}{68}},
  \bibinfo{pages}{056502}.

\bibitem[{\citenamefont{Penetrante}
  \emph{et~al.}(1992)\citenamefont{Penetrante, Bardsley, Wood, Siders, and
  Downer}}]{Penetrante:josab:9:2032}
\bibinfo{author}{\bibnamefont{Penetrante}, \bibfnamefont{B.~M.}},
  \bibinfo{author}{\bibfnamefont{J.~N.} \bibnamefont{Bardsley}},
  \bibinfo{author}{\bibfnamefont{W.~M.} \bibnamefont{Wood}},
  \bibinfo{author}{\bibfnamefont{C.~W.} \bibnamefont{Siders}}, and
  \bibinfo{author}{\bibfnamefont{M.~C.} \bibnamefont{Downer}},
  \bibinfo{year}{1992}, \bibinfo{journal}{J.\ Opt.\ Soc.\ Am.\ B}
  \textbf{\bibinfo{volume}{9}}, \bibinfo{pages}{2032}.

\bibitem[{\citenamefont{P{\'e}pin} \emph{et~al.}(2001)\citenamefont{P{\'e}pin,
  Comtois, Vidal, Chien, Desparois, Johnston, Kieffer, LaFontaine, Martin,
  Rizk, Potvin, Couture} \emph{et~al.}}]{Pepin:pop:8:2532}
\bibinfo{author}{\bibnamefont{P{\'e}pin}, \bibfnamefont{H.}},
  \bibinfo{author}{\bibfnamefont{D.}~\bibnamefont{Comtois}},
  \bibinfo{author}{\bibfnamefont{F.}~\bibnamefont{Vidal}},
  \bibinfo{author}{\bibfnamefont{C.~Y.} \bibnamefont{Chien}},
  \bibinfo{author}{\bibfnamefont{A.}~\bibnamefont{Desparois}},
  \bibinfo{author}{\bibfnamefont{T.~W.} \bibnamefont{Johnston}},
  \bibinfo{author}{\bibfnamefont{J.-C.} \bibnamefont{Kieffer}},
  \bibinfo{author}{\bibfnamefont{B.}~\bibnamefont{LaFontaine}},
  \bibinfo{author}{\bibfnamefont{F.}~\bibnamefont{Martin}},
  \bibinfo{author}{\bibfnamefont{F.~A.~M.} \bibnamefont{Rizk}},
  \bibinfo{author}{\bibfnamefont{C.}~\bibnamefont{Potvin}},
  \bibinfo{author}{\bibfnamefont{P.}~\bibnamefont{Couture}}, \emph{et~al.},
  \bibinfo{year}{2001}, \bibinfo{journal}{Phys.\ Plasmas}
  \textbf{\bibinfo{volume}{8}}, \bibinfo{pages}{2532}.

\bibitem[{\citenamefont{Perelomov and Popov}(1967)}]{Perelomov:spjetp:25:336}
\bibinfo{author}{\bibnamefont{Perelomov}, \bibfnamefont{A.~M.}}, and
  \bibinfo{author}{\bibfnamefont{V.~S.} \bibnamefont{Popov}},
  \bibinfo{year}{1967}, \bibinfo{journal}{Sov.\ Phys.\ JETP}
  \textbf{\bibinfo{volume}{25}}, \bibinfo{pages}{336}.

\bibitem[{\citenamefont{Perelomov} \emph{et~al.}(1966)\citenamefont{Perelomov,
  Popov, and Terent'ev}}]{Perelomov:spjetp:23:924}
\bibinfo{author}{\bibnamefont{Perelomov}, \bibfnamefont{A.~M.}},
  \bibinfo{author}{\bibfnamefont{V.~S.} \bibnamefont{Popov}}, and
  \bibinfo{author}{\bibfnamefont{M.~V.} \bibnamefont{Terent'ev}},
  \bibinfo{year}{1966}, \bibinfo{journal}{Sov.\ Phys.\ JETP}
  \textbf{\bibinfo{volume}{23}}, \bibinfo{pages}{924}.

\bibitem[{\citenamefont{Perelomov} \emph{et~al.}(1967)\citenamefont{Perelomov,
  Popov, and Terent'ev}}]{Perelomov:spjetp:24:207}
\bibinfo{author}{\bibnamefont{Perelomov}, \bibfnamefont{A.~M.}},
  \bibinfo{author}{\bibfnamefont{V.~S.} \bibnamefont{Popov}}, and
  \bibinfo{author}{\bibfnamefont{M.~V.} \bibnamefont{Terent'ev}},
  \bibinfo{year}{1967}, \bibinfo{journal}{Sov.\ Phys.\ JETP}
  \textbf{\bibinfo{volume}{24}}, \bibinfo{pages}{207}.

\bibitem[{\citenamefont{Perry}
  \emph{et~al.}(1988{\natexlab{a}})\citenamefont{Perry, Landen, Sz{\"o}ke, and
  Campbell}}]{Perry:pra:37:747}
\bibinfo{author}{\bibnamefont{Perry}, \bibfnamefont{M.~D.}},
  \bibinfo{author}{\bibfnamefont{O.~L.} \bibnamefont{Landen}},
  \bibinfo{author}{\bibfnamefont{A.}~\bibnamefont{Sz{\"o}ke}}, and
  \bibinfo{author}{\bibfnamefont{E.~M.} \bibnamefont{Campbell}},
  \bibinfo{year}{1988}{\natexlab{a}}, \bibinfo{journal}{Phys.\ Rev.\ A}
  \textbf{\bibinfo{volume}{37}}, \bibinfo{pages}{747}.

\bibitem[{\citenamefont{Perry}
  \emph{et~al.}(1988{\natexlab{b}})\citenamefont{Perry, Szoke, Landen, and
  Campbell}}]{Perry:prl:60:1270}
\bibinfo{author}{\bibnamefont{Perry}, \bibfnamefont{M.~D.}},
  \bibinfo{author}{\bibfnamefont{A.}~\bibnamefont{Szoke}},
  \bibinfo{author}{\bibfnamefont{O.~L.} \bibnamefont{Landen}}, and
  \bibinfo{author}{\bibfnamefont{E.~M.} \bibnamefont{Campbell}},
  \bibinfo{year}{1988}{\natexlab{b}}, \bibinfo{journal}{Phys.\ Rev.\ Lett.}
  \textbf{\bibinfo{volume}{60}}, \bibinfo{pages}{1270}.

\bibitem[{\citenamefont{Petit} \emph{et~al.}(2000)\citenamefont{Petit,
  Talebpour, Proulx, and Chin}}]{Petit:oc:175:323}
\bibinfo{author}{\bibnamefont{Petit}, \bibfnamefont{S.}},
  \bibinfo{author}{\bibfnamefont{A.}~\bibnamefont{Talebpour}},
  \bibinfo{author}{\bibfnamefont{A.}~\bibnamefont{Proulx}}, and
  \bibinfo{author}{\bibfnamefont{S.~L.} \bibnamefont{Chin}},
  \bibinfo{year}{2000}, \bibinfo{journal}{Opt.\ Commun.}
  \textbf{\bibinfo{volume}{175}}, \bibinfo{pages}{323}.

\bibitem[{\citenamefont{Petrov} \emph{et~al.}(1998)\citenamefont{Petrov,
  Torner, Martorell, Vilaseca, Torres, and Cojocaru}}]{Petrov:ol:23:1444}
\bibinfo{author}{\bibnamefont{Petrov}, \bibfnamefont{D.~V.}},
  \bibinfo{author}{\bibfnamefont{L.}~\bibnamefont{Torner}},
  \bibinfo{author}{\bibfnamefont{J.}~\bibnamefont{Martorell}},
  \bibinfo{author}{\bibfnamefont{R.}~\bibnamefont{Vilaseca}},
  \bibinfo{author}{\bibfnamefont{J.~P.} \bibnamefont{Torres}}, and
  \bibinfo{author}{\bibfnamefont{C.}~\bibnamefont{Cojocaru}},
  \bibinfo{year}{1998}, \bibinfo{journal}{Opt.\ Lett.}
  \textbf{\bibinfo{volume}{23}}, \bibinfo{pages}{1444}.

\bibitem[{\citenamefont{Pfeifer} \emph{et~al.}(2006)\citenamefont{Pfeifer,
  Spielmann, and Gerber}}]{Pfeifer:rpp:69:443}
\bibinfo{author}{\bibnamefont{Pfeifer}, \bibfnamefont{T.}},
  \bibinfo{author}{\bibfnamefont{C.}~\bibnamefont{Spielmann}}, and
  \bibinfo{author}{\bibfnamefont{G.}~\bibnamefont{Gerber}},
  \bibinfo{year}{2006}, \bibinfo{journal}{Rep.\ Prog.\ Phys.}
  \textbf{\bibinfo{volume}{69}}, \bibinfo{pages}{443}.

\bibitem[{\citenamefont{Pietsch} \emph{et~al.}(1991)\citenamefont{Pietsch,
  Blaha, Laedke, and Kumar}}]{Pietsch:epl:15:173}
\bibinfo{author}{\bibnamefont{Pietsch}, \bibfnamefont{H.}},
  \bibinfo{author}{\bibfnamefont{R.}~\bibnamefont{Blaha}},
  \bibinfo{author}{\bibfnamefont{E.~W.} \bibnamefont{Laedke}}, and
  \bibinfo{author}{\bibfnamefont{A.}~\bibnamefont{Kumar}},
  \bibinfo{year}{1991}, \bibinfo{journal}{Europhys.\ Lett.}
  \textbf{\bibinfo{volume}{15}}, \bibinfo{pages}{173}.

\bibitem[{\citenamefont{Porras} \emph{et~al.}(2004)\citenamefont{Porras,
  Parola, Faccio, Dubietis, and {Di Trapani}}}]{Porras:prl:93:153902}
\bibinfo{author}{\bibnamefont{Porras}, \bibfnamefont{M.~A.}},
  \bibinfo{author}{\bibfnamefont{A.}~\bibnamefont{Parola}},
  \bibinfo{author}{\bibfnamefont{D.}~\bibnamefont{Faccio}},
  \bibinfo{author}{\bibfnamefont{A.}~\bibnamefont{Dubietis}}, and
  \bibinfo{author}{\bibfnamefont{P.}~\bibnamefont{{Di Trapani}}},
  \bibinfo{year}{2004}, \bibinfo{journal}{Phys.\ Rev.\ Lett.}
  \textbf{\bibinfo{volume}{93}}, \bibinfo{pages}{153902}.

\bibitem[{\citenamefont{Quigora-Teixeiro and
  Michinel}(1997)}]{Quigora-Teixeiro:josab:14:2004}
\bibinfo{author}{\bibnamefont{Quigora-Teixeiro}, \bibfnamefont{M.}}, and
  \bibinfo{author}{\bibfnamefont{H.}~\bibnamefont{Michinel}},
  \bibinfo{year}{1997}, \bibinfo{journal}{J.\ Opt.\ Soc.\ Am.\ B}
  \textbf{\bibinfo{volume}{14}}, \bibinfo{pages}{2004}.

\bibitem[{\citenamefont{Rae and Burnett}(1992)}]{Rae:pra:46:1084}
\bibinfo{author}{\bibnamefont{Rae}, \bibfnamefont{S.~C.}}, and
  \bibinfo{author}{\bibfnamefont{K.}~\bibnamefont{Burnett}},
  \bibinfo{year}{1992}, \bibinfo{journal}{Phys.\ Rev.\ A}
  \textbf{\bibinfo{volume}{46}}, \bibinfo{pages}{1084}.

\bibitem[{\citenamefont{Rairoux} \emph{et~al.}(2000)\citenamefont{Rairoux,
  Schillinger, Niedermeier, Rodriguez, Ronneberger, Sauerbrey, Stein, Waite,
  Wedekind, Wille, W{\"o}ste, and Ziener}}]{Rairoux:apb:71:573}
\bibinfo{author}{\bibnamefont{Rairoux}, \bibfnamefont{P.}},
  \bibinfo{author}{\bibfnamefont{H.}~\bibnamefont{Schillinger}},
  \bibinfo{author}{\bibfnamefont{S.}~\bibnamefont{Niedermeier}},
  \bibinfo{author}{\bibfnamefont{M.}~\bibnamefont{Rodriguez}},
  \bibinfo{author}{\bibfnamefont{F.}~\bibnamefont{Ronneberger}},
  \bibinfo{author}{\bibfnamefont{R.}~\bibnamefont{Sauerbrey}},
  \bibinfo{author}{\bibfnamefont{B.}~\bibnamefont{Stein}},
  \bibinfo{author}{\bibfnamefont{D.}~\bibnamefont{Waite}},
  \bibinfo{author}{\bibfnamefont{C.}~\bibnamefont{Wedekind}},
  \bibinfo{author}{\bibfnamefont{H.}~\bibnamefont{Wille}},
  \bibinfo{author}{\bibfnamefont{L.}~\bibnamefont{W{\"o}ste}}, and
  \bibinfo{author}{\bibfnamefont{C.}~\bibnamefont{Ziener}},
  \bibinfo{year}{2000}, \bibinfo{journal}{Appl.\ Phys.\ B: Lasers \& Optics}
  \textbf{\bibinfo{volume}{71}}, \bibinfo{pages}{573}.

\bibitem[{\citenamefont{Rambo} \emph{et~al.}(2001)\citenamefont{Rambo,
  Schwartz, and Diels}}]{Rambo:joapao:3:146}
\bibinfo{author}{\bibnamefont{Rambo}, \bibfnamefont{P.}},
  \bibinfo{author}{\bibfnamefont{J.}~\bibnamefont{Schwartz}}, and
  \bibinfo{author}{\bibfnamefont{J.-C.} \bibnamefont{Diels}},
  \bibinfo{year}{2001}, \bibinfo{journal}{J.\ Opt.\ A: Pure Appl.\ Opt.}
  \textbf{\bibinfo{volume}{3}}, \bibinfo{pages}{146}.

\bibitem[{\citenamefont{Ranka and Gaeta}(1998)}]{Ranka:ol:23:534}
\bibinfo{author}{\bibnamefont{Ranka}, \bibfnamefont{J.~K.}}, and
  \bibinfo{author}{\bibfnamefont{A.~L.} \bibnamefont{Gaeta}},
  \bibinfo{year}{1998}, \bibinfo{journal}{Opt.\ Lett.}
  \textbf{\bibinfo{volume}{23}}, \bibinfo{pages}{534}.

\bibitem[{\citenamefont{Rasmussen and Rypdal}(1986)}]{Rasmussen:ps:33:481}
\bibinfo{author}{\bibnamefont{Rasmussen}, \bibfnamefont{J.~J.}}, and
  \bibinfo{author}{\bibfnamefont{K.}~\bibnamefont{Rypdal}},
  \bibinfo{year}{1986}, \bibinfo{journal}{Phys.\ Scr.}
  \textbf{\bibinfo{volume}{33}}, \bibinfo{pages}{481}.

\bibitem[{\citenamefont{Rayner} \emph{et~al.}(2005)\citenamefont{Rayner,
  Naumov, and Corkum}}]{Rayner:oe:13:3208}
\bibinfo{author}{\bibnamefont{Rayner}, \bibfnamefont{D.~M.}},
  \bibinfo{author}{\bibfnamefont{A.}~\bibnamefont{Naumov}}, and
  \bibinfo{author}{\bibfnamefont{P.~B.} \bibnamefont{Corkum}},
  \bibinfo{year}{2005}, \bibinfo{journal}{Opt.\ Express}
  \textbf{\bibinfo{volume}{13}}, \bibinfo{pages}{3208}.

\bibitem[{\citenamefont{Reiss}(1980)}]{Reiss:pra:22:1786}
\bibinfo{author}{\bibnamefont{Reiss}, \bibfnamefont{H.~R.}},
  \bibinfo{year}{1980}, \bibinfo{journal}{Phys.\ Rev.\ A}
  \textbf{\bibinfo{volume}{22}}, \bibinfo{pages}{1786}.

\bibitem[{\citenamefont{Ripoche} \emph{et~al.}(1997)\citenamefont{Ripoche,
  Grillon, Prade, Franco, Nibbering, Lange, and
  Mysyrowicz}}]{Ripoche:oc:135:310}
\bibinfo{author}{\bibnamefont{Ripoche}, \bibfnamefont{J.~F.}},
  \bibinfo{author}{\bibfnamefont{G.}~\bibnamefont{Grillon}},
  \bibinfo{author}{\bibfnamefont{B.}~\bibnamefont{Prade}},
  \bibinfo{author}{\bibfnamefont{M.}~\bibnamefont{Franco}},
  \bibinfo{author}{\bibfnamefont{E.}~\bibnamefont{Nibbering}},
  \bibinfo{author}{\bibfnamefont{R.}~\bibnamefont{Lange}}, and
  \bibinfo{author}{\bibfnamefont{A.}~\bibnamefont{Mysyrowicz}},
  \bibinfo{year}{1997}, \bibinfo{journal}{Opt.\ Commun.}
  \textbf{\bibinfo{volume}{135}}, \bibinfo{pages}{310}.

\bibitem[{\citenamefont{Rodriguez} \emph{et~al.}(2004)\citenamefont{Rodriguez,
  Bourayou, M{\'e}jean, Kasparian, Yu, Salmon, Scholz, Stecklum, Eisl{\"o}ffel,
  Laux, Hatzes, Sauerbrey} \emph{et~al.}}]{Rodriguez:pre:69:036607}
\bibinfo{author}{\bibnamefont{Rodriguez}, \bibfnamefont{M.}},
  \bibinfo{author}{\bibfnamefont{R.}~\bibnamefont{Bourayou}},
  \bibinfo{author}{\bibfnamefont{G.}~\bibnamefont{M{\'e}jean}},
  \bibinfo{author}{\bibfnamefont{J.}~\bibnamefont{Kasparian}},
  \bibinfo{author}{\bibfnamefont{J.}~\bibnamefont{Yu}},
  \bibinfo{author}{\bibfnamefont{E.}~\bibnamefont{Salmon}},
  \bibinfo{author}{\bibfnamefont{A.}~\bibnamefont{Scholz}},
  \bibinfo{author}{\bibfnamefont{B.}~\bibnamefont{Stecklum}},
  \bibinfo{author}{\bibfnamefont{J.}~\bibnamefont{Eisl{\"o}ffel}},
  \bibinfo{author}{\bibfnamefont{U.}~\bibnamefont{Laux}},
  \bibinfo{author}{\bibfnamefont{A.~P.} \bibnamefont{Hatzes}},
  \bibinfo{author}{\bibfnamefont{R.}~\bibnamefont{Sauerbrey}}, \emph{et~al.},
  \bibinfo{year}{2004}, \bibinfo{journal}{Phys.\ Rev.\ E}
  \textbf{\bibinfo{volume}{69}}, \bibinfo{pages}{036607}.

\bibitem[{\citenamefont{Rodriguez} \emph{et~al.}(2002)\citenamefont{Rodriguez,
  Sauerbrey, Wille, W{\"o}ste, Fujii, Andr{\'e}, Mysyrowicz, Klingbeil,
  Rethmeier, Kalkner, kasparian, Salmon} \emph{et~al.}}]{Rodriguez:ol:27:772}
\bibinfo{author}{\bibnamefont{Rodriguez}, \bibfnamefont{M.}},
  \bibinfo{author}{\bibfnamefont{R.}~\bibnamefont{Sauerbrey}},
  \bibinfo{author}{\bibfnamefont{H.}~\bibnamefont{Wille}},
  \bibinfo{author}{\bibfnamefont{L.}~\bibnamefont{W{\"o}ste}},
  \bibinfo{author}{\bibfnamefont{T.}~\bibnamefont{Fujii}},
  \bibinfo{author}{\bibfnamefont{Y.-B.} \bibnamefont{Andr{\'e}}},
  \bibinfo{author}{\bibfnamefont{A.}~\bibnamefont{Mysyrowicz}},
  \bibinfo{author}{\bibfnamefont{L.}~\bibnamefont{Klingbeil}},
  \bibinfo{author}{\bibfnamefont{K.}~\bibnamefont{Rethmeier}},
  \bibinfo{author}{\bibfnamefont{W.}~\bibnamefont{Kalkner}},
  \bibinfo{author}{\bibfnamefont{J.}~\bibnamefont{kasparian}},
  \bibinfo{author}{\bibfnamefont{E.}~\bibnamefont{Salmon}}, \emph{et~al.},
  \bibinfo{year}{2002}, \bibinfo{journal}{Opt.\ Lett.}
  \textbf{\bibinfo{volume}{27}}, \bibinfo{pages}{772}.

\bibitem[{\citenamefont{Rohwetter} \emph{et~al.}(2003)\citenamefont{Rohwetter,
  Stelmaszczyk, M{\'e}jean, Yu, Salmon, Kasparian, Wolf, and
  W{\"o}ste}}]{Rohwetter:jaas:19:437}
\bibinfo{author}{\bibnamefont{Rohwetter}, \bibfnamefont{P.}},
  \bibinfo{author}{\bibfnamefont{K.}~\bibnamefont{Stelmaszczyk}},
  \bibinfo{author}{\bibfnamefont{G.}~\bibnamefont{M{\'e}jean}},
  \bibinfo{author}{\bibfnamefont{J.}~\bibnamefont{Yu}},
  \bibinfo{author}{\bibfnamefont{E.}~\bibnamefont{Salmon}},
  \bibinfo{author}{\bibfnamefont{J.}~\bibnamefont{Kasparian}},
  \bibinfo{author}{\bibfnamefont{J.-P.} \bibnamefont{Wolf}}, and
  \bibinfo{author}{\bibfnamefont{L.}~\bibnamefont{W{\"o}ste}},
  \bibinfo{year}{2003}, \bibinfo{journal}{J.\ Anal. \ Atom. \ Spectroscopy}
  \textbf{\bibinfo{volume}{19}}, \bibinfo{pages}{437}.

\bibitem[{\citenamefont{Rothenberg}(1992)}]{Rothenberg:ol:17:1340}
\bibinfo{author}{\bibnamefont{Rothenberg}, \bibfnamefont{J.~E.}},
  \bibinfo{year}{1992}, \bibinfo{journal}{Opt.\ Lett.}
  \textbf{\bibinfo{volume}{17}}, \bibinfo{pages}{1340}.

\bibitem[{\citenamefont{Rypdal and Rasmussen}(1989)}]{Rypdal:ps:40:192}
\bibinfo{author}{\bibnamefont{Rypdal}, \bibfnamefont{K.}}, and
  \bibinfo{author}{\bibfnamefont{J.~J.} \bibnamefont{Rasmussen}},
  \bibinfo{year}{1989}, \bibinfo{journal}{Phys.\ Scr.}
  \textbf{\bibinfo{volume}{40}}, \bibinfo{pages}{192}.

\bibitem[{\citenamefont{Rypdal} \emph{et~al.}(1985)\citenamefont{Rypdal,
  Rasmussen, and Thomsen}}]{Rypdal:pd:16:339}
\bibinfo{author}{\bibnamefont{Rypdal}, \bibfnamefont{K.}},
  \bibinfo{author}{\bibfnamefont{J.~J.} \bibnamefont{Rasmussen}}, and
  \bibinfo{author}{\bibfnamefont{K.}~\bibnamefont{Thomsen}},
  \bibinfo{year}{1985}, \bibinfo{journal}{Physica D}
  \textbf{\bibinfo{volume}{16}}, \bibinfo{pages}{339}.

\bibitem[{\citenamefont{Sali\`eres and
  Lewenstein}(2001)}]{Salieres:mst:12:1818}
\bibinfo{author}{\bibnamefont{Sali\`eres}, \bibfnamefont{P.}}, and
  \bibinfo{author}{\bibfnamefont{M.}~\bibnamefont{Lewenstein}},
  \bibinfo{year}{2001}, \bibinfo{journal}{Meas.\ Sci.\ Technol.}
  \textbf{\bibinfo{volume}{12}}, \bibinfo{pages}{1818}.

\bibitem[{\citenamefont{Schj{\o}dt-Eriksen}
  \emph{et~al.}(2001{\natexlab{a}})\citenamefont{Schj{\o}dt-Eriksen, Gaididei,
  and Christiansen}}]{Schjoedteriksen:pre:64:066614}
\bibinfo{author}{\bibnamefont{Schj{\o}dt-Eriksen}, \bibfnamefont{J.}},
  \bibinfo{author}{\bibfnamefont{Y.~B.} \bibnamefont{Gaididei}}, and
  \bibinfo{author}{\bibfnamefont{P.~L.} \bibnamefont{Christiansen}},
  \bibinfo{year}{2001}{\natexlab{a}}, \bibinfo{journal}{Phys.\ Rev.\ E}
  \textbf{\bibinfo{volume}{64}}, \bibinfo{pages}{066614}.

\bibitem[{\citenamefont{Schj{\o}dt-Eriksen}
  \emph{et~al.}(2001{\natexlab{b}})\citenamefont{Schj{\o}dt-Eriksen, Moloney,
  Wright, Feng, and Christiansen}}]{Schjoedteriksen:ol:26:78}
\bibinfo{author}{\bibnamefont{Schj{\o}dt-Eriksen}, \bibfnamefont{J.}},
  \bibinfo{author}{\bibfnamefont{J.~V.} \bibnamefont{Moloney}},
  \bibinfo{author}{\bibfnamefont{E.~M.} \bibnamefont{Wright}},
  \bibinfo{author}{\bibfnamefont{Q.}~\bibnamefont{Feng}}, and
  \bibinfo{author}{\bibfnamefont{P.~L.} \bibnamefont{Christiansen}},
  \bibinfo{year}{2001}{\natexlab{b}}, \bibinfo{journal}{Opt.\ Lett.}
  \textbf{\bibinfo{volume}{26}}, \bibinfo{pages}{78}.

\bibitem[{\citenamefont{Schroeder and Chin}(2004)}]{Schroeder:oc:234:399}
\bibinfo{author}{\bibnamefont{Schroeder}, \bibfnamefont{H.}}, and
  \bibinfo{author}{\bibfnamefont{S.~L.} \bibnamefont{Chin}},
  \bibinfo{year}{2004}, \bibinfo{journal}{Opt.\ Commun.}
  \textbf{\bibinfo{volume}{234}}, \bibinfo{pages}{399}.

\bibitem[{\citenamefont{Schroeder} \emph{et~al.}(2004)\citenamefont{Schroeder,
  Liu, and Chin}}]{Schroeder:oe:12:4768}
\bibinfo{author}{\bibnamefont{Schroeder}, \bibfnamefont{H.}},
  \bibinfo{author}{\bibfnamefont{J.}~\bibnamefont{Liu}}, and
  \bibinfo{author}{\bibfnamefont{S.~L.} \bibnamefont{Chin}},
  \bibinfo{year}{2004}, \bibinfo{journal}{Opt.\ Express}
  \textbf{\bibinfo{volume}{12}}, \bibinfo{pages}{4768}.

\bibitem[{\citenamefont{Schwarz} \emph{et~al.}(2001)\citenamefont{Schwarz,
  Rambo, and Diels}}]{Schwarz:apb:72:343}
\bibinfo{author}{\bibnamefont{Schwarz}, \bibfnamefont{J.}},
  \bibinfo{author}{\bibfnamefont{P.}~\bibnamefont{Rambo}}, and
  \bibinfo{author}{\bibfnamefont{J.-C.} \bibnamefont{Diels}},
  \bibinfo{year}{2001}, \bibinfo{journal}{Appl.\ Phys.\ B: Lasers \& Optics}
  \textbf{\bibinfo{volume}{72}}, \bibinfo{pages}{343}.

\bibitem[{\citenamefont{Schwarz} \emph{et~al.}(2000)\citenamefont{Schwarz,
  Rambo, Diels, Kolesik, Wright, and Moloney}}]{Schwarz:oc:180:383}
\bibinfo{author}{\bibnamefont{Schwarz}, \bibfnamefont{J.}},
  \bibinfo{author}{\bibfnamefont{P.}~\bibnamefont{Rambo}},
  \bibinfo{author}{\bibfnamefont{J.-C.} \bibnamefont{Diels}},
  \bibinfo{author}{\bibfnamefont{M.}~\bibnamefont{Kolesik}},
  \bibinfo{author}{\bibfnamefont{E.~M.} \bibnamefont{Wright}}, and
  \bibinfo{author}{\bibfnamefont{J.~V.} \bibnamefont{Moloney}},
  \bibinfo{year}{2000}, \bibinfo{journal}{Opt.\ Commun.}
  \textbf{\bibinfo{volume}{180}}, \bibinfo{pages}{383}.

\bibitem[{\citenamefont{Scrinzi} \emph{et~al.}(2006)\citenamefont{Scrinzi,
  Ivanov, Kienberger, and Villeneuve}}]{Scrinzi:jpb:39:1}
\bibinfo{author}{\bibnamefont{Scrinzi}, \bibfnamefont{A.}},
  \bibinfo{author}{\bibfnamefont{M.~Y.} \bibnamefont{Ivanov}},
  \bibinfo{author}{\bibfnamefont{R.}~\bibnamefont{Kienberger}}, and
  \bibinfo{author}{\bibfnamefont{D.~M.} \bibnamefont{Villeneuve}},
  \bibinfo{year}{2006}, \bibinfo{journal}{J.\ Phys.\ B: At.\ Mol.\ Opt.\ Phys.}
  \textbf{\bibinfo{volume}{39}}, \bibinfo{pages}{1}.

\bibitem[{\citenamefont{Sharma} \emph{et~al.}(2003)\citenamefont{Sharma, Lucey,
  Ghosh, Hubble, and Horton}}]{Sharma:saa:59:2391}
\bibinfo{author}{\bibnamefont{Sharma}, \bibfnamefont{S.~K.}},
  \bibinfo{author}{\bibfnamefont{P.~G.} \bibnamefont{Lucey}},
  \bibinfo{author}{\bibfnamefont{M.}~\bibnamefont{Ghosh}},
  \bibinfo{author}{\bibfnamefont{H.~W.} \bibnamefont{Hubble}}, and
  \bibinfo{author}{\bibfnamefont{K.~A.} \bibnamefont{Horton}},
  \bibinfo{year}{2003}, \bibinfo{journal}{Spectroch.\ Acta A}
  \textbf{\bibinfo{volume}{59}}, \bibinfo{pages}{2391}.

\bibitem[{\citenamefont{Shen}(1976)}]{Shen:rmp:1:48}
\bibinfo{author}{\bibnamefont{Shen}, \bibfnamefont{Y.~R.}},
  \bibinfo{year}{1976}, \bibinfo{journal}{Rev.\ Mod.\ Phys.}
  \textbf{\bibinfo{volume}{1}}, \bibinfo{pages}{48}.

\bibitem[{\citenamefont{Shen}(1984)}]{Shen:PNO:84}
\bibinfo{author}{\bibnamefont{Shen}, \bibfnamefont{Y.~R.}},
  \bibinfo{year}{1984}, \emph{\bibinfo{title}{The Principles of Nonlinear
  Optics}} (\bibinfo{publisher}{John Wiley \& Sons},
  \bibinfo{address}{New-York}).

\bibitem[{\citenamefont{Silberberg}(1990)}]{Silberberg:ol:15:1282}
\bibinfo{author}{\bibnamefont{Silberberg}, \bibfnamefont{Y.}},
  \bibinfo{year}{1990}, \bibinfo{journal}{Opt.\ Lett.}
  \textbf{\bibinfo{volume}{15}}, \bibinfo{pages}{1282}.

\bibitem[{\citenamefont{Skarka} \emph{et~al.}(2003)\citenamefont{Skarka,
  Aleksi{\'c}, and Berezhiani}}]{Skarka:pla:319:317}
\bibinfo{author}{\bibnamefont{Skarka}, \bibfnamefont{V.}},
  \bibinfo{author}{\bibfnamefont{N.~B.} \bibnamefont{Aleksi{\'c}}}, and
  \bibinfo{author}{\bibfnamefont{V.~I.} \bibnamefont{Berezhiani}},
  \bibinfo{year}{2003}, \bibinfo{journal}{Phys.\ Lett.\ A}
  \textbf{\bibinfo{volume}{319}}, \bibinfo{pages}{317}.

\bibitem[{\citenamefont{Skryabin and Firth}(1998)}]{Skryabin:pre:58:1252}
\bibinfo{author}{\bibnamefont{Skryabin}, \bibfnamefont{D.~V.}}, and
  \bibinfo{author}{\bibfnamefont{W.~J.} \bibnamefont{Firth}},
  \bibinfo{year}{1998}, \bibinfo{journal}{Phys.\ Rev.\ E}
  \textbf{\bibinfo{volume}{58}}, \bibinfo{pages}{R1252}.

\bibitem[{\citenamefont{Skupin and Berg{\'e}}(2006)}]{Skupin:pd:220:14}
\bibinfo{author}{\bibnamefont{Skupin}, \bibfnamefont{S.}}, and
  \bibinfo{author}{\bibfnamefont{L.}~\bibnamefont{Berg{\'e}}},
  \bibinfo{year}{2006}, \bibinfo{journal}{Physica D}
  \textbf{\bibinfo{volume}{220}}, \bibinfo{pages}{14}.

\bibitem[{\citenamefont{Skupin}
  \emph{et~al.}(2004{\natexlab{a}})\citenamefont{Skupin, Berg{\'e}, Peschel,
  and Lederer}}]{Skupin:prl:93:023901}
\bibinfo{author}{\bibnamefont{Skupin}, \bibfnamefont{S.}},
  \bibinfo{author}{\bibfnamefont{L.}~\bibnamefont{Berg{\'e}}},
  \bibinfo{author}{\bibfnamefont{U.}~\bibnamefont{Peschel}}, and
  \bibinfo{author}{\bibfnamefont{F.}~\bibnamefont{Lederer}},
  \bibinfo{year}{2004}{\natexlab{a}}, \bibinfo{journal}{Phys.\ Rev.\ Lett.}
  \textbf{\bibinfo{volume}{93}}, \bibinfo{pages}{023901}.

\bibitem[{\citenamefont{Skupin}
  \emph{et~al.}(2004{\natexlab{b}})\citenamefont{Skupin, Berg\'e, Peschel,
  Lederer, M\'ejean, Yu, Kasparian, Salmon, Wolf, Rodriguez, W{\"o}ste,
  Bourayou} \emph{et~al.}}]{Skupin:pre:70:046602}
\bibinfo{author}{\bibnamefont{Skupin}, \bibfnamefont{S.}},
  \bibinfo{author}{\bibfnamefont{L.}~\bibnamefont{Berg\'e}},
  \bibinfo{author}{\bibfnamefont{U.}~\bibnamefont{Peschel}},
  \bibinfo{author}{\bibfnamefont{F.}~\bibnamefont{Lederer}},
  \bibinfo{author}{\bibfnamefont{G.}~\bibnamefont{M\'ejean}},
  \bibinfo{author}{\bibfnamefont{J.}~\bibnamefont{Yu}},
  \bibinfo{author}{\bibfnamefont{J.}~\bibnamefont{Kasparian}},
  \bibinfo{author}{\bibfnamefont{E.}~\bibnamefont{Salmon}},
  \bibinfo{author}{\bibfnamefont{J.-P.} \bibnamefont{Wolf}},
  \bibinfo{author}{\bibfnamefont{M.}~\bibnamefont{Rodriguez}},
  \bibinfo{author}{\bibfnamefont{L.}~\bibnamefont{W{\"o}ste}},
  \bibinfo{author}{\bibfnamefont{R.}~\bibnamefont{Bourayou}}, \emph{et~al.},
  \bibinfo{year}{2004}{\natexlab{b}}, \bibinfo{journal}{Phys.\ Rev.\ E}
  \textbf{\bibinfo{volume}{70}}, \bibinfo{pages}{046602}.

\bibitem[{\citenamefont{Skupin}
  \emph{et~al.}(2006{\natexlab{a}})\citenamefont{Skupin, Nuter, and
  Berg{\'e}}}]{Skupin:pra:74:043813}
\bibinfo{author}{\bibnamefont{Skupin}, \bibfnamefont{S.}},
  \bibinfo{author}{\bibfnamefont{R.}~\bibnamefont{Nuter}}, and
  \bibinfo{author}{\bibfnamefont{L.}~\bibnamefont{Berg{\'e}}},
  \bibinfo{year}{2006}{\natexlab{a}}, \bibinfo{journal}{Phys.\ Rev.\ A}
  \textbf{\bibinfo{volume}{74}}, \bibinfo{pages}{043813}.

\bibitem[{\citenamefont{Skupin} \emph{et~al.}(2003)\citenamefont{Skupin,
  Peschel, Etrich, Leine, Lederer, and Michaelis}}]{Skupin:oqe:35:573}
\bibinfo{author}{\bibnamefont{Skupin}, \bibfnamefont{S.}},
  \bibinfo{author}{\bibfnamefont{U.}~\bibnamefont{Peschel}},
  \bibinfo{author}{\bibfnamefont{C.}~\bibnamefont{Etrich}},
  \bibinfo{author}{\bibfnamefont{L.}~\bibnamefont{Leine}},
  \bibinfo{author}{\bibfnamefont{F.}~\bibnamefont{Lederer}}, and
  \bibinfo{author}{\bibfnamefont{D.}~\bibnamefont{Michaelis}},
  \bibinfo{year}{2003}, \bibinfo{journal}{Opt.\ Quant.\ Electron.}
  \textbf{\bibinfo{volume}{35}}, \bibinfo{pages}{573}.

\bibitem[{\citenamefont{Skupin} \emph{et~al.}(2002)\citenamefont{Skupin,
  Peschel, Etrich, Leine, Michaelis, and Lederer}}]{Skupin:ol:27:1812}
\bibinfo{author}{\bibnamefont{Skupin}, \bibfnamefont{S.}},
  \bibinfo{author}{\bibfnamefont{U.}~\bibnamefont{Peschel}},
  \bibinfo{author}{\bibfnamefont{C.}~\bibnamefont{Etrich}},
  \bibinfo{author}{\bibfnamefont{L.}~\bibnamefont{Leine}},
  \bibinfo{author}{\bibfnamefont{D.}~\bibnamefont{Michaelis}}, and
  \bibinfo{author}{\bibfnamefont{F.}~\bibnamefont{Lederer}},
  \bibinfo{year}{2002}, \bibinfo{journal}{Opt.\ Lett.}
  \textbf{\bibinfo{volume}{27}}, \bibinfo{pages}{1812}.

\bibitem[{\citenamefont{Skupin}
  \emph{et~al.}(2006{\natexlab{b}})\citenamefont{Skupin, Stibenz, Berg{\'e},
  Lederer, Sokollik, Schn{\"u}rer, Zhavoronkov, and
  Steinmeyer}}]{Skupin:pre:74:056604}
\bibinfo{author}{\bibnamefont{Skupin}, \bibfnamefont{S.}},
  \bibinfo{author}{\bibfnamefont{G.}~\bibnamefont{Stibenz}},
  \bibinfo{author}{\bibfnamefont{L.}~\bibnamefont{Berg{\'e}}},
  \bibinfo{author}{\bibfnamefont{F.}~\bibnamefont{Lederer}},
  \bibinfo{author}{\bibfnamefont{T.}~\bibnamefont{Sokollik}},
  \bibinfo{author}{\bibfnamefont{M.}~\bibnamefont{Schn{\"u}rer}},
  \bibinfo{author}{\bibfnamefont{N.}~\bibnamefont{Zhavoronkov}}, and
  \bibinfo{author}{\bibfnamefont{G.}~\bibnamefont{Steinmeyer}},
  \bibinfo{year}{2006}{\natexlab{b}}, \bibinfo{journal}{Phys.\ Rev.\ E}
  \textbf{\bibinfo{volume}{74}}, \bibinfo{pages}{056604}.

\bibitem[{\citenamefont{Soto-Crespo}
  \emph{et~al.}(1992)\citenamefont{Soto-Crespo, Wright, and
  Akhmediev}}]{Sotocrespo:pra:45:3168}
\bibinfo{author}{\bibnamefont{Soto-Crespo}, \bibfnamefont{J.~M.}},
  \bibinfo{author}{\bibfnamefont{E.~M.} \bibnamefont{Wright}}, and
  \bibinfo{author}{\bibfnamefont{N.~N.} \bibnamefont{Akhmediev}},
  \bibinfo{year}{1992}, \bibinfo{journal}{Phys.\ Rev.\ A}
  \textbf{\bibinfo{volume}{45}}, \bibinfo{pages}{3168}.

\bibitem[{\citenamefont{Sprangle} \emph{et~al.}(1996)\citenamefont{Sprangle,
  Esarey, and Krall}}]{Sprangle:pre:54:4211}
\bibinfo{author}{\bibnamefont{Sprangle}, \bibfnamefont{P.}},
  \bibinfo{author}{\bibfnamefont{E.}~\bibnamefont{Esarey}}, and
  \bibinfo{author}{\bibfnamefont{J.}~\bibnamefont{Krall}},
  \bibinfo{year}{1996}, \bibinfo{journal}{Phys.\ Rev.\ E}
  \textbf{\bibinfo{volume}{54}}, \bibinfo{pages}{4211}.

\bibitem[{\citenamefont{Sprangle} \emph{et~al.}(2002)\citenamefont{Sprangle,
  Pe{\~n}ano, and Hafizi}}]{Sprangle:pre:66:046418}
\bibinfo{author}{\bibnamefont{Sprangle}, \bibfnamefont{P.}},
  \bibinfo{author}{\bibfnamefont{J.~R.} \bibnamefont{Pe{\~n}ano}}, and
  \bibinfo{author}{\bibfnamefont{B.}~\bibnamefont{Hafizi}},
  \bibinfo{year}{2002}, \bibinfo{journal}{Phys.\ Rev.\ E}
  \textbf{\bibinfo{volume}{66}}, \bibinfo{pages}{046418}.

\bibitem[{\citenamefont{Sprangle} \emph{et~al.}(2004)\citenamefont{Sprangle,
  Pe{\~n}ano, Hafizi, and Kapetanakos}}]{Sprangle:pre:69:066415}
\bibinfo{author}{\bibnamefont{Sprangle}, \bibfnamefont{P.}},
  \bibinfo{author}{\bibfnamefont{J.~R.} \bibnamefont{Pe{\~n}ano}},
  \bibinfo{author}{\bibfnamefont{B.}~\bibnamefont{Hafizi}}, and
  \bibinfo{author}{\bibfnamefont{C.~A.} \bibnamefont{Kapetanakos}},
  \bibinfo{year}{2004}, \bibinfo{journal}{Phys.\ Rev.\ E}
  \textbf{\bibinfo{volume}{69}}, \bibinfo{pages}{066415}.

\bibitem[{\citenamefont{Steinmeyer}
  \emph{et~al.}(1999)\citenamefont{Steinmeyer, Sutter, Gallmann, Matuschek, and
  Keller}}]{Steinmeyer:sc:286:1507}
\bibinfo{author}{\bibnamefont{Steinmeyer}, \bibfnamefont{G.}},
  \bibinfo{author}{\bibfnamefont{D.~H.} \bibnamefont{Sutter}},
  \bibinfo{author}{\bibfnamefont{L.}~\bibnamefont{Gallmann}},
  \bibinfo{author}{\bibfnamefont{N.}~\bibnamefont{Matuschek}}, and
  \bibinfo{author}{\bibfnamefont{U.}~\bibnamefont{Keller}},
  \bibinfo{year}{1999}, \bibinfo{journal}{Science}
  \textbf{\bibinfo{volume}{286}}, \bibinfo{pages}{1507}.

\bibitem[{\citenamefont{Stelmaszczyk}
  \emph{et~al.}(2004)\citenamefont{Stelmaszczyk, Rohwetter, M{\'e}jean, Yu,
  Salmon, Kasparian, Ackermann, Wolf, and
  W{\"o}ste}}]{Stelmaszczyk:apb:85:3977}
\bibinfo{author}{\bibnamefont{Stelmaszczyk}, \bibfnamefont{K.}},
  \bibinfo{author}{\bibfnamefont{P.}~\bibnamefont{Rohwetter}},
  \bibinfo{author}{\bibfnamefont{G.}~\bibnamefont{M{\'e}jean}},
  \bibinfo{author}{\bibfnamefont{J.}~\bibnamefont{Yu}},
  \bibinfo{author}{\bibfnamefont{E.}~\bibnamefont{Salmon}},
  \bibinfo{author}{\bibfnamefont{J.}~\bibnamefont{Kasparian}},
  \bibinfo{author}{\bibfnamefont{R.}~\bibnamefont{Ackermann}},
  \bibinfo{author}{\bibfnamefont{J.-P.} \bibnamefont{Wolf}}, and
  \bibinfo{author}{\bibfnamefont{L.}~\bibnamefont{W{\"o}ste}},
  \bibinfo{year}{2004}, \bibinfo{journal}{Appl.\ Phys.\ B: Lasers \& Optics}
  \textbf{\bibinfo{volume}{85}}, \bibinfo{pages}{3977}.

\bibitem[{\citenamefont{Stibenz} \emph{et~al.}(2006)\citenamefont{Stibenz,
  Zhavoronkov, and Steinmeyer}}]{Stibenz:ol:31:274}
\bibinfo{author}{\bibnamefont{Stibenz}, \bibfnamefont{G.}},
  \bibinfo{author}{\bibfnamefont{N.}~\bibnamefont{Zhavoronkov}}, and
  \bibinfo{author}{\bibfnamefont{G.}~\bibnamefont{Steinmeyer}},
  \bibinfo{year}{2006}, \bibinfo{journal}{Opt.\ Lett.}
  \textbf{\bibinfo{volume}{31}}, \bibinfo{pages}{274}.

\bibitem[{\citenamefont{Stuart} \emph{et~al.}(1996)\citenamefont{Stuart, Feit,
  Herman, Rubenchik, Shore, and Perry}}]{Stuart:prb:53:1749}
\bibinfo{author}{\bibnamefont{Stuart}, \bibfnamefont{B.~C.}},
  \bibinfo{author}{\bibfnamefont{M.~D.} \bibnamefont{Feit}},
  \bibinfo{author}{\bibfnamefont{S.}~\bibnamefont{Herman}},
  \bibinfo{author}{\bibfnamefont{A.~M.} \bibnamefont{Rubenchik}},
  \bibinfo{author}{\bibfnamefont{B.~W.} \bibnamefont{Shore}}, and
  \bibinfo{author}{\bibfnamefont{M.~D.} \bibnamefont{Perry}},
  \bibinfo{year}{1996}, \bibinfo{journal}{Phys.\ Rev.\ B}
  \textbf{\bibinfo{volume}{53}}, \bibinfo{pages}{1749}.

\bibitem[{\citenamefont{Suda} \emph{et~al.}(2005)\citenamefont{Suda, Hatayama,
  Nagasaka, and Midorikawa}}]{Suda:apl:86:11116}
\bibinfo{author}{\bibnamefont{Suda}, \bibfnamefont{A.}},
  \bibinfo{author}{\bibfnamefont{M.}~\bibnamefont{Hatayama}},
  \bibinfo{author}{\bibfnamefont{K.}~\bibnamefont{Nagasaka}}, and
  \bibinfo{author}{\bibfnamefont{K.}~\bibnamefont{Midorikawa}},
  \bibinfo{year}{2005}, \bibinfo{journal}{Appl.\ Phys.\ Lett.}
  \textbf{\bibinfo{volume}{86}}, \bibinfo{pages}{111116}.

\bibitem[{\citenamefont{Sudrie} \emph{et~al.}(2002)\citenamefont{Sudrie,
  Couairon, Franco, Lamouroux, Prade, Tzortzakis, and
  Mysyrowicz}}]{Sudrie:prl:89:186601}
\bibinfo{author}{\bibnamefont{Sudrie}, \bibfnamefont{L.}},
  \bibinfo{author}{\bibfnamefont{A.}~\bibnamefont{Couairon}},
  \bibinfo{author}{\bibfnamefont{M.}~\bibnamefont{Franco}},
  \bibinfo{author}{\bibfnamefont{B.}~\bibnamefont{Lamouroux}},
  \bibinfo{author}{\bibfnamefont{B.}~\bibnamefont{Prade}},
  \bibinfo{author}{\bibfnamefont{S.}~\bibnamefont{Tzortzakis}}, and
  \bibinfo{author}{\bibfnamefont{A.}~\bibnamefont{Mysyrowicz}},
  \bibinfo{year}{2002}, \bibinfo{journal}{Phys.\ Rev.\ Lett.}
  \textbf{\bibinfo{volume}{89}}, \bibinfo{pages}{186601}.

\bibitem[{\citenamefont{Sudrie} \emph{et~al.}(1999)\citenamefont{Sudrie,
  Franco, Prade, and Mysyrowicz}}]{Sudrie:oc:171:279}
\bibinfo{author}{\bibnamefont{Sudrie}, \bibfnamefont{L.}},
  \bibinfo{author}{\bibfnamefont{M.}~\bibnamefont{Franco}},
  \bibinfo{author}{\bibfnamefont{B.}~\bibnamefont{Prade}}, and
  \bibinfo{author}{\bibfnamefont{A.}~\bibnamefont{Mysyrowicz}},
  \bibinfo{year}{1999}, \bibinfo{journal}{Opt.\ Commun.}
  \textbf{\bibinfo{volume}{171}}, \bibinfo{pages}{279}.

\bibitem[{\citenamefont{Sudrie} \emph{et~al.}(2001)\citenamefont{Sudrie,
  Franco, Prade, and Mysyrowicz}}]{Sudrie:oc:191:333}
\bibinfo{author}{\bibnamefont{Sudrie}, \bibfnamefont{L.}},
  \bibinfo{author}{\bibfnamefont{M.}~\bibnamefont{Franco}},
  \bibinfo{author}{\bibfnamefont{B.}~\bibnamefont{Prade}}, and
  \bibinfo{author}{\bibfnamefont{A.}~\bibnamefont{Mysyrowicz}},
  \bibinfo{year}{2001}, \bibinfo{journal}{Opt.\ Commun.}
  \textbf{\bibinfo{volume}{191}}, \bibinfo{pages}{333}.

\bibitem[{\citenamefont{Sulem and Sulem}(1999)}]{Sulem:NLS:99}
\bibinfo{author}{\bibnamefont{Sulem}, \bibfnamefont{C.}}, and
  \bibinfo{author}{\bibfnamefont{P.-L.} \bibnamefont{Sulem}},
  \bibinfo{year}{1999}, \emph{\bibinfo{title}{The Nonlinear {Schr{\"o}dinger}
  Equation: Self-focusing and Wave collapse}}
  (\bibinfo{publisher}{Springer-Verlag}, \bibinfo{address}{New York}),
  \bibinfo{edition}{first} edition.

\bibitem[{\citenamefont{Talebpour} \emph{et~al.}(1999)\citenamefont{Talebpour,
  Yang, and Chin}}]{Talebpour:oc:163:29}
\bibinfo{author}{\bibnamefont{Talebpour}, \bibfnamefont{A.}},
  \bibinfo{author}{\bibfnamefont{J.}~\bibnamefont{Yang}}, and
  \bibinfo{author}{\bibfnamefont{S.~L.} \bibnamefont{Chin}},
  \bibinfo{year}{1999}, \bibinfo{journal}{Opt.\ Commun.}
  \textbf{\bibinfo{volume}{163}}, \bibinfo{pages}{29}.

\bibitem[{\citenamefont{Tamaki} \emph{et~al.}(1999)\citenamefont{Tamaki,
  Itatani, Nagata, Obara, and Midorikawa}}]{Tamaki:prl:82:1422}
\bibinfo{author}{\bibnamefont{Tamaki}, \bibfnamefont{Y.}},
  \bibinfo{author}{\bibfnamefont{J.}~\bibnamefont{Itatani}},
  \bibinfo{author}{\bibfnamefont{Y.}~\bibnamefont{Nagata}},
  \bibinfo{author}{\bibfnamefont{M.}~\bibnamefont{Obara}}, and
  \bibinfo{author}{\bibfnamefont{K.}~\bibnamefont{Midorikawa}},
  \bibinfo{year}{1999}, \bibinfo{journal}{Phys.\ Rev.\ Lett.}
  \textbf{\bibinfo{volume}{82}}, \bibinfo{pages}{1422}.

\bibitem[{\citenamefont{Tempea and
  Brabec}(1998{\natexlab{a}})}]{Tempea:ol:23:1286}
\bibinfo{author}{\bibnamefont{Tempea}, \bibfnamefont{G.}}, and
  \bibinfo{author}{\bibfnamefont{T.}~\bibnamefont{Brabec}},
  \bibinfo{year}{1998}{\natexlab{a}}, \bibinfo{journal}{Opt.\ Lett.}
  \textbf{\bibinfo{volume}{23}}, \bibinfo{pages}{1286}.

\bibitem[{\citenamefont{Tempea and
  Brabec}(1998{\natexlab{b}})}]{Tempea:ol:23:762}
\bibinfo{author}{\bibnamefont{Tempea}, \bibfnamefont{G.}}, and
  \bibinfo{author}{\bibfnamefont{T.}~\bibnamefont{Brabec}},
  \bibinfo{year}{1998}{\natexlab{b}}, \bibinfo{journal}{Opt.\ Lett.}
  \textbf{\bibinfo{volume}{23}}, \bibinfo{pages}{762}.

\bibitem[{\citenamefont{Th\'eberge}
  \emph{et~al.}(2005{\natexlab{a}})\citenamefont{Th\'eberge, Ak{\"o}zbek, Liu,
  Gravel, and Chin}}]{Theberge:oc:245:399}
\bibinfo{author}{\bibnamefont{Th\'eberge}, \bibfnamefont{F.}},
  \bibinfo{author}{\bibfnamefont{N.}~\bibnamefont{Ak{\"o}zbek}},
  \bibinfo{author}{\bibfnamefont{W.}~\bibnamefont{Liu}},
  \bibinfo{author}{\bibfnamefont{J.-F.} \bibnamefont{Gravel}}, and
  \bibinfo{author}{\bibfnamefont{S.~L.} \bibnamefont{Chin}},
  \bibinfo{year}{2005}{\natexlab{a}}, \bibinfo{journal}{Opt.\ Commun.}
  \textbf{\bibinfo{volume}{245}}, \bibinfo{pages}{399}.

\bibitem[{\citenamefont{Th\'eberge}
  \emph{et~al.}(2005{\natexlab{b}})\citenamefont{Th\'eberge, Liu, Luo, and
  Chin}}]{Theberge:apb:80:221}
\bibinfo{author}{\bibnamefont{Th\'eberge}, \bibfnamefont{F.}},
  \bibinfo{author}{\bibfnamefont{W.}~\bibnamefont{Liu}},
  \bibinfo{author}{\bibfnamefont{Q.}~\bibnamefont{Luo}}, and
  \bibinfo{author}{\bibfnamefont{S.~L.} \bibnamefont{Chin}},
  \bibinfo{year}{2005}{\natexlab{b}}, \bibinfo{journal}{Appl.\ Phys.\ B: Lasers
  \& Optics} \textbf{\bibinfo{volume}{80}}, \bibinfo{pages}{221}.

\bibitem[{\citenamefont{Theopold} \emph{et~al.}(2005)\citenamefont{Theopold,
  Wolf, and W{\"o}ste}}]{Theopold:DR:2005}
\bibinfo{author}{\bibnamefont{Theopold}, \bibfnamefont{F.~A.}},
  \bibinfo{author}{\bibfnamefont{J.-P.} \bibnamefont{Wolf}}, and
  \bibinfo{author}{\bibfnamefont{L.}~\bibnamefont{W{\"o}ste}},
  \bibinfo{year}{2005}, \emph{\bibinfo{title}{Dial revisited: Belinda and
  white-light femtoseocnd lidar in range-resolved optical sensing of the
  atmosphere}} (\bibinfo{publisher}{Springer Verlag}, \bibinfo{address}{New
  York}).

\bibitem[{\citenamefont{Tien} \emph{et~al.}(1999)\citenamefont{Tien, Backus,
  Kapteyn, Murnane, and Mourou}}]{Tien:prl:82:3883}
\bibinfo{author}{\bibnamefont{Tien}, \bibfnamefont{A.-C.}},
  \bibinfo{author}{\bibfnamefont{S.}~\bibnamefont{Backus}},
  \bibinfo{author}{\bibfnamefont{H.}~\bibnamefont{Kapteyn}},
  \bibinfo{author}{\bibfnamefont{M.}~\bibnamefont{Murnane}}, and
  \bibinfo{author}{\bibfnamefont{G.}~\bibnamefont{Mourou}},
  \bibinfo{year}{1999}, \bibinfo{journal}{Phys.\ Rev.\ Lett.}
  \textbf{\bibinfo{volume}{82}}, \bibinfo{pages}{3883}.

\bibitem[{\citenamefont{Tikhonenko}
  \emph{et~al.}(1996)\citenamefont{Tikhonenko, Christou, and
  Luther-Davies}}]{Tikhonenko:prl:76:2698}
\bibinfo{author}{\bibnamefont{Tikhonenko}, \bibfnamefont{V.}},
  \bibinfo{author}{\bibfnamefont{J.}~\bibnamefont{Christou}}, and
  \bibinfo{author}{\bibfnamefont{B.}~\bibnamefont{Luther-Davies}},
  \bibinfo{year}{1996}, \bibinfo{journal}{Phys.\ Rev.\ Lett.}
  \textbf{\bibinfo{volume}{76}}, \bibinfo{pages}{2698}.

\bibitem[{\citenamefont{Ting}
  \emph{et~al.}(2005{\natexlab{a}})\citenamefont{Ting, Alexeev, Gordon, Fisher,
  Kaganovitch, Jones, Briscoe, Pe{\~n}ano, Hubbard, and
  Sprangle}}]{Ting:pop:12:056705}
\bibinfo{author}{\bibnamefont{Ting}, \bibfnamefont{A.}},
  \bibinfo{author}{\bibfnamefont{I.}~\bibnamefont{Alexeev}},
  \bibinfo{author}{\bibfnamefont{D.}~\bibnamefont{Gordon}},
  \bibinfo{author}{\bibfnamefont{R.}~\bibnamefont{Fisher}},
  \bibinfo{author}{\bibfnamefont{D.}~\bibnamefont{Kaganovitch}},
  \bibinfo{author}{\bibfnamefont{T.}~\bibnamefont{Jones}},
  \bibinfo{author}{\bibfnamefont{E.}~\bibnamefont{Briscoe}},
  \bibinfo{author}{\bibfnamefont{J.}~\bibnamefont{Pe{\~n}ano}},
  \bibinfo{author}{\bibfnamefont{R.}~\bibnamefont{Hubbard}}, and
  \bibinfo{author}{\bibfnamefont{P.}~\bibnamefont{Sprangle}},
  \bibinfo{year}{2005}{\natexlab{a}}, \bibinfo{journal}{Phys.\ Plasmas}
  \textbf{\bibinfo{volume}{12}}, \bibinfo{pages}{056705}.

\bibitem[{\citenamefont{Ting}
  \emph{et~al.}(2005{\natexlab{b}})\citenamefont{Ting, Gordon, Briscoe,
  Pe{\~n}ano, and Sprangle}}]{Ting:ao:44:1474}
\bibinfo{author}{\bibnamefont{Ting}, \bibfnamefont{A.}},
  \bibinfo{author}{\bibfnamefont{D.~F.} \bibnamefont{Gordon}},
  \bibinfo{author}{\bibfnamefont{E.}~\bibnamefont{Briscoe}},
  \bibinfo{author}{\bibfnamefont{J.~R.} \bibnamefont{Pe{\~n}ano}}, and
  \bibinfo{author}{\bibfnamefont{P.}~\bibnamefont{Sprangle}},
  \bibinfo{year}{2005}{\natexlab{b}}, \bibinfo{journal}{Appl.\ Opt.}
  \textbf{\bibinfo{volume}{44}}, \bibinfo{pages}{1474}.

\bibitem[{\citenamefont{Tohmon} \emph{et~al.}(1989)\citenamefont{Tohmon,
  Mizuno, Ohki, Sasagane, Nagasawa, and Hama}}]{Tohmon:prb:39:1337}
\bibinfo{author}{\bibnamefont{Tohmon}, \bibfnamefont{R.}},
  \bibinfo{author}{\bibfnamefont{H.}~\bibnamefont{Mizuno}},
  \bibinfo{author}{\bibfnamefont{Y.}~\bibnamefont{Ohki}},
  \bibinfo{author}{\bibfnamefont{K.}~\bibnamefont{Sasagane}},
  \bibinfo{author}{\bibfnamefont{K.}~\bibnamefont{Nagasawa}}, and
  \bibinfo{author}{\bibfnamefont{Y.}~\bibnamefont{Hama}}, \bibinfo{year}{1989},
  \bibinfo{journal}{Phys.\ Rev.\ B} \textbf{\bibinfo{volume}{39}},
  \bibinfo{pages}{1337}.

\bibitem[{\citenamefont{Tong} \emph{et~al.}(2002)\citenamefont{Tong, Zhao, and
  Lin}}]{Tong:pra:66:033402}
\bibinfo{author}{\bibnamefont{Tong}, \bibfnamefont{X.~M.}},
  \bibinfo{author}{\bibfnamefont{Z.~X.} \bibnamefont{Zhao}}, and
  \bibinfo{author}{\bibfnamefont{C.~D.} \bibnamefont{Lin}},
  \bibinfo{year}{2002}, \bibinfo{journal}{Phys.\ Rev.\ A}
  \textbf{\bibinfo{volume}{66}}, \bibinfo{pages}{033402}.

\bibitem[{\citenamefont{Tosa} \emph{et~al.}(2003)\citenamefont{Tosa, Takahashi,
  Nabekawa, and Midorikawa}}]{Tosa:pra:67:063817}
\bibinfo{author}{\bibnamefont{Tosa}, \bibfnamefont{V.}},
  \bibinfo{author}{\bibfnamefont{E.}~\bibnamefont{Takahashi}},
  \bibinfo{author}{\bibfnamefont{Y.}~\bibnamefont{Nabekawa}}, and
  \bibinfo{author}{\bibfnamefont{K.}~\bibnamefont{Midorikawa}},
  \bibinfo{year}{2003}, \bibinfo{journal}{Phys.\ Rev.\ A}
  \textbf{\bibinfo{volume}{67}}, \bibinfo{pages}{063817}.

\bibitem[{\citenamefont{Towers} \emph{et~al.}(2001)\citenamefont{Towers,
  Buryak, Sammut, Malomed, Crasovan, and Mihalache}}]{Towers:pla:288:292}
\bibinfo{author}{\bibnamefont{Towers}, \bibfnamefont{I.}},
  \bibinfo{author}{\bibfnamefont{A.~V.} \bibnamefont{Buryak}},
  \bibinfo{author}{\bibfnamefont{R.~A.} \bibnamefont{Sammut}},
  \bibinfo{author}{\bibfnamefont{B.~A.} \bibnamefont{Malomed}},
  \bibinfo{author}{\bibfnamefont{L.-C.} \bibnamefont{Crasovan}}, and
  \bibinfo{author}{\bibfnamefont{D.}~\bibnamefont{Mihalache}},
  \bibinfo{year}{2001}, \bibinfo{journal}{Phys.\ Lett.\ A}
  \textbf{\bibinfo{volume}{288}}, \bibinfo{pages}{292}.

\bibitem[{\citenamefont{Trillo and Torruellas}(2001)}]{Trillo:SS:01}
\bibinfo{editor}{\bibnamefont{Trillo}, \bibfnamefont{S.}}, and
  \bibinfo{editor}{\bibfnamefont{W.}~\bibnamefont{Torruellas}} (eds.),
  \bibinfo{year}{2001}, \emph{\bibinfo{title}{Spatial Solitons}}
  (\bibinfo{publisher}{Springer}, \bibinfo{address}{Berlin}).

\bibitem[{\citenamefont{Trushin} \emph{et~al.}(2005)\citenamefont{Trushin,
  Panja, Kosma, Schmid, and Fuss}}]{Trushin:apb:80:399}
\bibinfo{author}{\bibnamefont{Trushin}, \bibfnamefont{S.~A.}},
  \bibinfo{author}{\bibfnamefont{S.}~\bibnamefont{Panja}},
  \bibinfo{author}{\bibfnamefont{K.}~\bibnamefont{Kosma}},
  \bibinfo{author}{\bibfnamefont{W.~E.} \bibnamefont{Schmid}}, and
  \bibinfo{author}{\bibfnamefont{W.}~\bibnamefont{Fuss}}, \bibinfo{year}{2005},
  \bibinfo{journal}{Appl.\ Phys.\ B: Lasers \& Optics}
  \textbf{\bibinfo{volume}{80}}, \bibinfo{pages}{399}.

\bibitem[{\citenamefont{Tzortzakis}
  \emph{et~al.}(2006)\citenamefont{Tzortzakis, Anglos, and
  Gray}}]{Tzortzakis:ol:31:1139}
\bibinfo{author}{\bibnamefont{Tzortzakis}, \bibfnamefont{S.}},
  \bibinfo{author}{\bibfnamefont{D.}~\bibnamefont{Anglos}}, and
  \bibinfo{author}{\bibfnamefont{D.}~\bibnamefont{Gray}}, \bibinfo{year}{2006},
  \bibinfo{journal}{Opt.\ Lett.} \textbf{\bibinfo{volume}{31}},
  \bibinfo{pages}{1139}.

\bibitem[{\citenamefont{Tzortzakis}
  \emph{et~al.}(2001{\natexlab{a}})\citenamefont{Tzortzakis, Berg{\'e},
  Couairon, Franco, Prade, and Mysyrowicz}}]{Tzortzakis:prl:86:5470}
\bibinfo{author}{\bibnamefont{Tzortzakis}, \bibfnamefont{S.}},
  \bibinfo{author}{\bibfnamefont{L.}~\bibnamefont{Berg{\'e}}},
  \bibinfo{author}{\bibfnamefont{A.}~\bibnamefont{Couairon}},
  \bibinfo{author}{\bibfnamefont{M.}~\bibnamefont{Franco}},
  \bibinfo{author}{\bibfnamefont{B.}~\bibnamefont{Prade}}, and
  \bibinfo{author}{\bibfnamefont{A.}~\bibnamefont{Mysyrowicz}},
  \bibinfo{year}{2001}{\natexlab{a}}, \bibinfo{journal}{Phys.\ Rev.\ Lett.}
  \textbf{\bibinfo{volume}{86}}, \bibinfo{pages}{5470}.

\bibitem[{\citenamefont{Tzortzakis}
  \emph{et~al.}(1999)\citenamefont{Tzortzakis, Franco, Andr\'e, Chiron,
  Lamouroux, Prade, and Mysyrowicz}}]{Tzortzakis:pre:60:3505}
\bibinfo{author}{\bibnamefont{Tzortzakis}, \bibfnamefont{S.}},
  \bibinfo{author}{\bibfnamefont{M.~A.} \bibnamefont{Franco}},
  \bibinfo{author}{\bibfnamefont{Y.-B.} \bibnamefont{Andr\'e}},
  \bibinfo{author}{\bibfnamefont{A.}~\bibnamefont{Chiron}},
  \bibinfo{author}{\bibfnamefont{B.}~\bibnamefont{Lamouroux}},
  \bibinfo{author}{\bibfnamefont{B.~S.} \bibnamefont{Prade}}, and
  \bibinfo{author}{\bibfnamefont{A.}~\bibnamefont{Mysyrowicz}},
  \bibinfo{year}{1999}, \bibinfo{journal}{Phys.\ Rev.\ E}
  \textbf{\bibinfo{volume}{60}}, \bibinfo{pages}{R3505}.

\bibitem[{\citenamefont{Tzortzakis}
  \emph{et~al.}(2000{\natexlab{a}})\citenamefont{Tzortzakis, Lamouroux, Chiron,
  Franco, Prade, Mysyrowicz, and Moustaizis}}]{Tzortzakis:ol:25:1270}
\bibinfo{author}{\bibnamefont{Tzortzakis}, \bibfnamefont{S.}},
  \bibinfo{author}{\bibfnamefont{B.}~\bibnamefont{Lamouroux}},
  \bibinfo{author}{\bibfnamefont{A.}~\bibnamefont{Chiron}},
  \bibinfo{author}{\bibfnamefont{M.}~\bibnamefont{Franco}},
  \bibinfo{author}{\bibfnamefont{B.}~\bibnamefont{Prade}},
  \bibinfo{author}{\bibfnamefont{A.}~\bibnamefont{Mysyrowicz}}, and
  \bibinfo{author}{\bibfnamefont{S.~D.} \bibnamefont{Moustaizis}},
  \bibinfo{year}{2000}{\natexlab{a}}, \bibinfo{journal}{Opt.\ Lett.}
  \textbf{\bibinfo{volume}{25}}, \bibinfo{pages}{1270}.

\bibitem[{\citenamefont{Tzortzakis}
  \emph{et~al.}(2001{\natexlab{b}})\citenamefont{Tzortzakis, Lamouroux, Chiron,
  Moustaizis, Anglos, Franco, Prade, and Mysyrowicz}}]{Tzortzakis:oc:197:131}
\bibinfo{author}{\bibnamefont{Tzortzakis}, \bibfnamefont{S.}},
  \bibinfo{author}{\bibfnamefont{B.}~\bibnamefont{Lamouroux}},
  \bibinfo{author}{\bibfnamefont{A.}~\bibnamefont{Chiron}},
  \bibinfo{author}{\bibfnamefont{S.~D.} \bibnamefont{Moustaizis}},
  \bibinfo{author}{\bibfnamefont{D.}~\bibnamefont{Anglos}},
  \bibinfo{author}{\bibfnamefont{M.}~\bibnamefont{Franco}},
  \bibinfo{author}{\bibfnamefont{B.}~\bibnamefont{Prade}}, and
  \bibinfo{author}{\bibfnamefont{A.}~\bibnamefont{Mysyrowicz}},
  \bibinfo{year}{2001}{\natexlab{b}}, \bibinfo{journal}{Opt.\ Commun.}
  \textbf{\bibinfo{volume}{197}}, \bibinfo{pages}{131}.

\bibitem[{\citenamefont{Tzortzakis}
  \emph{et~al.}(2002)\citenamefont{Tzortzakis, M\'echain, Patalano, Andr\'e,
  Prade, Franco, Munier, Mysyrowicz, Gheudin, Beaudin, and
  Encrenaz}}]{Tzortzakis:ol:27:1944}
\bibinfo{author}{\bibnamefont{Tzortzakis}, \bibfnamefont{S.}},
  \bibinfo{author}{\bibfnamefont{G.}~\bibnamefont{M\'echain}},
  \bibinfo{author}{\bibfnamefont{G.}~\bibnamefont{Patalano}},
  \bibinfo{author}{\bibfnamefont{Y.-B.} \bibnamefont{Andr\'e}},
  \bibinfo{author}{\bibfnamefont{B.}~\bibnamefont{Prade}},
  \bibinfo{author}{\bibfnamefont{M.}~\bibnamefont{Franco}},
  \bibinfo{author}{\bibfnamefont{J.-M.} \bibnamefont{Munier}},
  \bibinfo{author}{\bibfnamefont{A.}~\bibnamefont{Mysyrowicz}},
  \bibinfo{author}{\bibfnamefont{M.}~\bibnamefont{Gheudin}},
  \bibinfo{author}{\bibfnamefont{G.}~\bibnamefont{Beaudin}}, and
  \bibinfo{author}{\bibfnamefont{P.}~\bibnamefont{Encrenaz}},
  \bibinfo{year}{2002}, \bibinfo{journal}{Opt.\ Lett.}
  \textbf{\bibinfo{volume}{27}}, \bibinfo{pages}{1944}.

\bibitem[{\citenamefont{Tzortzakis}
  \emph{et~al.}(2003)\citenamefont{Tzortzakis, M\'echain, Patalano, Franco,
  Prade, and Mysyrowicz}}]{Tzortzakis:apb:76:609}
\bibinfo{author}{\bibnamefont{Tzortzakis}, \bibfnamefont{S.}},
  \bibinfo{author}{\bibfnamefont{G.}~\bibnamefont{M\'echain}},
  \bibinfo{author}{\bibfnamefont{G.}~\bibnamefont{Patalano}},
  \bibinfo{author}{\bibfnamefont{M.}~\bibnamefont{Franco}},
  \bibinfo{author}{\bibfnamefont{B.}~\bibnamefont{Prade}}, and
  \bibinfo{author}{\bibfnamefont{A.}~\bibnamefont{Mysyrowicz}},
  \bibinfo{year}{2003}, \bibinfo{journal}{Appl.\ Phys.\ B: Lasers \& Optics}
  \textbf{\bibinfo{volume}{76}}, \bibinfo{pages}{609}.

\bibitem[{\citenamefont{Tzortzakis}
  \emph{et~al.}(2000{\natexlab{b}})\citenamefont{Tzortzakis, Prade, Franco, and
  Mysyrowicz}}]{Tzortzakis:oc:181:123}
\bibinfo{author}{\bibnamefont{Tzortzakis}, \bibfnamefont{S.}},
  \bibinfo{author}{\bibfnamefont{B.}~\bibnamefont{Prade}},
  \bibinfo{author}{\bibfnamefont{M.}~\bibnamefont{Franco}}, and
  \bibinfo{author}{\bibfnamefont{A.}~\bibnamefont{Mysyrowicz}},
  \bibinfo{year}{2000}{\natexlab{b}}, \bibinfo{journal}{Opt.\ Commun.}
  \textbf{\bibinfo{volume}{181}}, \bibinfo{pages}{123}.

\bibitem[{\citenamefont{Tzortzakis}
  \emph{et~al.}(2001{\natexlab{c}})\citenamefont{Tzortzakis, Prade, Franco,
  Mysyrowicz, H{\"u}ller, and Mora}}]{Tzortzakis:pre:64:057401}
\bibinfo{author}{\bibnamefont{Tzortzakis}, \bibfnamefont{S.}},
  \bibinfo{author}{\bibfnamefont{B.}~\bibnamefont{Prade}},
  \bibinfo{author}{\bibfnamefont{M.}~\bibnamefont{Franco}},
  \bibinfo{author}{\bibfnamefont{A.}~\bibnamefont{Mysyrowicz}},
  \bibinfo{author}{\bibfnamefont{S.}~\bibnamefont{H{\"u}ller}}, and
  \bibinfo{author}{\bibfnamefont{P.}~\bibnamefont{Mora}},
  \bibinfo{year}{2001}{\natexlab{c}}, \bibinfo{journal}{Phys.\ Rev.\ E}
  \textbf{\bibinfo{volume}{64}}, \bibinfo{pages}{057401}.

\bibitem[{\citenamefont{Tzortzakis}
  \emph{et~al.}(2001{\natexlab{d}})\citenamefont{Tzortzakis, Sudrie, Franco,
  Prade, Mysyrowicz, Couairon, and Berg{\'e}}}]{Tzortzakis:prl:87:213902}
\bibinfo{author}{\bibnamefont{Tzortzakis}, \bibfnamefont{S.}},
  \bibinfo{author}{\bibfnamefont{L.}~\bibnamefont{Sudrie}},
  \bibinfo{author}{\bibfnamefont{M.}~\bibnamefont{Franco}},
  \bibinfo{author}{\bibfnamefont{B.}~\bibnamefont{Prade}},
  \bibinfo{author}{\bibfnamefont{A.}~\bibnamefont{Mysyrowicz}},
  \bibinfo{author}{\bibfnamefont{A.}~\bibnamefont{Couairon}}, and
  \bibinfo{author}{\bibfnamefont{L.}~\bibnamefont{Berg{\'e}}},
  \bibinfo{year}{2001}{\natexlab{d}}, \bibinfo{journal}{Phys.\ Rev.\ Lett.}
  \textbf{\bibinfo{volume}{87}}, \bibinfo{pages}{213902}.

\bibitem[{\citenamefont{Vakhitov and Kolokolov}(1975)}]{Vakhitov:rqe:16:783}
\bibinfo{author}{\bibnamefont{Vakhitov}, \bibfnamefont{N.~G.}}, and
  \bibinfo{author}{\bibfnamefont{A.~A.} \bibnamefont{Kolokolov}},
  \bibinfo{year}{1975}, \bibinfo{journal}{Radiophys.\ Quant.\ Electron.}
  \textbf{\bibinfo{volume}{16}}, \bibinfo{pages}{783}.

\bibitem[{\citenamefont{Verhoef} \emph{et~al.}(2006)\citenamefont{Verhoef,
  Seres, Schmid, Nomura, Tempea, Veisz, and Krausz}}]{Verhoef:apb:82:513}
\bibinfo{author}{\bibnamefont{Verhoef}, \bibfnamefont{A.~J.}},
  \bibinfo{author}{\bibfnamefont{J.}~\bibnamefont{Seres}},
  \bibinfo{author}{\bibfnamefont{K.}~\bibnamefont{Schmid}},
  \bibinfo{author}{\bibfnamefont{Y.}~\bibnamefont{Nomura}},
  \bibinfo{author}{\bibfnamefont{G.}~\bibnamefont{Tempea}},
  \bibinfo{author}{\bibfnamefont{L.}~\bibnamefont{Veisz}}, and
  \bibinfo{author}{\bibfnamefont{F.}~\bibnamefont{Krausz}},
  \bibinfo{year}{2006}, \bibinfo{journal}{Appl.\ Phys.\ B: Lasers \& Optics}
  \textbf{\bibinfo{volume}{82}}, \bibinfo{pages}{513}.

\bibitem[{\citenamefont{Vidal and Johnston}(1996)}]{Vidal:prl:77:1282}
\bibinfo{author}{\bibnamefont{Vidal}, \bibfnamefont{F.}}, and
  \bibinfo{author}{\bibfnamefont{T.~W.} \bibnamefont{Johnston}},
  \bibinfo{year}{1996}, \bibinfo{journal}{Phys.\ Rev.\ Lett.}
  \textbf{\bibinfo{volume}{77}}, \bibinfo{pages}{1282}.

\bibitem[{\citenamefont{Vidal and Johnston}(1997)}]{Vidal:pre:55:3571}
\bibinfo{author}{\bibnamefont{Vidal}, \bibfnamefont{F.}}, and
  \bibinfo{author}{\bibfnamefont{T.~W.} \bibnamefont{Johnston}},
  \bibinfo{year}{1997}, \bibinfo{journal}{Phys.\ Rev.\ E}
  \textbf{\bibinfo{volume}{55}}, \bibinfo{pages}{3571}.

\bibitem[{\citenamefont{Vin{\c{c}}otte and
  Berg{\'e}}(2004)}]{Vincotte:pra:70:061802}
\bibinfo{author}{\bibnamefont{Vin{\c{c}}otte}, \bibfnamefont{A.}}, and
  \bibinfo{author}{\bibfnamefont{L.}~\bibnamefont{Berg{\'e}}},
  \bibinfo{year}{2004}, \bibinfo{journal}{Phys.\ Rev.\ A}
  \textbf{\bibinfo{volume}{70}}, \bibinfo{pages}{061802(R)}.

\bibitem[{\citenamefont{Vin{\c{c}}otte and
  Berg\'e}(2005)}]{Vincotte:prl:95:193901}
\bibinfo{author}{\bibnamefont{Vin{\c{c}}otte}, \bibfnamefont{A.}}, and
  \bibinfo{author}{\bibfnamefont{L.}~\bibnamefont{Berg\'e}},
  \bibinfo{year}{2005}, \bibinfo{journal}{Phys.\ Rev.\ Lett.}
  \textbf{\bibinfo{volume}{95}}, \bibinfo{pages}{193901}.

\bibitem[{\citenamefont{Vlasov} \emph{et~al.}(1974)\citenamefont{Vlasov,
  Petrishchev, and Talanov}}]{Vlasov:rqe:14:1062}
\bibinfo{author}{\bibnamefont{Vlasov}, \bibfnamefont{S.~N.}},
  \bibinfo{author}{\bibfnamefont{V.~A.} \bibnamefont{Petrishchev}}, and
  \bibinfo{author}{\bibfnamefont{V.~I.} \bibnamefont{Talanov}},
  \bibinfo{year}{1974}, \bibinfo{journal}{Radiophys.\ Quant.\ Electron.}
  \textbf{\bibinfo{volume}{14}}, \bibinfo{pages}{1062}.

\bibitem[{\citenamefont{Vlasov} \emph{et~al.}(1989)\citenamefont{Vlasov,
  Piskunova, and Talanov}}]{Vlasov:spjetp:68:1125}
\bibinfo{author}{\bibnamefont{Vlasov}, \bibfnamefont{S.~N.}},
  \bibinfo{author}{\bibfnamefont{L.~V.} \bibnamefont{Piskunova}}, and
  \bibinfo{author}{\bibfnamefont{V.~I.} \bibnamefont{Talanov}},
  \bibinfo{year}{1989}, \bibinfo{journal}{Sov.\ Phys.\ JETP}
  \textbf{\bibinfo{volume}{68}}, \bibinfo{pages}{1125}.

\bibitem[{\citenamefont{Vuong} \emph{et~al.}(2006)\citenamefont{Vuong, Grow,
  Ishaaya, Gaeta, 't~Hooft, Eliel, and Fibich}}]{Vuong:prl:96:133901}
\bibinfo{author}{\bibnamefont{Vuong}, \bibfnamefont{L.~T.}},
  \bibinfo{author}{\bibfnamefont{T.~D.} \bibnamefont{Grow}},
  \bibinfo{author}{\bibfnamefont{A.}~\bibnamefont{Ishaaya}},
  \bibinfo{author}{\bibfnamefont{A.~L.} \bibnamefont{Gaeta}},
  \bibinfo{author}{\bibfnamefont{G.~W.} \bibnamefont{'t~Hooft}},
  \bibinfo{author}{\bibfnamefont{E.~R.} \bibnamefont{Eliel}}, and
  \bibinfo{author}{\bibfnamefont{G.}~\bibnamefont{Fibich}},
  \bibinfo{year}{2006}, \bibinfo{journal}{Phys.\ Rev.\ Lett.}
  \textbf{\bibinfo{volume}{96}}, \bibinfo{pages}{133901}.

\bibitem[{\citenamefont{Wagner} \emph{et~al.}(2004)\citenamefont{Wagner,
  Gibson, Popmintchev, Christov, Murnane, and Kapteyn}}]{Wagner:prl:93:173902}
\bibinfo{author}{\bibnamefont{Wagner}, \bibfnamefont{N.~L.}},
  \bibinfo{author}{\bibfnamefont{E.~A.} \bibnamefont{Gibson}},
  \bibinfo{author}{\bibfnamefont{T.}~\bibnamefont{Popmintchev}},
  \bibinfo{author}{\bibfnamefont{I.~P.} \bibnamefont{Christov}},
  \bibinfo{author}{\bibfnamefont{M.~M.} \bibnamefont{Murnane}}, and
  \bibinfo{author}{\bibfnamefont{H.~C.} \bibnamefont{Kapteyn}},
  \bibinfo{year}{2004}, \bibinfo{journal}{Phys.\ Rev.\ Lett.}
  \textbf{\bibinfo{volume}{93}}, \bibinfo{pages}{173902}.

\bibitem[{\citenamefont{Ward and Berg{\'e}}(2003)}]{Ward:prl:90:053901}
\bibinfo{author}{\bibnamefont{Ward}, \bibfnamefont{H.}}, and
  \bibinfo{author}{\bibfnamefont{L.}~\bibnamefont{Berg{\'e}}},
  \bibinfo{year}{2003}, \bibinfo{journal}{Phys.\ Rev.\ Lett.}
  \textbf{\bibinfo{volume}{90}}, \bibinfo{pages}{053901}.

\bibitem[{\citenamefont{Weinstein}(1983)}]{Weinstein:cmp:87:567}
\bibinfo{author}{\bibnamefont{Weinstein}, \bibfnamefont{M.~I.}},
  \bibinfo{year}{1983}, \bibinfo{journal}{Commun.\ Math.\ Phys.}
  \textbf{\bibinfo{volume}{87}}, \bibinfo{pages}{567}.

\bibitem[{\citenamefont{Wiens} \emph{et~al.}(2002)\citenamefont{Wiens,
  Arvidson, Cremers, Ferris, Blacic, and IV}}]{Wiens:jgrp:107:FIDO3-1-14}
\bibinfo{author}{\bibnamefont{Wiens}, \bibfnamefont{R.~C.}},
  \bibinfo{author}{\bibfnamefont{R.~E.} \bibnamefont{Arvidson}},
  \bibinfo{author}{\bibfnamefont{D.~A.} \bibnamefont{Cremers}},
  \bibinfo{author}{\bibfnamefont{M.~J.} \bibnamefont{Ferris}},
  \bibinfo{author}{\bibfnamefont{J.~D.} \bibnamefont{Blacic}}, and
  \bibinfo{author}{\bibfnamefont{F.~P.~S.} \bibnamefont{IV}},
  \bibinfo{year}{2002}, \bibinfo{journal}{J. \ Geophys. \ Res. \ Planets}
  \textbf{\bibinfo{volume}{107(E11)}}, \bibinfo{pages}{FIDO3}.

\bibitem[{\citenamefont{Wille} \emph{et~al.}(2002)\citenamefont{Wille,
  Rodriguez, Kasparian, Mondelain, Yu, Mysyrowicz, Sauerbrey, Wolf, , and
  W{\"o}ste}}]{Wille:epjap:20:183}
\bibinfo{author}{\bibnamefont{Wille}, \bibfnamefont{H.}},
  \bibinfo{author}{\bibfnamefont{M.}~\bibnamefont{Rodriguez}},
  \bibinfo{author}{\bibfnamefont{J.}~\bibnamefont{Kasparian}},
  \bibinfo{author}{\bibfnamefont{D.}~\bibnamefont{Mondelain}},
  \bibinfo{author}{\bibfnamefont{J.}~\bibnamefont{Yu}},
  \bibinfo{author}{\bibfnamefont{A.}~\bibnamefont{Mysyrowicz}},
  \bibinfo{author}{\bibfnamefont{R.}~\bibnamefont{Sauerbrey}},
  \bibinfo{author}{\bibfnamefont{J.~P.} \bibnamefont{Wolf}}, , and
  \bibinfo{author}{\bibfnamefont{L.}~\bibnamefont{W{\"o}ste}},
  \bibinfo{year}{2002}, \bibinfo{journal}{Eur.\ Phys.\ Jour.\ - Appl. Phys.}
  \textbf{\bibinfo{volume}{20}}, \bibinfo{pages}{183}.

\bibitem[{\citenamefont{Wolf}(2000)}]{Wolf:EAC:2000:2226}
\bibinfo{author}{\bibnamefont{Wolf}, \bibfnamefont{J.-P.}},
  \bibinfo{year}{2000}, \emph{\bibinfo{title}{Ultraviolet/visible light
  detection and ranging Applications in air monitoring}}, in
  \cite{Meyers:EAC:2000}, p. \bibinfo{pages}{2226}.

\bibitem[{\citenamefont{W{\"o}ste} \emph{et~al.}(1997)\citenamefont{W{\"o}ste,
  Wedekind, Wille, Rairoux, Stein, Nikolov, Werner, Niedermeier, Ronneberger,
  Schillinger, and Sauerbrey}}]{Woste:lo:29:51}
\bibinfo{author}{\bibnamefont{W{\"o}ste}, \bibfnamefont{L.}},
  \bibinfo{author}{\bibfnamefont{C.}~\bibnamefont{Wedekind}},
  \bibinfo{author}{\bibfnamefont{H.}~\bibnamefont{Wille}},
  \bibinfo{author}{\bibfnamefont{P.}~\bibnamefont{Rairoux}},
  \bibinfo{author}{\bibfnamefont{B.}~\bibnamefont{Stein}},
  \bibinfo{author}{\bibfnamefont{S.}~\bibnamefont{Nikolov}},
  \bibinfo{author}{\bibfnamefont{C.}~\bibnamefont{Werner}},
  \bibinfo{author}{\bibfnamefont{S.}~\bibnamefont{Niedermeier}},
  \bibinfo{author}{\bibfnamefont{F.}~\bibnamefont{Ronneberger}},
  \bibinfo{author}{\bibfnamefont{H.}~\bibnamefont{Schillinger}}, and
  \bibinfo{author}{\bibfnamefont{R.}~\bibnamefont{Sauerbrey}},
  \bibinfo{year}{1997}, \bibinfo{journal}{Laser \& Optoelektron.}
  \textbf{\bibinfo{volume}{29}}, \bibinfo{pages}{51}.

\bibitem[{\citenamefont{Wyller}(2001)}]{Wyller:pd:157:90}
\bibinfo{author}{\bibnamefont{Wyller}, \bibfnamefont{J.}},
  \bibinfo{year}{2001}, \bibinfo{journal}{Physica D}
  \textbf{\bibinfo{volume}{157}}, \bibinfo{pages}{90}.

\bibitem[{\citenamefont{Xi} \emph{et~al.}(2006)\citenamefont{Xi, Lu, and
  Zhang}}]{Xi:prl:96:025003}
\bibinfo{author}{\bibnamefont{Xi}, \bibfnamefont{T.-T.}},
  \bibinfo{author}{\bibfnamefont{X.}~\bibnamefont{Lu}}, and
  \bibinfo{author}{\bibfnamefont{J.}~\bibnamefont{Zhang}},
  \bibinfo{year}{2006}, \bibinfo{journal}{Phys.\ Rev.\ Lett.}
  \textbf{\bibinfo{volume}{96}}, \bibinfo{pages}{025003}.

\bibitem[{\citenamefont{Xu} \emph{et~al.}(2006{\natexlab{a}})\citenamefont{Xu,
  Daigle, Luo, and Chin}}]{Xu:apb:82:655}
\bibinfo{author}{\bibnamefont{Xu}, \bibfnamefont{H.~L.}},
  \bibinfo{author}{\bibfnamefont{J.~F.} \bibnamefont{Daigle}},
  \bibinfo{author}{\bibfnamefont{Q.}~\bibnamefont{Luo}}, and
  \bibinfo{author}{\bibfnamefont{S.~L.} \bibnamefont{Chin}},
  \bibinfo{year}{2006}{\natexlab{a}}, \bibinfo{journal}{Appl.\ Phys.\ B: Lasers
  \& Optics} \textbf{\bibinfo{volume}{82}}, \bibinfo{pages}{655}.

\bibitem[{\citenamefont{Xu} \emph{et~al.}(2006{\natexlab{b}})\citenamefont{Xu,
  Liu, and Chin}}]{Xu:ol:31:1540}
\bibinfo{author}{\bibnamefont{Xu}, \bibfnamefont{X.~H.}},
  \bibinfo{author}{\bibfnamefont{W.}~\bibnamefont{Liu}}, and
  \bibinfo{author}{\bibfnamefont{S.~L.} \bibnamefont{Chin}},
  \bibinfo{year}{2006}{\natexlab{b}}, \bibinfo{journal}{Opt.\ Lett.}
  \textbf{\bibinfo{volume}{31}}, \bibinfo{pages}{1540}.

\bibitem[{\citenamefont{Yablonovitch}(1974)}]{Yablonovitch:pra:10:1888}
\bibinfo{author}{\bibnamefont{Yablonovitch}, \bibfnamefont{E.}},
  \bibinfo{year}{1974}, \bibinfo{journal}{Phys.\ Rev.\ A}
  \textbf{\bibinfo{volume}{10}}, \bibinfo{pages}{1888}.

\bibitem[{\citenamefont{Yablonovitch and
  Bloembergen}(1972)}]{Yablonovitch:prl:29:907}
\bibinfo{author}{\bibnamefont{Yablonovitch}, \bibfnamefont{E.}}, and
  \bibinfo{author}{\bibfnamefont{N.}~\bibnamefont{Bloembergen}},
  \bibinfo{year}{1972}, \bibinfo{journal}{Phys.\ Rev.\ Lett.}
  \textbf{\bibinfo{volume}{29}}, \bibinfo{pages}{907}.

\bibitem[{\citenamefont{Yang and Shen}(1984)}]{Yang:ol:9:510}
\bibinfo{author}{\bibnamefont{Yang}, \bibfnamefont{G.}}, and
  \bibinfo{author}{\bibfnamefont{Y.~R.} \bibnamefont{Shen}},
  \bibinfo{year}{1984}, \bibinfo{journal}{Opt.\ Lett.}
  \textbf{\bibinfo{volume}{9}}, \bibinfo{pages}{510}.

\bibitem[{\citenamefont{Yang} \emph{et~al.}(2003)\citenamefont{Yang, Zhang,
  Zhang, Zhao, Li, Teng, Li, Wang, Chen, Wei, Ma, Yu}
  \emph{et~al.}}]{Yang:pre:67:015401}
\bibinfo{author}{\bibnamefont{Yang}, \bibfnamefont{H.}},
  \bibinfo{author}{\bibfnamefont{J.}~\bibnamefont{Zhang}},
  \bibinfo{author}{\bibfnamefont{J.}~\bibnamefont{Zhang}},
  \bibinfo{author}{\bibfnamefont{L.~Z.} \bibnamefont{Zhao}},
  \bibinfo{author}{\bibfnamefont{Y.~J.} \bibnamefont{Li}},
  \bibinfo{author}{\bibfnamefont{H.}~\bibnamefont{Teng}},
  \bibinfo{author}{\bibfnamefont{Y.~T.} \bibnamefont{Li}},
  \bibinfo{author}{\bibfnamefont{Z.~H.} \bibnamefont{Wang}},
  \bibinfo{author}{\bibfnamefont{Z.~L.} \bibnamefont{Chen}},
  \bibinfo{author}{\bibfnamefont{Z.~Y.} \bibnamefont{Wei}},
  \bibinfo{author}{\bibfnamefont{J.~X.} \bibnamefont{Ma}},
  \bibinfo{author}{\bibfnamefont{W.}~\bibnamefont{Yu}}, \emph{et~al.},
  \bibinfo{year}{2003}, \bibinfo{journal}{Phys.\ Rev.\ E}
  \textbf{\bibinfo{volume}{67}}, \bibinfo{pages}{015401(R)}.

\bibitem[{\citenamefont{Yu} \emph{et~al.}(2001)\citenamefont{Yu, Mondelain,
  Ange, Volk, Niedermeier, Wolf, Kasparian, and Sauerbrey}}]{Yu:ol:26:533}
\bibinfo{author}{\bibnamefont{Yu}, \bibfnamefont{J.}},
  \bibinfo{author}{\bibfnamefont{D.}~\bibnamefont{Mondelain}},
  \bibinfo{author}{\bibfnamefont{G.}~\bibnamefont{Ange}},
  \bibinfo{author}{\bibfnamefont{R.}~\bibnamefont{Volk}},
  \bibinfo{author}{\bibfnamefont{S.}~\bibnamefont{Niedermeier}},
  \bibinfo{author}{\bibfnamefont{J.-P.} \bibnamefont{Wolf}},
  \bibinfo{author}{\bibfnamefont{J.}~\bibnamefont{Kasparian}}, and
  \bibinfo{author}{\bibfnamefont{R.}~\bibnamefont{Sauerbrey}},
  \bibinfo{year}{2001}, \bibinfo{journal}{Opt.\ Lett.}
  \textbf{\bibinfo{volume}{26}}, \bibinfo{pages}{533}.

\bibitem[{\citenamefont{Yu} \emph{et~al.}(2003)\citenamefont{Yu, Mondelain,
  Kasparian, Salmon, Geffroy, Favre, Boutou, and Wolf}}]{Yu:ao:42:7117}
\bibinfo{author}{\bibnamefont{Yu}, \bibfnamefont{J.}},
  \bibinfo{author}{\bibfnamefont{D.}~\bibnamefont{Mondelain}},
  \bibinfo{author}{\bibfnamefont{J.}~\bibnamefont{Kasparian}},
  \bibinfo{author}{\bibfnamefont{E.}~\bibnamefont{Salmon}},
  \bibinfo{author}{\bibfnamefont{S.}~\bibnamefont{Geffroy}},
  \bibinfo{author}{\bibfnamefont{C.}~\bibnamefont{Favre}},
  \bibinfo{author}{\bibfnamefont{V.}~\bibnamefont{Boutou}}, and
  \bibinfo{author}{\bibfnamefont{J.-P.} \bibnamefont{Wolf}},
  \bibinfo{year}{2003}, \bibinfo{journal}{Appl.\ Opt.}
  \textbf{\bibinfo{volume}{42}}, \bibinfo{pages}{7117}.

\bibitem[{\citenamefont{Zakharov and Kuznetsov}(1986)}]{Zakharov:spjetp:64:773}
\bibinfo{author}{\bibnamefont{Zakharov}, \bibfnamefont{V.~E.}}, and
  \bibinfo{author}{\bibfnamefont{E.~A.} \bibnamefont{Kuznetsov}},
  \bibinfo{year}{1986}, \bibinfo{journal}{Sov.\ Phys.\ JETP}
  \textbf{\bibinfo{volume}{64}}, \bibinfo{pages}{773}.

\bibitem[{\citenamefont{Zakharov and Rubenchik}(1974)}]{Zakharov:spjetp:38:494}
\bibinfo{author}{\bibnamefont{Zakharov}, \bibfnamefont{V.~E.}}, and
  \bibinfo{author}{\bibfnamefont{A.~M.} \bibnamefont{Rubenchik}},
  \bibinfo{year}{1974}, \bibinfo{journal}{Sov.\ Phys.\ JETP}
  \textbf{\bibinfo{volume}{38}}, \bibinfo{pages}{494}.

\bibitem[{\citenamefont{Zeng} \emph{et~al.}(2003)\citenamefont{Zeng, Li, Yu,
  and Xu}}]{Zeng:pra:67:013815}
\bibinfo{author}{\bibnamefont{Zeng}, \bibfnamefont{Z.}},
  \bibinfo{author}{\bibfnamefont{R.}~\bibnamefont{Li}},
  \bibinfo{author}{\bibfnamefont{W.}~\bibnamefont{Yu}}, and
  \bibinfo{author}{\bibfnamefont{Z.}~\bibnamefont{Xu}}, \bibinfo{year}{2003},
  \bibinfo{journal}{Phys.\ Rev.\ A} \textbf{\bibinfo{volume}{67}},
  \bibinfo{pages}{013815}.

\bibitem[{\citenamefont{Zharova} \emph{et~al.}(2003)\citenamefont{Zharova,
  Litvak, and Mironov}}]{Zharova:jetp:96:643}
\bibinfo{author}{\bibnamefont{Zharova}, \bibfnamefont{N.~A.}},
  \bibinfo{author}{\bibfnamefont{A.~G.} \bibnamefont{Litvak}}, and
  \bibinfo{author}{\bibfnamefont{V.~A.} \bibnamefont{Mironov}},
  \bibinfo{year}{2003}, \bibinfo{journal}{J.\ Exp.\ Theor.\ Phys.}
  \textbf{\bibinfo{volume}{96}}, \bibinfo{pages}{643}.

\bibitem[{\citenamefont{Zozulya} \emph{et~al.}(1999)\citenamefont{Zozulya,
  Diddams, {Van Engen}, and Clement}}]{Zozulya:prl:82:1430}
\bibinfo{author}{\bibnamefont{Zozulya}, \bibfnamefont{A.~A.}},
  \bibinfo{author}{\bibfnamefont{S.~A.} \bibnamefont{Diddams}},
  \bibinfo{author}{\bibfnamefont{A.~G.} \bibnamefont{{Van Engen}}}, and
  \bibinfo{author}{\bibfnamefont{T.~S.} \bibnamefont{Clement}},
  \bibinfo{year}{1999}, \bibinfo{journal}{Phys.\ Rev.\ Lett.}
  \textbf{\bibinfo{volume}{82}}, \bibinfo{pages}{1430}.

\end{thebibliography}

\end{document}